\documentclass[preprints,article,accept,pdftex,oneauthors]{Definitions/mdpi} 

\usepackage{xcolor}

\usepackage{enumitem}

\firstpage{1} 
\pubvolume{1}
\issuenum{1}
\articlenumber{0}
\pubyear{2026}
\copyrightyear{2026}
\datereceived{ } 
\daterevised{ } 
\dateaccepted{ } 
\datepublished{ } 

\DeclareGraphicsExtensions{ps,eps,eps.gz,bmp}

\usepackage[english]{babel}
\usepackage[utf8]{inputenc}

\usepackage{titlesec}
\titleformat{\subsection}{\normalfont\bfseries}{\thesubsection}{1em}{}

\usepackage{multicol}

\newcommand{\ar}{\to}
\newcommand{\al}{\gets}

	\newcommand{\nW}{\rotatebox[origin=c]{180}{$W$}}
	
	\newcommand{\diff}{\mathop{}\!\mathrm{d}}

	    \newcommand{\calN}{{\lambda}}

	\newcommand{\Tr}{{\mathrm{Tr}}}
	\newcommand{\underinequality}[2]{{\underset{\substack{ {}_{#2}}}{\, #1\, }}}
	
	\newcommand{\smallsum}{{\textstyle \sum}}
	
	\newcommand{\ind}[1]{{}}
	
	\newcommand{\lst}[2]{\ensuremath{#1_1, #1_2,\discretionary{}{}{}\ldots, #1_#2}}

\def\Boltz{{k_\text{\scriptsize B}}}
	
	\def\bn{{\boldsymbol{n}}}
	\def\bbeta{{\boldsymbol{\beta}}}
	\def\bmu{{\boldsymbol{\mu}}}
	\def\by{{\boldsymbol{y}}}
	\def\bx{{\boldsymbol{x}}}
	\def\bz{{\boldsymbol{z}}}
				\def\bj{{\boldsymbol{j}}}
	\def\Eu{{E\!u}}

\newcommand{\Sirr}{S_{\rm irr}}
\newcommand{\Sdotirr}{\dot S_{\rm irr}}

\newcommand{\hparder}[3]{\left(\partial #1 /\partial #2\right)_{\!#3}}
\newcommand{\vparder}[3]{\left(\frac{\partial #1}{\partial #2}\right)_{\!\!#3}}

\Title{Universal Foundations of Thermodynamics:
	Entropy and Energy Beyond Equilibrium and Without Extensivity}

	\Author{Gian Paolo Beretta $^{1}$\orcidA{}}

	\AuthorNames{Gian Paolo Beretta}

	\address{%
	$^{1}$ \quad Department of Mechanical and Industrial Engineering, University of Brescia, 25123 Brescia, Italy; gianpaolo.beretta@unibs.it}

\corres{Correspondence: gianpaolo.beretta@unibs.it}

\abstract{Thermodynamics is commonly presented as a theory of macroscopic systems in stable equilibrium, built upon assumptions of extensivity and scaling with system size. In this paper, we present a universal formulation of the elementary foundations of thermodynamics, in which entropy and energy are defined and employed beyond equilibrium and without assuming extensivity. The formulation applies to all systems --- large and small, with many or few particles --- and to all states, whether equilibrium or nonequilibrium, by relying on carefully stated operational definitions and existence principles rather than macroscopic idealizations.
	Key thermodynamic concepts, including adiabatic availability and available energy, are developed and illustrated using the energy–entropy diagram representation of nonequilibrium states, which provides geometric insight into irreversibility and the limits of work extraction for systems of any size. A substantial part of the paper is devoted to the analysis of entropy transfer in non-work interactions, leading to precise definitions of heat interactions and heat-and-diffusion interactions of central importance in mesoscopic continuum theories of nonequilibrium behavior in simple and complex solids and fluids.
	As a direct consequence of this analysis, Clausius inequalities and the Clausius statement of the second law are derived in forms explicitly extended to nonequilibrium processes. The resulting framework presents thermodynamics as a universal theory whose concepts apply uniformly to all systems, large and small, and provides a coherent foundation for both teaching and modern applications.}

\keyword{Foundations of thermodynamics;	
	Nonequilibrium entropy;	
	Entropy as a state property;	
	Thermodynamics of small systems;	
	Entropy transfer mechanisms;	
	Heat interaction;	
	Heat-and-diffusion interaction;	
	Energy–entropy diagram;	
	Clausius inequality (nonequilibrium);	
	Adiabatic availability and available energy}

\begin{document}
	
\begingroup
\scriptsize
\setlength{\columnsep}{0.5cm}
\begin{multicols}{2}
\renewcommand{\baselinestretch}{0.8}\selectfont
\makeatletter
\renewcommand{\tableofcontents}{%
	\section*{\contentsname}%
	\vspace{-0.1\baselineskip}%
	\@starttoc{toc}%
}
\makeatother
\makeatletter
\renewcommand*\l@section[2]{%
	\ifnum \c@tocdepth >\z@
	\addvspace{2pt}%
	\noindent
	\hangindent=\cftsecnumwidth
	\hangafter=1
	\bfseries\scriptsize\raggedright
	#1\par
	\fi}
\renewcommand*\l@subsection[2]{%
	\ifnum \c@tocdepth >1\relax
	\addvspace{1pt}%
	\noindent
	\hspace*{2.5em}%
	\hangindent=\cftsubsecnumwidth
	\hangafter=1
	\normalfont\scriptsize\raggedright
	#1\par
	\fi}
\makeatother
\tableofcontents
\end{multicols}
\endgroup

	\section{Introduction}
	
Thermodynamics is one of the most mature and successful theories in physical science. Its principles govern phenomena across an extraordinary range of systems and scales, from macroscopic energy conversion devices and chemical reactors to multicomponent transport, biological processes, and, increasingly, nanoscale and quantum technologies. Yet, despite more than two centuries of development and application, the elementary foundations of thermodynamics continue to invite scrutiny, reinterpretation, and reformulation.

This sustained foundational activity is not accidental. Classical thermodynamics was historically shaped to describe macroscopic systems in stable equilibrium, guided by phenomenological observations and practical engineering needs \cite{Fermi1956,Callen1985,VanWylenSonntag1978,MoranShapiro1988}. As a result, many traditional formulations incorporate, often implicitly, assumptions of extensivity and scale separation that are valid only for systems with many particles. While these assumptions are extraordinarily effective in their domain of applicability, they are not intrinsic to thermodynamics itself and become increasingly restrictive as the theory is extended to nonequilibrium phenomena, multicomponent systems, and systems of arbitrary size.

Among the fundamental concepts of thermodynamics, entropy occupies a uniquely central position. It appears as a property of state, a criterion of equilibrium, a measure of irreversibility, and a generator of transport laws and variational principles. At the same time, its definition, operational meaning, and domain of validity are often taken for granted --- particularly outside the realm of stable equilibrium. As a consequence, entropy is frequently invoked with different meanings in different contexts, sometimes without explicit acknowledgment of the modeling assumptions involved.

The present paper provides a modern exposition of the elementary foundations of thermodynamics, with particular emphasis on the definition and role of entropy for nonequilibrium states. The objective is not to replace existing formulations, but to clarify their logical structure, identify minimal assumptions, and articulate a coherent conceptual framework in which all thermodynamic concepts apply uniformly to systems of arbitrary size, particle number, and degree of disequilibrium, without reliance on extensivity or macroscopic idealizations.

\subsection{Conceptual motivations and historical background}

In many traditional expositions, thermodynamics is introduced through the distinction between heat and work, often motivated by mechanical analogies or heuristic pictures of microscopic motion. Classic examples include the description of heat as energy associated with random molecular motion \cite{Feynman1963,Tisza1966} or as that part of an energy change not attributable to work \cite{LandauLifshitz1980}. While intuitively appealing, such descriptions conceal fundamental conceptual difficulties.

First, neither heat nor work is a property stored in a system; each is a mode of energy transfer between systems. Second, and more fundamentally, the very distinction between heat and work cannot be justified within mechanics alone. As emphasized by Hatsopoulos and Keenan \cite{HatsopoulosKeenan1965}, without the second law of thermodynamics and the existence of entropy as a property of matter, heat and work would be indistinguishable. Any formulation that treats heat as a primitive concept therefore encounters logical circularities when entropy is later defined in terms of heat and temperature.

Several refinements of the traditional approach have been proposed. Heat has been defined as energy transfer driven by a temperature difference \cite{Keenan1941}, or as that part of an energy exchange not accounted for by mechanical work \cite{Guggenheim1967}. These definitions are operationally effective in restricted settings, but they implicitly assume the existence of subsystems in stable equilibrium at the interfaces between interacting systems. As a result, they restrict the domain of validity of entropy from the outset, effectively confining it to equilibrium or local-equilibrium states.

This limitation becomes particularly problematic when thermodynamics is applied to nonequilibrium processes, multicomponent transport, or systems that depart significantly from macroscopic idealizations. In such contexts, the initial intuitive appeal of heat-based definitions gives way to ambiguity, and thermodynamics is sometimes perceived --- incorrectly --- as intrinsically vague or logically inconsistent.

\subsection{Entropy as a fundamental property}

An alternative route to the foundations of thermodynamics has been developed over several decades within the Keenan–Hatsopoulos tradition and further elaborated by Gyftopoulos, Beretta, and collaborators \cite{HatsopoulosKeenan1965,GyftopoulosBeretta1991,ZanchiniBeretta2010,BerettaZanchini2011}. In this approach, thermodynamics is formulated as an autonomous physical theory, complementary to mechanics, based on carefully worded operational definitions of all the basic concepts, starting from system, state, property, and process.

Within this framework, energy is introduced as a consequence of the first law expressed as a postulate asserting the adiabatic interconnectability of all pair of states with fixed composition and constraints. From this postulate follows the principle of energy conservation and the additivity of energy. 
Entropy is introduced independently of heat, empirical temperature, and calorimetric constructions. The second law is expressed as a postulate asserting that any given state of  a system with fixed composition and constraints and an admissible value of the energy can be adiabatically and reversibly interconnected to a stable equilibrium state. From this postulate follow, as consequences rather than assumptions, the principle of entropy nondecrease and the additivity of entropy.

A central conceptual shift inherent in this formulation is that entropy is a property of all states, not merely of stable equilibrium states. Stable equilibrium states are not the foundation of thermodynamics, but rather a distinguished subset of states characterized by extremal properties. Only for such states does entropy admit a fundamental equation whose derivatives define potentials such as temperature, pressure, and chemical potentials. Nonequilibrium states, by contrast, possess well-defined entropy values but no associated potentials --- a distinction essential for both logical clarity and physical interpretation.

This perspective aligns with, while remaining distinct from, other axiomatic developments, including Carathéodory’s formulation \cite{Caratheodory1909} and the order-theoretic approach of Lieb and Yngvason \cite{LiebYngvason1999,LiebYngvason2013,LiebYngvason2014,ZanchiniBeretta2014}. Across these approaches, a common insight emerges: entropy must be defined independently of heat if thermodynamics is to achieve full logical coherence and generality.

\subsection{Structure of the present exposition}

The first part of the paper is devoted to the careful formulation of operational definitions and fundamental postulates. These definitions are deliberately chosen so as not to rely on extensivity, macroscopic homogeneity, or large-system limits, and are therefore applicable without modification to systems with few particles as well as to macroscopic systems.

Building on this foundation, several key thermodynamic concepts --- most notably adiabatic availability and available energy --- are developed and illustrated using the energy–entropy diagram representation of nonequilibrium states introduced by our school of thermodynamics. This geometric representation treats equilibrium and nonequilibrium states on equal footing and provides direct insight into irreversibility, dissipation, and the limits of work extraction, independently of system size.

A substantial portion of the latter part of the paper is devoted to the analysis of entropy transfer in non-work interactions. This analysis leads to precise operational definitions of heat interactions and of more general heat-and-diffusion interactions, concepts that are essential for mesoscopic continuum theories of nonequilibrium behavior in simple and complex solids and fluids, including systems with internal structure and nonlocal effects. The formulation emphasizes entropy balance relations that remain valid beyond equilibrium and beyond macroscopic idealizations.

An important byproduct of this analysis is the derivation of Clausius inequalities and of the Clausius statement of the second law in forms explicitly extended to nonequilibrium processes.

\subsection{Nonequilibrium states, small systems, and nanothermodynamics}

A central point emphasized throughout this paper is not merely the applicability of thermodynamic concepts to small systems, but their universality: all concepts defined and employed here apply to all systems, independently of size, particle number, or degree of extensivity. No assumption of extensiveness is made at the foundational level. Properties such as entropy and energy are defined operationally for individual systems in individual states, rather than inferred from scaling arguments or ensemble limits. We purposely avoid to discuss here any instance or assumption of extensivity, because extensivity is not a prerequisite for thermodynamic reasoning, but  a contingent feature of certain classes of systems and states that, using the terminology introduced in \cite[Ch. 16]{GyftopoulosBeretta1991},  can be modeled under the ``simple-system model'' approximation, also known as macroscopic limit.   This emphasis is increasingly relevant in the context of what is often termed nanothermodynamics \cite{Hill1964,Hill2001,Chamberlin2015,Chamberlin2024,bedeaux2024}, where traditional macroscopic idealizations may fail, yet many thermodynamic results remain valid and indispensable.

The approach adopted here does not rely on statistical ensembles or fluctuation arguments. Instead, it emphasizes operational definitions and existence principles that hold for systems of arbitrary size, provided the states under consideration are appropriately defined (for example, as separable and uncorrelated). In this sense, the present treatment clarifies the thermodynamic content that underlies --- and constrains --- statistical and microscopic descriptions, rather than replacing them, as illustrated in the last part of the paper (from Section \ref{quantumexamples} on).

This perspective also provides a coherent conceptual foundation for near-equilibrium theories, such as Onsager’s reciprocal relations and entropy production principles, as well as for ongoing efforts to extend thermodynamics to far-from-equilibrium phenomena.

\subsection{Role of thermal reservoirs}

Unlike some earlier developments --- including our own work aimed at eliminating the notion of thermal reservoirs from the operational definition of entropy \cite{BerettaZanchini2019}  --- the present exposition does make explicit use of the concept of a thermal reservoir, carefully defined and employed within a clearly delimited scope.

This choice is motivated by considerations of simplicity, transparency, and pedagogical effectiveness. When rigorously defined, thermal reservoirs provide an efficient operational tool for relating entropy changes to measurable energy exchanges in a broad class of processes. Their use in this paper represents a deliberate modeling choice rather than a claim of fundamental necessity, and its assumptions and limitations are made explicit throughout.

\subsection{Relation to teaching and applications}

This paper also serves an explicitly pedagogical purpose. It is intended as a written companion to a graduate-level course on advanced thermodynamics developed over many years and recently made available through MIT OpenCourseWare \cite{Beretta2024MITOCW}. That course emphasizes precise thermodynamic language, explicit modeling assumptions, and a logically coherent progression from elementary definitions to advanced nonequilibrium and small-systems applications.

Accordingly, the exposition is structured so that subsets of the material can support instruction at different levels, while remaining fully consistent with applications in energy systems, environmental and climate engineering, separation processes, and coupled energy–mass–charge transport phenomena.

\subsection{Perspective and scope}

The guiding philosophy of this work is that many of the conceptual difficulties traditionally associated with thermodynamics are not intrinsic to the theory itself, but stem from historical choices in the order and manner in which its basic concepts are introduced. By revisiting these foundations with explicit attention to logical structure, operational meaning, and domain of validity, thermodynamics emerges as a coherent and universal physical theory --- applicable without qualification to equilibrium and nonequilibrium states, to macroscopic and few-particle systems alike. 

By formulating thermodynamics without assuming extensivity and by treating nonequilibrium states on the same conceptual footing as equilibrium states, the present work aims to clarify what is essential, what is contingent, and what is universal in the foundations of the theory.

\subsection{Proofs are omitted}

In this paper, several results are stated without explicit proofs. Unless otherwise indicated, complete proofs are available in Ref.~\cite{GyftopoulosBeretta1991}. The use of expressions such as \emph{it follows that\dots} or \emph{it can be proved that\dots} without explicitly presenting the proof is a deliberate expedient intended to streamline the exposition, allowing emphasis on the structure of the arguments, the precise formulation of the concepts involved, and their range of applicability. The same choice has also proven effective in instructional settings, where it facilitates concentration on the essential conceptual content.

The logical rigor of the presentation rests on the existence of complete and detailed proofs supporting each assertion. Their availability ensures that the foundational framework employed here is sound, unambiguous, and applicable whenever the stated modeling assumptions and definitions are satisfied.

		\section{What is Thermodynamics?}
	
Thermodynamics has survived all major scientific revolutions and advances in physics, chemistry, and engineering. Far from being an obsolete discipline, it has experienced a renewed centrality in recent decades, driven in particular by the study of nonequilibrium phenomena and systems with few degrees of freedom. This resurgence stands in contrast with a period, not so long ago, during which thermodynamics was sometimes regarded as a closed or exhausted subject. Yet, despite its longevity and pervasive influence, a precise and universally accepted answer to the seemingly simple question --- \emph{what is thermodynamics about?} --- remains surprisingly elusive. Even experts in the field often hold distinct, and evolving, personal definitions, and may find it difficult to articulate them unambiguously or to reconcile them with alternative viewpoints.

Nonetheless, it is useful to attempt a clear statement of perspective. From the present point of view, \emph{applied thermodynamics} may be regarded as the art of modeling the kinematics and dynamics of physical systems by selecting an appropriate level of description for a given application of interest, and by enforcing the general principles, rules, and constraints that such models must satisfy in order to provide a faithful representation of physical reality. The application of interest may arise in engineering, chemistry, physics, biology, or cosmology; the same logical structure can, in principle, be extended to other domains --- such as economics or social systems --- whenever a well-defined ``plane of perceptions'' in the sense of Margenau \cite{Margenau1950} can be identified.

\emph{Foundational thermodynamics}, by contrast, is concerned with the inverse problem: the extraction, distillation, and identification of the most general and unifying principles from the successes and failures of diverse modeling efforts aimed at rationalizing experimental observations. Its objective is not the construction of specific models, but the clarification of the structural constraints (such as the great conservation principles \cite{Feynman1964great}) that any admissible model of physical systems must satisfy, independently of scale, composition, or degree of disequilibrium.

In this sense, thermodynamics may be described as the science that studies the instantaneous condition of material systems and the evolution in time of such conditions, whether the evolution occurs spontaneously or as a result of interactions with other systems. Viewed from this perspective, thermodynamics constitutes a genuine extension --- indeed, a generalization --- of mechanics. The meaning and implications of this statement will become progressively clearer as the exposition unfolds, and will eventually acquire a direct geometric interpretation through the energy--entropy diagram representation introduced in Sec.~\ref{ESdiagram} and thereafter.

Given the breadth and generality of its scope, thermodynamics requires the unequivocal definition of a number of basic concepts upon which the theory is founded. Some of these concepts are inherited from mechanics and will be assumed known. Others --- such as system, property, state, process, equilibrium, stable equilibrium, energy, entropy, temperature, and pressure --- must be defined with particular care. In the present work,  following \cite{GyftopoulosBeretta1991}, these concepts are redefined  not only to eliminate ambiguity, but also to extend their validity and operational meaning beyond the traditional confines of macroscopic, extensive systems and stable equilibrium.

	\section{The loaded meaning of the word ``system''}\label{system}
	
	Matter is composed of particles, either free or bound together to form nuclei, atoms, molecules, ions, and other structures, as well as the electromagnetic field. Depending on the phenomenon to be described, it is appropriate to identify a model of reality that is as simple as possible, which limits itself to a level of simplified description that completely ignores aspects (of subatomic or submolecular structure, nuclear reactions, chemical reactions, radioactive decay, etc.) that, although potentially active in principle, have negligible effects on the study of the phenomenon of interest.
	
	For example, if the effects of chemical reactions are not relevant, we can study the properties of water by assuming that the $\rm H_2O$ molecules are indivisible, or the properties of oxygen by assuming that the $\rm O_2$ molecules are indivisible. The choice of the appropriate level of description and the ``indivisible constituents''\ind{constituents} are the first steps in defining what we call a 
	``system.''\ind{system} However, in thermodynamics, in order to talk about a system, a precise condition regarding the forces acting on these indivisible constituents must also be satisfied. 
	The condition is that none of the forces acting on the set of constituents of interest depends on the coordinates of other constituents external to the object of study and that none of the outcomes of measurements performed on the set of constituents of interest should be statistically correlated to the outcomes of measurements performed on external constituents.
	 However, these forces can depend on geometric parameters (such as the shape of a container limiting the available space) or on fields generated by ``static'' sets of external constituents or control devices.
	
A ``system''\ind{system} is, therefore, a set of constituents that is ``separable'' and ``uncorrelated'' from ``external'' constituents, defined by the following specifications: (a) the type or types of ``constituents''\ind{constituents}, for example, water molecules, or a mixture of oxygen molecules and nitrogen molecules; (b) the  ``parameters''\ind{parameters of external forces} that characterize all external forces, i.e., constraints and forces exerted from the outside on the constituents, for example, specifications describing the geometric shape of a sealed container or the volume of the container itself, or an electric, magnetic, or gravitational field; (c) the nature of the 
	``internal forces''\ind{internal forces} between the constituents that are to be considered in the model, such as intermolecular forces or the condition that some or all chemical reactions within the system are inhibited; and (d) the nature of any 
	``internal constraints''\ind{internal constraints} that characterize the interconnections between separate parts and define the internal structure of the model, such as a fixed or mobile wall that divides the volume available to the constituents into two partitions. Again, ``external'' constituents are those not included in the set under examination; ``separable'' refers to the condition that none of the external forces depends on the coordinates of external constituents;  ``uncorrelated''  refers to the condition that no measurement done on the constituents of interest is statistically dependent on measurements done on external constituents. Anything external to the system and therefore excluded from it is called the 
	``system's environment'' or simply the 
	``environment.''\ind{environment}  In principle, the environment should represent a model of the ``rest of the universe''; in practice, however, it is sufficient to adopt a simplified model that includes only those external constituents that effectively constrain and interact with the system's constituents over the time interval of interest.
	
	For a system consisting of $r$ different types of constituents,\ind{constituents} we indicate their amounts using the variables \lst{n}{r} where $n_i$ stands for the number of units (molecules, atoms, or particles) of the $i$-th constituent.\footnote{In the following, the bold symbol, $\bn$, will denote the set of all amounts of constituents, $\bn=\{n_1,n_2,\dots,n_r\}$.}
	
	The unit of measurement in the International System for the amount of a constituent is the 
	``mole,''\ind{mole}\ind{gram-mole} indicated by the symbol 
	``mol'' (sometimes also called the 
	``gram-mole'' and indicated by the symbol 
	``gmol''), defined as the number of units (molecules, atoms, or particles) equal to  Avogadro's number,\ind{Avogadro's number of} $N_{\rm Av} = 6.02214076 \times 10^{23}$, and thus 1 mol (= 1 gmol) = $N_{\rm Av}$ particles. Of course, the International System prefixes for multiples are applicable. For example, for the kilomole, 1 kmol = $10^3$ mol.
	
	The ratio $M_i$ between the mass $m_i$ and the amount $n_i$ of the $i$-th constituent, $M_i = m_i/n_i$, is called the ``molecular (or atomic) mass,''\ind{mass!molecular}\ind{mass!atomic} and is normally expressed in g/mol or kg/kmol.
	
	Internal forces\ind{internal forces} can be of various types. For example, a system that is typically studied in detail in introductory engineering courses consists only of $\rm H_2O$ molecules subject only to intermolecular internal forces and external forces that confine them to a region of space with volume $V$. Another standard example of a system consists of three species, $\rm H_2$, $\rm O_2$, and $\rm H_2O$, subject, in addition to the intermolecular forces between all three types of molecules, to internal forces that control the chemical reaction mechanism $\rm H_2 + \frac{1}{2}\, O_2 = H_2O$.
	
	For a system with constraints and external forces\ind{external forces} dependent on $s$ parameters, we indicate the parameters\ind{parameters of external forces} with the symbols \lst{\beta}{s}.\footnote{The bold symbol, $\bbeta$, will denote the set of all parameters, $\bbeta=\{\beta_1, \beta_2,\dots, \beta_r\}$.} For example, the parameters can be the sides $\ell_1$, $\ell_2$, $\ell_3$, and the volume $V=\ell_1\ell_2\ell_3$ of a parallelepiped-shaped region enclosed by walls (understood as impenetrable barriers) of a container that separates the constituents of the system from others that are external and therefore do not belong to it.
	
	Other parameters can be provided by external force fields,\ind{field!gravitational}\ind{field!electric}\ind{field!magnetic} such as the gravitational field ${\bf G}_e$, electric field ${\bf E}_e$, and magnetic field ${\bf H}_e$ generated by stationary distributions of mass\ind{density!mass}\ind{density!charge}\ind{density!current} density $\rho$, charge density $\rho_e$, and current density $\bj$ outside the region of space occupied by the constituents of the system. Here, ${\bf G}_e$, ${\bf E}_e$, and ${\bf H}_e$ represent the values of the fields that these distributions $\rho$, $\rho_e$, $\bj$ would generate in the region of space of the system in the absence of its constituents, with the understanding that outside this region, the distributions of electric and magnetic dipole or multipole moments are zero. 
	
	The principles we state in this paper and the results that follow are valid for both ``macroscopic systems,''\ind{system!macroscopic} composed of large amounts of constituents (such as 1 kg of H$_2$O or an entire thermal power plant), and ``microscopic systems,''\ind{system!microscopic} composed of small amounts of constituents (such as a single molecule of H$_2$ or a structureless point particle confined in a box). It is important to note that this observation is rarely acknowledged in thermodynamics textbooks, which immediately restrict their treatment to the simple-system model, in our opinion missing the opportunity to expose the universal aspects of thermodynamic theory and clarify its relations with mechanics.
	
	It is worth noting that the given definition of a system, which generally coincides with the one adopted (although often only implicitly) in physics, is made rather restrictive by the conditions of  statistical independence of measurement results and independence of  external  forces from  coordinates of external objects. 	
	Therefore, not always does a material object or, better, a model of a material object constitute a well-defined system. 
	
	In particular, separability requires that the forces acting on the system's constituents must all be either internal or external. For example, an electron can constitute a system if it is free or if it is immersed in an external electrostatic field but not if it is subject to interaction with other electrons in the bonding of a molecule or with the nucleus of an atom. The restriction is significant, and it will be well to keep it in mind because the principles we will state and the results that follow are only valid for well-defined systems, defined in the manner just described.\footnote{Anticipating ideas that we develop in what follows, a fundamental consequence of the laws of thermodynamics is that energy and entropy are additive properties defined for all systems in every possible state. However, these properties do not apply to material objects subject to external forces dependent on external coordinates or to correlations with external constituents. For such models that do not satisfy our formal definition of a system, energy and entropy are not merely non-additive --- they are not defined.}

	In the case of constituents immersed in a gravitational\ind{field!gravitational}\ind{field!electric}\ind{field!magnetic} field ${\bf G}_e$, electric field ${\bf E}_e$, or magnetic field ${\bf H}_e$, the independence of external forces from the coordinates of external objects, necessary for the system to be well-defined, is guaranteed only if these fields are generated by stationary distributions, i.e., time-invariant ones, of\ind{density!mass}\ind{density!charge}\ind{density!current} mass density $\rho$, charge density $\rho_e$, and current density $\bj$ that generate them, and if outside the volume of the system, the electric and magnetic dipole or multipole moments are zero.
	
	\section{The loaded meaning of the word 	``property''}\label{properties}
	
	The experimental method involves studying the behavior of a system subjected to 
	``measurement procedures.''\ind{measurement procedure} Each measurement procedure is associated with a 
	``physical observable''\ind{physical observable} representing the system's response to the procedure. Each procedure leads to the determination of a result, generally expressible in numerical terms: the 
	``value of the physical observable.''
	
	An important subclass of physical observables is properties.\ind{properties} A 
	``property'' $P$ is defined by a measurement procedure that, when applied to a system at time $t$, provides a numerical result $P(t)$, the value of the property at that instant, which must be independent of: details of the measurement devices, other systems in the environment, and instants of time different from $t$.
	
	This definition is quite restrictive, and there are numerous examples of measurement procedures that do not satisfy it and therefore, while defining physical observables, do not define properties. For example, the distance traveled by a particular molecule in a given finite interval of time divided by the interval itself is not a property because the measurement procedure for its value necessarily depends on the results of two position measurements at different times. However, if the time interval is made to tend to zero, then the limit value depends only on the initial time, and the procedure defines a property well known in mechanics: velocity. Examples of procedures that satisfy the definition of a property just given are well-known measurement procedures in mechanics that define instantaneous position, instantaneous velocity, and instantaneous acceleration of a particular molecule of a constituent.
	
	The procedure for counting the number of particles, atoms, or molecules of the $i$-th type present in the system at time $t$, defining the amount $n_i$, satisfies the definition of a property and provides the value $n_i(t)$. The same applies to the measurement procedure for the volume available to the constituents of the system at time $t$, defining the parameter $V$ of external forces, which satisfies the definition of a property and provides the value $V(t)$. The same holds for the other parameters of external forces.

	\section{What exactly do we mean by ``state'' of a system?}\label{state}

	To completely characterize\ind{state} a system at a given instant of time $t$, one must specify how it responds to all possible measurement procedures it can undergo. Therefore, in particular, it is necessary to specify the values of all amounts of constituents, all parameters of external forces, and all other properties. This set of values, which is generally an infinite list of numbers, defines the 
	``state'' of the system at that instant,
	\begin{equation} A(t) = \{n_1(t),\dots,n_r(t),\beta_1(t),\dots,\beta_s(t),P_1(t),P_2(t),\dots\},\label{stateEq}\end{equation}
	where $A(t)$ denotes the state of system $A$ at time $t$, and $P_1(t)$, $P_2(t)$, $\dots$ represent the values at time $t$ of various properties.
	To say that the state of the system is known means that the values of all properties, amounts, and parameters are known.\footnote{In some treatments, properties are also referred to as 
		``state variables'' or 
		``state functions,'' and it is stated that the value of each property depends exclusively on the state of the system. In reality, we have seen that it is the values of the properties that define the state, not the other way around.}\ind{state variables}\ind{state functions}
	
	The definition of a state does not impose any restrictions on the number of properties that contribute to defining it. In general, this number is infinite even for the most elementary systems. For example, for a single material point confined to a given region of space, it is known from quantum mechanics that at least one property is defined for each geometric point in the space available: the probability that the material point is in that position following a position measurement. This infinite set of numerical values (one per point), which contributes to defining the state, can be represented, for a particular subclass of states, by the so-called 
	``wave function.''
	
	The fact that the list of values defining the state of a system at a given instant of time is infinite, means that a given system admits a vast multitude of possible states. Two identical systems are in two different states if the values of at least one property are different for the two systems: the two infinite lists of values defining the states of the two identical systems differ in at least one of the corresponding values.
	
	This great variety of states is rarely mentioned in thermodynamics textbooks, which immediately restrict their treatment to stable equilibrium states, thereby also missing the opportunity to clarify the relations between mechanics and thermodynamics and to extend the treatment to nonequilibrium states, which are by far the most numerous and also the most interesting in terms of applications, as we will see.

			\subsection{Representations of states across nonequilibrium thermodynamic frameworks}\label{frameworks}

	As noted in the Introduction, the profound significance of the laws of thermodynamics lies in their universal applicability across any level or framework of description chosen to model empirical reality. This holds provided the model possesses a fundamental mathematical structure and satisfies certain reasonable conditions --- ensuring, for instance, that concepts like separability and statistical independence between the system and its surroundings are precisely defined. 
		
	Throughout two centuries of thermodynamic history, numerous frameworks have successfully modeled nonequilibrium phenomena. While these approaches vary significantly in their choice of independent properties defining the state space (in the sense of Eq.~\ref{stateEq}), most share common geometrical features within their mathematical structures, as captured for example by the  different iterations of the GENERIC (General Equation for the Non-Equilibrium Reversible-Irreversible Coupling) construction~\cite{Pavelka2018, Ottinger2005,Grmela2026} or variational formulations~\cite{Gyarmati1970,Van2020}.

			The choice of independent properties used to define a ``state'' is not merely a matter of mathematical convenience; it reflects the physical scale, the degree of rarefaction, and the distance from equilibrium under consideration. Although these frameworks differ significantly in their mathematical formalisms, a common thread emerges: most modern approaches represent the state through various forms of probability distributions. Whether describing the likelihood of microstates in phase space, the distribution of internal mesoscopic configurations, or the statistical populations and coherences encoded in a density operator, these probabilistic representations provide a bridge between microscopic fluctuations and macroscopic observables.
			
		The following list provides a minimal summary of the typical sets of independent properties adopted for state description in several prominent frameworks. The comparison highlights how different theories prioritize either a finite set of macroscopic fields or a more elaborate distribution function in order to capture the relevant physics. To ground these abstract descriptions in a concrete application, Section \ref{quantumexamples} presents an explicit example of a state formulation within the framework of quantum thermodynamics, illustrating how these principles manifest at the interface between information and energy.

	\begin{itemize}[labelindent=0em,labelsep=0.5em,leftmargin=*,wide = 0pt]
		\item \textit{Classical Statistical Mechanics}~\cite{Gibbs1902, LandauLifshitz1980}: Probability distributions (densities) over classical phase space, representing the possible microstates defined by the positions and momenta of all particles.
		
		\item \textit{Quantum Statistical Mechanics}~\cite{Tolman1938, Pathria2021}: Probability distributions over a set of pure quantum states, typically expressed in terms of the eigenvectors of the system’s Hamiltonian operator.
		
		\item \textit{Information-Theoretic Thermodynamics}~\cite{Jaynes1957, Parrondo2015}: Probability distributions over a discrete or continuous set of microstates or events, representing the observer’s uncertainty or informational description.
		
		\item \textit{Stochastic Thermodynamics}~\cite{Seifert2012, Sekimoto2010, Risken1996}: Probability distributions over fluctuating trajectories and path-dependent variables (such as stochastic work and heat) that characterize the energetics of individual realizations.
		
		\item \textit{Macroscopic Nonequilibrium Thermodynamics (CIT)}~\cite{Onsager1931, DeGroot1962, Ziegler1983}: Local-equilibrium field variables --- functions of position and time --- representing the macroscopic properties of continuum parcels assumed to be in locally stable equilibrium states.

		\item \textit{Gradient (Non-local) Thermodynamics}~\cite{Cahn1958,Antanovskii1996,Van2003,Mauri2013}: A set of local field variables augmented by their spatial gradients, representing the state of non-uniform systems where the local energy density depends on the neighboring environment, such as in the description of diffuse interfaces and phase separation.

		\item \textit{Continuum Mechanics}~\cite{Malvern1969, Truesdell2004, Gurtin2010}: A finite set of local field variables, such as displacement, deformation gradient, and mass, momentum, and total energy densities, representing the macroscopic properties of continuum parcels under the local-equilibrium hypothesis.
		
	\item \textit{Internal Variable Theories}~\cite{Berezovski2017, Van2020}: A finite set of local field variables supplemented by hidden or internal variables that describe the microstructural state of a material (e.g., in thermoelasticity), often derived via variational principles to ensure thermodynamic consistency.

		\item \textit{Extended Nonequilibrium Thermodynamics (EIT)}~\cite{Jou2010}: A finite set of local field variables that includes classical densities augmented by dissipative fluxes (e.g., heat flux, viscous stress) treated as independent state variables in order to account for memory effects and finite signal propagation speeds away from equilibrium.
		
		\item \textit{Mesoscopic Nonequilibrium Thermodynamics}~\cite{Mazur1998, Rubi2012, Anisimov2024}: Probability distributions over a set of local mesovariables describing internal configurations, such as the position of a Brownian particle, molecular orientation, cluster size, or degree of protein folding.
		
		\item \textit{Rational Extended Thermodynamics (RET)}~\cite{Muller1998, Ruggeri2015}: A finite set of local fields representing moments of the velocity distribution function, such as mass, momentum, and energy densities, as well as higher-order moments including heat flux and momentum flux.

		\item \textit{Small-Scale and Rarefied Gas Dynamics}~\cite{Cercignani1988,Struchtrup2005}: The local particle velocity distribution function, representing the probability of finding a molecule with a specified velocity at a given position and time, typically governed by the Boltzmann equation.
		
		\item \textit{Chemical Kinetics}~\cite{Prigogine1967, Beretta2012, Kondepudi2014}: A set of bulk or local macroscopic variables, such as species concentrations and reaction coordinates, sufficient to describe the chemical evolution of a reactive mixture toward stable equilibrium.
		
		\item \textit{Quantum Thermodynamics}~\cite{Alicki2007, Deffner2013, Binder2018}: The density operator, serving as a generalized probability distribution that represents both classical populations (probabilities) and quantum coherences.
		
		\item \textit{Nanothermodynamics}~\cite{Hill1964,Hill2001,Chamberlin2015,Chamberlin2024,bedeaux2024}: Statistical distributions over a large number of independent, small-scale replicas of a system, where the state is defined by finite-size parameters and internal degrees of freedom that account for non-extensive energy contributions and fluctuations.
	\end{itemize}

	While connecting these ideas to the extensive literature on nonequilibrium thermodynamics is valuable, acknowledging the many pioneers of the various approaches --- and summarizing the different attempts at unification,  construction --- would require a level of discussion beyond the scope of this paper. Rather than claiming exhaustiveness, the preceding list is intended to illustrate the broad range of state representations to which the general thermodynamic concepts developed in this work may apply.
	
	This diverse landscape of state representations illustrates the remarkable versatility of nonequilibrium thermodynamics across different physical scales and modeling priorities. It is important to emphasize that the foundational approach presented in this work is intended to be compatible with, and applicable within, any of these specific frameworks. By providing a rigorous set of operational definitions --- grounded in the first and second laws as they apply to well-defined (separable and uncorrelated) systems --- our construction establishes a self-consistent logical basis that helps clarify the limits of applicability of the assumptions underlying any chosen set of state variables. In particular, this grounding makes explicit the conditions of separability and absence of correlations required for the valid application of the energy balance and the principle of entropy non-decrease, thereby reducing the risk of conceptual misinterpretations when analyzing complex or interacting composite systems whose components become correlated during interaction.

	\section{Time evolution, interactions, and the concept of ``process''}\label{statechange}
	
	The state of a system can evolve spontaneously,\ind{time} driven by its internal dynamics,\ind{internal dynamics}\ind{spontaneous evolution}\ind{state!change in} or as a result of interactions with its environment. An \textit{isolated system}\ind{isolated system}, one that cannot interact with the environment and hence cannot cause any change in the environment's state, can only undergo spontaneous evolutions. Non-isolated systems interact\ind{interaction} in various ways, resulting in the \textit{flow} (or \textit{transfer} or \textit{exchange}) of certain properties from one system to another. This involves a decrease in the value of a property in one system, accompanied by a simultaneous increase of the same value in the other system. For example, during the interaction between two systems undergoing an elastic collision, there is a transfer of momentum and kinetic energy from one system to the other.
	
	The equation that describes the evolution of the state of a system over time is called the \textit{equation of motion}\ind{equation of motion}\ind{dynamics}\ind{motion law}\ind{temporal evolution}\ind{state!change in} of the system. To illustrate this, we can say that it will have a structure similar to
	\begin{equation}  \frac{\diff A(t)}{\diff t}= f(A(t),{\rm internal\ forces}(t), {\rm external\ forces}(t))\end{equation}
	where the function $f$ depends on the nature of the system's constituents.
	Given the function $f$ and the state $A(t_0)$ of system $A$ at time $t_0$, integration of the equation of motion allows us to calculate the state $A(t)$ of the system at any other instant in time, whether earlier (past) or later (future) than $t_0$.\footnote{In some models, a notable example being the quantum theory of open systems, the existence of the solution of the initial value problem for the equation of motion is granted only forward in time, and often this mathematical irreversibility  is confused with the idea of thermodynamic irreversibility.}
	
	This problem, as formulated, poses an enormous mathematical complexity due, on the one hand, to the fact that the state $A(t)$ is a mathematical object representing an infinity of numbers, and on the other hand, to the fact that for many systems and for the most important states in thermodynamics, the general equation of motion is still a subject of research. Thus, the chosen approach is necessarily different. We limit ourselves to verifying that the temporal evolution of the system's state is consistent with the two main implications of the equation of motion, recognized as universally valid and therefore to be respected by all systems. These implications are initially expressed in a non-mathematical form as the statements of the \textit{first law}\ind{first law of thermodynamics} and the \textit{second law}\ind{second law of thermodynamics} of thermodynamics and are then translated into two relations that must always be satisfied: the energy and entropy balance equations. In practice, we give up the possibility of determining how the state changes over time by solving the system's equation of motion and instead limit ourselves to determining how only the values of some main properties, defined for all states of all systems (amounts of constituents, energy, entropy), change over time.
	
	In other words, the temporal evolution is characterized (in an incomplete way) by: (a) the description of the initial state $A(t_1)$ and the final state $A(t_2)$ of the system; (b) the description of the interactions that occur during the change of state, which cause the flow or exchange of certain properties between the system and its environment; and (c) verification that the change is compatible with the first law\ind{first law of thermodynamics} and the second law\ind{second law of thermodynamics} of thermodynamics, or with the main implications of the general equation of motion, including the principles of energy conservation and entropy non-decrease.
	
	For simplicity, in the following, we use the notation $A_1$ to represent the state of system $A$ at time $t_1$ instead of $A(t_1)$.
	
	The first law\ind{first law of thermodynamics} and the second law\ind{second law of thermodynamics} of thermodynamics are stated as laws or principles, i.e., as unprovable postulates. However, as we have seen, in an approach that postulates a general equation of motion valid for all systems, these principles would emerge as theorems, consequences of the equation of motion. Therefore, it remains a fact that the laws (the principles) of thermodynamics express general consequences of the equation of motion, i.e., of the dynamics of all (well-defined) systems.

	\begin{figure}[!ht]
		\begin{center}
				\includegraphics[scale=0.45]{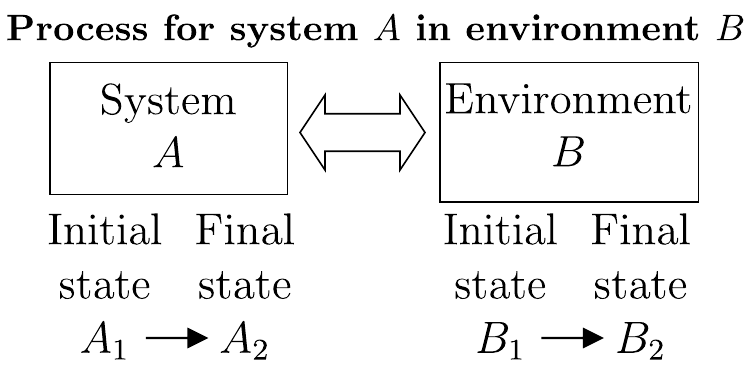}
				\caption{\label{Process}The term ``process'' refers to the description of the initial state, the final state, and the effects caused on the environment, related to a given temporal evolution of the state of a system.}
		\end{center}
	\end{figure}

	The descriptions\ind{process}\ind{interaction} of the initial state, the final state, and the effects caused by the interactions on the values of the main properties (amounts of constituents, energy, entropy) related to a given temporal evolution specify a \textit{process} (Figure \ref{Process}). Processes can be classified based on the effects they have on the system's surroundings, i.e., changes in state induced by interactions with its environment. For example, a process is called \textit{spontaneous}\ind{process!spontaneous} if it is not accompanied by any external effects. As previously seen in Section \ref{statechange}, an \textit{isolated system}\ind{isolated system} can only undergo spontaneous processes since it cannot induce changes in the state of other systems or be affected by them.
	
	The notion of a process is also compatible with a time evolution in which the separation between the constituents of the system and those of the environment is temporarily broken, provided that both the separation and the statistical independence are reestablished at the end of the evolution. In other words, it suffices that the system and the environment must be well defined --- i.e., separable and independent --- at both the initial and final instants of time.\footnote{In a quantum description, this requirement is highly restrictive and substantially limits the applicability of the thermodynamic foundations discussed in this paper. Consider two subsystems. By definition they are initially separable and uncorrelated. Separability may be temporarily broken by turning on an interaction Hamiltonian between the corresponding subsets of constituents, and later restored by turning it off. However, the interaction generally generates correlations that persist after the interacting constituents are separated. As a result, the condition of statistical independence of measurements performed on the two subsets is lost, and the resulting entities no longer satisfy the requirements of the present definition of a system, i.e., they are no longer subsystems. See, e.g., Refs.~\cite{BerettaZanchini2019,RayBeretta2025}.}
	
	\section{Weight processes and adiabatic accessibility}
	
\begin{figure}[!ht]
		\begin{center}
				\includegraphics[scale=0.45]{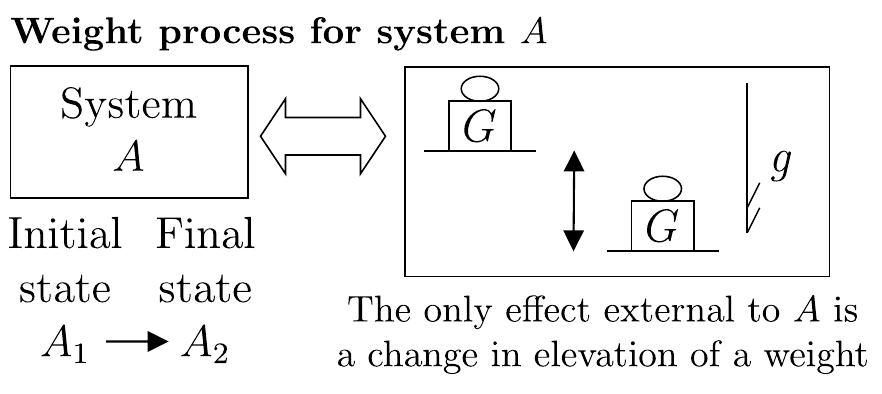}
				\caption{\label{WeightProcess}A process is a weight process if the only external effect caused by the system's interactions is a change in the elevation of a weight.}
		\end{center}
	\end{figure}

	A process\ind{process!weight} in which interactions result in the only external effect being a change in the elevation of a weight (or another equivalent mechanical effect) is called a \textit{weight process} (Figure \ref{WeightProcess}). Weight processes are conceptually important in developing the foundations of thermodynamics because, as we will see, they provide a clear presentation as an extension of mechanics.\ind{mechanics} Two states that may be interconnected via a weight process are said to be \textit{adiabatically accessible}.\footnote{Further discussions of the role of adiabatic accessibility are found in Refs. \cite{Zanchini1986,Zanchini1988,LiebYngvason1999,LiebYngvason2013,LiebYngvason2014,ZanchiniBeretta2014}.}

		\section{Reversible versus irreversible processes}
	
\begin{figure}[!ht]
		\begin{center}
			\includegraphics[scale=0.45]{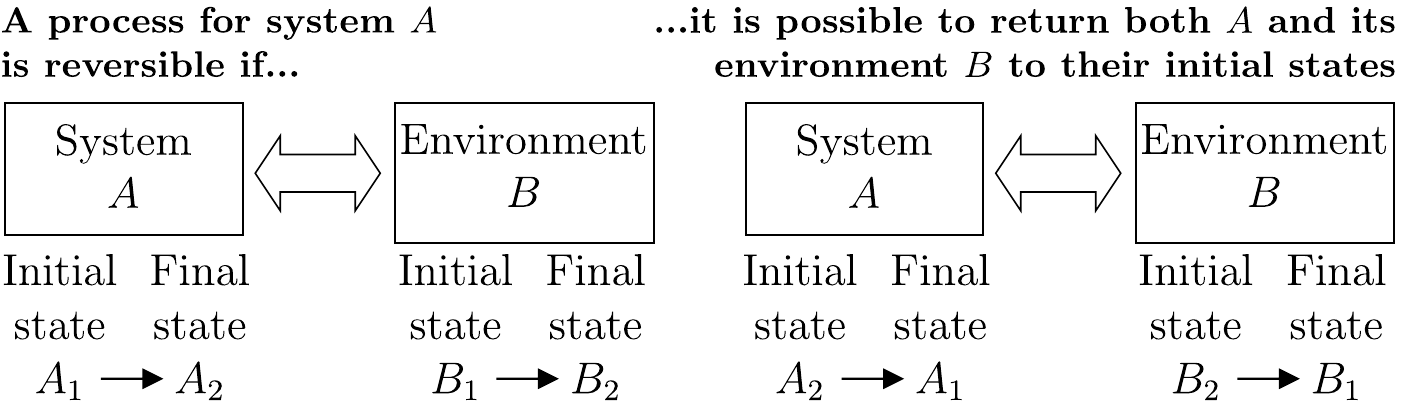}
			\caption{\label{ReversibleProcess}A process is \textit{reversible} if there is a way to return both the system and its environment to their respective initial states.}
		\end{center}
	\end{figure}

	Another important classification of processes is based on the possibility of reversing their effects. Thus, a process\ind{process!reversible} is called \textit{reversible} if there exists, i.e., if it is physically conceivable,\footnote{Permitted by the equation of motion.} a way to return both the system and its environment to their respective initial states (Figure \ref{ReversibleProcess}), i.e., if all the effects of the process (including those external to the system) are reversible. Otherwise, the process is \textit{irreversible}, meaning that there is no physically conceivable way to return the system and its environment to their respective initial states.

	In the particular case of an isolated system, the reversibility of a process implies the existence of a reverse process capable of returning the system to its initial state. The reverse process need not retrace, in reverse order, the same sequence of states traversed by the forward process.
	
	Directly verifying, based on the definition given, whether a process is reversible or not is impractical, as it would require testing all conceivable ways to return the system to its initial state and, for each one, checking whether the environment also returns to its initial state. Nevertheless, the given definition is valid and allows, as we will see, an indirect criterion (necessary and sufficient condition for the reversibility of a process) based on the simple balance of the property entropy, which we will define.
	
	\section{Statement of the First Law}\label{FirstLaw}
	
	The \textit{first law of thermodynamics} consists of two assertions: the first is  that \textit{any pair of states $A_1$ and $A_2$ with compatible values of the amounts of constituents and the parameters of a (well-defined) system $A$  can always be interconnected by means of a weight process}. Indicating with $z_2-z_1$ the change in height produced by the weight process on a mass $m$ in a uniform gravitational field with acceleration $g$, the second assertion is that \textit{the product $mg\,(z_2-z_1)$ assumes the same value for all weight processes that connect the two given states $A_1$ and $A_2$}.

	\section{Energy: definition, additivity, conservation, exchangeability}
	
	The main consequence of the first law is that for every system $A$ in any of its states $A_1$, a property called \textit{energy}\ind{energy} is defined, indicated by the symbol $E_1$.
	The \textit{energy}\ind{energy!definition} $E_1$ of system $A$ in state $A_1$ is defined by the following measurement procedure: an auxiliary weight process is realized that connects state $A_1$ to a reference state $A_0$ (of the same system $A$) chosen once and for all, to which a reference value $E_0$ is assigned. It is then stated that \begin{equation} E_1 = E_0  -mg\,(z_1 - z_0) \label{defEnergy}\end{equation}where $m$ is the mass of the weight, $g$ is the gravitational acceleration, and $z_1-z_0$ is the change in height of the weight, which is the only external effect in the weight process.

	Due to the dimensional homogeneity of Eq.\ 
	\ref{defEnergy}, $E_1$ and $E_0$ must have the same dimensions as $mgz$, i.e., energy. Therefore, the dimensions of $E$ are [mass]$\times$[length]$^2\times$[time]$^{-2}$, and the unit of measurement in the International System is the joule, J.\ind{Joule!unit of measure}
	
	The first law guarantees that the measurement procedure\ind{energy!definition}
	described defines $E$ as a property of system $A$.
	In fact, it follows from the first law's statement that the weight process connecting states $A_1$ and $A_0$ exists\footnote{The first law does not specify in which direction the weight process is possible, but it does allow us to conclude that if it is not possible from $A_1$ to $A_0$, it is definitely possible in the opposite direction.} and that the value of $mg\,(z_1 -
	z_0)$ and therefore $E_1$ is: (a) independent of the instruments used for measurement, i.e., of the details of the interaction between system $A$ and the weight; (b) independent of other systems in the environment, i.e., of the weight's details, which is the only other system affected by the weight process; and (c) independent
	of measurements made at other instants of time, i.e., of the details of the changes in $A$ during the process. It is important to note that the measurement procedure does not imply any restrictions on the nature of state $A_1$ and system $A$. Thus, {\it  energy is property, defined for all states of all systems}.\footnote{This is the most important consequence of the first
		law: the definition of energy for all states of all systems could not be taken for granted, nor is it valid if the word system is not loaded with the restrictive meaning we defined in Section \ref{system}.}
	
	The definition given extends what was already seen in {\it mechanics} to states that are not covered by mechanics. This will become clearer as we proceed. One fact, already known from mechanics, remains valid in general: for given values of $\bn$ and $\bbeta$, the value of a system's energy is bounded from below, i.e.,
	\begin{equation} E\ge E_{\rm min}(\bn,\bbeta) \end{equation}
	
	Since $E$ is a property, it contributes to defining the state of the
	system, together with the values of all the other properties. If the state of the system is known, then the value of energy is also known. The change $E_2-E_1$ corresponding to a given change in state from $A_1$ to $A_2$ depends only on states $A_1$ and $A_2$ and therefore not on the interaction methods with other systems, nor on the forces that induced it. The same change in state from $A_1$ to $A_2$ can be achieved by many (infinite)
	different processes, each with the same value of $E_2-E_1$.

	\begin{figure}[!ht]
		\begin{center}
			\includegraphics[scale=0.45]{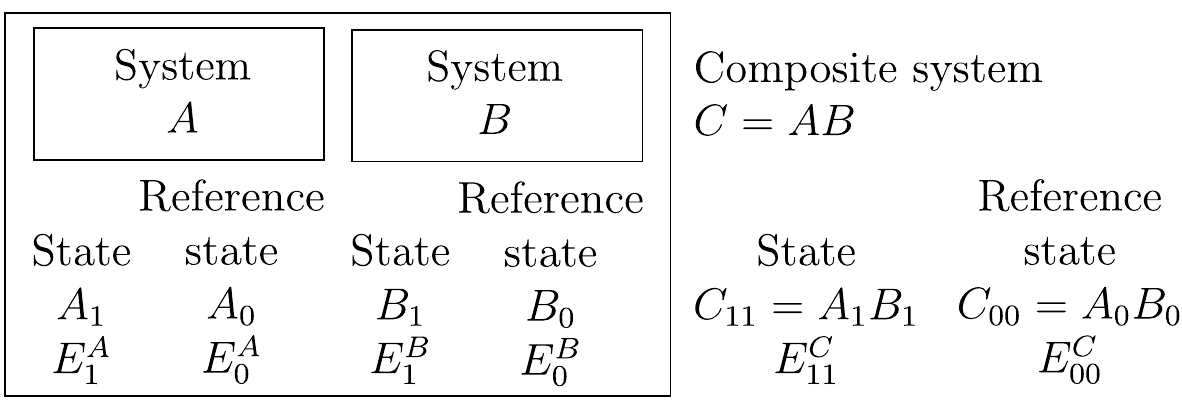}
			\caption{\label{EnergyAdditivity}Energy differences are additive, $E_{11}^C-E_{00}^C=(E_1^A-E_0^A)+(E_1^B-E_0^B)$. Energy values can be made additive by selecting reference values for composite systems so that $E_{00}^C=E_0^A+E_0^B$. As a result, $E_{11}^C=E_1^A+E_1^B$.}
		\end{center}
	\end{figure}
	
From the first law it  also follows that energy is an {\it additive}\ind{energy!additivity} property in the following sense. With the help of Figure \ref{EnergyAdditivity}, consider two systems $A$ and $B$ in states $A_1$ and $B_1$, respectively, and consider the {\it composite system}\ind{composite system} $C$ formed by the set of the two systems, $C=AB$, of which the state is indicated by the symbol $C_{11}$, where the double index refers to the respective states of the two subsystems ($C_{11}=A_1B_1$). Applying the procedure that defines energy for systems $A$, $B$, and $C$, after choosing the respective reference states $A_0$, $B_0$, and $C_{00}$ once and for all (this is the state of $C$ in which subsystem $A$ is in state $A_0$ and subsystem $B$ is in state $B_0$) and the respective reference values $E^A_0$, $ E^B_0$, and $ E^C_{00}$, the respective values of energy in states $A_1$, $B_1$, and $C_{11}$ are indicated by $E^A_1$, $ E^B_1$, and $ E^C_{11}$. Recalling the second assertion of the first law, it is easy to prove that\footnote{It is sufficient to consider: (1) the weight process for $A$ between $A_1$ and $A_0$ and the weight process for $C$ between $C_{11}$ and $C_{01}$ and observe that they have identical external effects; (2) the weight process for $B$ between $B_1$ and $B_0$ and the weight process for $C$ between $C_{01}$ and $C_{00}$ and observe that they have identical external effects; (3) the sequence of the two weight processes for $C$ between $C_{11}$ and $C_{01}$ and between $C_{01}$ and $C_{00}$, which is effectively a weight process between $C_{11}$ and $C_{00}$.}  \begin{equation} E^A_1- E^A_0+ E^B_1 -E^B_0= E^C_{11}- E^C_{00} \end{equation}
and therefore that {\it if the reference value for the composite system $C$ is chosen as the sum of the reference values of the subsystems that compose it}, i.e., if $ E^C_{00}=E^A_0+ E^B_0$, then 
\begin{equation} E^C_{11}=E^A_1 + E^B_1 \end{equation}
	
Finally, the first law also implies that the value of the energy {\it is conserved} every time the system undergoes a process without net external effects, such as a spontaneous process, for example. The conservation\ind{energy!conservation} of energy is of fundamental theoretical and practical importance.\footnote{Note that although in some formulations it is presented as the {\it principle of conservation of energy}, the conservation of energy here emerges not as a principle but as a consequence of the first law, as is the very existence of energy as a property.}
	This importance derives from the additivity of energy (or rather of energy differences) and from the fact that
	any process can always be considered part of a process without net
	external effects of a larger system that includes all interacting systems, for which energy --- the sum of the energies of the
	subsystems --- remains unchanged.

	\begin{figure}[!h]
		\begin{center}
				\includegraphics[scale=0.45]{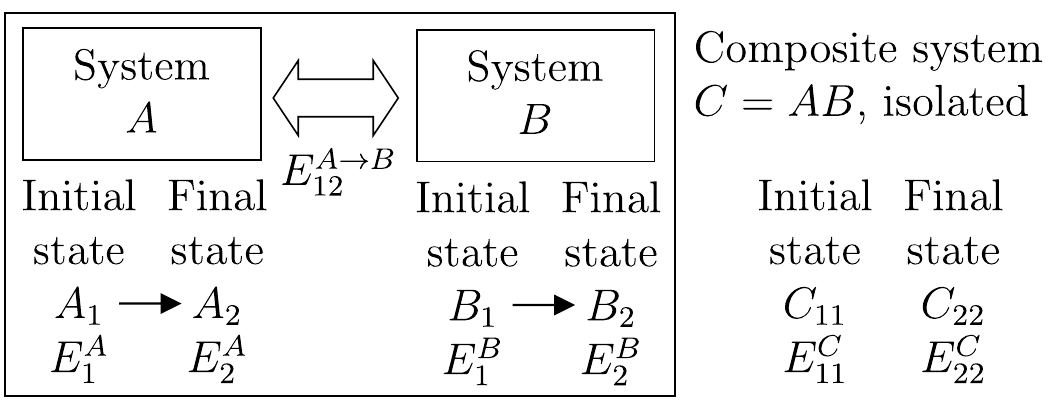}
				\caption{\label{EnergyTransfer}Energy can be exchanged between two systems $A$ and $B$ through interaction. In this example, the composite  system $C=AB$ is isolated.}
		\end{center}
	\end{figure}
	
	\section{Notation for energy exchange and the energy balance equation}
	
	From the conservation and additivity of energy, it follows that energy can be transferred (exchanged) between interacting systems.\ind{energy!exchange of} Using Figure \ref{EnergyTransfer} as an example, consider a system $C$ composed of subsystems $A$ and $B$ and a spontaneous process (without external effects) in which the state of $A$ changes from $A_1$ to $A_2$ and that of $B$ changes from $B_1$ to $B_2$. Since there are no external effects on $C$, the value of $C$'s energy remains unchanged, i.e., $E^C_{22}= E^C_{11}$. Due to the additivity of energy differences, this means that 
	$E^A_2- E^A_1+ E^B_2 -E^B_1=0$, or in other words, $E^A_2- E^A_1=-(E^B_2 -E^B_1)$. The change in energy of $A$ is equal and opposite to that of $B$. This justifies the notion of {\it energy exchange}\ind{exchange!of energy}, such that if the energy of $B$ increases, we say that $B$ receives energy from $A$, and consequently, the energy of $A$ decreases by an equal amount.

	\begin{figure}[!ht]
		\begin{center}
				\includegraphics[scale=0.45]{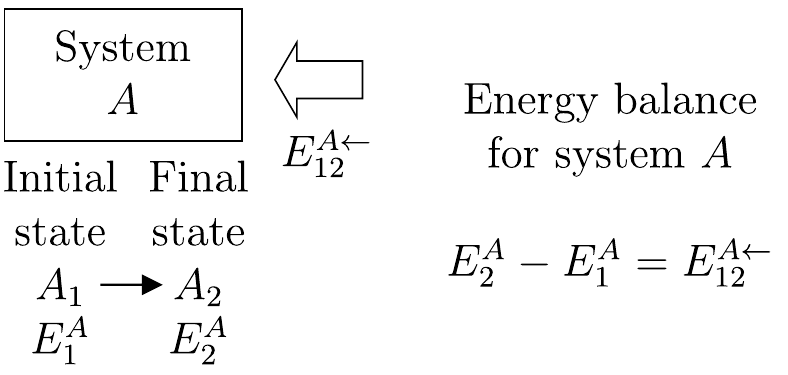}
				\caption{\label{EnergyBalance}Energy balance for system $A$ for a process in which the state of $A$ changes from $A_1$ at time $t_1$ to $A_2$ at time $t_2$, and the net effect of the interaction between $A$ and its environment includes an energy transfer $E^{A\gets}_{12}$ (positive if in the direction of the arrows, i.e., if received by $A$, negative if in the opposite direction).}
		\end{center}
	\end{figure}
	
	The notion of energy transferred from $A$ to $B$ is of practical importance, and it is convenient to adopt special notation to indicate it. The symbol used is
	\begin{equation} E^{A\rightarrow B}_{12} \end{equation}
	where the subscript 12 indicates that it is the quantity of energy transferred between time $t_1$ and time $t_2$ during the process in which the state of system $A$ changes from $A_1$ to $A_2$ of $A$ and that of its environment $B$ from $B_1$ to $B_2$. Thus, we have
	\begin{equation} E^A_2- E^A_1=-(E^B_2 -E^B_1)=- E^{A\rightarrow B}_{12} \end{equation}
	Equivalently, we can indicate the quantity of energy transferred from $B$ to $A$ using the symbol 
	$E^{A\leftarrow B}_{12}$, which leads to
	\begin{equation} E^A_2- E^A_1=-(E^B_2 -E^B_1)= E^{A\leftarrow B}_{12} \end{equation}
	Therefore, the two introduced symbols are not independent, and we have
	\begin{equation} E^{A\rightarrow B}_{12}=- E^{A\leftarrow B}_{12} \end{equation}
	
	If we consider that $B$ is the environment of $A$, we can simplify the notation by omitting the subscript $B$ and writing\footnote{When there is no ambiguity about the system under consideration, the notation can be further simplified by omitting the subscript $A$ and writing $E^{\rightarrow}_{12}=-E^{\leftarrow}_{12}$. When there is no ambiguity about the process being considered, it is possible to omit the subscript $12$ and write $E^{\rightarrow}=-E^{\leftarrow}$. It is worth noting that when $E^{A\leftarrow}$ takes a negative value, it means that system $A$ has released energy. For example, if $E^{A\leftarrow}=-5$ J, from Relation \ref{segniE}, we can infer that $E^{A\rightarrow}=5$ J, indicating that system $A$ has released 5 joules of energy.}
	\begin{equation} E^{A\rightarrow}_{12}=-E^{A\leftarrow}_{12}\label{segniE}\end{equation}
	 The relations just seen can be written in the form of the \textit{energy balance equation},.\ind{energy balance equation}\ind{balance equation!energy}
	\begin{equation} E^A_2 - E^A_1 = E^{A\leftarrow}_{12} \qquad\text{
	or, equivalently,}\qquad 
	 E^A_2 - E^A_1 = -E^{A\rightarrow}_{12} \end{equation}
	This important consequence of energy additivity and conservation  states that the change in energy $E^A_2-E^A_1$ of system $A$ following a process from $A_1$ to $A_2$ must be equal to the (net) amount of energy $E^{A\leftarrow}_{12}$ transferred into  system $A$ due to interactions with its environment.\ind{first law of thermodynamics!and energy balance}\footnote{In some texts, the energy balance equation is succinctly referred to as the 
		``first law of thermodynamics,'' so much so that the jargon 
		``writing the first law'' is used to mean 
		``writing the energy balance equation.'' We have already emphasized that the first law of thermodynamics leads to many important conclusions beyond the energy balance equation, and therefore, the jargon mentioned can be misleading in addition to being reductive with respect to the role and other implications of the first law.}
	
	It is useful to remember that the energy balance equation, like the first law itself from which it derives, is an expression of the laws of \textit{dynamics}. The time variable\ind{time} does not appear explicitly, but it is strongly present: recall that
	$A_1$ indicates the state of system $A$ at time $t_1$ and $A_2$ at time $t_2$. To make the dynamic nature of the equation more explicit, we can express it in the following alternative form:
	\begin{equation} \diff E^A/ \diff t= \dot{E}^{A\leftarrow} \end{equation}
	This form is obtained when $t_1=t$ and $t_2=t+\diff t$, with $ \diff E^A =E^A_{t+\diff t}-E^A_t$, and $\dot E^{A\leftarrow}= \delta E^{A\leftarrow}_{t,t+\diff t}/\diff t$ (energy per unit of time, power, transferred into $A$ from its environment).\ind{infinitesimals, notation}\ind{notation for infinitesimal quantities} We adopt the convention of using the prefix $\delta$ rather than $\diff$ for infinitesimal quantities of observables that are not properties. For example, the energy transferred in the time interval $t,t+\diff t$ is not a property because it depends on two instants of time, $t$ and $t+\diff t$. Therefore, it is indicated with $\delta E^{A\leftarrow}_{t,t+\diff t}$.
	
	To apply the energy balance equation, we will develop various concepts necessary to express, on one hand, the change in energy $E^A_2 - E^A_1$ as a function of the composition of  system $A$ and the nature of states $A_1$ and $A_2$, and on the other hand, the quantity of energy exchanged $E^{A\leftarrow}_{12}$ as a function of the types of interactions experienced by  system $A$ in the process from $A_1$ to $A_2$.
	
	For instance, later on, we will see that the interaction between a system $A$ and a weight in a weight process corresponds to an example of what we will call {\it work interaction}.\ind{interaction!work} The energy exchanged in an interaction of this type will be called work and denoted by the symbol $W^{A\leftarrow}$ instead of the generic $E^{A\leftarrow}$, so that for a weight process, we will write the energy balance as
	\begin{equation} E^A_2 - E^A_1 = W^{A\leftarrow}  \qquad\text{
			or, equivalently,}\qquad E^A_2 - E^A_1 = -W^{A\rightarrow}  \label{bilEW}\end{equation}

		\section{Steady and unsteady states. Equilibrium and nonequilibrium states}
	
	Since the number\ind{state!types of}\ind{types of states} of independent properties of a system is very large --- even for a single particle --- and since many properties can vary over an infinite range of values, the number of possible states of a system is infinite. The classification of states based on the type of temporal evolution of each state highlights some important aspects of thermodynamics. States of a system can be classified into four types: unsteady, steady, nonequilibrium, and equilibrium states. Equilibrium states can be further classified into three subtypes: unstable, metastable, and stable.

	If a system is subject to interactions with other systems that have non-zero effects, its state usually changes over time, and it is called a \textit{unsteady state}.\ind{state!unsteady} However, in practical engineering, interactions with other systems are often regulated and balanced in such a way that their net effects on the state of the system are zero. In this case, the state is called \textit{steady} or, sometimes,  \textit{stationary}.
	
	Another important way to classify states is based on the behavior dictated solely by the internal dynamics of the system, i.e., the behavior the system would exhibit if all interactions of the system with other systems were 
	``turned off'' or 
	``frozen,'' and only the internal dynamics remained active. If the state changes over time due to the internal dynamics alone, i.e., spontaneously, it is called \textit{nonequilibrium}.\ind{state!nonequilibrium} If it does not change, it is called \textit{equilibrium}.\ind{state!equilibrium}\ind{equilibrium}
	
	There are various types of equilibrium states. An \textit{unstable equilibrium} is a state that can be induced to spontaneously evolve toward  different often distant) state by some vanishingly small and brief interaction --- an infinitesimal perturbation. Such a perturbation has only a temporary, negligible effect on the environment, leaving no permanent net external change. In contrast, a  \textit{metastable equilibrium} state is stable under small perturbations, but can be changed without leaving permanent effects on the environment by means of larger perturbations.
	
	\section{Stable equilibrium states}
	
	Lastly, a \textit{stable equilibrium state} (abbreviated SES) is an equilibrium state that \textit{can only be modified through interactions that leave nonzero net effects on the  environment of the system}. In other words, a stable equilibrium state cannot be altered without leaving net effects in the system's environment, i.e., without changing the state of the environment.
	
	We will see that starting from a non-stable equilibrium state (i.e., either a nonequilibrium state or an  unstable or metastable equilibrium state), it is always possible for the system to evolve by means of a weight process that results in the lifting of a weight and does not leave any other changes in the state of the environment.

		\section{Statement of the Second Law}\label{SecondLaw}

	We shall adopt as statement of the second law a refinement of the one due to  Hatsopoulos and Keenan \cite{HatsopoulosKeenan1965} because from this statement we can derive all other traditional statements, namely, those by Kelvin-Planck, Clausius, and Carath\'eodory.

	We have already seen that the number of possible states of a system is infinite. Among all these states, consider the subset of all the states that share a given set of values of  amounts of constituents \lst{n}{r}, parameters \lst{\beta}{s}, and energy $E$. Also the number of states in this subset is usually infinite. The \textit{second law of thermodynamics}\ind{second law of thermodynamics!Hatsopoulos-Keenan statement}\ind{Hatsopoulos-Keenan statement of the second law} consist of two assertions: the first is that \textit{in the subset of states of a system compatible with given values of $\bn$ and $\bbeta$, there is always one and only one stable equilibrium state for each value of the energy $E$}.\footnote{The word ``compatible'' can be omitted in the absence of reaction mechanisms (such as chemical reactions capable of spontaneously altering the amounts of constituents), internal constraints (such as movable internal walls capable of spontaneously altering the parameters), and other transition restrictions (such as selection rules among quantized energy levels or the constraint of strictly unitary quantum evolution). To fix ideas and for simplicity, we proceed by assuming that it can be omitted, so that the first part of the second-law statement reduces to the simple assertion that \textit{among all the states that share a given set of values of energy, amounts of constituents, and parameters, one and only one is a stable equilibrium state}. For the discussion of systems with chemical reaction mechanisms,  internal constraints, and transition restrictions see, e.g., \cite{GyftopoulosBeretta1991,HatsopoulosKeenan1965}.}
	
	It is important to contrast this assertion with its analogous in the domain of mechanics: for given values of $\bn$ and $\bbeta$, the set of states considered in \textit{mechanics} contains only one stable equilibrium state, which is the one with the minimum energy, $E_{\rm min}(\bn,\bbeta)$. By contrast, for given values of $\bn$ and $\bbeta$, \textit{thermodynamics} considers a much broader set of states, where there is one (and only one) stable equilibrium state for each value of energy $E$.\footnote{The representation on the energy-entropy plane discussed in Section \ref{ESdiagram} will illustrate  also graphically the significance of this important statement.} It is in this sense\ind{thermodynamics vs mechanics}\ind{mechanics vs thermodynamics} that thermodynamics is an extension of mechanics.\footnote{The so-called paradoxes of thermodynamics, often mentioned in popular science, history and philosophy of science books, originated from the mistaken belief, still prevalent, that the conclusions of mechanics and those of thermodynamics should be compared, assuming that both theories contemplate the same set of states for a given system. In this case, the paradox would be that for given values of $\bn$ and $\bbeta$, mechanics has only one stable equilibrium state, while thermodynamics has infinitely many, one for each value of energy. The paradox is resolved by admitting that the set of states considered by thermodynamics contains, yes, but only as a subset, the states considered by mechanics. Thus, both statements are valid, each in its own context.}

	The second assertion of the  statement of the second law is that \textit{starting from any initial state of the system, it is always possible, through a reversible weight process, to reach a stable equilibrium state with values of $\bn$ and $\bbeta$ arbitrarily fixed among those compatible with the initial state.}

	\section{Proof of the Kelvin-Planck statement of impossibility of a perpetual motion machine of the second kind}

\begin{figure}[!ht]
		\begin{center}
				\includegraphics[scale=0.45]{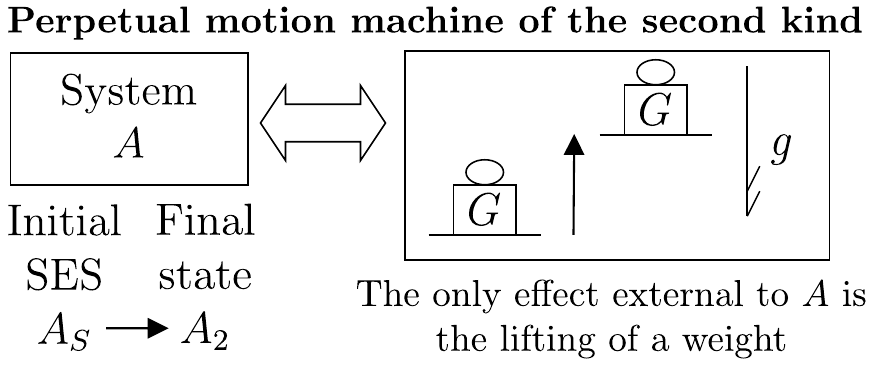}
				\caption{\label{PMM2} Perpetual motion of the second kind refers to the possibility of extracting mechanical energy (lifting a weight) without any other effects (weight process) from a system that initially is in a stable equilibrium state.}
		\end{center}
	\end{figure}

	The first important consequence of the second law is that, in general, not all of a system's energy can be transferred to a weight through a weight process, i.e., not all states of a system can reach via a weight process a state  with the minimum energy for the same or compatible values of  $\bn$ and $\bbeta$. In particular, if a system is in a stable equilibrium state, it cannot transfer energy to a weight through a weight process that does not alter the set of compatible values of $\bn$ and $\bbeta$. This statement is also known as the \textit{impossibility of perpetual motion of the second kind}\ind{perpetual motion of the second kind} (Figure \ref{PMM2}) and is referred to as the \textit{Kelvin-Planck statement of the second law of thermodynamics} (1897).\ind{Kelvin, Kelvin-Planck statement}\ind{Planck, Kelvin-Planck statement}\ind{second law of thermodynamics!Kelvin-Planck statement}\ind{Clausius, Clausius statement}\ind{second law of thermodynamics!Clausius statement}\footnote{Another 
		``historical'' statement of the second law is Clausius' statement (1850). We will see later (Section \ref{ClausiusStatement}), after defining temperature, that also Clausius' statement is a consequence (theorem) of the more general second-law statement we have adopted.}
	
	Since the impossibility of perpetual motion of the second kind leads to a fundamental inequality for all that follows, it is worth outlining the proof. We proceed by contradiction; suppose that a perpetual motion machine of the second kind is possible: it would be a system $A$ (Figure \ref{PMM2}) initially in a stable equilibrium state $A_S$ that can lift a weight as a result of a weight process; after the energy is transferred to the weight, the system $A$ would end up in a state $A_2$ with lower energy than the initial state. Now, suppose that the system $A$ is made up of at least two parts;\footnote{This is a more than acceptable simplification that makes the proof much simpler than the complete one \cite[Sec.4.5]{GyftopoulosBeretta1991}.} then it is possible to create a second process in which the weight returns all the energy it received to system $A$, causing the two parts of $A$ to move relative to each other. The new state $A_3$ of system $A$, thus determined, is certainly different from the initial state $A_S$ because the two parts of $A$ are in relative motion. We therefore succeeded in changing the state of system $A$ from $A_S$ to $A_3\ne A_S$ while leaving no net effects in the environment. Since this contradicts the definition of stable equilibrium, we conclude that the assumption is absurd and, therefore, a perpetual motion machine of the second kind is impossible.
	
	From what has been proven, it follows that when a system $A$ is initially in a stable equilibrium state with energy  $E^A_S$, in a weight process it can only reach  states with higher energy $E^A_2$. Using the notation introduced earlier,
	\begin{equation} E^A_2 - E^A_S = -W^{A\ar} > 0 \label{impPM2}\end{equation}

	\section{Adiabatic availability: definition}\label{AvailabilityAdiabatic}
	
	The historical formulations of the principles of thermodynamics originated from a careful examination of the technical question that, with the terminology we have developed so far, can be formulated as follows: 
	``How much of the energy $E_1$ of a system $A$ in a given state $A_1$ can be transferred to a weight through a weight process?''

	The answer identifies, for each system $A$ and any state $A_1$, a property called \textit{adiabatic availability},\ind{adiabatic availability} denoted by the symbol $\Psi$. It consists of evaluating, for system $A$ in state $A_1$, the maximum amount of energy $W^{A\rightarrow G}_{\rm max}$, denoted as $\Psi_1$, that can be transferred to a weight in a weight process without altering the (set of compatible) values of $\bn$ and $\bbeta$.

	\begin{figure}[!ht]
		\begin{center}
			\includegraphics[scale=0.45]{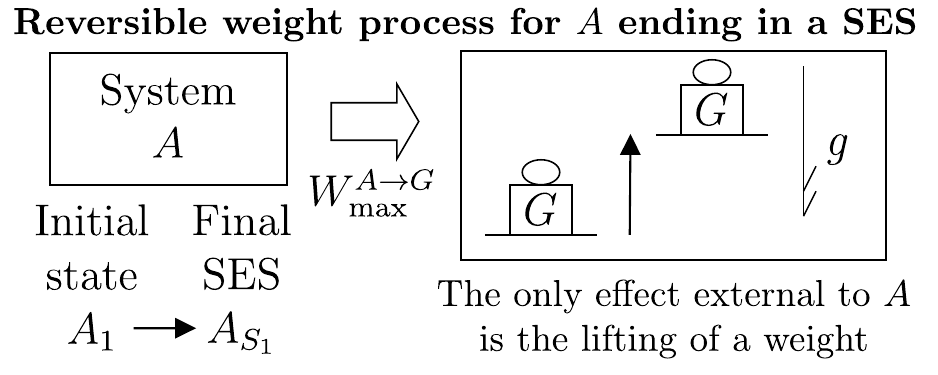}
			\caption{\label{AdiabaticAvailability} Adiabatic availability $\Psi^A_1$ measures the maximum amount of energy that can be transferred out of system $A$ initially in state $A_1$ by means of a weight process. Such maximum is obtained when the weight process for $A$ is reversible and the system ends in a stable equilibrium state. The final stable equilibrium state $A_{S_1}$ is uniquely determined by the initial state $A_1$.}		
		\end{center}
	\end{figure}

	The measurement procedure that defines it is sketched in Figure \ref{AdiabaticAvailability}. It can be proven that $W^{A\rightarrow G}_{\rm max}$ obtains when the weight process for $A$ is reversible and changes state $A_1$ into a stable equilibrium state $A_{S_1}$ with values of  $\bn$ and $\bbeta$ compatible with state $A_1$. The existence of such process is guaranteed directly by the second law (second assertion). It is also proven that state $A_{S_1}$ is uniquely determined by $A_1$, being the only stable equilibrium state with the compatible values of amounts and parameters that can be reached from $A_1$ through a reversible weight process. 
	
	From the energy balance we have
	\begin{equation} \Psi_1 = E_1 - E_{S_1} \label{adiabatic}\end{equation}
	Clearly, adiabatic availability has the same dimensions as energy and is measured in joules, J, in the International System of Units.  
	
	From the impossibility of perpetual motion of the second kind (Eq.~\ref{impPM2}), it immediately follows that for any stable equilibrium state, the adiabatic availability is zero. If the state is not  stable equilibrium, the adiabatic availability is nonzero and certainly positive.
	
	Adiabatic availability $\Psi$ has another important utility: it provides an operational criterion to ascertain the reversibility\ind{process!reversible} of a weight process.  In fact, it is shown that a given weight process for system $A$ from state $A_1$ to state $A_2$ is reversible if and only if \begin{equation} E_2 - \Psi_2 = E_1 - \Psi_1 \label{mechrev}\end{equation}while it is irreversible if and only if \begin{equation} E_2 - \Psi_2 > E_1 - \Psi_1\label{mechirr} \end{equation}These results determine the \textit{direction} in which the weight process is possible. If \begin{equation} E_2 - \Psi_2 < E_1 - \Psi_1 \end{equation}then the weight process in the direction from $A_1$ to $A_2$ is impossible. In this case, the first law guarantees that the process is possible in the opposite direction.
	
	From these results, the important connection between irreversibility and the loss of the ability to produce useful effects emerges. It can be seen that for weight processes, the irreversibility of the process increases the value of $E - \Psi$, which is the difference between energy and adiabatic availability. This is the portion of energy in the system that is \textit{not adiabatically available}, meaning it cannot be transferred to a weight in a weight process. The fact that $E - \Psi$ cannot decrease in any weight process leads to the conclusion that the unavailable portion of a system's energy cannot decrease in a weight process: it remains unchanged if the process is reversible and increases if the process is irreversible.
	
	Although adiabatic availability allows for such general and important conclusions and provides an operational and quantitative criterion to verify the reversibility of weight processes, it has a 
	``defect'' that makes it unsuitable for practical use: it is \textit{not} an additive property. This can be easily seen from a simple example, which we will consider after introducing the notion of mutual equilibrium.

	\section{Mutual stable equilibrium: definition}
	
	Two systems are said to be \textit{in mutual stable equilibrium} (MSE), or simply \textit{in mutual equilibrium},\ind{equilibrium!mutual} if their respective states are such that the composite system is in a stable equilibrium state.\footnote{It should be noted that a necessary condition for the mutual equilibrium of two systems is that each is in a stable equilibrium state. However, vice versa is not sufficient; it is not enough for two systems to be in stable equilibrium states for them to be in mutual equilibrium.}
	
	Now, consider two systems, $A$ and $B$, that are not in mutual equilibrium, even though each is in a stable equilibrium state. Taken individually, each of the two systems has zero adiabatic availability. However, the composite system, not being in a stable equilibrium state, has nonzero adiabatic availability. Therefore, the adiabatic availability of the composite system is not equal to the sum of the adiabatic availabilities of the subsystems.
	
	It is possible, however, to define, for each system, a monotonic function of $E - \Psi$ in a way that the new property resulting from it is additive. We will call this important property \textit{entropy}\ind{entropy} and denote it by the symbol $S$.\footnote{From a mathematical point of view, the definition of $S$ is already unambiguously, but implicitly, defined by the condition that for every system $A$ (and $B$) and for every pair of states $A_1$ and $A_2$ (and $B_1$ and $B_2$), there exists a function $S(Y)$, where $Y$ denotes $E - \Psi$, such that \begin{equation} S^A(Y^A_2) > S^A(Y^A_1) \quad \text{if and only if} \quad Y^A_2 > Y^A_1 \end{equation}\begin{equation} S^{AB}(Y^{AB}_2) - S^{AB}(Y^{AB}_1) = S^A(Y^A_2) - S^A(Y^A_1) + S^B(Y^B_2) - S^B(Y^B_1) \end{equation}For a mathematical formulation along these lines see \cite{LiebYngvason2014}. However, the explicit definition we propose in the next sections is pedagogically preferable as it is much more concrete, even though it requires some non-trivial reasoning.} In order to arrive at a clear and explicit operational definition of entropy, next we introduce the notion of  thermal reservoirs and a measurement procedure that characterizes them by direct comparison. 
	
	\section{Thermal reservoir: definition}

	We call a system $R$ that approximately satisfies the following  limiting condition a \textit{thermal reservoir}\ind{thermal reservoir} or simply a \textit{reservoir}: \textit{In any of its stable equilibrium states with given values of  amounts of constituents  and parameters it is in mutual stable equilibrium  with a  given system $C$ in a fixed given state $C_R$.}

	\begin{figure}[!ht]
		\begin{center}
				\includegraphics[scale=0.45]{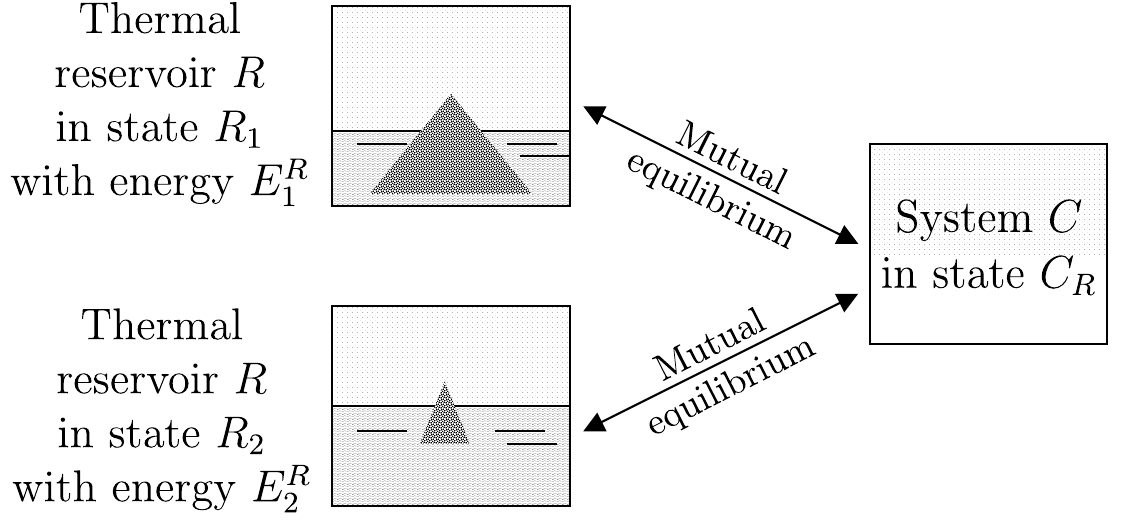}
				\caption{\label{Reservoir}A practical approximation of a thermal reservoir can be obtained with $\rm H_2 O$ at the triple point. In any state, $R_1$, $R_2$, \dots, in which the solid, liquid, and vapor phases coexist in stable equilibrium, even though they have different energy values, the reservoir $R$ is always in mutual equilibrium with a system $C$ in state $C_R$, also containing $\rm H_2 O$ at the triple point.}
		\end{center}
	\end{figure}

	A ``practical'' reservoir can be easily created in any laboratory, as illustrated in Figure \ref{Reservoir}, by placing $\rm H_2 O$ in a container under conditions (referred to as the ``triple point'')\ind{triple point} such that some is in the solid phase (ice), some is in the liquid phase (water), and some is in the vapor phase. In the range of  stable equilibrium states in which the three phases (solid, liquid, and vapor) are all present in finite and not microscopic amounts, this system behaves as a reservoir.
	
The concept of a thermal reservoir is an idealized limiting abstraction, providing a pedagogical framework that simplifies both practical modeling and theoretical development. But it is important to note that its defining condition provides only an approximate description of physical reality, that holds with good accuracy  for systems containing a  large number of particles (exceedingly good for very large numbers, such as the triple-point model), but fails for systems with few particles.\footnote{	As a consequence, when a thermal reservoir $R$ is employed in the measurement procedure used to define the entropy difference between two states $A_1$ and $A_2$ 
		of a system $A$ (see Sec.~\ref{entropy}), the procedure requires measuring the energy change of 
		$R$ in a reversible weight process of the composite system $AR$, during which the state of 
		$A$ changes from $A_1$ to $A_2$. If system $A$
		contains only a few particles while $R$ can only be realized with a  large number of particles, the corresponding energy change of 		$R$ is exceedingly small and, in practice, difficult to detect.
		For this reason, in Ref.~\cite{BerettaZanchini2019} we developed a more elaborate --- though pedagogically less practical --- alternative formulation that avoids the use of thermal reservoirs as ``entropymeters'' and is better aligned with recent technological advances in the domain of small systems; see, e.g., Ref.~\cite{PekolaKarimi2024}.} Moreover, its strict validity  would constitute a violation of the second law.\footnote{To see this, consider two identical reservoirs $R'$ and $R''$, initially in the same state $R'_1=R''_1$. Since they are both in MSE with $C$ in state $C_R$, they are in MSE with one another. Hence, state $R'_1R''_1$ of the composite system $R'R''$ is a stable equilibrium state with energy $E^{R'R''}_{11}=E_1^{R'}+E_1^{R''}$. Now, consider states $R'_2$ and $R''_2$ such that $E_2^{R'}=E_1^{R'}+\Delta E$ and $E_2^{R''}=E_1^{R''}-\Delta E$. State $R'_2R''_2$ is another  stable equilibrium state for the composite system $R'R''$. But $E^{R'R''}_{22}=E^{R'R''}_{11}$, so we would have two stable equilibrium states with the same values of energy, amounts, and parameters, in direct violation of the uniqueness asserted by the second law.} 

	\section{Available energy with respect to a thermal reservoir}\label{AvailableEnergy}

\begin{figure}[!ht]
		\begin{center}
			\includegraphics[scale=0.45]{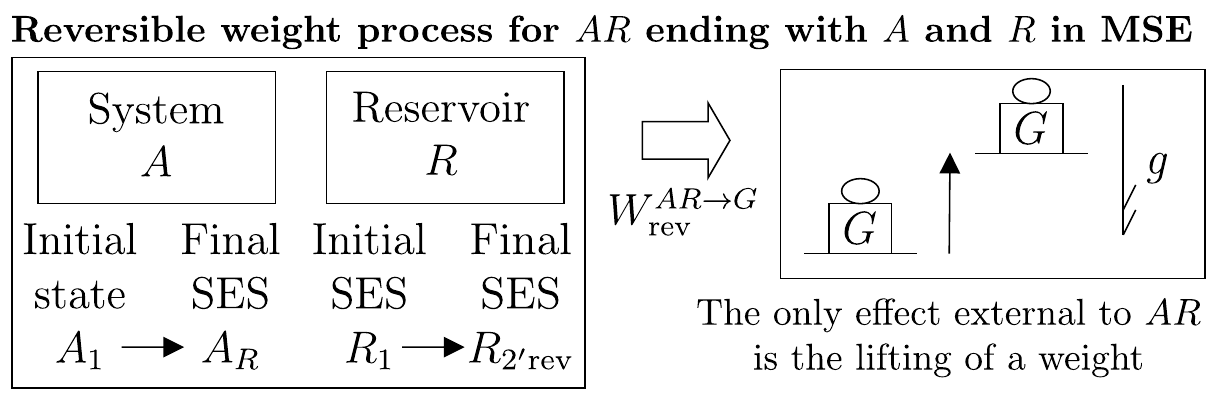}
			\caption{\label{RevWeightProcessForAR} Available energy $(\Omega^R)_1^A$ with respect to thermal reservoir $R$ measures the maximum amount of energy that can be transferred out of system $A$ initially in state $A_1$ by means of a weight process for the composite system  $AR$. Such maximum is obtained independently of the initial stable equilibrium state $R_1$ of the reservoir when the weight process for $AR$ is reversible and the composite system $AR$ ends in a stable equilibrium state, i.e., $A$ and $R$ end in mutual stable equilibrium.}
		\end{center}
	\end{figure}

	The second assertion of the second law guarantees that starting from any initial state $A_1$ of any system $A$ that can interact with a reservoir $R$ initially in state $R_1$, it is always possible, through a reversible weight process for the composite system $AR$ (see Figure \ref{RevWeightProcessForAR}), to reach a stable equilibrium state for $AR$ with values of $\bn$ and $\bbeta$ arbitrarily fixed among those compatible with the initial states. In the final state of $AR$, the system and the reservoir are in mutual equilibrium. The final state $A_R$ of system  $A$ is uniquely determined by the chosen reservoir $R$ and the chosen values of $\bn$ and $\bbeta$. The final state $R_{2'\text{rev}}$ of $R$ is uniquely determined by the chosen initial states $A_1$ and $R_1$  and the chosen values of $\bn$ and $\bbeta$. 
	
	Note that the energy transferred to the weight in this process, $W^{AR\to G}_{\rm rev}$,  corresponds to the adiabatic availability $\Psi^{AR}_{11}$ of the composite system $AR$ in state $A_1R_1$. It can be shown that it is independent of the initial state $R_1$ of the reservoir, i.e., it depends only the state $A_1$ of system $A$ and the chosen reservoir $R$.

	Therefore, we can view the process just described (Figure \ref{RevWeightProcessForAR}) as the measurement procedure that, with respect to a chosen reservoir $R$, defines a property of system $A$ that we call \textit{available energy with respect to thermal reservoir $R$}, denoted by the symbol  $\Omega^R$. For state $A_1$ of system $A$ and the chosen reservoir $R$, regardless of the initial state $R_1$, we denote its value by $ (\Omega^R)^A_1=\Psi^{AR}_{11}$ or simply $\Omega^R_1$. Like adiabatic availability, it has the same dimensions as energy and is measured in joules, J, in the International System of Units.
	
	Unlike adiabatic availability, however, the conditions that define thermal reservoirs make $\Omega^R$ an additive property (with respect to a fixed $R$), in the sense that $(\Omega^R)^{AB}_1=(\Omega^R)^A_1+(\Omega^R)^B_1$ for all states $A_1$ and $B_1$ of all systems $A$ and $B$. From the results already seen for adiabatic availability, it follows that the available energy with respect to reservoir $R$ takes nonzero and positive values for all states of system $A$ except for the stable equilibrium state $A_R$ in which system $A$ is in mutual equilibrium with reservoir $R$, in which case it is zero, i.e., $(\Omega^R)^A_R = 0$.
	
	Like adiabatic availability, available energy also gives rise to a quantitative criterion to determine whether a given weight process is reversible or not: the weight process for system $A$ from state $A_1$ to state $A_2$ is reversible if and only if \begin{equation} E^A_2-(\Omega^R)^A_2 = E^A_1-(\Omega^R)^A_1 \label{mechrevR}\end{equation}
	while it is irreversible if and only if \begin{equation} E^A_2-(\Omega^R)^A_2 > E^A_1-(\Omega^R)^A_1 \label{mechirrR}\end{equation}
	and this holds for any choice of the reservoir $R$ used to measure $\Omega^R$ [as long as, of course, the same $R$ is used for $(\Omega^R)^A_1$ and $(\Omega^R)^A_2$].
	
	Like the \textit{adiabatically unavailable energy}, $E-\Psi$, the \textit{unavailable energy with respect to reservoir $R$}, $E-\Omega^R$, is conserved, meaning it remains constant over time, in reversible weight processes;\ind{process!reversible} it increases if the weight process is irreversible.
	
Like energy and adiabatic availability, also available energy is property defined for \textit{all} states of a system, including stationary and non-stationary states, equilibrium and nonequilibrium states, not just for stable equilibrium states.
	
	In Section \ref{FirstLaw}, we saw that energy $E$ is an additive property that can be transferred between systems through interactions. These characteristics make it ideal for the analysis of processes in complex systems, as the system can be schematized as composed of various subsystems, and energy is the sum of the energies of the subsystems. Adiabatic availability $\Psi$ is not suitable for this purpose because it is not additive. Available energy $\Omega^R$ is additive, but it depends on the choice of a specific reservoir, so it does not measure a property of the system \textit{itself} but of the composite system, system-reservoir. The next step will be to define a characteristic of reservoirs that will finally allow us to define the property of entropy, which, as we will see, is additive and independent of the reservoir chosen to measure it.

\begin{figure}[!ht]
	\begin{center}
			\includegraphics[scale=0.45]{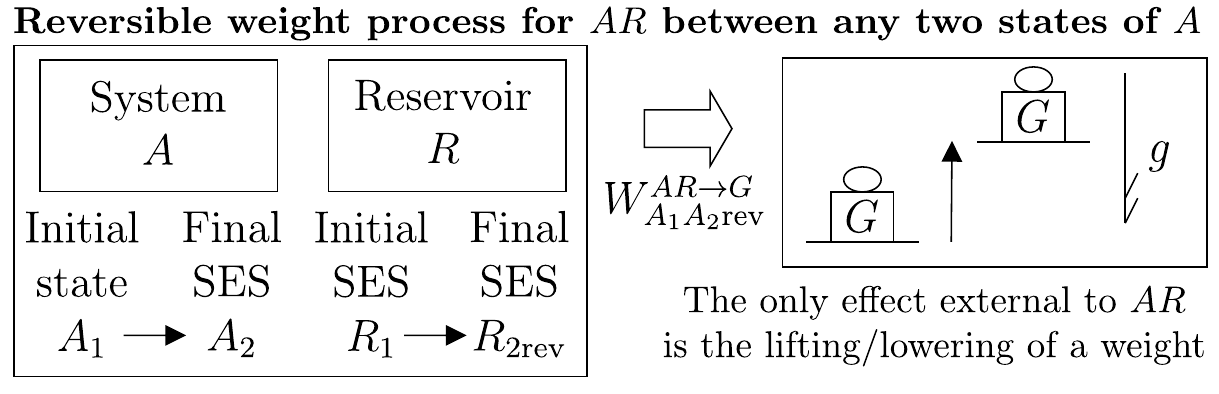}
			\caption{\label{RevWeightProcessAuxiliary}If system $A$ can interact with a thermal reservoir $R$, any pair of states $A_1$ and $A_2$ can be interconnected by means of reversible weight process for the composite system $AR$.} 
	\end{center}
\end{figure}

A direct consequence of the results stated in this section, is that starting from any initial state $A_1$ of any system $A$ that can interact with a reservoir $R$ (initially in stable equilibrium state $R_1$), it is always possible, through a reversible weight process for the composite system $AR$ (see Figure \ref{RevWeightProcessAuxiliary}), to reach a final state for $AR$ with values of $\bn$ and $\bbeta$ arbitrarily fixed among those compatible with the initial states and with system $A$ in an arbitrarily chosen final state $A_2$. In this case, in general, the system and the reservoir do not reach mutual equilibrium. The final state $R_{2\text{rev}}$ of $R$ is uniquely determined by the chosen initial and final states  $A_1$ and $A_2$ of the system, the initial state $R_1$,  and the chosen values of $\bn$ and $\bbeta$. 

	It is easy to show that the energy $W^{AR\to G}_{A_1A_2\rm rev}$ transferred to the weight in this process is equal to the  difference $(\Omega^R)^A_1-(\Omega^R)^A_2$ in the adiabatic availabilities of the composite system $AR$, i.e., of the available energies of system $A$ with respect to reservoir $R$.  Therefore, the energy balance for  system $AR$, 
		\begin{equation} E^{AR}_2-E^{AR}_1=  -W^{AR\to G}_{A_1A_2\rm rev} \end{equation}
using energy additivity, can be rewritten as	
		\begin{equation} (E^R_{2\rm rev}-E^R_1) +(E^A_2-E^A_1)=  -(\Omega^R)^A_1+(\Omega^R)^A_2 \end{equation}
or, equivalently, adding subscripts summarizing the process details described in Figure \ref{RevWeightProcessAuxiliary},
	\begin{equation}\label{DeltaER} (E^R_{2\rm rev}-E^R_1)_{A_1R_1\underset{\text{w,rev}}{\Longrightarrow} A_2R_{2\rm rev}} = [ E^A_1-(\Omega^R)^A_1]-[E^A_2-(\Omega^R)^A_2] \end{equation}
The setup of Figure \ref{RevWeightProcessAuxiliary} is important because it supports the measurement procedures that we discuss next and that lead to the general operational definition of entropy.

	\section{Temperature of a thermal reservoir: definition}
	
\begin{figure}[!ht]
		\begin{center}
			\includegraphics[scale=0.45]{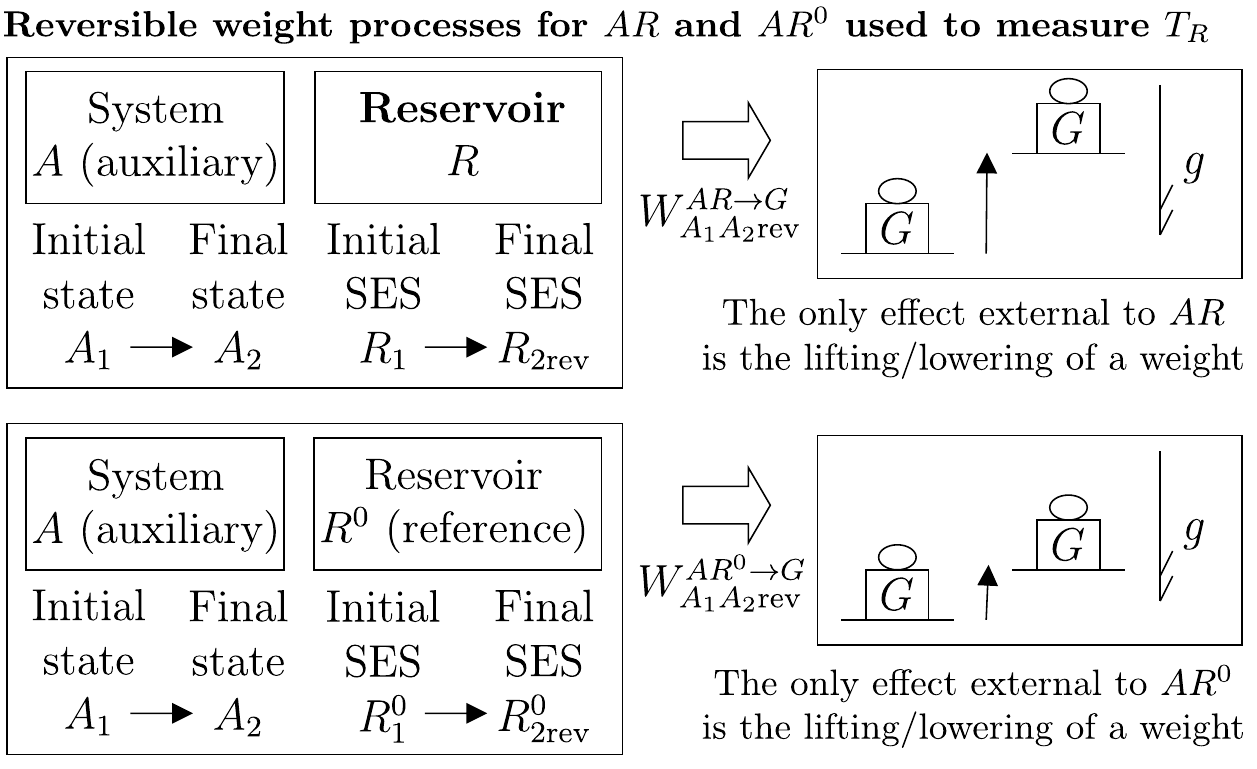}
			\caption{\label{TRdefinition}Visualization of the measurement procedure defining the temperature $T_R$ of thermal reservoir $R$ by comparison with the reference thermal reservoir $R^0$. System $A$ and its states $A_1$ and $A_2$ are chosen arbitrarily and  play only an auxiliary role in the procedure, by determining uniquely the final stable equilibrium states $R_{2\rm rev}$ and $R^0_{2\rm rev}$ of the two reservoirs. The objective  of the procedure is to measure the  energy changes of $R$ and $R^0$ in these two reversible weight processes and compute the dimensionless ratio $(E_{2\rm rev}^R-E_1^R)/(E_{2\rm rev}^{R^0}-E_1^{R^0})$.}
		\end{center}
	\end{figure}

	We  define the \textit{temperature of a reservoir}\ind{temperature!of a reservoir}, denoted as $T_R$, through the following measurement procedure, sketched in Figure \ref{TRdefinition}. 

First, choose a reference reservoir $R^0$, an arbitrary stable equilibrium state $R_1^0$, an arbitrary auxiliary system $A$, and two arbitrary states $A_1$ and $A_2$. Consider a reversible weight process for $AR^0$ in which $A$ goes from $A_1$ to $A_2$, and measure the change in energy of $R^0$, $(E^{R^0}_{2\rm rev}-E^{R^0}_1)_{A_1R^0_1\underset{\text{w,rev}}{\Longrightarrow} A_2R^0_{2\rm rev}}$. Then, take the reservoir $R$ to be measured in an arbitrary stable equilibrium state $R_1$ and consider a reversible weight process for $AR$ in which $A$ goes from $A_1$ to $A_2$, and measure the change in energy of $R$, $(E^{R}_{2\rm rev}-E^{R}_1)_{A_1R_1\underset{\text{w,rev}}{\Longrightarrow} A_2R_{2\rm rev}}$. Finally, calculate
\begin{equation}\label{TR1} T_R=T_{R^0}\, \frac{ (E^{R}_{2\rm rev}-E^{R}_1)_{A_1R_1\underset{\text{w,rev}}{\Longrightarrow} A_2R_{2\rm rev}}}{(E^{R^0}_{2\rm rev}-E^{R^0}_1)_{A_1R^0_1\underset{\text{w,rev}}{\Longrightarrow} A_2R^0_{2\rm rev}}} \end{equation}
where $ T_{R^0}$ is an arbitrarily assigned reference value for the reference reservoir $R^0$. 

It should be noted that $T_{R^0}$ is chosen once and for all. 	The reservoir realized with water at the triple point (Figure \ref{Reservoir}) can be chosen as the reference reservoir $R^0$ to be used in the measurement procedure just defined. It is a secondary standard reservoir easily realized in all laboratories, to which the reference value $T_{R^0}=273.16$ K is conventionally assigned, where K stands for the {\it kelvin},\ind{kelvin, unit of measure} the unit of temperature in the International System. Note that the measurement procedure defining $T_R$ implies a comparison between reservoir $R$ and a reference reservoir $R^0$, and therefore, $T_R$ is a fundamental property that cannot be expressed in terms of other fundamental properties of mechanics (length, time, and mass) or electromagnetism (current).

	It can be shown that the value $T_R$ defined in this way is independent of  the choice of the auxiliary system $A$ and of its states $A_1$ and $A_2$, i.e.,  the role of  system $A$ in the procedure is purely auxiliary. Moreover, $T_R$ is constant for a given reservoir. This means that the measurement procedure defined here always results in the same value, regardless of the initial stable equilibrium state $R_1$ of reservoir $R$, and the initial stable equilibrium state $R^0_1$ of the reference reservoir $R^0$.  Finally, it can be shown that two reservoirs $R$ and $R'$ in mutual equilibrium have the same temperature, $T_R=T_{R'}$.
	
	Equation \ref{TR1} can be rewritten using Eq.~\ref{DeltaER} as
	\begin{equation}\label{TR2} T_R=T_{R^0}\, \frac{ [ E^A_1-(\Omega^R)^A_1]-[E^A_2-(\Omega^R)^A_2] }{[ E^A_1-(\Omega^{R^0})^A_1]-[E^A_2-(\Omega^{R^0})^A_2] } \end{equation}
	 which shows that, once the auxiliary system $A$ and the two states $A_1$ and $A_2$ are chosen, an alternative to the direct measurement procedure outlined in Figure \ref{TRdefinition} is to use  the measurement procedures  previously defined for energy and for available energy with respect to $R$ and substitute the results in Eq.~\ref{TR2}. 
	
	It is important to note that what we gave in this section is not the definition of temperature for systems that are not reservoirs. 
	 That will be done later (Section \ref{MutualEquilibriumConditions}) and is entirely different from the one just described, although when applied to a reservoir, it, of course, will provide the same value as $T_R$ defined above.

	We have finally defined everything we need to define entropy.

	\section{Entropy: definition (valid also for nonequilibrium states)}\label{entropy}

\begin{figure}[!ht]
		\begin{center}
			\includegraphics[scale=0.45]{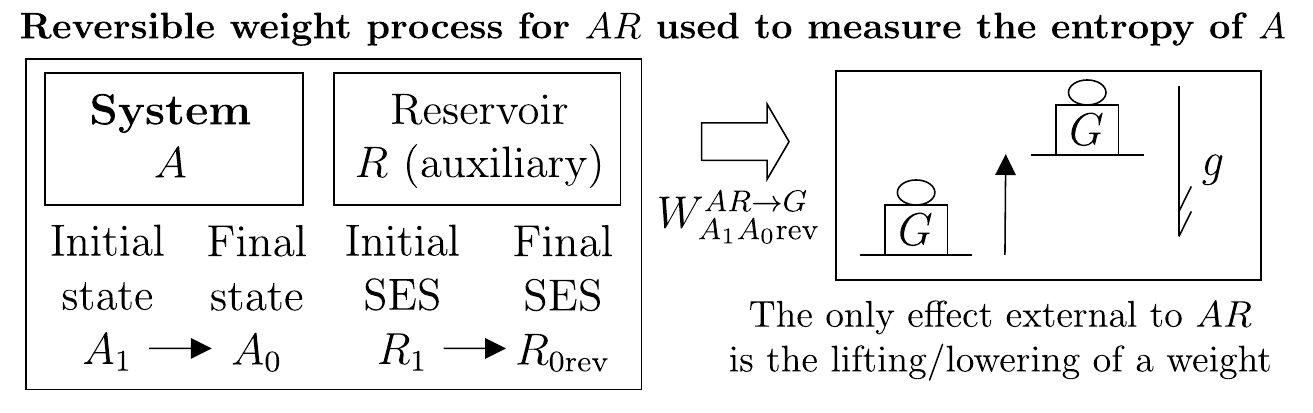}
			\caption{\label{EntropyDefinition}Visualization of the measurement procedure defining the entropy of system $A$ with respect to an arbitrarily chosen reference state $A^0$. Reservoir $R$ and its initial state $R_1$  are chosen arbitrarily and play only an auxiliary role in the procedure, by determining uniquely the final state $R_{0\rm rev}$ of the reservoir. The objective of the procedure is to measure the  energy change of $R$  in the  reversible weight process for $AR$ in order to compute the ratio $(E_{0\rm rev}^R-E_1^R)/T_R$.}
		\end{center}
	\end{figure}

	The entropy\ind{entropy!definition} $S_1$ of any system $A$ in state $A_1$ is defined by the following measurement procedure, sketched in Figure \ref{EntropyDefinition}. 
	
	First, choose a reference state $A_0$ (of system $A$) to which you assign the reference value $S_0$. Measure the energy $E_0$ of this state using the corresponding procedure. Second, select a thermal reservoir $R$ and measure its temperature $T_R$ using the procedure discussed earlier. Consider a reversible weight process for $AR$ in which $A$ goes from $A_1$ to $A_0$, and measure the change in energy of $R$, $(E^{R}_{0\rm rev}-E^{R}_1)_{A_1R_1\underset{\text{w,rev}}{\Longrightarrow} A_0R_{2\rm rev}}$. Finally calculate
		\begin{equation} S_1 = S_0 + \frac{(E^{R}_{0\rm rev}-E^{R}_1)_{A_1R_1\underset{\text{w,rev}}{\Longrightarrow} A_0R_{2\rm rev}}}{T_R} \label{defentropy1}\end{equation}
		The dimensions of $S_1$ and $S_0$ are the same as those of $E/T_R$, which are [energy]/[temperature], and the International System unit of measurement for entropy is the joule per kelvin, J/K.

		Equation \ref{defentropy1} can be rewritten using Eq.~\ref{DeltaER} as
	\begin{equation}\label{defentropy2}  S_1 = S_0 + \frac{(E_1-\Omega^R_1 ) - (E_0- \Omega^R_0)}{T_R} \end{equation}
	which shows that, once the auxiliary reservoir $R$ has been chosen and its temperature $T_R$ measured, an alternative to the direct measurement procedure outlined in Figure \ref{EntropyDefinition} is to use  the measurement procedures  previously defined for energy and for available energy with respect to $R$ and substitute the results in Eq.~\ref{defentropy2}. 
	
	It can be shown\footnote{Rewriting Equation \ref{TR1} for two reservoirs $R$ and $R'$ with the state $A_0$ instead of $A_2$, it is easy to derive that 
		\begin{equation} \frac{1}{T_R}(E^{R}_{0\rm rev}-E^{R}_1)_{A_1R_1\underset{\text{w,rev}}{\Longrightarrow} A_0R_{2\rm rev}}=\frac{1}{T_{R'}}(E^{R'}_{0\rm rev}-E^{R'}_1)_{A_1R'_1\underset{\text{w,rev}}{\Longrightarrow} A_0R'_{2\rm rev}} \end{equation}
		which shows that this ratio, for the given pair of states $A_0$ and $A_1$, is equal for all reservoirs. Hence,  the value of $S_1-S_0$ in Equation \ref{defentropy1} is independent of the choice of reservoir $R$. Since $S_0$ is chosen for system $A$ independently of $R$, it follows that the entropy value $S_1$ is entirely independent of the choice of the reservoir $R$ used to measure it.} that the entropy value $S_1$ resulting from this definition is independent of the choice of the reservoir $R$, which plays a purely auxiliary role in the measurement procedure. This implies that entropy $S$, like energy $E$, is a property of system $A$ in and of itself. In particular, it does not depend on the reservoir $R$ chosen for measurement. 
	
	It is important to note that since the properties $E$ and $\Omega^R$ are defined for all states of all (well defined) systems, including  nonequilibrium states and small systems, the given definition of property $S$ is valid for any state and any system.\footnote{In traditional thermodynamics textbooks, it is not uncommon to find the assertion that entropy, and thermodynamics as a whole, is defined only for stable equilibrium states. However, this is due to the fact that the definition given there for entropy is inherently limited to stable equilibrium states, as it is based on temperature, which, as we will see in Section \ref{MutualEquilibriumConditions}, is only defined for stable equilibrium states.}

		\section{Practical meaning of entropy}

	To emphasize the physical and technical significance of entropy, it is interesting to note from Eq.~\ref{defentropy2} that, apart from the constants $S_0$, $E_0$, and $\Omega^R_0$ related to the choice of reference state $A_0$, entropy $S$ is proportional to the 
	``unavailable'' energy with respect to reservoir $R$, $E-\Omega^R$. For example, the change in unavailable energy with respect to $R$ is equal to the change in entropy of the system multiplied by the temperature of reservoir $R$,\ind{energy!unavailable with respect to a reservoir}
	\begin{equation} (E_2- \Omega^R_2) - (E_1- \Omega^R_1) = T_R \,(S_2-S_1) \label{TRS}\end{equation}
	In this sense, the thermal reservoir can be viewed to play the role of an ``entropymeter.''
	
	From what has been discussed, we can also derive the expression that allows us to calculate the available energy with respect to a reservoir $R$. We have already observed that the state $A_R$, in which system $A$ is in mutual equilibrium with reservoir $R$, has available energy with respect to $R$ equal to zero, $(\Omega^R)^A_R = 0$. From Equation \ref{TRS} with state $A_R$ replacing state $A_2$, we can derive the two equivalent expressions
	\begin{equation} \Omega^R_1 = E_1-E_R- T_R \,(S_1-S_R) \label{OmegaR}\end{equation}
		\begin{equation} S_1=\frac{E_1-\Omega^R_1}{T_R}+S_R-\frac{E_R}{ T_R} \label{Smeaning}\end{equation}
	where $E_R$ and $S_R$ are the energy and entropy of system $A$ in the stable equilibrium state $A_R$ of mutual equilibrium with $R$. As we will demonstrate in Section \ref{MutualEquilibriumConditions}, the temperature $T^A_R$ of the stable equilibrium state $A_R$ is equal to the temperature $T_R$ of the reservoir.
	
	Eq.~\ref{Smeaning} provides an explicit interpretation of the practical (engineering) meaning of entropy. Apart from the constant combination $S_R - E_R/T_R$, which is independent of state $A_1$, the entropy is proportional to $E_1-\Omega^R_1$, the  unavailable energy with respect to reservoir $R$, the constant of proportionality being the inverse of the reservoir's temperature. 
	
	Since $E$ and $S$ are properties, the differences $E_2-E_1$ and $S_2-S_1$ corresponding to a given change in state from $A_1$ to $A_2$ depend only on these states and not on the mode of interaction with other systems or the type of process or the forces or reactions that induced the change. The same change in state, from $A_1$ to $A_2$, can be obtained with many (infinite) modes of interaction, but they all yield the same values for $E_2-E_1$ and $S_2-S_1$.
	
	\section{Principle of entropy non-decrease in weight processes}
	
	The criteria for reversibility\ind{process!reversible} of a weight process that we have derived in terms of adiabatic availability, Eqs.~\ref{mechrev} and \ref{mechirr}, and in terms of available energy, Eqs.~\ref{mechrevR} and \ref{mechirrR}), can be reformulated in terms of entropy. The weight process that takes system $A$ from state $A_1$ to state $A_2$ is reversible if and only if\begin{equation} S_2=S_1 \label{nondecrescita=}\end{equation}while it is irreversible if and only if \begin{equation} S_2>S_1 \label{nondecrescita>}\end{equation}
	Equations \ref{nondecrescita=} and \ref{nondecrescita>} are known as the \textit{principle of non-decrease of entropy in weight processes}. These equations can  be rewritten in a single form, valid only for weight processes
	\begin{equation} (S_2-S_1)_{A_1\underset{\text{w}}{\Longrightarrow} A_2} = (\Sirr)_{12}\qquad  (\Sirr)_{12}\ge 0\label{nondecrease}\end{equation}
	with the condition that $(\Sirr)_{12}=0$ if the weight process is reversible and $(\Sirr)_{12}>0$ if the weight process is irreversible. During an irreversible weight process, the system loses some of its ability to transfer energy to a weight, and the entropy of the system increases. This increase, $(\Sirr)_{12}$, is called \textit{entropy produced} (or \textit{generated} or \textit{created}) in the system \textit{due to irreversibility}\ind{entropy!produced due to irreversibility}\ind{entropy!generation of}\ind{entropy!creation of} or simply \textit{entropy production} (or \textit{generation} or \textit{creation}).
	
	While energy is conserved in weight processes without net external effects, entropy is conserved in reversible weight processes. For example, for an isolated system, energy always remains constant, while entropy remains constant if the process is reversible and increases if the process is irreversible.

	\section{Entropy: additivity, non-decrease, exchangeability}

	Like energy $E$ and available energy $\Omega^R$, entropy $S$ is also an additive property.\ind{entropy!additivity}\footnote{Just like for energy, additivity holds in general for differences in entropy, i.e., \begin{equation} S^A_1- S^A_0+ S^B_1 -S^B_0= S^C_{11}- S^C_{00} \end{equation}for $C=AB$. To make it hold for absolute values as well, \begin{equation} S^C_{11}=S^A_1 + S^B_1\,, \end{equation} we need the reference values for  composite systems to be chosen such that $ S^C_{00}=S^A_0+ S^B_0$.}

\begin{figure}[!h]
		\begin{center}
				\includegraphics[scale=0.45]{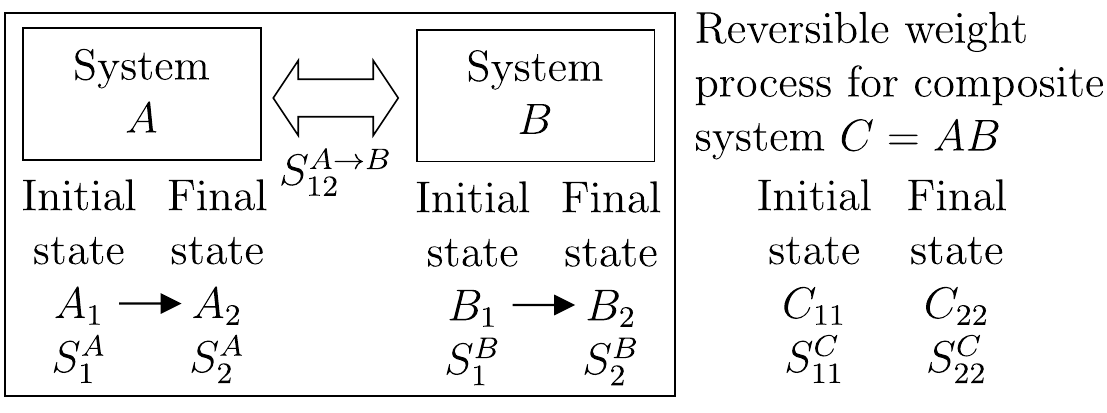}
				\caption{\label{EntropyTransfer}Entropy can be exchanged between two systems $A$ and $B$ through interaction. In this example, the composite  system $C=AB$ undergoes a reversible weight process.}
		\end{center}
	\end{figure}

	From the principle of non-decrease in weight processes and the additivity of entropy, it follows that entropy can be transferred (exchanged) between interacting systems.\ind{entropy!exchange of} Using Figure \ref{EntropyTransfer}, consider a system $C$ composed of subsystems $A$ and $B$ and a reversible weight process in which the state of $A$ changes from $A_1$ to $A_2$ and that of $B$ changes from $B_1$ to $B_2$. Since the process for $C$ is reversible, the value of the entropy of $C$ remains unchanged. In fact,  Eq.~\ref{nondecrease} for system $C$ yields $S^C_{22}- S^C_{11}=(\Sirr)^C_{12}=0$. Due to the additivity of differences in entropy, this means that $(S^A_2- S^A_1)+ (S^B_2 -S^B_1)=0$, or in other words, $(S^A_2- S^A_1)=-(S^B_2 -S^B_1)$. The change in entropy of $A$ is equal and opposite to that of $B$. This justifies the notion of \textit{entropy transfer} or \textit{exchange}\ind{exchange!of entropy}, meaning that if the entropy of $B$ increases, we say that $B$ receives entropy from $A$, as the entropy of $A$ decreases by an equal amount.

	\section{Notation for entropy exchange and the entropy (im)balance equation}
	
	For the quantity of entropy transferred from $A$ to $B$, we use the symbol\begin{equation} S^{A\rightarrow B}_{12} \end{equation}and, consequently, for a reversible weight process for the composite system $C=AB$, \begin{equation} S^A_2- S^A_1=-(S^B_2 -S^B_1)=- S^{A\rightarrow B}_{12} \end{equation}Equivalently, we can indicate the quantity of entropy transferred from $B$ to $A$ using the symbol $ S^{A\leftarrow B}_{12}$ which leads to\begin{equation} S^A_2- S^A_1=-(S^B_2 -S^B_1)= S^{A\leftarrow B}_{12} \end{equation}Therefore, the two introduced symbols are not independent, and we have\begin{equation} S^{A\rightarrow B}_{12}=- S^{A\leftarrow B}_{12} \end{equation}
	
	If we consider $B$ as the environment of $A$, we can simplify the notation by omitting the subscript $B$ and writing\footnote{When there is no ambiguity about the system under consideration, the notation can be further simplified by omitting the subscript $A$ and writing $S^{\ar}_{12}=-S^{\al}_{12}$. When there is no ambiguity about the process under consideration, the subscript $12$ can be omitted as well, writing $S^{A\ar}=-S^{A\al}$ or even $S^{\ar}=-S^{\al}$. It is important to note that when $S^{A\al}$ takes a negative value, it means that system $A$ has released entropy. For example, if $S^{A\al}=-5$ J/K, according to Equation \ref{segniS}, $S^{A\ar}=5$ J/K, indicating that system $A$ has released 5 joules per kelvin of entropy.}\begin{equation} S^{A\ar}=-S^{A\al}\label{segniS}\end{equation}for the quantity of entropy transferred between environment $B$ and system $A$ as a result of all interactions in the process that changes the state of $A$ from $A_1$ to $A_2$.
	
	These relations, valid \textit{if the process is reversible}, can be written in the form\ind{entropy!balance}\ind{balance!of entropy}\begin{equation} S^A_2 - S^A_1 = S^{A\al} \qquad\text{or, equivalently,}\qquad S^A_2 - S^A_1 = -S^{A\ar} \end{equation}This important consequence of entropy additivity and the principle of
	entropy conservation in reversible processes can be generalized using the principle of non-decrease of entropy in weight processes. In this case, for system $C$, $S^C_{22}- S^C_{11}=(\Sirr)^C_{12}>0$, and due to entropy additivity, we can write $(S^A_2- S^A_1)+ (S^B_2 -S^B_1)= (\Sirr)^A_{12}+(\Sirr)^B_{12}$, where we have decomposed the entropy created due to irreversibility in system $C$ into two non-negative contributions, one from each subsystem. It is important to distinguish between the entropy transferred $S^{A\al B}_{12}$ and the entropy produced due to irreversibility, as follows:\begin{eqnarray} S^A_2 - S^A_1 =  S^{A\al B}_{12}+ (\Sirr)^A_{12} \qquad (\Sirr)^A_{12}\ge 0\\ S^B_2 - S^B_1 =  -S^{A\al B}_{12}+ (\Sirr)^B_{12} \qquad (\Sirr)^B_{12}\ge 0 \end{eqnarray} These relations generalize the entropy balance to processes, even non-weight ones, characterized by entropy exchange between $A$ and $B$, as well as entropy generation by irreversibility in both subsystems.
	
	\begin{figure}[!ht]
		\begin{center}
				\includegraphics[scale=0.45]{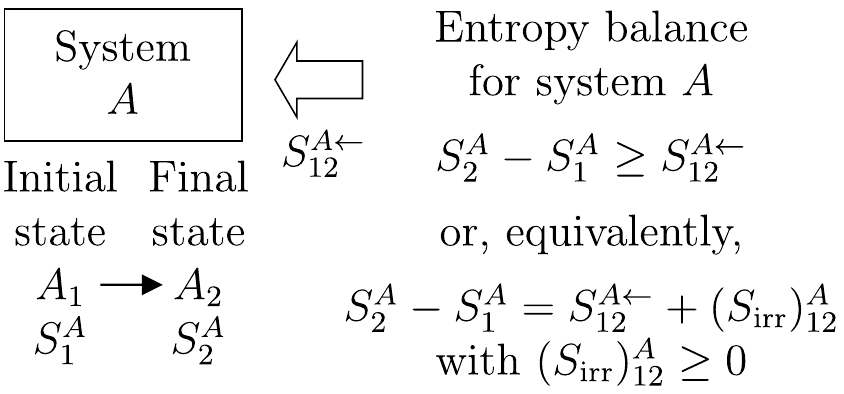}
				\caption{\label{EntropyBalance}Entropy balance for system $A$ for a process in which the state of $A$ changes from $A_1$ at time $t_1$ to $A_2$ at time $t_2$, and the net effect of the interaction between $A$ and its environment includes an entropy transfer $S^{A\gets}_{12}$ (positive if in the direction of the arrow, i.e., if received by $A$, negative if in the opposite direction).}
		\end{center}
	\end{figure}

	With a focus on system $A$ (Figure \ref{EntropyBalance}), the \textit{entropy balance equation}\footnote{In some texts, the entropy balance equation is briefly referred to as the 
		``second law of thermodynamics.'' Therefore, the jargon 
		``writing the second law'' is used to mean 
		``writing the entropy balance equation.'' We have already pointed out that the second law of thermodynamics leads to many other important conclusions beyond the entropy balance equation. Thus, the mentioned jargon can be misleading and also fails to acknowledge the broader role and other implications of the second law, such as the state principle, the maximum entropy principle, and the numerous relations that derive from them.} (by some authors also named ``imbalance equation'')\begin{equation} S^A_2 - S^A_1 =  S^{A\al}_{12}+ (\Sirr)^A_{12}\qquad (\Sirr)^A_{12}\ge 0\end{equation}imposes that the change in entropy $S^A_2 - S^A_1$ resulting from a process for $A$ from $A_1$ to $A_2$ is greater (if the process is irreversible) or equal (if the process is reversible) to the net quantity  of entropy $S^{A\al}_{12}$ transferred to system $A$ as a result of interactions with its environment.
	
	As with the energy balance equation, it is important to remember that the entropy balance equation, as well as the first and second principles from which it derives, is an expression of the laws of \textit{dynamics}. The variable time\ind{time} does not appear explicitly but is strongly present: recall that $A_1$ denotes the state of system $A$ at time $t_1$ and $A_2$ at time $t_2$. To make the dynamic nature of the equation more explicit, it can be expressed in the following alternative form, which is useful for the analysis of continuous processes\begin{equation} \diff S^A/ \diff t= \dot S^{A\al}+ (\Sdotirr)^A \end{equation}This form is obtained when $t_1=t$ and $t_2=t+\diff t$, leading to $ \diff S^A =S^A_{t+\diff t}-S^A_t$, $\dot S^{A\al}= \delta S^{A\al}_{\{t\}-\{t+\diff t\}}/\diff t$ (entropy per unit time transferred to $A$ from its environment), and $(\Sdotirr)^A= \delta (\Sirr)^A_{\{t\}-\{t+\diff t\}}  /\diff t $ (entropy per unit time generated in system $A$ due to irreversibility).
	
	As we have seen, assessing changes in entropy is important, as they are directly related  to changes in the unavailable energy with respect to a reservoir (Equation \ref{TRS}) and, i.e., dissipation of available energy due to irreversibility.

		Stable equilibrium states play a special role in the statement of the second law of thermodynamics and derive from it formal characteristics that make them are easier to  study than other states of a system.  In the following Sections \ref{maxEntropyPrinciple}--\ref{pressure}, we outline the main features of this subset of states of a system.

	\section{Maximum entropy and minimum energy principles}\label{maxEntropyPrinciple}

	From the definition of a stable equilibrium state, as well as from the statement of the second law, another important result follows: the entropy  of every stable equilibrium state is greater (strictly greater) than the entropy  of any other state with the same value of $E$ and compatible values of $\bn$, and $\bbeta$. This result is known as the 
	``maximum entropy principle.''\footnote{Although the term `principle' is improper, as this is another theorem that derives from the statement of the second law.}
	
	It is useful to sketch the proof of this `principle' because it allows us to invoke various fundamental principles and definitions. Consider any system $C$ and the stable equilibrium state $C_0$ with energy $E_0^C$, amounts $\bn_0^C$, and parameters $\bbeta_0^C$. Then consider any other state, $C_1$, different from $C_0$, but with the same value of the energy, $E_0^C$, and amounts and parameters compatible with   $\bn_0^C$ and $\bbeta_0^C$. The first law guarantees that there exists a weight process for $C$ that interconnects the two states $C_0$ and $C_1$, but it does not specify the direction. Since the two states have the same energy, this weight process has no net external effect. It follows that the direction cannot be from $C_0$ to $C_1$ because, by definition, $C_0$, being a stable equilibrium state, cannot be altered without leaving external effects. Therefore, the weight process is in the direction from $C_1$ to $C_0$ and is irreversible.\footnote{We have just seen that it has no external effects and it is not possible in the opposite direction.} From the principle of entropy non-decrease in weight processes (Relation \ref{nondecrease}), it follows that \begin{equation} S_0 > S_1\label{maxS}\end{equation}which is what we wanted to prove. Among all the states with given values of $E$, $\bn$, and $\bbeta$, the stable equilibrium state has the maximum entropy, and all other states have lower entropy.
	
	\section{State principle}\label{StatePrinciple}
	From the statement of the second law of thermodynamics, in particular from the assertion that the values of $E$, $\bn$, and $\bbeta$ uniquely determine one and only one stable equilibrium state for any system, it directly follows that every stable equilibrium state of a system is uniquely determined by the values of $E$, $\bn$, and $\bbeta$. But if the state is determined, by the definition of 	``state,'' the values of all the properties of the system are determined.
	
	This one-to-on connection between the values of $E$, $\bn$, $\bbeta$, and the value of any other property $P$ at stable equilibrium is equivalent to the existence of the mathematical relation \begin{equation} P = P(E, n_1, n_2, \ldots, n_r, \beta_1, \beta_2, \ldots, \beta_s) \label{Pstate}\end{equation}This result, valid for all properties but only for stable equilibrium states, is known as the 
	``state principle,''\footnote{Although `principle' is not but rather a `theorem,' which, as just seen, directly derives from the general (Hatsopoulos-Keenan) statement of the second law we adopted.} and it expresses a general characteristic of the stable equilibrium states of all systems: it implies the existence of interrelations between the properties of this family of states. For all other states, those that are not stable equilibrium, Eq.~\ref{Pstate} is not valid, and such interrelations do not generally exist.

	\section{Fundamental stable equilibrium state relation}
	\ind{fundamental relation}Applying Relation \ref{Pstate} to the property entropy, $S$, we obtain that for the stable equilibrium states of any (well-defined) system, there exists the relation \begin{equation} S = S(E,n_1,n_2, \ldots, n_r, \beta_1, \beta_2, \ldots, \beta_s) \label{fundamentalS}\end{equation}which implies specific interrelations among the values of $S$, $E$, $\bn$, and $\bbeta$. This relation is characteristic of the system, meaning that its functional form varies from system to system and is known as the 
	``fundamental stable equilibrium relation of the system in entropy form'' or simply as the 
	``fundamental relation of the system'' or 
	``fundamental relation in entropy form.''
	
	In general, Relation \ref{fundamentalS} has partial derivatives of all orders,\footnote{Except for some states within the framework of the simple system model \cite[Ch.16]{GyftopoulosBeretta1991}, where some second derivatives are discontinuous due to the many-particle limiting approximations (simplifications) introduced by the model. These states usually define the boundary between regions where different states of matter prevail, which we  refer to as 
		``phases.'' However, it is only in the directions that cross this boundary that discontinuities occur, while in other directions, all derivatives are generally continuous.} so any difference between the entropies of two stable equilibrium states can always be expressed in the form of a Taylor series in terms of the differences in their values of  $E$, $\bn$, and $\bbeta$.
	
	For example, considering two stable equilibrium states  with identical values for all amounts of constituents and parameters but different energy values, $E_1$ and $E_0$, we can write
	\begin{eqnarray} S_0&=&S(E_0,\bn,\bbeta)\\ S_1&=&S(E_1,\bn,\bbeta)\nonumber\\
	&=&S(E_0+(E_1-E_0),\bn,\bbeta)\nonumber\\
	&=&S_0+\left.\frac{\partial S(E,\bn,\bbeta)}{\partial E}\right|_0
	(E_1-E_0) + \frac{1}{2}\left.\frac{\partial^2
		S(E,\bn,\bbeta)}{\partial E^2}\right|_0 (E_1-E_0)^2 + \cdots
	\end{eqnarray}
	where the symbol $\big|_0$ indicates that the partial derivative must be evaluated at the values $E_0,\bn,\bbeta$, as it is in the vicinity of these values that the series expansion of the fundamental relation with respect to the variable $E$ has been performed.
	
	We call \textit{normal systems} those for which  energy values have no upper bound, such as when they consist of constituents with translational degrees of freedom, i.e., the great majority of practical systems relevant to engineering.
	\footnote{However, in the realm of quantum thermodynamics, some systems (qubits, qutrits, spin, N-level atoms, such as  models of three-level atoms useful for understanding the operation of lasers) are assumed to have a finite set of energy levels, so that the energy values they can take are limited by an upper bound. For these \textit{special systems} Relation \ref{fundamentalS} is not monotone and therefore it can be  inverted only over two restricted energy intervals, yielding two relations of the form of Eq.~\ref{fundamentalE}, respectively called the \textit{positive temperature branch} and the  \textit{negative temperature branch}. See later Fig. \ref{ESdiagram_NegativeT}. }
		For normal systems, the fundamental relation \ref{fundamentalS} is strictly monotonic in the variable $S$ and, therefore, 
	it can be inverted by expressing $E$ as a function of $S$, $\bn$, and $\bbeta$, thus obtaining the relation
	\begin{equation} E =  E(S,n_1,n_2, \ldots, n_r, \beta_1, \beta_2, \ldots, \beta_s) \label{fundamentalE}\end{equation}
	called the 
	``fundamental  stable equilibrium  relation in energy form'' or simply the 
	``fundamental relation in energy form.'' 
	
	\subsection{Notation for partial derivatives and differentials}\label{notation}
	A peculiar tradition in thermodynamics is to indicate partial derivatives in the following way. Given the relation $z=z(x,y)$, the symbol is introduced as \begin{equation} \vparder{z}{x}{y } =
	\frac{\partial z(x,y)}{\partial x}\end{equation}which is sometimes (inelegantly) read as the derivative of $z$ with respect to $x$ 
	``at constant $y$.''
	The utility of this notation lies in the fact that the symbol of the partial derivative contains all the information about which variables are the independent variables of the function subject to the derivative. For example, if we have the functions $z=z(x,y)$ and $z=z(x,w)$, the symbol $\partial z/\partial x$ is ambiguous because it is not clear which of the two functions is the subject of the derivative, whereas the symbols $\hparder{z}{x}{y }$ and $\hparder{z}{x}{ w}$ leave no room for ambiguity.
	
	If the relation $z=z(x,y)$ can be 
	``solved'' with respect to either variable $y$, yielding the relation $y=y(x,z)$, or with respect to variable $x$, yielding the relation $x=x(y,z)$, it is obvious that the three obtained relations represent the same surface in the $x$-$y$-$z$ space. For each of them, the differential evaluated at a given 
	``point'' ($x$,$y$,$z$) represents the tangent plane to that surface, for which we have the three equivalent expressions \begin{align}
\diff z&=\vparder{z}{x}{y}\diff x+\vparder{z}{y}{x}\diff y
	\label{zxy}\\
\diff y&=\vparder{y}{x}{z}\diff x+\vparder{y}{z}{x}\diff z
	\label{yxz}\\
	\diff x&=\vparder{x}{y}{z}\diff y+\vparder{x}{z}{y}\diff z
	\label{xyz}\end{align} Since these expressions represent the same tangent plane, the various partial derivatives that appear in them are not independent of each other. For example, by solving expression \ref{xyz} for $\diff y$, we obtain \begin{equation}
	\diff y=\frac{1}{\hparder{x}{y}{z}}
	\diff x-\frac{\hparder{x}{z}{y}}{\hparder{x}{y}{z}}\diff z \label{dy}\end{equation} and by comparing Eq.~\ref{dy} with Eq.~\ref{yxz}, we deduce the relations \begin{equation} \vparder{y}{x}{z}=1\bigg/ \displaystyle
		\vparder{x}{y}{z} \label{inversexyz}\qquad\text{and}\qquad
	\vparder{y}{z}{x}=-\frac{\hparder{x}{z}{y}}{\hparder{x}{y}{z}}\end{equation}which can be rewritten in the following forms, the first of
	which is  called  
	``cyclic relation,''\ind{cyclic relation} \begin{equation}
	\vparder{x}{y}{z}\vparder{z}{x}{y}\vparder{y}{z}{x}=-1  \qquad
	{\rm or}\qquad
	\vparder{x}{y}{z}=-\vparder{x}{z}{y}\vparder{z}{y}{x}
	\label{cyclicRelation}\end{equation}

	For example, the fundamental relation in energy form, $E=
	E(S,\bn,\bbeta)$, (Eq.~\ref{fundamentalE}), is obtained from the fundamental relation in entropy form, $S=S(E,\bn,\bbeta) $,
	(Eq.~\ref{fundamentalS}). Both represent the same surface in the $E$-$S$-$\bn$-$\bbeta$ space. Therefore, Relation \ref{inversexyz} implies that \begin{equation}
	\vparder{S}{E}{\bn,\bbeta}=1\bigg/ \displaystyle
		\vparder{E}{S}{\bn,\bbeta} \label{inverseT}\end{equation}
	
	Finally, writing  the fundamental Relation \ref{fundamentalS} in the compact form   $S=S(\bx)$, with $\bx=(E,\bn,\bbeta)$, we can approximate the difference in its values between two neighboring stable equilibrium states with values $\bx$ and $\bx\pm\diff\bx$ by the Taylor series
	\begin{equation}\label{Taylor}
		S(\bx\pm\diff\bx)-S(\bx)=\pm\diff S|_\bx + \frac{1}{2}\diff^2 S|_\bx + \cdots
	\end{equation}
where $|_\bx$ means  that the first and second-order differentials are ``evaluated at state $\bx$,'' i.e.,
	\begin{equation}\label{differentials}\diff S|_\bx=\left.\frac{\partial S}{\partial\bx}\right|_{\bx}\cdot\diff\bx \qquad\text{and}\qquad \diff^2 S|_\bx = \diff\bx\cdot\left.\frac{\partial^2 S}{\partial\bx\,\partial\bx}\right|_{\bx} \cdot\diff\bx
	\end{equation}
	
	\section{Temperature, total potentials, pressure}\label{MutualEquilibriumConditions}
	Each of the first-order derivatives of the fundamental relation in entropy form, $S(E,\bn,\bbeta)$, or of the one in energy form, $E(S,\bn,\bbeta)$, defines a property of the family of stable equilibrium states of the system.
	
	The \textit{absolute temperature},\ind{temperature} or simply temperature $T$, is defined as \begin{equation} T =
	\vparder{E}{S}{\bn,\bbeta}= 1\bigg/ \vparder{S}{E}{\bn,\bbeta}
	\label{defT}\end{equation}where we used Eq.~\ref{inverseT}. For dimensional consistency, temperature has units of [energy]/[entropy], and in the International System of Units, it is measured in kelvin, K.\ind{kelvin, unit of measure}
	
	The \textit{total potential of the $i$-th constituent},\ind{potential!total} denoted as $\mu_i$, is defined as \begin{equation}
	\mu_i = \vparder{E}{n_i}{S,\bn,\bbeta} = -T
	\vparder{S}{n_i}{E,\bn,\bbeta}\label{defmu} \end{equation}For dimensional consistency, it has units of [energy]/[amount of constituent], and in SI, it is measured in joule/mole, J/mol.
	
	The \textit{generalized force conjugate to the $j$-th parameter},\ind{generalized force} denoted as $f_j$, is defined as \begin{equation}
	f_j = \vparder{E}{\beta_j}{S,\bn,\bbeta} = -T
	\vparder{S}{\beta_j}{E,\bn,\bbeta}\label{defgenforce} \end{equation}When the volume $V$ is a parameter, the generalized force conjugate to $V$, with a sign change, is called \textit{pressure}\ind{pressure} $p$, and it is given by \begin{equation} p =
	-\vparder{E}{V}{S,\bn,\bbeta} = T \vparder{S}{V}{E,\bn,\bbeta}
	\label{defp}\end{equation}For dimensional consistency, pressure has units of [energy]/[volume], and in SI, it is measured in joule/m$^3=$ newton/m$^2=$ pascal, Pa.\ind{pascal, unit of measure}
	
	These derivatives are defined and measurable for stable equilibrium states and are, therefore, properties. They play an important role in determining the conditions for mutual equilibrium between systems and the spontaneous tendency for systems in stable equilibrium but not in mutual equilibrium to exchange energy, entropy, amounts of constituents, and additive parameters. It is evident that each of these properties is defined only for the stable equilibrium states of the system: for other states, the fundamental relation does not exist, and, consequently, its derivatives do not exist either.
	
	\subsection{Necessary conditions for mutual equilibrium}
	
	It can be shown (see below) that the equality of temperatures of two systems is a necessary condition for the two systems that can exchange energy to be in mutual equilibrium. The practical importance of this result arises from the fact that it allows for the indirect measurement of the temperature  of a system $A$ (a partial derivative of its fundamental relation) by measuring the temperature of another system $B$ in mutual equilibrium with $A$. A \textit{thermometer} is a system for which the temperature is easily  measurable and the result readily displayed. By placing a thermometer $B$ in contact with system $A$ and waiting for mutual equilibrium to be reached, the temperature reading of the thermometer also provides the measurement of the temperature of $A$.\footnote{It should be noted that while a thermometer placed in contact with a system in a nonequilibrium state will provide a reading, this `apparent temperature' is a property of the thermometer's own stable equilibrium state  after the  system+thermometer composite reaches a specific steady-state or transient interaction, rather than an intrinsic property of the nonequilibrium system. In the present framework, we strictly distinguish between such interaction-dependent readings --- which depend on the nature of the contact and the thermometer's own properties ---  and the  temperature, which is defined exclusively for stable equilibrium states via the entropy-energy relation.} 
	
	Similarly, the other equalities (total potentials and pressures) necessary for mutual equilibrium between systems can be proven.
	
	The equality of the total potentials of a common constituent in two systems is a necessary condition for the mutual equilibrium of the two systems if they can exchange that constituent, for example, through a semipermeable membrane or simply through an opening or conduit that connects them. Pressure equality is a necessary condition for the mutual equilibrium of two systems when they can exchange volume, for example, if they are separated by a movable partition.
	
	\subsection{Proof of temperature and potential equality at mutual equilibrium}
	
\begin{figure}[!ht]
		\begin{center}
				\includegraphics[scale=0.70]{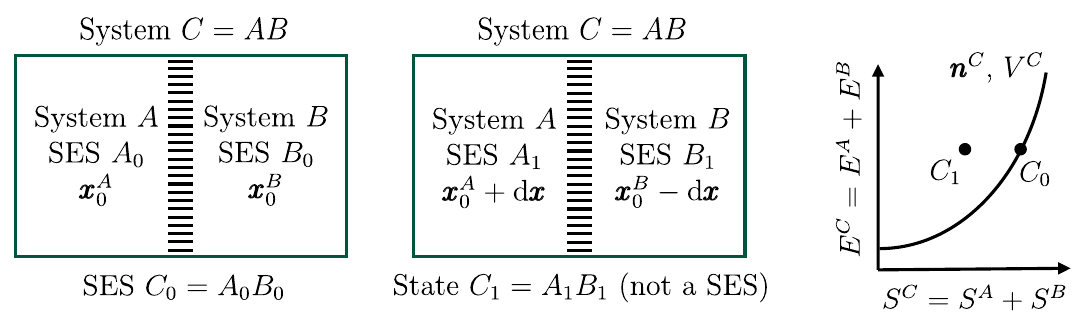}
				\caption{\label{MEPconditions}If $A$ and $B$ are in mutual equilibrium in states $A_0$ and $B_0$, then $C_0=A_0B_0$ is a stable equilibrium state. The maximum entropy principle implies that any other state 
					$C_1$ with the same or compatible values of $\bx^C=(E^A+E^B,\bn^A+\bn^B,V^A+V^B)$ cannot be a stable equilibrium state and, therefore, $S^C_1<S^C_0$. Compatibility depends on the interactions between systems $A$ and $B$ allowed by the partition that separates them. For example, it can allow them to exchange only energy, or energy and only one of the constituents, or energy and volume, and so on. }
		\end{center}
	\end{figure}

	We provide this proof here to show how the necessary conditions for mutual equilibrium can be derived from the maximum entropy principle. With the help of Fig. \ref{MEPconditions}, consider the two states $C_1$ and $C_0$ of the composite system $C=AB$ defined as follows. In state $C_0=A_0B_0$, systems $A$ and $B$ are in mutual equilibrium, hence $C_0$ is a stable equilibrium state. Assume, for simplicity that $A$ and $B$ have volume as the only parameter of the external forces, so that, all variables $\bx=(E,\bn,V)$ are additive, i.e., for any state $C_1=A_1B_1$, $\bx^C_1=\bx^A_1+\bx^B_1$. State $C_1$ is chosen so that $A_1$ is the stable equilibrium state with values $\bx^A_1=\bx^A_0+\diff\bx$ and  $B_1$ is the stable equilibrium state with values $\bx^B_1=\bx^B_0-\diff\bx$. As a result, $\bx^C_1 =\bx^C_0$ and clearly $C_1\ne C_0$, therefore, the maximum entropy principle implies that  $S^C_1<S^C_0$. Using the additivity of entropy, we can write this condition as
		\begin{equation}
		S^C_1-S^C_0=(S^A_1+S^B_1)-(S^A_0+S^B_0) =(S^A_1-S^A_0)+(S^B_1-S^B_0)<0  \label{MEPcondition}
	\end{equation}
		
Since $A_0$ and $A_1$ are stable equilibrium states,  the fundamental relation for $A$ yields $S_0^A=S_A(\bx_0^A)$ and  $S_1^A=S_A(\bx_0^A+\diff\bx)$, and similarly for $B$, $S_0^B=S_B(\bx_0^B)$ and $S_1^B=S_B(\bx_0^B-\diff\bx)$. Substituting in  Relation \ref{MEPcondition} and  using Eq.~\ref{Taylor}, yields the condition
		\begin{equation}
	S^C_1-S^C_0=\diff S_A|_{\bx^A_0} - \diff S_B|_{\bx^B_0}  + \frac{1}{2}\diff^2 S_A|_{\bx^A_0} +  \frac{1}{2}\diff^2 S_B|_{\bx^B_0} + \cdots <0  \label{MEPcondition2}
\end{equation}
which, using Eq.~\ref{differentials}, becomes
			\begin{equation}
		\left[\left.\frac{\partial S_A}{\partial\bx}\right|_{\bx^A_0} - \left.\frac{\partial S_B}{\partial\bx}\right|_{\bx^B_0}\right]\cdot\diff\bx + \frac{1}{2}\diff^2 S^A|_{\bx^A_0} + \frac{1}{2}\diff^2 S^B|_{\bx^B_0} + \cdots <0  \label{MEPcondition3}
	\end{equation}
This inequality must hold for all  choices of $\diff\bx$ compatible with the allowed interactions between systems $A$ and $B$. For example, if $A$ and $B$ can exchange  energy, but not constituents nor volume, i.e., the rigid partition in Fig. \ref{MEPconditions} is impermeable and fixed, than the only compatible choices are  $\diff\bx =(\diff E,0,0)$ so that the first term in the lhs of  Relation \ref{MEPcondition3} reduces to $[(1/T^A_0)-(1/T^B_0)]\diff E$. Then, the inequality can be satisfied for arbitrary values (positive and negative) of $\diff E$ only if the term in  brackets is zero, i.e., if $T^A_0=T^B_0$: temperature equality.  If $A$ and $B$ can exchange  energy and the $i$-th constituent, but not the other constituents nor volume, i.e., the rigid partition in Fig. \ref{MEPconditions} is semi-permeable (only to constituent $i$) and fixed, than the only compatible choices are  $\diff\bx =(\diff E,\diff n_i,0,0)$ and the first term in the lhs of  Relation \ref{MEPcondition3} reduces to $[(1/T^A_0)-(1/T^B_0)]\diff E-[(\mu_{i0}^A/T^A_0)-(\mu_{i0}^B/T^B_0)]\diff n_i$. Then, the inequality can be satisfied for arbitrary values (positive and negative) of $\diff E$ and $\diff n_i$ only if the terms in the brackets are zero, i.e., if $T^A_0=T^B_0$ and $\mu_{i0}^A=\mu_{i0}^B$: equality of temperature and $i$-th total potential. Again, if $A$ and $B$ can exchange also volume (movable partition) then mutual equilibrium requires also pressure equality.

	\section{Concavity of the fundamental relation}\label{concave}
	
Since the   first order term in the lhs of  Relation \ref{MEPcondition3} is necessarily zero, the strict inequality stems from second order terms or, if  they vanish (e.g., for thermal reservoirs), from  higher order terms in the Taylor expansion. Consider the particular case in which systems $A$ and $B$ are identical and are allowed to exchange energy, all constituents, and volume. Then, when they are in mutual equilibrium they share the same temperature, total potentials, and pressure. Relation \ref{MEPcondition3} reduces to 
	$\diff^2 S^A+\cdots<0$ which implies that in general, for any system in any stable equilibrium state, $	\diff^2 S\le 0$, i.e.,  the fundamental relation is \textit{concave} in all its variables. In other words, recalling Eq.~\ref{differentials}, the Hessian of the fundamental relation, $\partial^2 S/\partial\bx\,\partial\bx$, is a negative semi-definite matrix.

In particular, for any system,\footnote{If the system is a thermal reservoir, or behaves as such in a given set of states, then $\left(\partial^2 S/\partial
	E^2\right)_{\bn,\bbeta} =0$.} \begin{equation} \left(\frac{\partial^2 S}{\partial
		E^2}\right)_{\bn,V} =  \left(\frac{\partial (1/T)}{\partial
		E}\right)_{\bn,V}=-\frac{1}{T^2}\left(\frac{\partial T}{\partial
		E}\right)_{\bn,V} \le 0\label{concaveS}\end{equation}
where we used Eq.~\ref{defT}. This shows that the  temperature $T$ and the negative of its inverse, $-1/T$,  are increasing functions of the energy. 

	 Since it can be shown that, except where $1/T=0$, i.e., for finite temperatures, \begin{equation} \left(\frac{\partial^2 E}{\partial
		S^2}\right)_{\bn,V}=-T^3\left(\frac{\partial^2 S}{\partial
		E^2}\right)_{\bn,V}\label{convexE} \end{equation}it also follows that for a normal system the fundamental relation in energy form, $E=E(S,\bn,V)$, is \textit{convex} with respect to the variable $S$.\footnote{For a special system, Eq.~\ref{convexE} shows that the positive-temperature branch of the fundamental relation in energy form is convex, but the  negative-temperature branch  is concave.}

	\section{Gibbs relation}
	
	By differentiating the fundamental relation in energy form, $E=E(S,\bn,\bbeta)$, and using the definitions of $T$, $p$, $\mu_i$, and $f_j$ as just described, we can express the interrelations between differences in energy, $\diff E$, entropy, $\diff S$, volume,\footnote{For convenience, let's consider volume as the first parameter, $\beta_1=V$, and denote the set of other parameters as $\bbeta'=\{\beta_2,\dots,\beta_s\}$.} $\diff V$, other parameters, $\diff \beta_2$, $\diff \beta_3$, \dots, $\diff \beta_s$, and amounts of constituents, $\diff n_1$, $\diff n_2$, \dots, $\diff n_r$ between  neighboring stable equilibrium states as follows
	\begin{equation}  \diff E = T\diff S - p\diff V + \sum_{i=1}^r
	\mu_i\diff n_i + \sum_{j=2}^s f_j\diff \beta_j  \label{GibbsRelation}\end{equation}
	This relation, known as the \textit{Gibbs relation}, expresses the condition that must be satisfied if by varying the values of $E$, $S$, $V$, the $n_i$'s, and the $\beta_j$'s  we want the state of the system to shift along the stable-equilibrium-states manifold. The Gibbs relation represents the tangent plane to the stable-equilibrium-states manifold.
	
		\section{Pressure and force per unit area}\label{pressure}
	
	\begin{figure}[!ht]
		\begin{center}
				\includegraphics[scale=0.45]{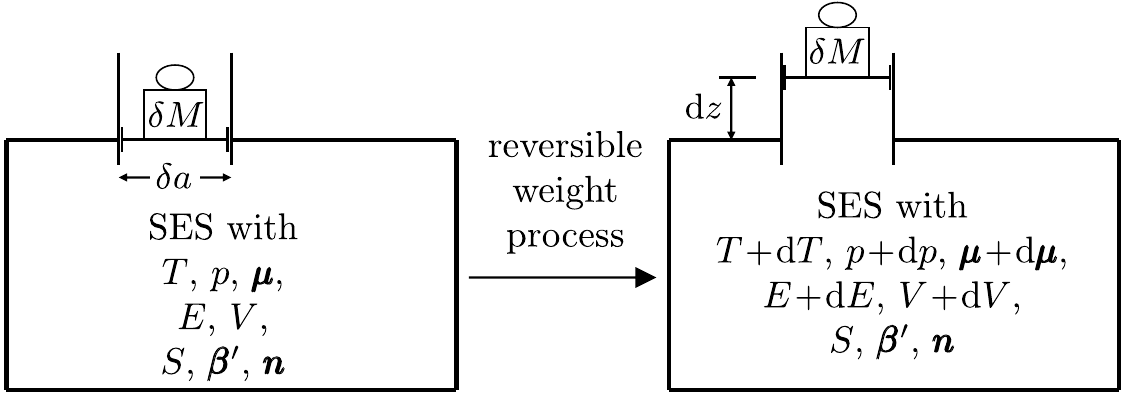}
				\caption{\label{Pressure}The pressure $p$ (defined at stable equilibrium by Eq.~\ref{defp}) is equal to the force per unit area exerted by the system on the walls confining its constituents in volume $V$.}
		\end{center}
	\end{figure}
	
	As an application of the Gibbs relation, it is useful to prove why the pressure $p$, defined at stable equilibrium by Eq.~\ref{defp} for a system with volume $V$ as one of the external parameters, is equal to the force per unit area exerted by the constituents of the system on the walls that confine them in the container of volume $V$. As shown in Figure \ref{Pressure}, we replace any segment of the wall with a small piston of area $\delta a$, which is mobile but sealed. We apply a force $\delta F=g\,\delta M$ to it using a mass $\delta M$ that is exactly needed to keep the piston in the same position as the replaced wall segment when the system is in a stable equilibrium state with values $T$, $p$, $\bmu$, $S$, $E$, $V$, $\bn$, $\bbeta'$. Now consider the reversible weight process ($\diff S=0$, Equation \ref{nondecrease}) that brings the system to an adjacent stable equilibrium state with values $S$, $E+\diff E$, $V+\diff V$, $\bn$, $\bbeta'$, having as its only external effect the displacement $\diff z$ of the piston on which the mass rests. From the energy balance \ref{bilEW}, we have \begin{equation}
\diff E=-\delta W^\ar=-g\,\delta M\diff z=-\delta F\diff z \label{dEdW}\end{equation}
	while from the Gibbs relation \ref{GibbsRelation} with $\diff S=0$, $\diff \bn=0$, $\diff \bbeta'=0$, we have \begin{equation} \diff E= -p\diff V \label{dEpV}\end{equation}
	Comparing \ref{dEdW} with \ref{dEpV}, it follows that $\delta F\diff z=p\diff V$. In other words, since $\diff V=\delta a\diff z$, \begin{equation}
	\frac{g\,\delta M}{\delta a}=\frac{\delta F}{\delta a}=p \end{equation} The force per unit area required to maintain the piston in position by counterbalancing the action of the system's constituents in a stable equilibrium state with pressure $p$ is equal to the pressure itself. Such force per unit area is exerted at every point on the surface confining the system's constituents.\footnote{Note that if the state is not a stable equilibrium state, the scheme illustrated in Figure \ref{Pressure} can result in different values of $\delta F/\delta a$ from point to point on the wall, as the Eq.~\ref{dEpV} ceases to hold, and pressure is not even defined.}
	
	\section[Energy vs entropy diagrams]{\label{ESdiagram}Energy vs entropy diagrams to represent nonequilibrium states and visually illustrate processes and summarize basic principles}
	
	In this section, we introduce the $E$--$S$ diagram representation, which is very useful to visualize states and processes of a system. We use it to visually illustrate and summarize the basic concepts and principles discussed so far. It is important to note that this representation is different from the  state diagrams used in traditional expositions of thermodynamics to represent the properties of the stable equilibrium states under the simple-system model. In contrast, the $E$--$S$ diagram represents not only stable equilibrium states but all other states, most of which are nonequilibrium. Moreover, the representation is valid for all systems, large and small, with many or few particles, even for a single quantum particle. This diagram is particularly effective for graphically depicting the relations between energy and entropy, and adiabatic availability and available energy.
	
	\subsection{Construction of the $E$--$S$  diagram}\label{diagrammiES}
	
\begin{figure}[!ht]
		\begin{center}
		\includegraphics[scale=0.42]{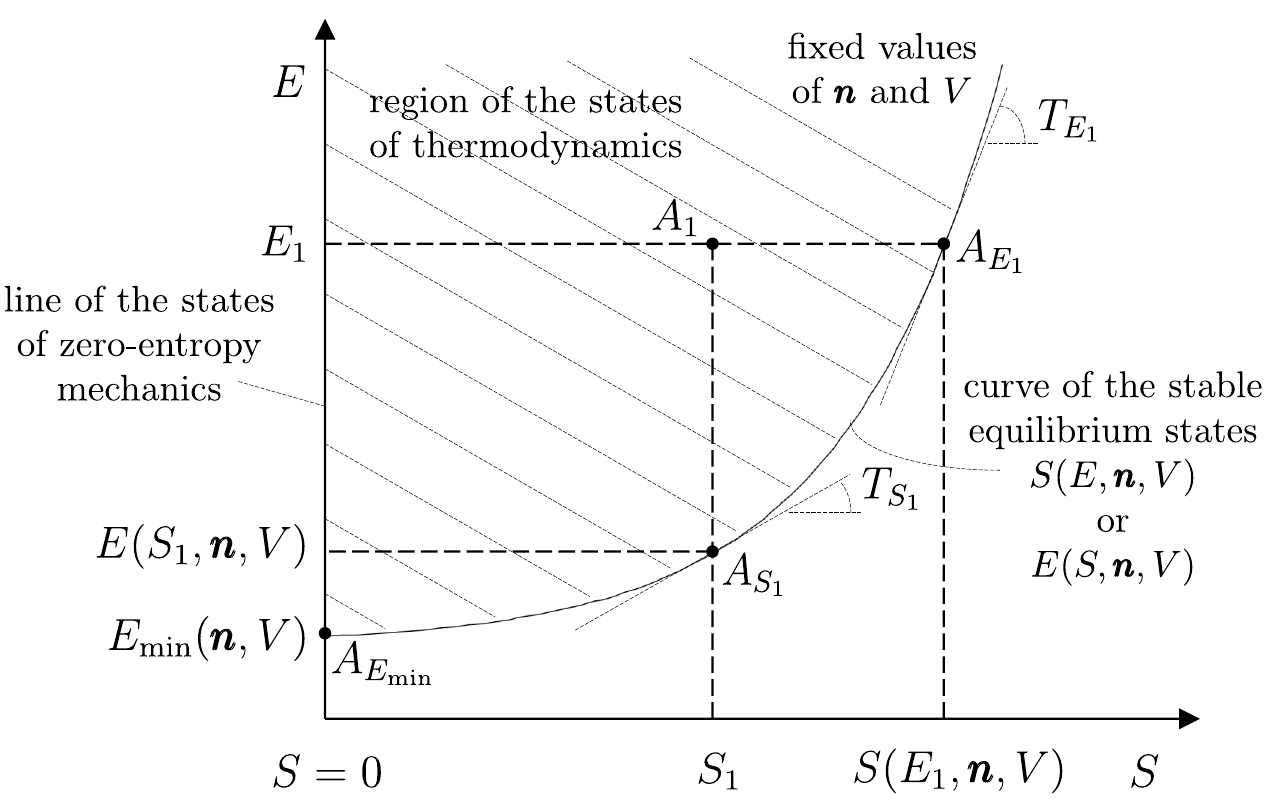}	
		\caption{\label{ESdiagramBase}Projection onto the $E$--$S$ plane of the multidimensional geometric space with an axis for each amount of constituents, parameter, and independent property, restricted to states that lie in the subspace corresponding to fixed values of  amounts and parameters, assuming for simplicity that volume $V$ is the only parameter and the system is normal and has non-degenerate ground-energy levels.}
		\end{center}
	\end{figure}

	Recall (Section \ref{state}) that the state $A_1=A(t_1)$ of a system at time $t_1$ is defined by the values of the amounts of constituents $\bn$, of the parameters $\bbeta$, and of all\footnote{A complete set of independent properties is sufficient, given which all other properties can be calculated.} the properties $P_1$, $P_2$, \ldots at time $t_1$. States can, in principle, be represented as points in a multidimensional geometric space with an axis for each amount of constituents, parameter, and independent property. However, such a presentation would not be particularly useful because the number of independent properties in a complete set is almost always infinite. Nevertheless, useful information can be obtained by intersecting this multidimensional geometric space with a plane (hyperplane) corresponding to fixed values of amounts of constituents and parameters. Subsequently, we can project this subspace onto the two-dimensional energy-entropy plane. For a system with volume $V$ as the only parameter, these states are projected inside the shaded area in Figure \ref{ESdiagramBase},\ind{energy-entropy diagram} bounded on the left by the vertical line of zero entropy states (mechanical states) and on the right by the curve defined by the restriction  of stable-equilibrium-state manifold  to the given set of values of $\bn$ and $\bbeta$. For simplicity, but without loss of generality, we proceed by assuming that volume $V$ is the only external parameter.

	A point located within the shaded area or on the vertical line $S=0$ generally represents the projection of an infinite number of states. All these states have the same values of amounts $\bn$, parameters $V$, energy $E$, and entropy $S$ but different values of other properties. They can be of any type, but not  stable equilibrium state. By contrast, a point on the convex curve of stable equilibrium states represents a single state, not a multiplicity of states. For each of the points (states) on this curve, the values of all properties are uniquely determined (state principle) by the values of $\bn$,  $V$, and $E$.

	\subsection{Maximum entropy and minimum energy principles}
	
	Every stable equilibrium state is the state of maximum entropy among all those with the same values of $E$, $\bn$, and $V$. It is also the state of minimum energy among all those with the same values of $S$, $\bn$, and $V$.
	
Figure \ref{ESdiagramBase}, shows that the set of states with values $E_1$, $\bn$, and $V$ projects onto a horizontal segment between $S=0$ and $S_{E_1}=S(E_1,\bn,V)$. The point at the far right of this segment (constrained-maximum entropy) represents the state $A_{E_1}$, the unique stable equilibrium state with values $E_1$, $\bn$, and $V$. There are no states to the right of $A_{E_1}$ with energy $E=E_1$ and the same values of $\bn$ and $V$. Moreover, as demonstrated in Section \ref{maxEntropyPrinciple}, no other state (of any kind)  with the same values of $E_1$, $\bn$, $V$ projects onto this point.
	
	In an isolated system (system-environment interactions cannot affect their respective states), every state $A_1$ on the $E = E_1$ segment, if not a metastable or unstable equilibrium, is pushed by  internal (so-called dissipative) dynamics toward states with increasing entropy until it reaches the stable equilibrium state $A_{E_1}$. This spontaneous process is irreversible because, in the absence of effects of interactions with the environment, the increase in entropy can only be generated internally  by the dissipative dynamics of the system. By definition, it is impossible to return from the stable equilibrium state $A_{E_1}$ back to the state $A_1$ without leaving effects in the environment.
	
		\subsection{Zero-entropy subspace: mechanical states}
		
	Figure \ref{ESdiagramBase} refers to a normal system (no upper bound in the energy) with non-degenerate ground-energy levels and  shows that the set of states with values $S_1$, $\bn$, and $V$ projects onto a vertical half-line with the lower endpoint at $E_{S_1}=E(S_1,\bn,V)$. The point at this endpoint (minimum energy) represents the state $A_{S_1}$, which for a normal system is the unique stable equilibrium state with values $S_1$, $\bn$, and $V$. There are no states below $A_{S_1}$ with entropy $S=S_1$ and the same values of $\bn$ and $V$.
	
	If we consider the set of states with values $S=0$, $\bn$, and $V$ (mechanical states), the half-line has an endpoint at $E=E(0,\bn,V)=E_{\rm min}(\bn,V)$ corresponding to the absolute minimum value that energy can take for the given values of $\bn$ and $V$.  The point at this endpoint represents the state $A_{E_{\rm min}}$. For systems with non-degenerate ground-energy levels, this is the unique stable equilibrium state with values $S=0$, $\bn$, and $V$, and the only equilibrium state (with values $\bn$ and $V$) considered in mechanics.

	The zero-entropy line represents all the states considered in mechanics (classical or quantum). As previously observed, the energy-entropy diagram clearly shows how mechanics emerges in this general presentation as a particular branch of thermodynamics, namely, its   restriction to the zero-entropy states.
	
	\subsection{Maximum-entropy subspace: thermodynamic equilibrium states}
	
	Similarly, the  thermodynamics of equilibrium states (so-called  \textit{thermostatics}), which considers only stable equilibrium states (so-called  \textit{thermodynamic equilibrium states}) and processes that occur exclusively through sequences of stable equilibrium states (so-called \textit{quasi-static processes}), emerges as another particular branch of thermodynamics, namely, its restriction to the (constrained)-maximum-entropy states.

	\subsection{Fundamental relation and temperature}
	
	The stable-equilibrium-state curve on the $E$-$S$ diagram represents, for fixed values of $\bn$ and volume $V$, the fundamental relation $S=S(E,\bn,V)$ in entropy form or, equivalently, the positive-temperature branch of its inversion into the energy form, $E=E(S,\bn,V)$ (the only branch for a normal system). The slope of the tangent line to this curve, $\hparder{E}{ S}{\bn,V}=1\big/\hparder{S}{ E}{\bn,V}$, coincides (Eq.~\ref{defT}) with the temperature $T$ of the stable equilibrium state represented by the point where the line is tangent. Temperature is not defined for states that are not stable equilibrium states because the fundamental relation does not hold for them, and in general, $E$ depends on more variables than just $S$, $\bn$, and $V$. 
	
	\subsection{Third Law. Zero-temperature at ground-energy stable equilibrium states}\label{ThirdLaw}
	
		For a normal system (not behaving as a thermal reservoir), the fundamental relation in energy form, $E=E(S,\bn,V)$, is convex  in the variable $S$ (Eqs.~\ref{concaveS} and \ref{convexE}), $\hparder{{}^2 E}{ S^2}{\bn,V}>0$. The temperature is positive except for the ground-energy stable equilibrium states, i.e., those with  minimal energy $E_{\rm min}(\bn,V)$ for the given values of $\bn$ and $V$. The assertion that the temperature of the ground-energy stable equilibrium state is zero is known as the \textit{third  law of thermodynamics}. It is not a  consequence of the first and the second laws, and we cover it here only marginally.\footnote{For a discussion on the quantum foundations of this assertion in the context of the present method of exposition of thermodynamics,   see \cite{BerettaGyftopoulos2015}, where we also provide an improved statement of the second   law compatible with the possibility, not excluded by the third law, of system's models with degenerate ground-energy levels, for which the zero-temperature ground-energy stable equilibrium states need not have zero entropy.}

	As already seen, the states considered in mechanics all have zero entropy, and for systems with non-degenerate ground-energy levels the state of minimum energy is a stable equilibrium state, as depicted in Figure \ref{ESdiagramBase}. Due to the convexity of the fundamental relation, temperature is an increasing function of energy, so the stable equilibrium state of minimum energy also has the minimum temperature for the given values of $\bn$ and $\bbeta$.

	However, convexity and the statements of the first and second laws of thermodynamics  do not exclude that the value of $T_{E_{\rm min}}$ be finite. That it is zero emerges within the formulations of quantum and statistical models.  To avoid resorting to such formalisms, in introductory expositions it suffices to assume an additional law, the \textit{third law of thermodynamics}, by asserting that \textit{all the ground-energy stable equilibrium states have zero temperature}.
	
	\begin{figure}[!ht]
		\begin{center}
				\includegraphics[scale=0.42]{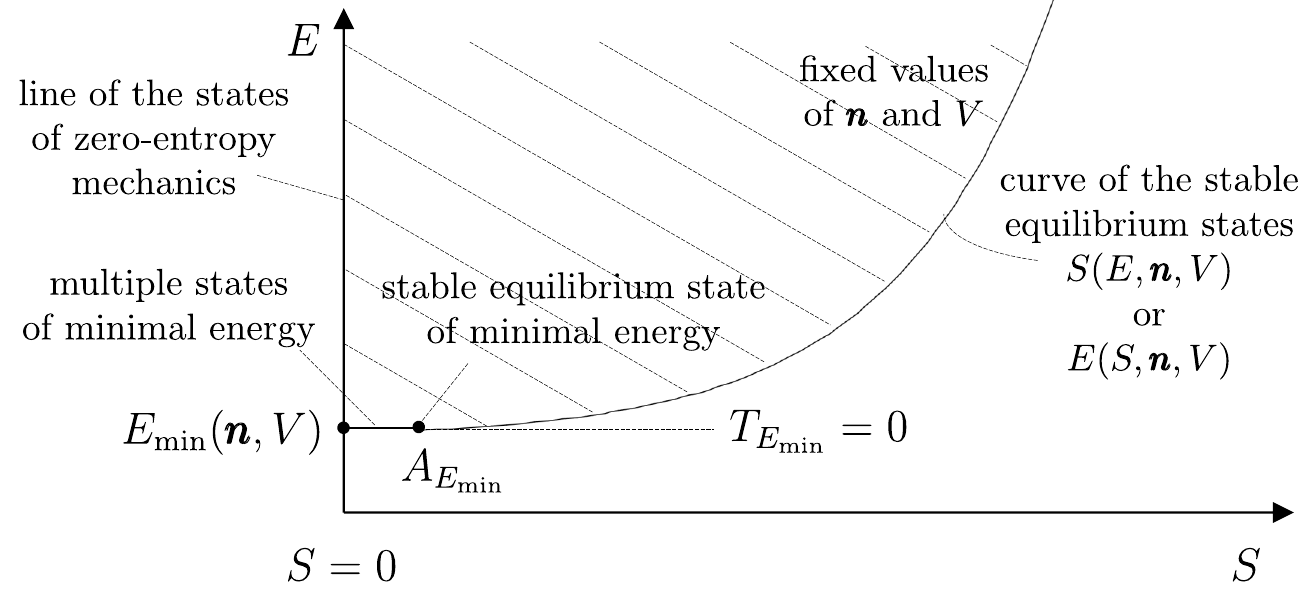}	
				\caption{\label{ESdiagram_third_law}$E$--$S$ diagram for a system with degenerate ground-energy levels. The stable equilibrium state corresponding to the minimum energy $E_{\rm min}(\bn,V)$ does not have zero entropy, but the third law asserts it has zero temperature.}
		\end{center}
	\end{figure}

	This statement is also compatible with the possibility, considered in the context of quantum theory, that the stable equilibrium state of minimum energy has nonzero entropy, as depicted in Figure \ref{ESdiagram_third_law}, with an entropy value given by
	\begin{equation}
		S(E_{\rm min},\bn,V)=\Boltz \ln g_1(\bn,V),
	\end{equation}
	where $\Boltz $ is the Boltzmann constant, and $g_1(\bn,V)$ is the \textit{multiplicity} of the minimum energy states for the given values of $\bn$ and $V$. For such systems, the minimum-energy states in mechanics, those with zero entropy, are not stable equilibrium states. However, if viewed from a restricted perspective that only considers states of mechanics, subject to a non-dissipative equation of motion valid only in this restricted domain of states, they typically appear as `partially' stable equilibrium states. They are considered `partially' stable because they are stable only with respect to perturbations that keep entropy equal to zero.

	\subsection{Adiabatic availability}
	
	In a weight process, every state $A_1$ at the intersection of the $S=S_1$ segment with the $E=E_1$ segment, if not a metastable or unstable equilibrium, is pushed by internal dynamics toward states with increasing entropy. It can also be subject to interactions that result in energy exchange (with the external weight). In such a process, there is a competition between the internal dynamics, whose dissipative part tends to generate entropy with a characteristic timescale, and the interactions with the external weight designed, for example,  to extract as much energy as possible from the system. Since the characteristic timescale of irreversible part of the system's internal dynamics typically that depends on the system's structure construction details, the designer will be able to extract more energy from the system the faster the action of the interactions with the external weight. This ensures that energy is extracted so rapidly that the internal dynamics have the least possible time to generate entropy. In the limit, if the weight process is reversible, the entropy will remain unchanged, and the state will move along the $S=S_1$ segment. In this case, it is possible to bring the system to the state of minimum energy $A_{S_1}$, and therefore, the extracted energy will be equal to $E_1-E_{S_1}$, which is the adiabatic availability of the system in the state $A_1$ (Eq.~\ref{adiabatic}),
$\Psi_1 = E_1 - E_{S_1}$.

	\begin{figure}[!ht]
	\begin{center}
			\includegraphics[scale=0.42]{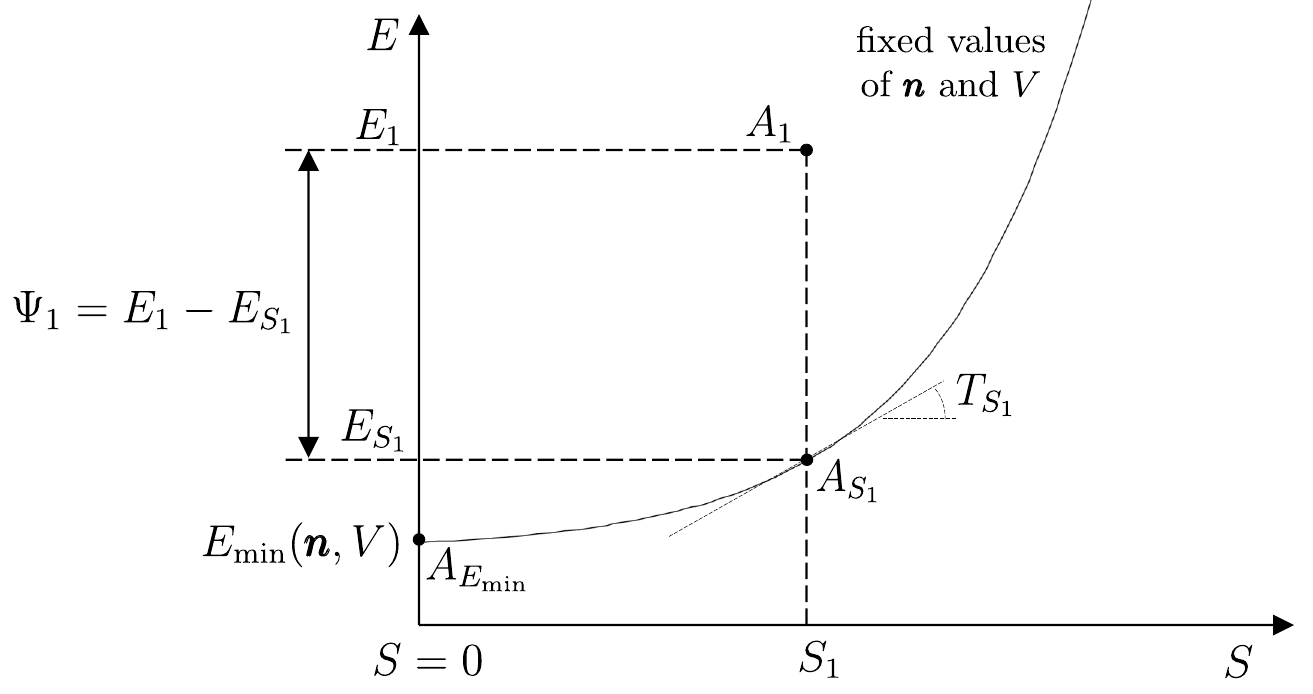}
			\caption{\label{ESdiagram_AdiabaticAvailability}Graphical representation on the $E$-$S$ diagram  of the adiabatic availability of state $A_1$ of system $A$.}
	\end{center}
\end{figure}

	\subsection{Available energy with respect to a thermal reservoir}
	
\begin{figure}[!ht]
		\begin{center}
				\includegraphics[scale=0.42]{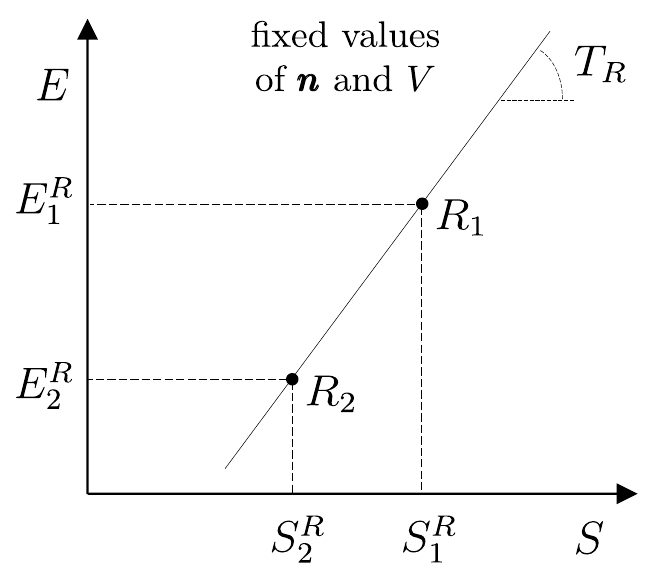}
				\caption{\label{ESdiagram_Reservoir} $E$--$S$ diagram for a thermal reservoir $R$ showing that the stable equilibrium state curve has $\left(\partial^2 S/\partial
					E^2\right)_{\bn,V} =0$ and constant slope $T_R$, i.e., it is a straight line. }
		\end{center}
	\end{figure}

	The $E$--$S$ diagram for a reservoir $R$ is shown in Figure  \ref{ESdiagram_Reservoir}. The stable equilibrium state curve is  a simple straight line with a slope equal to $T_R$,\footnote{Note that no system can behave like a thermal reservoir in all of its states because the condition that the temperature $T_R$ is equal for all stable equilibrium states is incompatible with the third law, which requires it to  vanish at the minimum energy.} 
	\begin{equation}
		E^R_2-E^R_1=T_R\,(S^R_2-S^R_1)  \label{FundRelR}
	\end{equation}

\begin{figure}[!ht]
		\begin{center}
				\includegraphics[scale=0.42]{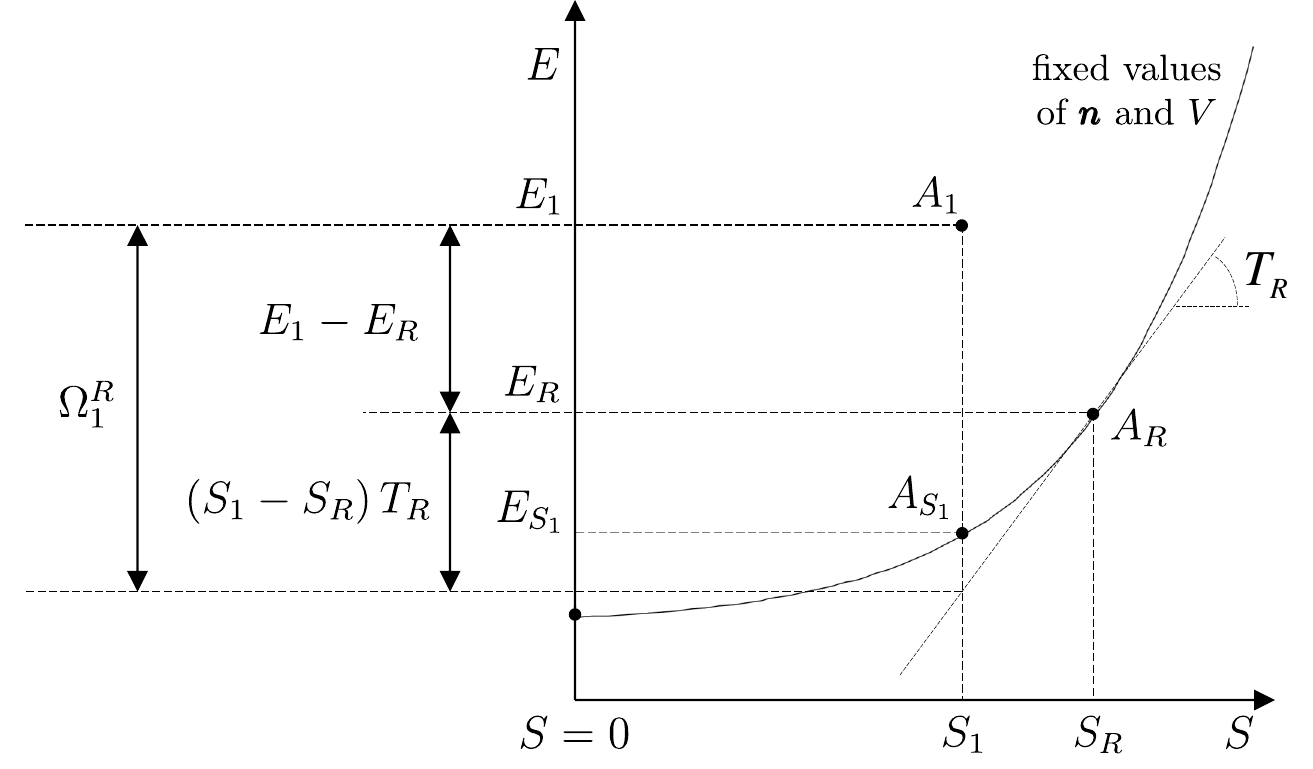}
				\caption{\label{ESdiagram_Omega} Graphical representation on the $E$--$S$ diagram of the available energy of state $A_1$ with respect to a thermal reservoir $R$ with temperature $T_R$.}
		\end{center}
	\end{figure}

	Now, consider  the available energy of a system $A$ in state $A_1$ with respect to reservoir $R$ and recall Eq.~\ref{OmegaR},
	\begin{equation}
		\Omega^R_1 = E_1-E_R- T_R \,(S_1-S_R)
	\end{equation}
	where $E_1$ and $S_1$ are the energy and entropy of state $A_1$ of $A$, $E_R$ and $S_R$ are the energy and entropy of $A$ in the stable equilibrium state $A_R$ with temperature $T_R$, i.e., the state in which $A$ is in mutual equilibrium with reservoir $R$. Figure \ref{ESdiagram_Omega} shows the graphical representation of $\Omega^R_1$ on the $E$--$S$ diagram for system $A$. The two terms $E_1-E_R$ and $T_R \,(S_R-S_1)$ are represented separately. Remember that available energy is the energy transferred to the weight in a reversible weight process  for the composite system $AR$ in which the state of $A$ changes from state $A_1$ to state $A_R$.  Therefore, the change in entropy of  $A$, $S_R-S_1$, must be accompanied by an equal and opposite change in the entropy of $R$, $S^R_2-S^R_1=-(S_R-S_1)$ which, as visualized in Fig.~\ref{ESdiagram_Reservoir},  requires a change in reservoir energy of $E^R_2-E^R_1=T_R\,(S^R_2-S^R_1)=-T_R\,(S_R-S_1)$. This change is essential in the energy balance for the composite system $AR$ and ensures that the overall energy transferred to the weight is indeed $(E_1-E_R)+ T_R \,(S_R-S_1)$. The two contributions are visualized  in Fig.~\ref{ESdiagram_Omega}.

	\subsection{Pressure and chemical potentials}
	
\begin{figure}[!ht]
		\begin{center}
				\includegraphics[scale=0.42]{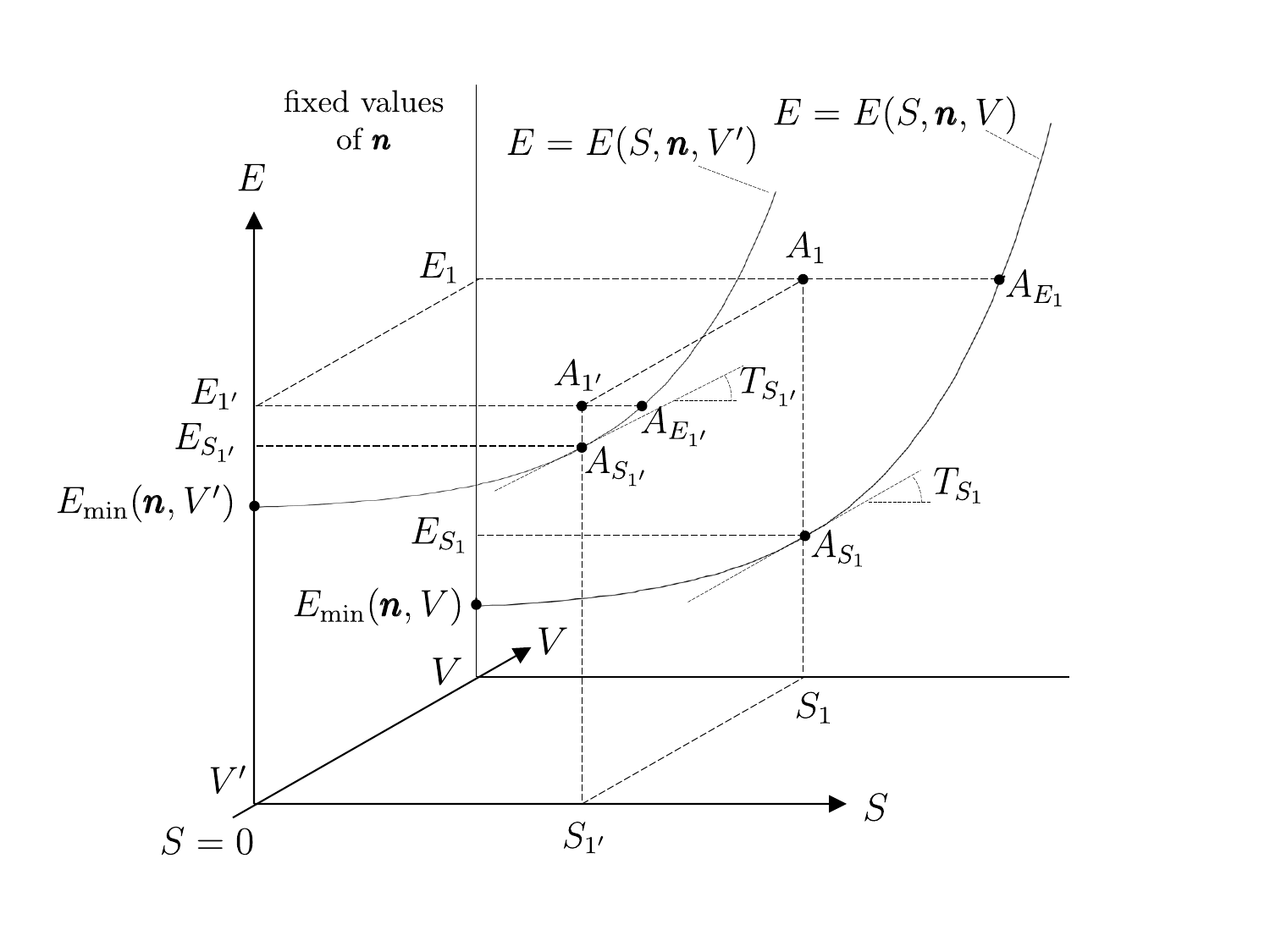}
				\caption{\label{ESdiagram_VariableVolume} Graphical representation on an $E$--$S$--$V$ diagram of two states $A_1$ and $A_{1'}$ with equal energy ($E_1=E_{1'}$) and entropy ($S_1=S_{1'}$) but different volumes and, therefore, different adiabatic availability ($E_1-E_{S_1}\ne E_{1'}-E_{S_{1'}}$).}
		\end{center}
	\end{figure}

	Figure \ref{ESdiagram_VariableVolume} hints at constructing a three-dimensional graph by adding an axis corresponding to the volume $V$. For simplicity, only the stable-equilibrium-state curves corresponding to two values $V$ and $V'$ are drawn. Geometrically, in this $E$--$S$--$V$ diagram, the stable equilibrium states fall on a surface obtained by projecting the points representing states onto a multidimensional geometric space with one axis for each amount of constituents, parameter, and independent property, restricted  to those lying on a subspace corresponding to fixed values of the amounts (and other parameters, if any, excluding volume).
	
	Two states, $A_1$ and $A_{1'}$, with equal energy ($E_1=E_{1'}$) and entropy ($S_1=S_{1'}$) but different volumes, are represented. It is noted that the adiabatic availability of the two states is also different, as $E_1-E_{S_1}\ne E_{1'}-E_{S_{1'}}$.
	
	The slope of the tangent plane to stable-equilibrium-state  $E$--$S$--$V$ surface   in the direction of constant $S$, $\hparder{E}{V}{S,\bn}$, coincides with the negative of the pressure, $-p$, of the stable equilibrium state where the plane is tangent.
	
	A similar three-dimensional diagram can be constructed by adding an axis corresponding not to volume but to the amount $n_i$ of one of the constituents. This results in the $E$--$S$--$n_i$ diagram, in which the slope of the tangent plane to the stable-equilibrium-state  $E$--$S$--$n_i$ surface in the direction of constant $S$, $\hparder{E}{n_i}{S,\boldsymbol{n'},V}$, coincides with the chemical potential $\mu_i$ (or total potential if there are other parameters besides volume) of the $i$-th constituent in the stable equilibrium state where the plane is tangent.

	\subsection{Special systems. Negative temperatures}\label{negative-temperatures}

	Almost\ind{negative temperature}\ind{fundamental relation!concavity of}\ind{concavity of the fundamental relation} all systems of practical interest are characterized by the ability to accommodate unlimited amounts of energy, which can be distributed among translational, rotational, vibrational, and electronic degrees of freedom of the molecules and/or atoms that constitute them. For all these systems, the $E$--$S$ diagram is as shown in Figure \ref{ESdiagramBase}: the fundamental relation $S=S(E,\bn,\bbeta)$ is monotonically increasing in energy, and therefore, its inversion with respect to $E$ yields the energy function $E=E(S,\bn,\bbeta)$, a single-valued function with a convex shape, $\hparder{{}^2E}{S^2}{\bn,\bbeta}>0$; the temperature $\hparder{ E}{ S}{\bn,\bbeta}$ is a non-negative function increasing with energy (starting from zero for the minimum energy state).

\begin{figure}[!ht]
	\begin{center}
			\includegraphics[scale=0.45]{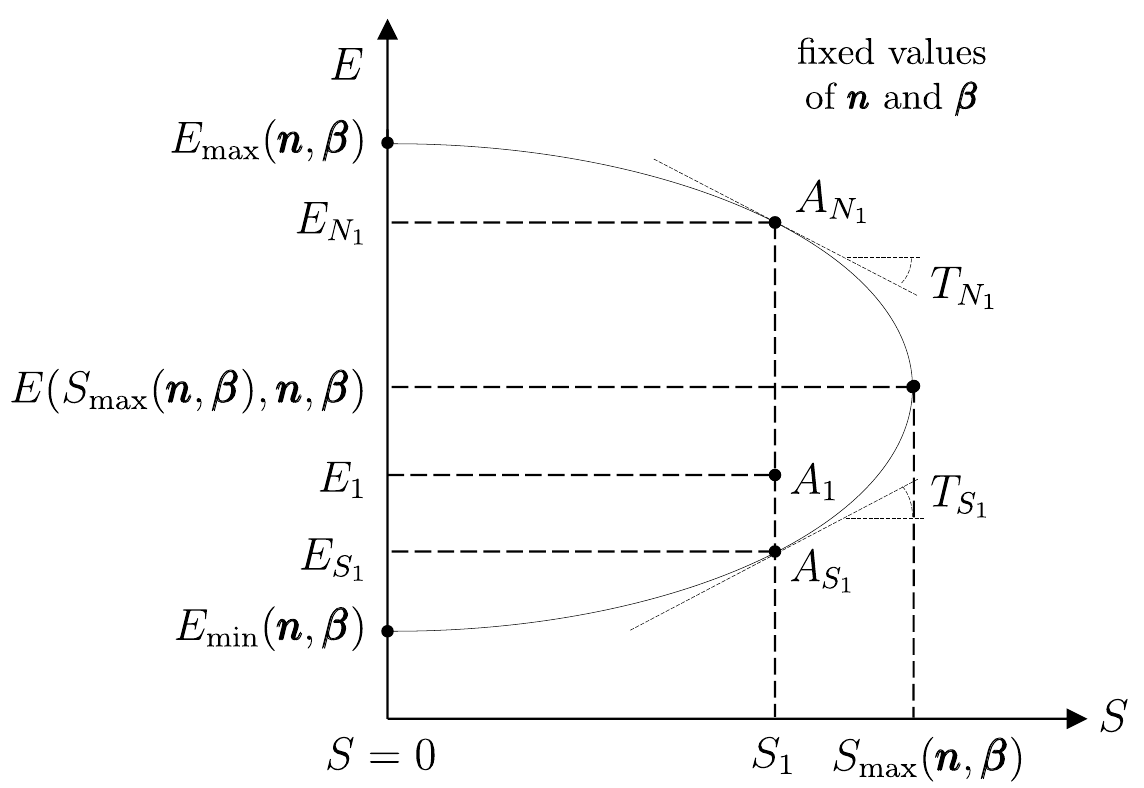}
			\caption{\label{ESdiagram_NegativeT} $E$--$S$ diagram for a special system that, for fixed values of $\bn$ and $\bbeta$, has energy values bounded between a minimum and a maximum. Stable equilibrium states with energy higher  than $E_{S_{\rm max}}$ have negative temperatures.}
	\end{center}
\end{figure}

	However, there are some special systems of quantum interest whose models require the existence of both a minimum and a maximum energy value for fixed amounts and parameters. For example, the model of an electron's spin in a magnetic field, the three-level atom model used to understand the operation of some lasers, and many others are systems characterized by a finite range of energy values existence between a lower and upper limit for energy values (levels).
	
	Such special systems, like all others, still adhere to the laws of thermodynamics we have described. However, the fundamental relation $S=S(E,\bn,\bbeta)$ is not monotonically increasing in energy, and therefore,  its inversion with respect to $E$ does not yield a single-valued function  of $(S,\bn,\bbeta)$. Figure \ref{ESdiagram_NegativeT} shows the $E$--$S$ diagram for a special system. 	
	The fundamental relation $S=S(E,\bn,\bbeta)$ maintains a concave shape, $\hparder{{}^2S}{E^2}{\bn,\bbeta}<0$. The negative of the inverse  temperature $-1/T=-\hparder{S}{E}{\bn,\bbeta}$ is an increasing function with respect to energy, ranging from $-\infty$ for the minimum energy state to $+\infty$ for the maximum energy state, passing through zero at the state with the maximum entropy $S_{\rm max}(\bn,\bbeta)$. Therefore, in addition to `normal' equilibrium states with positive temperatures ($-1/T$ between  $-\infty$ and zero), the system allows for `special' stable equilibrium states with negative temperatures for energies greater than the value $E_{S_{\rm max}}$ (where  $-1/T=0$), up to the stable equilibrium state with maximum energy (maximum for the given values of $\bn$ and $\bbeta$), where $-1/T=+\infty$ and hence the temperature is again zero. It is noteworthy that  $-1/T$ is  well-defined for all stable equilibrium states and  changes smoothly from $-\infty$ to $+\infty$ passing through zero. By contrast the temperature $T$ has a discontinuity at the stable equilibrium state with $E_{S_{\rm max}}$ where it jumps from  $+\infty$ to  $-\infty$. This state is not the hottest stable equilibrium state of the system for the given $\bn$ and $\bbeta$. Later, in Section \ref{ClausiusStatement} we  define what we mean by ``hot'' and  ``cold,'' and show that  the stable equilibrium states  with negative temperature are all hotter than the positive-temperature stable equilibrium states.
	
	\subsection{Energy--entropy constraints on energy conversion. The role of entropy sinks}
	
\begin{figure}[!ht]
	\begin{center}
			\includegraphics[scale=0.42]{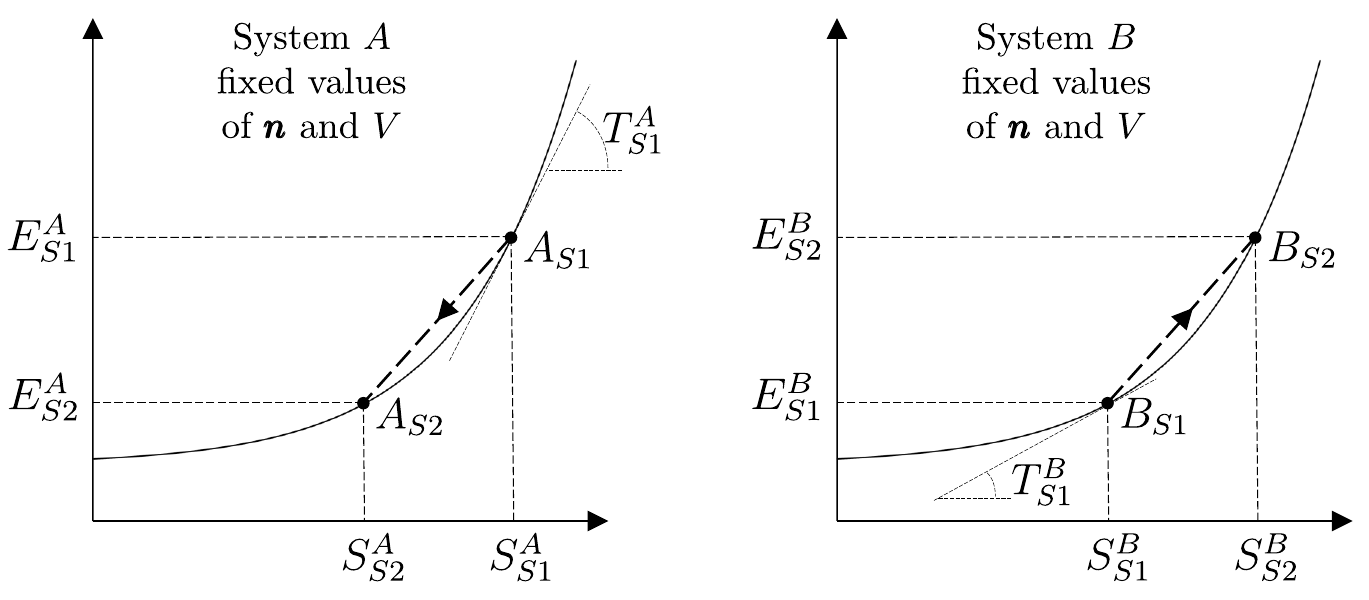}
			\caption{\textbf{Left}: To extract energy from a system $A$ initially in state  $A_{S1}$ we must reduce its entropy. \textbf{Right}: To transfer entropy into a system $B$ initially in state  $B_{S1}$  we must increase its energy. } \label{energy_entropy_constraints}
	\end{center}
\end{figure}

Figure \ref{energy_entropy_constraints} illustrates a fundamental constraint on energy extraction from systems initially in stable equilibrium, expressed most transparently with the help of the $E$–$S$ diagram. Consider a system $A$ initially in a stable equilibrium state $A_{S1}$ with energy $E^A_{S1}$, entropy $S^A_{S1}$, and temperature $T^A_{S1}$.

	A transition to a lower-energy state $A_{S2}$ cannot occur without a simultaneous decrease in entropy. In particular, the extraction of an energy amount $E_{12}^{A\ar}=E^A_{S1}-E^A_{S2}$ requires the extraction of at least an entropy amount $S_{12}^{A\ar}$ satisfying
	\begin{equation}\label{conversionA}
		S_{12}^{A\ar}= S^A_{S1}-S^A_{S2} +S^A_{\rm irr} > \frac{E^A_{S1}-E^A_{S2}}{T^A_{S1}} +S^A_{\rm irr}=\frac{E_{12}^{A\ar}}{T^A_{S1}} +S^A_{\rm irr}
	\end{equation}
where we used the entropy balance equation for $A$ and the inequality is illustrated in Figure \ref{energy_entropy_constraints}.Left.
	A purely vertical downward displacement in the $E$–$S$ diagram --- corresponding to energy extraction without entropy extraction --- is  impossible, as it would violate the second law and amount to a perpetual motion machine of the second kind.
	
	The necessary entropy extraction requires the presence in the environment of $A$ of an auxiliary system $B$  capable of accepting entropy. In practical energy conversion applications, this role is typically played by an external system such as a river, a lake, the atmosphere, the sea.\footnote{In physics and chemistry applications this role is often assigned to a so-called ``heat bath'' or a thermostatic bath, modeled by conditions essentially equivalent to our definition of a thermal reservoir.}  Such systems function as ``entropy sinks.''
	
	Assume the auxiliary system $B$ is initially in a stable equilibrium state $B_{S1}$ with energy $E^B_{S1}$, entropy $S^B_{S1}$, and temperature $T^B_{S1}$. A transition to a higher-entropy state  $B_{S2}$ cannot occur without a simultaneous increase in energy. Therefore, no entropy can be transferred to system $B$ without an accompanying transfer of energy. To accept an entropy amount $S_{12}^{B\al}$, system $B$ must also receive at least an energy amount satisfying
	\begin{equation}\label{conversionB}
	E_{12}^{B\al}=	E_{S2}-E_{S1} > (S_{S2}-S_{S1})\,T^B_{S1} =(S_{12}^{B\al}+S^B_{\rm irr} )\,T^B_{S1}
	\end{equation}
where the inequality is illustrated in Figure \ref{energy_entropy_constraints}.Right.
	This energy transfer to the environment of $A$ is often described, misleadingly, as wasted energy. In fact, it performs an essential thermodynamic function: it enables the disposal of entropy required for useful energy extraction from the system of interest. The presence of an entropy sink is therefore not a source of inefficiency, but a necessary condition for the operation of any energy-conversion device.
	
	The true sources of inefficiency in practical energy systems arise instead from internal irreversibilities within system $A$ and  $B$, as well as the  machinery $X$  used to accomplish the energy and entropy transfers and the energy conversion, i.e., from $S^A_{\rm irr}+S^X_{\rm irr}+S^B_{\rm irr}$.  Combining Eqs.~\ref{conversionA} and \ref{conversionB} assuming the machinery $X$ undergoes a cyclic process, i.e., $X_2=X_1$, so that $S_{12}^{B\al}=S_{12}^{A\ar} +S^X_{\rm irr}$, yields
	\begin{equation}\label{conversionM}
	E_{12}^{B\al}> \frac{T^B_{S1}}{T^A_{S1}}E_{12}^{A\ar} +(S^A_{\rm irr}+S^X_{\rm irr}+S^B_{\rm irr} )\,T^B_{S1}\qquad\text{that is}\qquad  \left.E_{12}^{B\al}\right|_\text{min}> \frac{T^B_{S1}}{T^A_{S1}}E_{12}^{A\ar}
\end{equation}
This result can be expressed also by saying that of the energy $E_{12}^{A\ar}$ extracted from $A$, only the energy amount $E_{12}^{A\ar}-E_{12}^{B\al}$ is available for performing useful tasks, because the energy 	$E_{12}^{B\al}$ must be used to accomplish the disposal   into system $B$ of the entropy that must be removed from system $A$ in order to achieve the energy extraction, therefore, at best (i.e., even in the absence of irreversibility) the fraction of extracted energy that remains  available for  useful tasks is bounded by\footnote{The rhs of Eq.~\ref{conversionCarnot} is referred to as the \textit{Carnot coefficient} between the two temperatures $T^B_{S1}$ and $T^A_{S1}$. The lhs is often referred to as the \textit{ energy efficiency of  the conversion}, but this language is misleading because, as already explained, it suggests that $E_{12}^{B\al}$ is avoidable, while it is not. }
	\begin{equation}\label{conversionCarnot}
		\frac{E_{12}^{A\ar}-E_{12}^{B\al}}{E_{12}^{A\ar}} <1-\frac{T^B_{S1}}{T^A_{S1}}
\end{equation}

Figure \ref{energy_entropy_constraints}  thus provides a direct geometric interpretation of the need for entropy sinks which rules the design of energy conversion devices.
	
		\section{\label{Interactions}Modes of interaction between systems}
		
	The foregoing discussion has progressed substantially --- including the definition of entropy and several other key results --- without invoking the notion of \emph{heat}. In doing so, we have developed all the conceptual and analytical tools required to introduce a rigorous definition of heat and to generalize it to \emph{heat-and-diffusion}. This is the subject of the next several sections, which are devoted to characterizing the various modes of interaction between systems.
	
	Because our modeling approach almost invariably begins with balances of energy, entropy, amounts of constituents, and volume, particular attention is devoted to the exchanges of these quantities across the frontiers separating interacting systems. The nature of these exchanges provides the basis for a precise classification of interactions.
	
	Interactions that involve exchanges of energy and volume only, without any exchange of entropy or constituents, are termed \emph{work interactions}\ind{work interaction}. A paradigmatic example is the interaction between a system and a weight in a weight process.
	
	Other interactions involve exchanges of both energy and entropy, with or without exchanges of constituents and volume. These are termed \emph{non-work interactions}\ind{non-work interaction}. As will be shown, \emph{heat} and \emph{heat-and-diffusion} interactions are special subclasses of non-work interactions, for which explicit relations can be established between the exchanged amounts of energy, entropy, constituents, and volume.
	
	Interactions generally drive the interacting systems into nonequilibrium states. If the interaction is momentary, these nonequilibrium states subsequently evolve spontaneously toward stable equilibrium, thereby inducing further changes in nonconserved properties. In particular, the spontaneous and irreversible evolution from a nonequilibrium state toward stable equilibrium entails the spontaneous generation of entropy within the system. Accordingly, interactions may change the entropy of a system both directly, through entropy exchange with other systems, and indirectly, through entropy generation associated with irreversible internal dynamics.
	
	Distinguishing between changes in properties due to exchanges with other systems and those due to spontaneous internal generation is essential for both understanding and engineering processes. For example, in an energy-conversion device, minimizing the spontaneous generation of entropy within its boundaries is a primary objective in improving efficiency. Conversely, in compact heat-transfer devices, maximizing the ratio of transferred energy to device volume may require accepting high rates of spontaneous entropy generation within the device.
	
	In general, when two systems begin to interact, they temporarily lose their separability and therefore, according to the present definitions, cease to be systems in their own right. Their individual energies are no longer defined, and only the energy of the composite system can be meaningfully specified. Part of this energy is associated directly with the interaction itself and cannot be unambiguously attributed to either collection of constituents. A simple illustration is provided by the collision of two molecules: as they approach, electrostatic interactions build up, temporarily storing energy in the interaction field; when the molecules separate again, this contribution vanishes. Once separated, the molecules return to being well-defined systems only if the internal dynamics has eliminated the correlations generated during the interaction.
	
	An important exception arises when the interaction is produced by a controlled variation of an external parameter common to both systems. In a weight process, for example, a rigid coupling can be engineered between a system parameter and the elevation of a weight in a gravitational field. Since the weight has a single independent property, no elastic or field-mediated energy storage external to the systems is involved. As a result, the system and the weight remain continuously separable and uncorrelated, and thus qualify as bona fide systems throughout the process.
	
	In the modeling of complex energy systems, it is essential to identify subsystems in a manner that allows the contributions of each to the overall entropy generation by irreversibility to be clearly identified. This analysis is carried out through energy and entropy balances, which require explicit specification of the types of interactions through which subsystems exchange energy and entropy. The classification of interactions into categories such as work, heat, diffusion, heat-and-diffusion, and radiative interactions is therefore instrumental and can be achieved only through precise and restrictive definitions.
	
	In particular, the concepts of work and heat provide a quantitative means to distinguish entropy generated by irreversibility from entropy exchanged through interaction. As will be shown, these concepts enable the precise identification of opportunities to reduce entropy generation and thereby improve the energy performance of thermodynamic systems.

	\section{\label{Work}Work  interactions}

	Work interactions  involve exchanges of energy and volume only, without any exchange of entropy or constituents.
	The energy transferred between system $A$ and $B$ by means of a work interaction is  called {\it work} and denoted with the symbol $W^{A\ar B}$,\ind{work interaction}\ind{work} which assumes  positive values if the energy is transferred {\it from $A$  to $B$} and negative if the transfer is in the opposite direction. We use the symbol $\delta W^{A\ar B}$ when the amount of energy transferred is infinitesimal and $\dot W^{A\ar B}$ for the rate of transfer in a continuous process. When the context allows it, and the focus is on system $A$ (and $B$ is its environment),  the symbols may be simplified to $W^{A\ar}$, $\delta W^{A\ar}$, and $\dot W^{A\ar}$ or even $W^{\ar}$, $\delta W^{\ar}$, and $\dot W^{\ar}$. Notice the identity $W^{A\ar B}=-W^{A\al B}$, i.e., reversing the arrow on the symbol is equivalent to changing its sign: a negative value of $W^{A\al B}$ means that the transfer is from $A$ to $B$, opposite to the direction of the arrow. 
	
	A process in which a system undergoes only  work-type interactions is called {\it adiabatic process}.\ind{adiabatic!process}
	If system $A$ changes from state $A_1$ to state $A_2 $ in an adiabatic process, the energy exchange $E^{A\al}$ is equal to the opposite of the work done on the environment $W^{A\ar}$, and the entropy exchange $S^{A\al} = 0$. Denoting the entropy generated within system $A$ by $S^A_{\rm irr} $, the energy and entropy balances \textit{for an adiabatic process} take the alternative forms
	\begin{align}
		E^A_2 - E^A_1 &= - W^{A\ar}&S^A_2 - S^A_1 &= S^A_{\rm irr}\\
		\diff E^A &= - \delta W^{A\ar} &\diff S^A &=  \delta S^A_{\rm irr} \\
		\diff E^A/\diff t &= - \dot W^{A\ar}&  \diff S^A /\diff t&= \dot S^A_{\rm irr}
	\end{align} 

\begin{figure}[!ht]
	\begin{center}
		\includegraphics[scale=0.35]{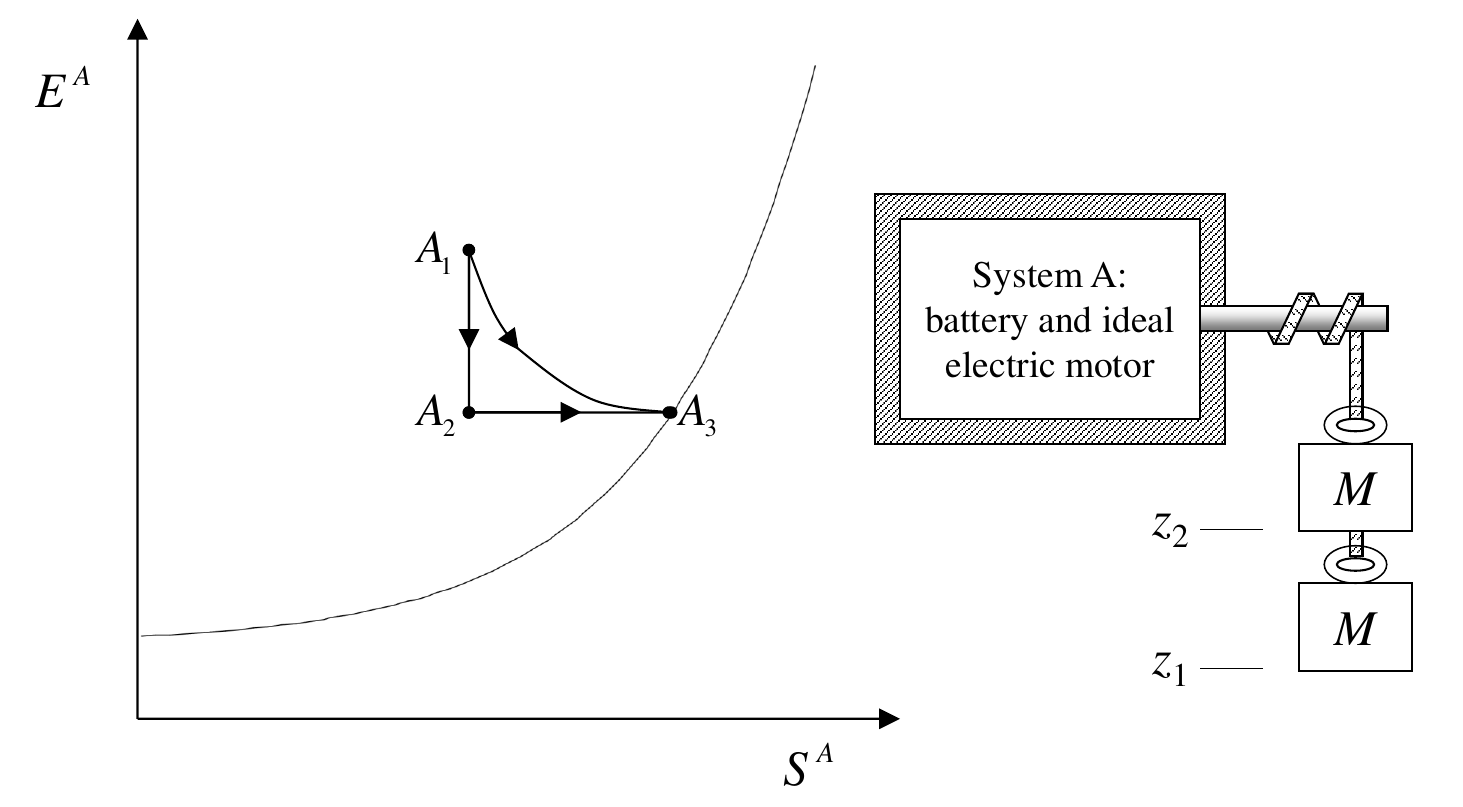} 
		\caption{\label{LavoroES4}$E$-$S$ diagram for a system $A$ containing an initially charged battery and an ideal electric motor connected  to a weight $B$ via a rope wound on its shaft, showing different paths in state space that may result depending on how rapid is the internal battery discharge with respect to the work interaction with the weight.  }
	\end{center}
\end{figure}
	
	The $E$--$S$ diagram allows  a graphical illustration of these ideas.
	Consider first the example of a system $A$  consisting of a battery and an ideal electric motor  on whose shaft a weight $B$ hangs as shown in Figure \ref{LavoroES4}. At time $t_1$ the battery is charged and the state is  $A_1$. Between $t_1$ and $t_2$ the motor, connected to the battery terminals, is activated, the weight is raised and the system reaches  state $A_2$. Between $t_2$ and $t_3$ the battery is disconnected from the engine and  system $A$ remains perfectly isolated. However the battery discharges internally and the system reaches the state $A_3$ in which the battery is completely discharged. 
	It is clear that the mechanism that causes the internal discharge of the battery is always active. If its speed  is much lower than the speed with which the lifting of the weight occurs, then the sequence of states is the broken one that passes through  state $A_2$. If instead  the internal discharge and the weight lifting  occur at comparable speeds and therefore proceed simultaneously, then the states between $A_1$ and $A_3$ follow a curved path,  as shown in Figure \ref{LavoroES4}.

\begin{figure}[!ht] 
		\begin{center} \includegraphics[scale=0.3]{ 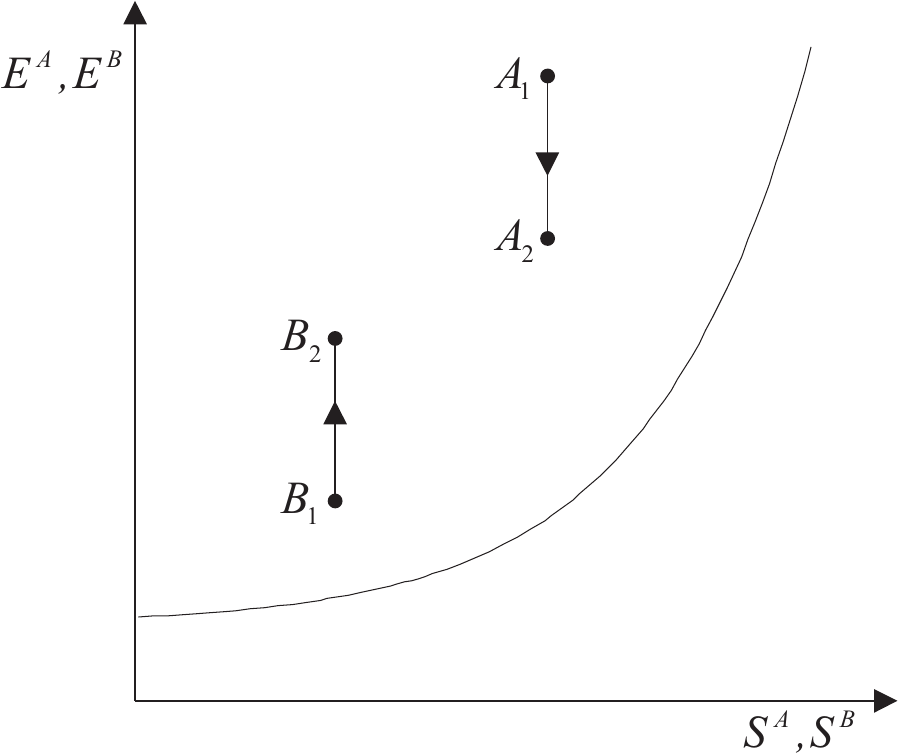}\hskip0.1mm \includegraphics[scale=0.3]{ 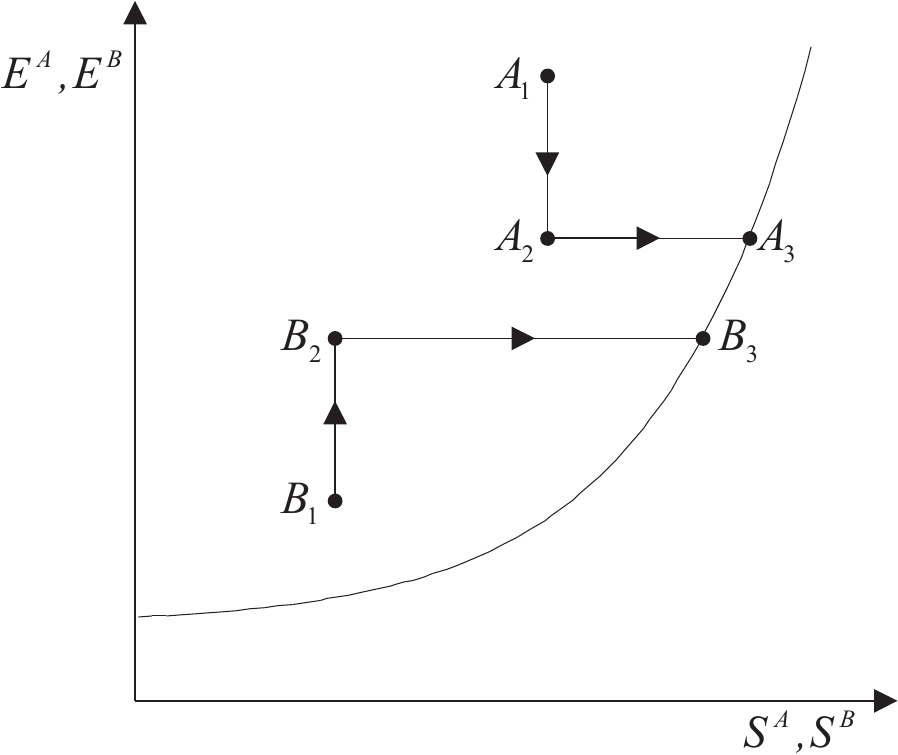}\hskip0.1mm \includegraphics[scale=0.3]{ 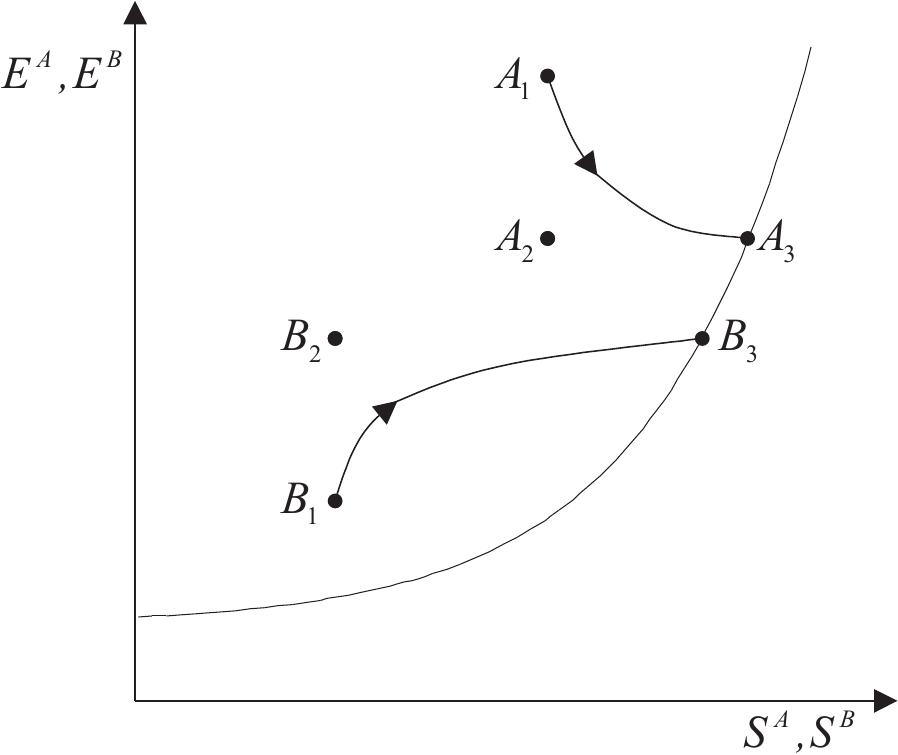} \caption{\label{LavoroES123}Superposed $E$--$S$ diagrams for two identical systems $A$ and $B$ exchanging energy by a work interaction. \textbf{Left}: the process is reversible and the systems end in nonequilibrium states. \textbf{Center}: the reversible process is followed by a spontaneous irreversible relaxation of each system to stable equilibrium state. \textbf{Right}: the  spontaneous irreversible relaxation toward stable equilibrium state starts and takes place simultaneously to the work interaction. } 
	\end{center} 
\end{figure}

	Consider a work interaction between two identical systems $A$ and $B$ with identical values of the amounts of constituents and parameters (volume, etc.). With this particular choice, the stable equilibrium state curves of the two systems are identical and we can superpose their $E$--$S$ diagrams on a single plot. Assume (Figure \ref{LavoroES123}) that states $A_1$ and $A_2$ have the same entropy, $S_2^A=S_1^A$, and the same holds for states $B_1$ and $B_2$, $S_2^B=S_1^B$. The entropy balances, $S^{A\ar}=S^A_{\rm irr}\ge 0$ for  system $A$ and  $S^{B\ar}=S^B_{\rm irr}\ge 0$ for  system $B$, imply that, if  $A$ and $B$ interact only with each other and not with other systems so that $ S^{B\ar}=S^{A\al}=-S^{A\ar}$, then $S^{B\ar}=S^{A\ar}=0$, i.e., the exchange of energy between the two systems is not accompanied by any exchange of entropy. It is a work interaction, with $W^{A\ar B}=E^{A\ar}= E^{B\al}$. 
	Graphically, the work is represented by the equal length of the vertical segments $A_1A_2$ and $B_1B_2$ on the diagram in Figure \ref{LavoroES123}. If the final states are $A_2$ and $B_2$ (as in Figure \ref{LavoroES123}-Left) the entropy balances also imply that  the  process is reversible (for both systems, $S^{\ar}=0$ and $S_2=S_1$ imply $S_{\rm irr}= 0$).

	However, since $A_2$ and $B_2$ are not stable equilibrium states, they will evolve spontaneously towards stable equilibrium  thus causing an irreversible generation of entropy.  For example, in Figure \ref{LavoroES123}-Center the spontaneous evolutions, start after the work interaction has ended, as the nonequilibrium states $A_2$ and $B_2$ relax towards the stable equilibrium states $A_3$ and $B_3$, respectively.  
	But the change of state from $A_1$ to $A_3$  can occur in many other ways, represented by different paths on these diagrams. The curved paths $A_1A_3$ and $B_1B_3$ in Figure \ref{LavoroES123}-Right show the possible paths when  the spontaneous relaxations towards stable equilibrium  occur in both systems simultaneously to the energy exchange by work interaction.

	We already noted that all weight processes are also adiabatic, since a weight has zero entropy and cannot accomodate any entropy transfer. Not all adiabatic processes, however, are weight processes. For example, if system $A$ has a work interaction with system $B$, as a result of which entropy is generated within $B$, the process for system $A$ is adiabatic but not a weight process, since the effects external to $A$ are not only mechanical. However, it can be shown that given any non-mechanical adiabatic process there always exists a weight process with the same initial and final states. 
	
	
	\section{\label{Nonwork}Non-work interactions}
	
	To begin the discussion of non-work interactions, let us introduce the symbol $\nW$ to denote non-work, i.e., the energy transferred  by means of a non-work interaction, and  recall that we call non-work any interaction in which in addition to energy transfer there is also an entropy transfer. The balance equations for system $A$ if it experiences a  {\it non-adiabatic process} with both work and non-work interactions,  become
	\begin{align}
		E^A_2 - E^A_1 &= W^{A\al}-\nW^{A\ar}&S^A_2 - S^A_1 &=-S^{A\ar}+ S^A_{\rm irr}\\
		\diff E^A &=  \delta W^{A\al}-\delta \nW^{A\ar} &\diff S^A &= -\delta S^{A\ar} + \delta S^A_{\rm irr} \\
		\diff E^A/\diff t &=  \dot W^{A\al}-\dot \nW^{A\ar} & \diff S^A /\diff t&=-\dot S^{A\ar}+ \dot S^A_{\rm irr}
	\end{align}
	In the next subsections we show that in non-work interactions the initial states of the interacting systems determine the range of  values of the entropy transfer that allow a given energy transfer. 
	Notice that for a cyclic process ($E_2-E_1=0$ and $S_2-S_1=0$) we have $\nW^{A\ar}= W^{A\al}$ and  $S^{A\ar}= S^A_{\rm irr}$. Similarly, at steady state ($\diff E^A/\diff t=0$ and $\diff S^A /\diff t=0$) we have $\dot \nW^{A\ar}=\dot W^{A\al}$ and $\dot S^{A\ar}=\dot S^A_{\rm irr}$.   In these special cases  the conditions $ S^A_{\rm irr}\ge 0$ and $\dot S^A_{\rm irr}\ge 0$, imply that
	\begin{equation}\label{ClausiusInequalitiesNonwork}
		S^{A\ar}\Big|_
		{\substack{\rm cyclic\\ \rm process}}\ge 0\qquad  \dot S^{A\ar}\Big|_{\substack{\rm steady\\ \rm state}}\ge 0	 
	\end{equation}
	These relations are general forms of the so-called \textit{Clausius inequality} (we will see its traditional forms in Section \ref{Clausius_inequalities}).
	
	\section{Entropy transfer bounds in non-work interactions}\label{Exchange_S}

\begin{figure}[!ht]
		\begin{center}
			\includegraphics[scale=0.45]{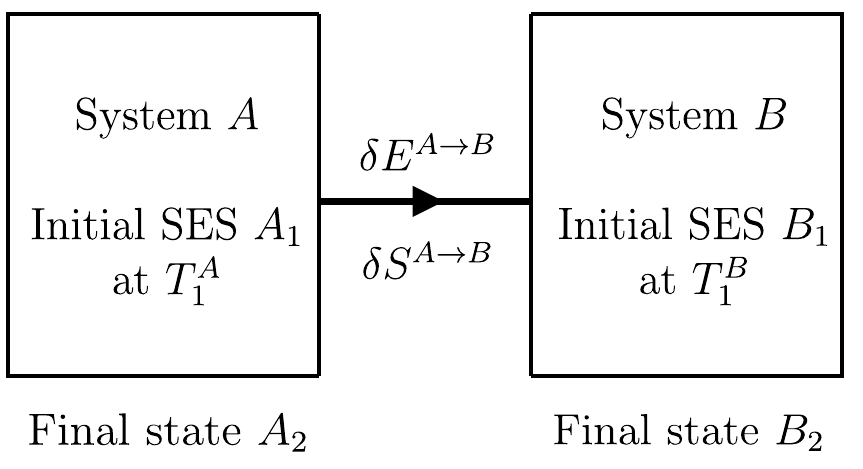}\\
			\includegraphics[scale=0.56]{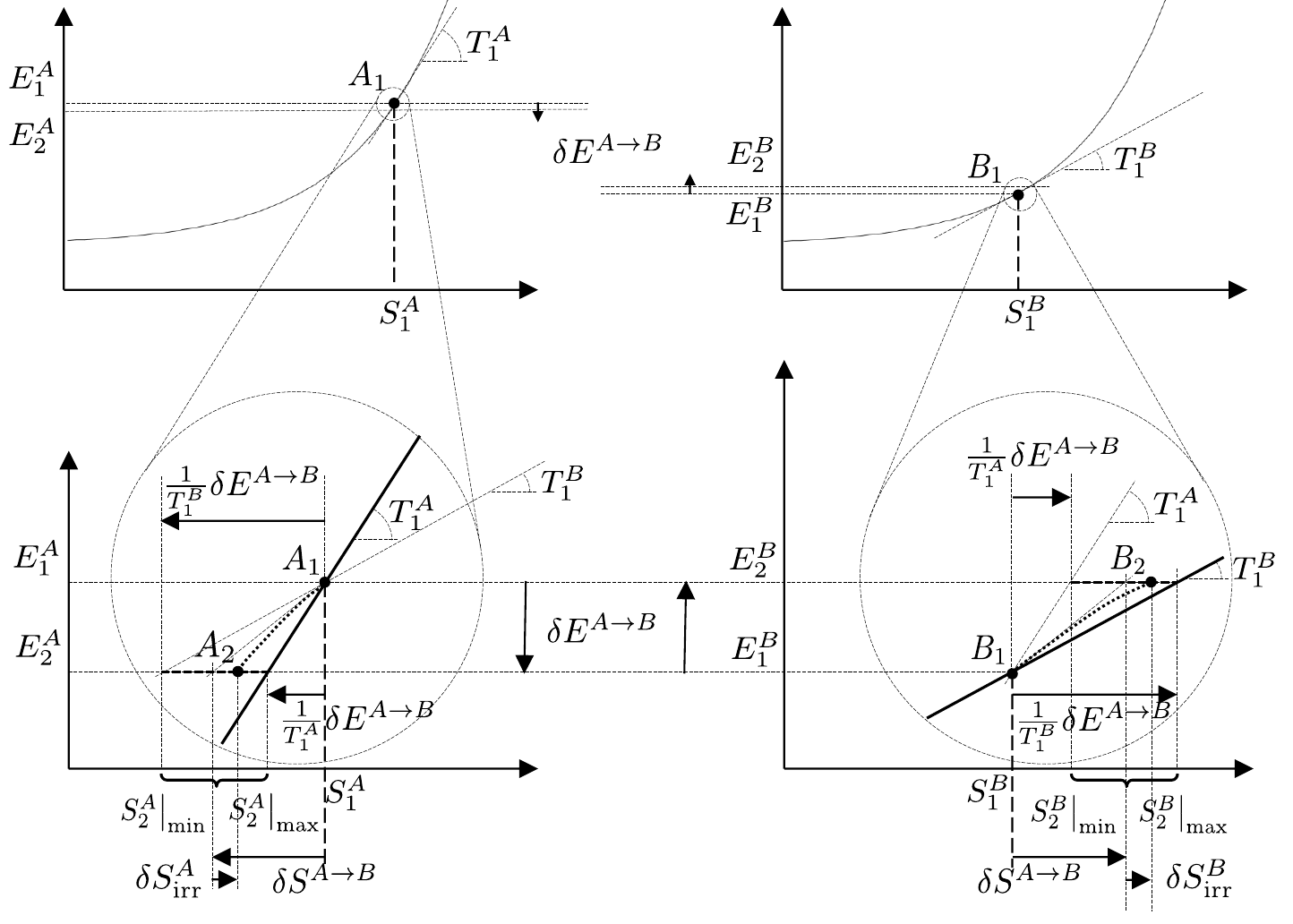}
			\caption{\label{HeatNoTest}Systems $A$ and $B$ are initially in stable equilibrium  and interact with each other (without leaving net  effects external to $AB$) by exchanging an infinitesimal amount $\delta E^{A\to B}$ of energy. Such exchange can occur only if  $\delta S^{A\to B}$ satisfies Relation \ref{Sscamcalore}. The $E$--$S$ diagrams in this Figure are quite complex and full of details, but once the derivation in this section is understood, they provide a graphical illustration of its various elements. The dashed lines represent the range of possible final states $A_2$ and $B_2$, respectively, while the dotted paths from $A_1$ to $A_2$ and from $B_1$ to $B_2$ represent one particular realization, compatible with Rel.~ \ref{Sscamcalore}, in which the systems, simultaneously to their energy and entropy exchange, relax toward stable equilibrium but at time $t_2$ have not reached yet the stable equilibrium state.  }
		\end{center}
	\end{figure}

	Before proceeding with the precise definition of heat interactions, let us  clarify an important point by considering two systems, $A$ and $B$ (Figure \ref{HeatNoTest}-Top), initially at different temperatures $T_1^A$ and $T_1^B$, which interact with each other directly (or are made to interact indirectly through some cyclic machinery $X$, but without leaving net  effects external to $AB$) in such a way as to exchange an amount of energy equal to $\delta E^{A\to B}$.

	For systems $A$ and $B$, the energy and entropy balances are
	\begin{align}
		\diff E^A&= -\delta E^{A\to B}&\quad
		\diff S^A&= -\delta S^{A\to B}+\delta\Sirr^A&\quad\delta\Sirr^A&\ge_{{}_{1A}} 0\\
		\diff E^B&= \delta E^{A\to B}&\quad
		\diff S^B&= \delta S^{A\to B}+\delta\Sirr^B&\quad\delta\Sirr^B&\ge_{{}_{1B}} 0
	\end{align}
	Moreover, the maximum entropy principle implies the inequalities\footnote{Rel.~\ref{maxSA} for system $A$ is proven as follows. The initial state  $A_1$ is a stable equilibrium state with energy $E^A_1$ and entropy $S^A_1$. If the energy changes by $\diff E^A$, the final state $A_2$ has energy $E^A_2=E^A_1+\diff E^A$. Among all the states with this energy, the stable equilibrium state has the maximum entropy, i.e., (assuming for simplicity that   parameters, $V^A$, and amounts of constituents, $\bn^A$, remain unchanged) $S^A_2=S_1^A+\diff S^A\le S^A_{2,\rm max}=S^A_{\rm SES}(E^A_1+\diff E^A,V^A,\bn^A)$  where the strict equality  holds only if also the final state $A_2$ is a stable equilibrium state. By Taylor expansion, $S^A_{2,\rm max}=S_1^A+\diff E^A/T_1^A+\hparder{{}^2 S_{\rm SES}}{E^2}{\bn,V}|^A_1(\diff E^A)^2/2+\cdots \le S_1^A+\diff E^A/T_1^A$, where the strict equality  holds only if the second derivative  of the fundamental relation (otherwise always negative) vanishes, such as for a thermal reservoir or two-phase states. Combining these relations yields Rel.~\ref{maxSA} for $A$, and similarly,  Rel.~\ref{maxSB} for $B$.}
	\begin{align}
	\diff S^A &\underinequality{\le}{2A} \frac{\diff E^A}{T_1^A} + \frac{1}{2}\left.\vparder{{}^2 S_{\rm SES}}{E^2}{\bn,V}\right|^A_1(\diff E^A)^2 \,\underinequality{\le}{3A} \frac{\diff E^A}{T_1^A} \label{maxSA}\\  \diff S^B &\underinequality{\le}{2B} \frac{\diff E^B}{T_1^B} + \frac{1}{2}\left.\vparder{{}^2 S_{\rm SES}}{E^2}{\bn,V}\right|^B_1(\diff E^B)^2\, \underinequality{\le}{3B} \frac{\diff E^B}{T_1^B} \label{maxSB}
	\end{align}

	Combining these relations (by eliminating $\diff E^A$, $\diff S^A$, $\diff E^B$, $\diff S^B$), yields
	\begin{equation}
	-\delta S^{A\to B}+\delta\Sirr^A  \underinequality{\le}{2A,3A} -\delta E^{A\to B}/T_1^A \qquad \text{and}\qquad
	\delta S^{A\to B}+\delta\Sirr^B  \underinequality{\le}{2B,3B} \delta E^{A\to B}/T_1^B
	\end{equation}
	and, solving for $\delta S^{A\to B}$, we obtain the following important train of inequalities\footnote{Strict equality 2A,3A holds if the final state $A_2$ of $A$ is a stable equilibrium state (2A) and if $\hparder{{}^2 S_{\rm SES}}{E^2}{\bn,V}|^A_1=0$ (3A), for example if $A$ is a thermal reservoir.  Strict equality 1A holds if $\delta\Sirr^A=0$, i.e., no   entropy generation within system $A$.  Strict equality 1B holds if $\delta\Sirr^B=0$, i.e., no   entropy generation within system $B$. Strict equality 2B,3B holds if the final state $B_2$ of $B$ is a stable equilibrium state (2B) and if $\hparder{{}^2 S_{\rm SES}}{E^2}{\bn,V}|^B_1=0$ (3B), for example if $B$ is a thermal reservoir. Clearly, for $T_1^A\ne T_1^B$ it is impossible that all equal signs apply. For example, if both $A$ and $B$ are thermal reservoirs and their final states $A_2$ and $B_2$ are stable equilibrium states, Relations \ref{Sscamcalore} imply that either $\delta\Sirr^A>0$ or $\delta\Sirr^B>0$, or both.}
	\begin{equation}
	\frac{\delta E^{A\to B}}{T_1^A}\, \underinequality{\le}{2A,3A} \, \delta S^{A\to B}-\delta\Sirr^A \,\underinequality{\le}{1A}\, \delta S^{A\to B} \,\underinequality{\le}{1B} \,\delta S^{A\to B}+\delta\Sirr^B\, \underinequality{\le}{2B,3B} \,\frac{\delta E^{A\to B}}{T_1^B} \label{Sscamcalore}
	\end{equation}
	from which it is observed that, short of additional conditions, there is no unique relationship between the exchanged entropy $\delta S^{A\to B}$, the exchanged energy $\delta E^{A\to B}$, and the initial temperatures $T_1^A$ and $T_1^B$. In other words, $\delta S^{A\to B}$ can range from $\delta E^{A\to B}/T_1^A$ to $\delta E^{A\to B}/T_1^B$.  
	
	The $E$--$S$ diagrams in  Figure \ref{HeatNoTest}  provide a graphical illustration of the various elements of the above derivation. The dashed lines represent the range of possible final states $A_2$ and $B_2$, respectively, while the dotted paths (from $A_1$ to $A_2$ and from $B_1$ to $B_2$) represent one particular realization, compatible with Rel.~\ref{Sscamcalore}, in which the systems, simultaneously to their energy and entropy exchange, relax toward stable equilibrium but at time $t_2$ have not  reached yet the stable equilibrium state. 
	
	It is important to note that in the very special limiting cases in which the initial temperatures of $A$ and $B$ are very close, i.e., for $T_1^A\to T_1^B$, the range of possible values of $\delta S^{A\to B}$ defined by Rel.~\ref{Sscamcalore}  shrinks to a single value. Furthermore, in order for all equality signs to hold in this limit, it is necessary for both $A$ and $B$ to end up in stable equilibrium states, and $\delta\Sirr^A = \delta\Sirr^B= 0$. 		In Section \ref{Heat} we will prove   that this  limiting situation is important and is precisely what characterizes a heat interaction, because then and only then the non-work interaction is entirely distinguishable from work.
	But before that, we discuss other important results that follow from  Relation \ref{Sscamcalore}. 
	
	For example, if $\delta E^{A\to B}$ is negative, using the identities $\delta E^{A\to B}=-\delta E^{A\gets B}$ and $\delta S^{A\to B}=-\delta S^{A\gets B}$ Rel.~\ref{Sscamcalore} is more conveniently rewritten in the equivalent form
    \begin{equation}
	\frac{\delta E^{A\gets B}}{T_1^A} \,\underinequality{\ge}{2A,3A} \, \delta S^{A\gets B}+\delta\Sirr^A \,\underinequality{\ge}{1A}\, \delta S^{A\gets B}\, \underinequality{\ge}{1B} \,\delta S^{A\gets B}-\delta\Sirr^B \,\underinequality{\ge}{2B,3B}\, \frac{\delta E^{A\gets B}}{T_1^B} \label{Sscamcaloreneg}
	\end{equation}
	The direct reading of either  Rel.~\ref{Sscamcalore} or \ref{Sscamcaloreneg} yields the following general conclusion (theorem): 
	\textit{two systems initially in stable equilibrium states with different temperatures that interact with each other and nothing else cannot exchange energy without a simultaneous exchange of entropy, unless their temperatures have opposite signs}. For positive temperatures, say $T_1^A>T_1^B>0$, to accomplish  a given energy transfer $\delta E^{A\to B}$ the interaction must produce also an entropy transfer $\delta S^{A\to B}$, at least equal $\delta E^{A\to B}/T_1^A$  but not more than $\delta E^{A\to B}/T_1^B$. Said differently, a work interaction ($\delta S^{A\to B}=0$) between $A$ and $B$ under these conditions is impossible for temperatures of the same sign, whereas for initial temperatures of opposite signs can occur only if the work is in the direction from the system with higher value of $-1/T$ into the one with lower value.

	\section{Clausius statement of the second law (proof)}\label{ClausiusStatement}
	
	The train of inequalities in Rel.~\ref{Sscamcalore} provides, among other things, the `proof'\footnote{Often, in traditional expositions, one assumes Clausius' statement as the statement of the second law, after having introduced heuristic definitions of temperature and heat. Clearly, in any axiomatic exposition, the postulated statement of the second law cannot be proved, as it is taken as the starting point of the deductive structure. In our approach,  the statement of the second law  that we postulate as the starting point is the Hatsopoulos-Keenan statement. In our context, the Clausius' statement emerges as a theorem, and here we provide the rigorous proof, without using the definition of heat.} of Clausius' statement (1850) of the second law of thermodynamics\ind{Clausius!second law statement}\ind{second law of thermodynamics!Clausius statement}, which states that \textit{a process that has as its only effect the transfer of energy from a system in a stable equilibrium state with positive temperature to another at a higher temperature is  not possible, not even if the energy transfer is infinitesimal}.

	The proof follows directly from Rel.~\ref{Sscamcalore} or the equivalent \ref{Sscamcaloreneg}. By focusing on the extreme sides of these inequalities and collecting $\delta E^{A\to B}$ ,we obtain
	\begin{equation}\label{Clausius1}
	\bigg(\frac{1}{T_1^A}-\frac{1}{T_1^B}\bigg)\,\delta E^{A\to B} \le 0 
	\qquad \mbox{or} \qquad 
	\bigg(\frac{1}{T_1^A}-\frac{1}{T_1^B}\bigg)\,\delta E^{A\gets B} \ge 0 
	\end{equation}
	From this follows that the interaction with $\delta E^{A\to B}>0$ is possible only if the temperatures are such that $-1/T_1^A \ge -1/T_1^B$ , i.e., only if either $ T_1^A\ge T_1^B\ge 0$ (both $A$ and $B$ are normal systems, i.e., their stable equilibrium states have positive temperatures) or $ T_1^A\le T_1^B\le 0$ (both $A$ and $B$ are  special systems and are both in stable equilibrium states with negative temperatures) or  $ T_1^A\le 0$ and  $T_1^B\ge 0$  ($A$ is a special system in a negative-temperature stable equilibrium state, while $B$ is in a positive-temperature stable equilibrium state). 
	
	For positive temperatures, which is the only possibility for almost all practical systems of engineering interest,  these inequalities simplify to
	\begin{equation}\label{Clausius2}
	(T_1^A- T_1^B)\,\delta E^{A\to B} \ge 0 \qquad \mbox{or}  \qquad (T_1^B- T_1^A)\, \delta E^{A\gets B} \ge 0
	\end{equation}
	from which it is easier to see that $\delta E^{A\to B}$ can be positive, and thus the flow of energy can be in the direction from system $A$ to system $B$ only  if $T_1^A\ge T_1^B$, i.e., if $A$ is ``warmer'' than $B$.
	From this result, it emerges that, in the realm of normal systems, \textit{temperature measures the tendency of a system in a stable equilibrium state to give up energy}.\footnote{More generally, without assuming that temperatures are positive, Rel.~\ref{Clausius1} implies that $-1/T$ can be interpreted as a `potential' that measures the tendency to give up energy. In quantum information technologies, it is common to have to deal with systems with a finite number of energy levels, such as spin systems or polarized photons which, because the energy is upper bounded, have both positive ad negative temperature stable equilibrium states.  For these systems, the potential  $-1/T$ ranges from $-\infty$ (zero absolute positive temperature, the ``coldest'' stable equilibrium state) to $+\infty$ (zero absolute negative temperature,  the ``hottest'' stable equilibrium state), passing through zero (the maximal-entropy stable equilibrium state).}\ind{quantum thermodynamics}\ind{negative temperatures}\ind{temperature!negative}  
	
	The $E$--$S$ diagrams in  Figure \ref{HeatNoTest}  provide a graphical illustration of the reasons why the Clausius' statement holds true. The dashed lines in the diagrams show the range of possible final states $A_2$ and $B_2$. For  $\delta E^{A\to B}>0$, i.e., to transfer energy out of system $A$, we must transfer out of $A$ (and therefore into $B$) also at least $\delta E^{A\to B}/T_1^A$ of entropy. But the maximum entropy that $B$ can accomodate is $\delta E^{A\to B}/T_1^B$.
	
\begin{figure}[!ht]
	\begin{center}
		\includegraphics[scale=0.57]{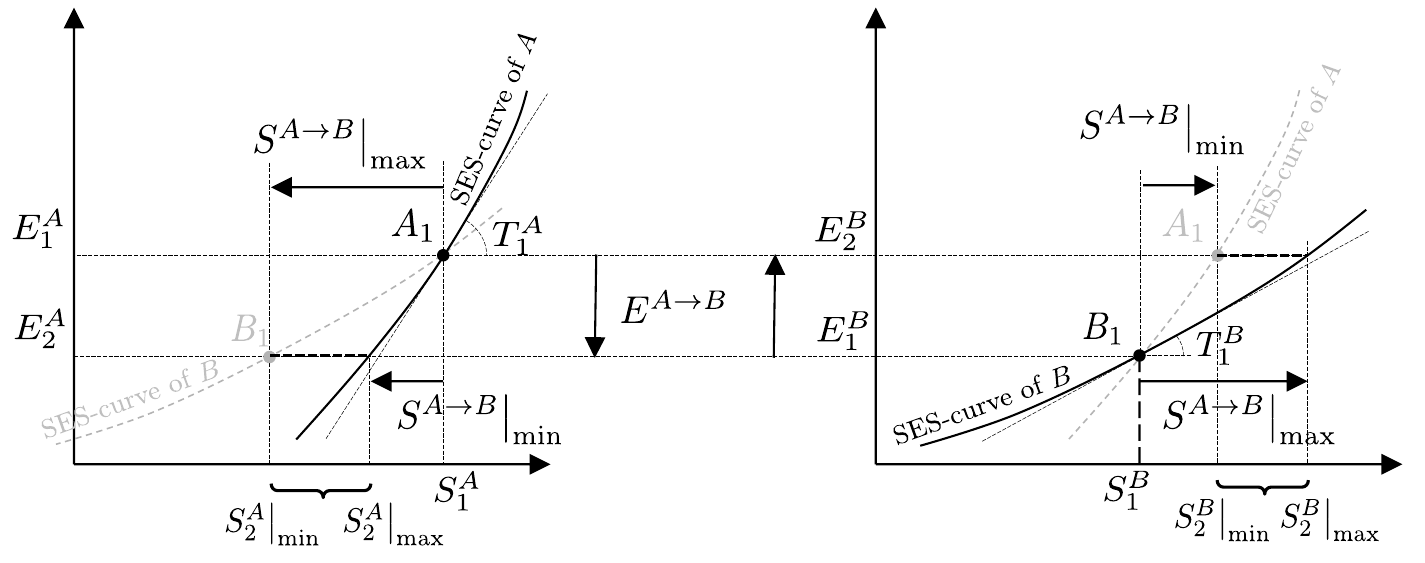}
		\caption{\label{NonWorkESdiagramsFinite}Systems $A$ and $B$ are initially in stable equilibrium states and interact with each other (without leaving net  effects external to $AB$) by exchanging a finite amount $E^{A\to B}$ of energy. Such exchange can occur only if there is also an entropy transfer $ S^{A\to B}$, at least $S^{A\to B}|_{\rm min}$ but no more than $S^{A\to B}|_{\rm max}$. } 
		\end{center}
	\end{figure}
	
	Similar limitations, but more restrictive, apply in general if the energy transfer is to be finite, $E^{A\to B}$. The $E$--$S$ diagrams in Figure \ref{NonWorkESdiagramsFinite} show the graphical constraints, which depend on the fundamental relations of the two systems. In fact, to  transfer energy out of system $A$, we must transfer out of $A$ (and therefore into $B$) also at least the amount  of entropy needed to reduce its entropy at or below the maximum value possible for its final energy $E_2^A=E_1^A-E^{A\to B}$. At most, system $B$ can accomodate the amount  of entropy needed to end in the stable equilibrium state with its final energy $E_2^B=E_1^B+E^{A\to B}$. Therefore, the ranges of possible final states are as shown by the dashed lines in the Figure. The generalization  of Relation \ref{Sscamcalore} to this case is (dropping superscripts on $\bn$ and $V$ for compactness)
	\begin{equation}\label{SscamcaloreFinite}
	S_{\rm SES}^A(E_1^A,V,\bn)-S_{\rm SES}^A(E_1^A-E^{A\to B},V,\bn)
	\le S^{A\to B}\le 
	S_{\rm SES}^B(E_1^B+E^{A\to B},V,\bn)-S_{\rm SES}^B(E_1^B,V,\bn)
	\end{equation}
	which entails, albeit implicitly through the fundamental relations of the two interacting systems, a restriction on how much energy $E^{A\to B}$ the two systems can exchange for the given initial stable equilibrium states as well as the lower and upper bounds on the entropy $ S^{A\to B}$ that must and can be transferred for a given energy transfer $E^{A\to B}$. Note that for a normal system both entropy bounds have the same sign as $E^{A\to B}$  and, therefore, also $ S^{A\to B}$ has the same sign, meaning that the  entropy transfer is in the same direction as the energy transfer.

	\section{Clausius statement of the second law extended to nonequilibrium}\label{ClausiusStatement_NE}
	
	The foregoing result allows us to extend Clausius' statement of the second law to  a process that has as its only effect the transfer of  energy and entropy  between subsystems $A$ and $B$ that start in nonequilibrium states $A_1$ and $B_1$. Repeating the same procedure we used to derive  Relation \ref{SscamcaloreFinite}, it is easy to show that for a finite transfer of energy  $E^{A\to B}$ we end up again with the same inequalities
	\begin{equation}
	S_1^A-S^A_{\rm SES}(E^A_1-E^{A\to B},V^A,\bn^A) \le S^{A\to B} \le S^B_{\rm SES}(E^B_1+E^{A\to B},V^B,\bn^B) - S_1^B\label{SscamcaloreNES1}
	\end{equation}
	where, however, the left hand side can be negative and therefore the direction of net entropy transfer may be zero or even opposite to that of the energy transfer.  
	This may occur when the state of the system that yields energy to the other is sufficiently far from stable equilibrium. More precisely, for $E^{A\to B}>0$, this occurs when
	\begin{equation}
	D_1^A \ge  S^A_{\rm SES}(E^A_1,V^A,\bn^A) - S^A_{\rm SES}(E^A_1-E^{A\to B},V^A,\bn^A)\label{verydistantNESA}
	\end{equation}
	or, for $E^{A\to B}<0$, when 
	\begin{equation}
	D_1^B \ge  S^B_{\rm SES}(E^B_1,V^B,\bn^B) - S^B_{\rm SES}(E^B_1+E^{A\to B},V^A,\bn^A)\label{verydistantNESB}
	\end{equation}
	where  $D_1^A$ and $D_1^B$ denote
	the (nonnegative) ``distances from stable equilibrium'' of the initial nonequilibrium states  $A_1$ and $B_1$, respectively defined by
	\begin{equation}
	D_1^A = S^A_{\rm SES}(E^A_1,V^A,\bn^A) - S_1^A\qquad\text{and}\qquad D_1^B =  S^B_{\rm SES}(E^B_1,V^B,\bn^B) - S_1^B\label{distanceNES}
	\end{equation}
	Clearly, if the net entropy transfer may be zero, it means that the energy exchange  $E^{A\to B}$ can be done by means of a work interaction. 
	
	If the transfer of energy is infinitesimal, Rel.~\ref{SscamcaloreNES1} becomes
	\begin{equation}
	\frac{\delta E^{A\to B}}{T^A_{\rm SES}(E^A_1,V^A,\bn^A)}-D_1^A \le   S^{A\to B} \le \frac{\delta E^{A\to B}}{T^B_{\rm SES}(E^B_1,V^B,\bn^B)}+D_1^B \label{SscamcaloreNES2}
	\end{equation}
	which of course, reduces to Rel.~\ref{Sscamcalore} if the initial states are stable equilibrium  ($D_1^A=D_1^B=0$). But in general, if the distances from stable equilibrium are finite, the infinitesimal terms on the rhs and lhs can be neglected, leaving 
	\begin{equation}
	-D_1^A \le S^{A\to B} \le D_1^B
	\end{equation}
	which involves no approximation if the energy transfer is exactly zero and  means that  in principle it is possible to achieve ``pure entropy transfer interactions'' ranging from the two extremes whereby on the one end $A$ is placed in a stable equilibrium state by receiving from $B$ an amount $S^{A\gets B}=D_1^A$, and on the other end  $B$  is placed in a stable equilibrium state by receiving from $A$ an amount $S^{A\to B}=D_1^B$. 
	
	Similar conclusions can be drawn from Rel.~\ref{SscamcaloreNES1} for finite values of $E^{A\to B}$. Focusing on the extreme sides of those inequalities we obtain (dropping the dependence on $\bn$ and $V$ for compactness)
	\begin{equation}
	S_1^A-S^A_{\rm SES}(E^A_1-E^{A\to B})  \le S^B_{\rm SES}(E^B_1+E^{A\to B}) - S_1^B\label{SscamcaloreNES3}
	\end{equation}
	or equivalently, in terms of $D_1^A$ and $D_1^B$,
	\begin{equation}
	S^A_{\rm SES}(E^A_1)-S^A_{\rm SES}(E^A_1-E^{A\to B})-D_1^A \le S^B_{\rm SES}(E^B_1+E^{A\to B})-S^B_{\rm SES}(E^B_1)+D_1^B \label{Sscamcalorenonequilibrium states4}
	\end{equation}
	This represents the extension of Clausius' statement of the second law to nonequilibrium states. Indirectly, through the fundamental stable-equilibrium-state relations of the two systems, it entails a bound on the amount of energy that can be transferred from $A$ to $B$ for the given initial nonequilibrium states  $A_1$ and $B_1$.

	\section{Heat interactions. Definition}\label{Heat}
	
	A special limiting class of non-work interactions between two systems that are initially in stable equilibrium states is one that can be \textit{completely distinguished from a work interaction}.  Here we will define precisely what we mean by this and prove that it may happen only in the limiting situation in which the difference in the initial temperatures of the interacting systems vanishes. In such limit,  the ratio of the exchanged energy to the exchanged entropy  equals the initial temperature of either system. This is called a \textit{heat interaction},\ind{interaction!heat}\ind{heat!definition} and the resulting energy exchanged is called \textit{heat}.

	\begin{figure}[!ht]
		\begin{center}
			\includegraphics[scale=0.45]{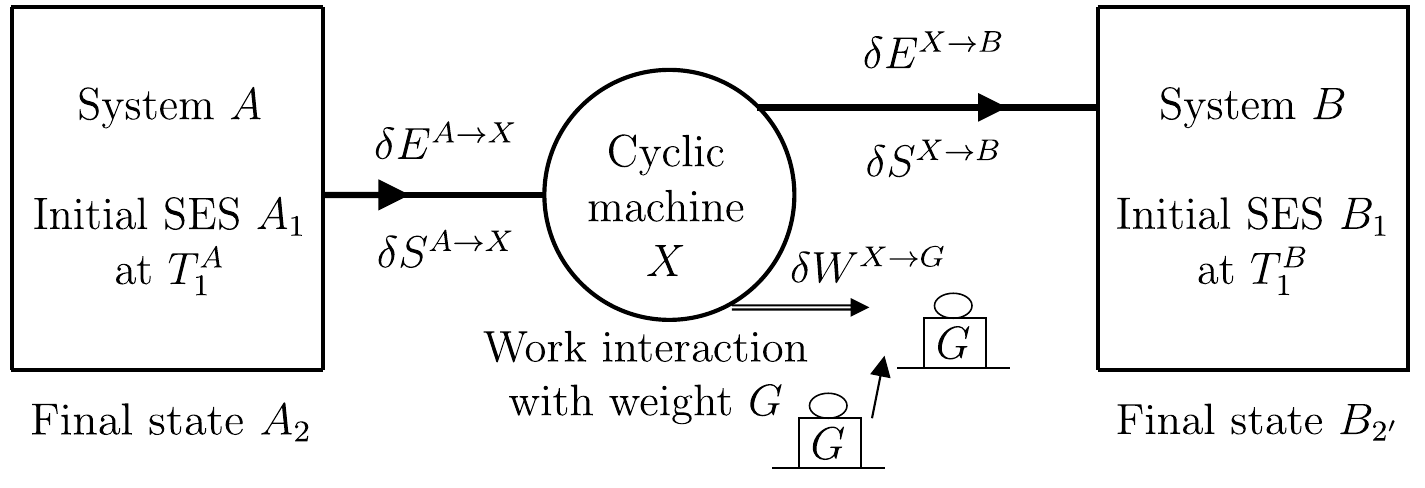}
			\caption{\label{HeatTest1}The cyclic machine $X$ interposed between interacting systems $A$ and $B$ intercepts the energy and entropy they exchange and attempts to channel as much energy as possible into lifting a weight $G$. This lifting  becomes impossible in the limit as $T_1^A\to T_1^B$. In this limit, the non-work interaction between $A$ and $B$ is a heat interaction.}
		\end{center}
	\end{figure}
	
	\begin{figure}[!ht]
			\begin{center}
			\includegraphics[scale=0.45]{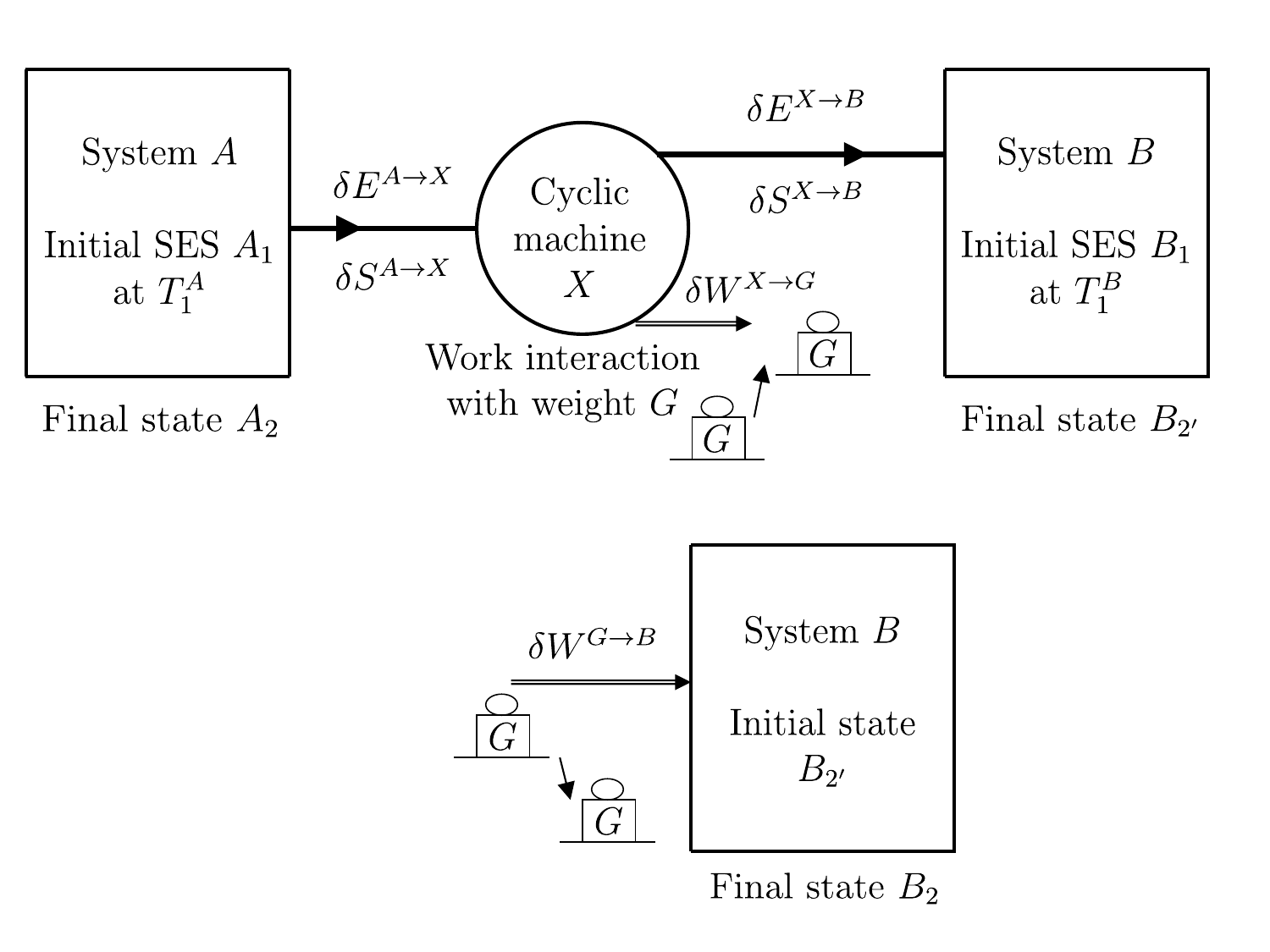}
			\caption{\label{HeatTest2} If the initial conditions of $A$ and $B$ are such that the machine $X$ in Figure \ref{HeatTest1} can transfer  a non-negligible amount of energy $\delta W^{X\to G}$ to the weight $G$, the interaction between $A$ and $B$ is not heat. If that energy is then given to $B$ by means of a work interaction, the final effects on $B$ are the same as in Figure \ref{HeatNoTest}. The energy received by $B$ from the weight is clearly identifiable as work. Therefore, the  machine $X$ has been able to split the energy transferred from $A$ to $B$ so that a finite fraction is work. When this is possible, the non-work interaction between $A$ and $B$ is \textit{not} heat.}
			\end{center}
	\end{figure}

	To do this, consider again the interaction between $A$ and $B$ sketched in Figure \ref{HeatNoTest} and suppose that we operate it as the result of the sequence of two separate processes with the assistance of a \textit{stationary} or \textit{cyclic} machine\ind{machine!stationary}\ind{machine!cyclic}\footnote{A \textit{stationary machine} is a system that, while interacting with other systems, always remains in the same state. A \textit{cyclic machine} is a system that, while interacting with other systems, periodically returns to its initial state. Note that the definition of a cyclic machine includes the case of a stationary machine.} $X$ interposed between $A$ and $B$ as sketched in Figure \ref{HeatTest1}.\footnote{It is important to emphasize that the cyclic machine
		introduced here is not an `engine' in the sense of a practical application. Rather, it serves as a conceptual measuring device --- a test system --- that allows us to operationally determine if a  non-work interaction can be decomposed into a work part and an irreducible non-work part.} In the first process, machine $X$ receives from $A$ the amounts of energy $\delta E^{A\to X}$ and entropy $\delta S^{A\to X}$ respectively equal to the amounts $\delta E^{A\to B}$ and $\delta S^{A\to B}$ that in Figure \ref{HeatNoTest} pass directly  from $A$ to $B$. Machine $X$ uses them to attempt to separate part of the energy received from $A$ and to store  it temporarily by lifting a weight $G$. System $A$ ends up in the same state $A_2$ as in Figure \ref{HeatNoTest}, while system $B$ ends up in a different state $B_{2'}$. In the second process (Figure \ref{HeatTest2}), the weight transfers to $B$ the energy it received from the machine, and $B$ ends up in the same state $B_2$ as in Figure \ref{HeatNoTest}. In the end, the sequence of  two processes has the same net effects and, therefore, is equivalent to  the direct exchange process in Figure \ref{HeatNoTest}. However, if the initial conditions of $A$ and $B$ are such that the amount of energy passed to the weight $G$ is non-negligible, we must conclude that a finite fraction of the energy received by $B$ (specifically, the energy it receives from the weight in the second process in Figure \ref{HeatTest2}) is clearly identifiable as work. In this case, the non-work interaction between $A$ and $B$ is \textit{not} heat.
	
	To calculate what fraction of the energy $\delta E^{A\to X}$ the machine $X$ can transfer to the weight $G$, we write the energy balance for  $X$ and the entropy balance for the composite system $AXB$ in the process of Figure \ref{HeatTest1}, recalling that by definition $\diff S^X=0$, 
	\begin{equation}\label{heatdef1}
	0 = \delta E^{A\to X} - \delta W^{X\to G} - \delta E^{X\to B}
	\qquad
\diff S^A + \diff S^B = \delta \Sirr^{AXB} \ge 0
	\end{equation}
	The energy balances for system $A$ and system $B$ (before receiving the work $\delta W^{G\to B}$) are
	\begin{equation}\label{heatdef2}
	\diff E^A = -\delta E^{A\to X} \qquad \text{and} \qquad \diff E^B = \delta E^{X\to B}
	\end{equation}
	Furthermore, as previously seen, the principle of maximum entropy and the fact that initially $A$ and $B$ are is stable equilibrium states requires that
	\begin{equation}\label{heatdef3}
	\diff S^A \le \frac{\diff E^A}{T_1^A} \qquad \text{and} \qquad \diff S^B \le \frac{\diff E^B}{T_1^B}
	\end{equation}
	where the strict equalities hold only if they also end in stable equilibrium states. Combining these relations, we obtain
	\begin{equation}\label{WXtoG}
	\delta W^{X\to G} \le  \bigg(1 - \frac{T_1^B}{T_1^A}\bigg)\delta E^{A\to X} - T_1^B\, \delta \Sirr^{AXB}
	\end{equation}
	where the equality sign holds only if both $A$ and $B$ end in stable equilibrium states. In the best-case scenario, namely,  if the machine $X$ operates reversibly ($\delta\Sirr^{AXB} = 0$) and both $A$ and $B$ end in stable equilibrium states, the fraction of the energy that $X$ receives  from $A$ manages to send to $G$ is equal to
	\begin{equation}\label{WXtoGmax}
	\frac{\delta W^{X\to G}_{\rm max}}{\delta E^{A\to X}} = 1 - \frac{T_1^B}{T_1^A}
	\end{equation}
This result is interesting in itself, because it proves the famous Carnot expression for the maximum work. But here, its importance is in showing that if $T_1^A \neq T_1^B$, this fraction assumes finite values and therefore we conclude that a finite fraction of the exchanged energy $\delta E^{A\to B}$ can be separated as work.
	
	However, in the limit as $T_1^A\to T_1^B$, we have
	\begin{equation}
	\lim_{T_1^A\to T_1^B} \frac{\delta W^{X\to G}_{\rm max}}{\delta E^{A\to B}} = 0
	\end{equation}
	In this limit, i.e., in practical terms when the temperatures $T_1^A$ and $T_1^B$ differ by at most an infinitesimal amount, $(T_1^A - T_1^B) \ll T_1^B$, machine $X$, even under the best conditions, cannot possibly spit the energy transferred between $A$ and $B$ so that a finite fraction is work. These are the limiting conditions that define a  ``heat interaction,'' i.e., a non-work interaction entirely distinguishable from work, in which the interacting systems  exchange energy and entropy, but no constituents nor volume. The extension of this same logic and definition to interactions that in addition to   energy and entropy exchange also volume and constituents is discussed in the following sections. 
	
Rel.~\ref{Sscamcalore} implies that in the limit as  $T_1^A\to T_Q$ and  $T_1^B\to T_Q$ the difference between the upper and lower bounds to allowed values of $\delta S^{A\to B}$ for a given value of $\delta E^{A\to B}$ becomes vanishing, therefore, the range of possible values of $\delta S^{A\to B}$ squeezes to a single value
	\begin{equation}\label{squeezed}
	\frac{\delta E^{A\to B}}{T_Q}=\lim_{T_1^A\to T_Q}\frac{\delta E^{A\to B}}{T_1^A} \le \delta S^{A\to B}\le \lim_{T_1^B\to T_Q} \frac{\delta E^{A\to B}}{T_1^B} =  \frac{\delta E^{A\to B}}{T_Q}
	\end{equation}
	In this limit the relationship between the energy and entropy exchanged from $A$ to $B$ is uniquely determined by the temperature $T_Q$ at which the heat interaction occurs, where $T_Q$ denotes the nearly common value of the initial temperatures $T_1^A$ and $ T_1^B$ of the two interacting systems. The  exchanged energy is called ``heat'' and is traditionally denoted by the symbol $Q^\to$ instead of $E^\to$, i.e., $\delta Q^{A\to B}$ instead of $\delta E^{A\to B}$, so that Rel.~\ref{squeezed} reduces to the famous relation
	\begin{equation}\label{entropyHeat}
	\delta S^{A\to B} = \frac{\delta Q^{A\to B}}{T_Q}
	\end{equation}
	The ratio of the energy and entropy exchanged in a heat interaction is equal to the  temperature at which the interaction occurs. 
	
\begin{figure}[!ht]
		\begin{center}
			\includegraphics[scale=0.45]{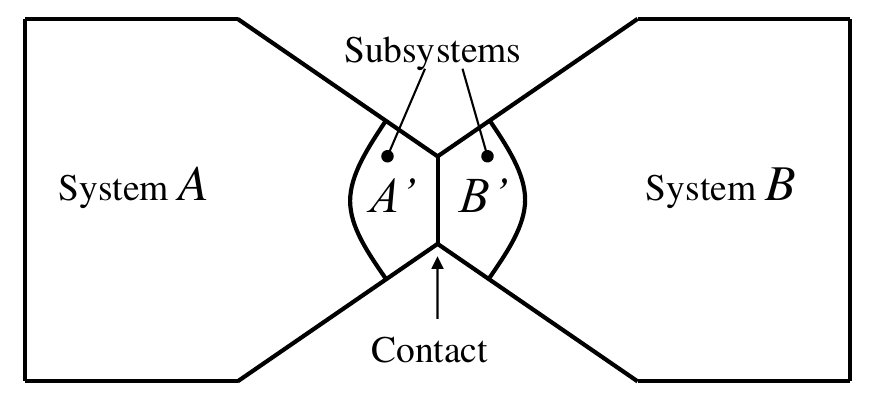}
			\caption{\label{Heat3}The interacting systems $A$ and $B$ are not in stable equilibrium state, but their respective subsystems $A'$ and $B'$ are in contact and have nearly identical temperatures, $T^{A'}\approx T^Q\approx T^{B'}$.}
		\end{center}
	\end{figure}

	Often in practical applications, as sketched in Figure \ref{Heat3}, a system $A$ may not be in a stable equilibrium state but can be modeled as the composition of multiple subsystems, one of which, $A'$, is in (or near) a stable equilibrium state  at temperature $T_Q$. Similarly, also system $B$ may consist of multiple subsystems, one of which, $B'$, is in (or near) a stable equilibrium state at a temperature that differs from $T_Q$ by an infinitesimal amount, $T_Q\pm \diff T$. If the two subsystems $A'$ and $B'$ undergo a heat interaction at temperature $T_Q$ then we generalize the definition given above and say that  systems $A$ and $B$  undergo a heat interaction at temperature $T_Q$ across their contact through their respective subsystems $A'$ and $B'$, even if $A$ and $B$ do not start in stable equilibrium states.
	
	\section{The role of heat interactions in heat transfer modeling}
	Among the conditions that identify heat interactions, the most restrictive one appears to be the requirement that the temperature difference between the interacting systems be infinitesimal. In fact, the common notion of heat seems in contradiction with  this idea, since in everyday language we always refers to heat as the energy exchanged between systems at \textit{different} temperatures. However, when two bodies at different temperatures come into contact, the phenomenology is quite complex because the contact brings them into nonequilibrium states. The discipline that studies this ubiquitous phenomenon is traditionally called \textit{heat transfer}.
	
These nonequilibrium states are usually modeled by making the continuum, local quasi-equilibrium, and simple-system assumptions. We do not discuss these assumptions here in any detail, but it is well known that they allow to represent the two interacting bodies as composite systems made up of many small (infinitesimal volume) subsystems (fluid parcels, in fluid mechanics; material points, in solid mechanics), each in a nonequilibrium state not too far from stable equilibrium  so that, even if they are nonequilibrium states, some of their  properties can be approximated with those of the unique stable equilibrium state with the same energy and compatible amounts of constituents, that each small volume would  spontaneously reach if it were isolated from the adjacent volumes. Among these ``local'' properties is the temperature and as a result the model defines a generally continuous field of temperature that may vary with time. The interaction between adjacent small volumes through their surface of contact is well approximated by a heat interaction. Indeed, by continuity the temperatures assigned to adjacent infinitesimal elements of the continuum may differ only infinitesimally, thus fulfilling the very restrictive limiting condition that defines a heat interaction. 

\begin{figure}[!ht]
	\begin{center}
			\includegraphics[scale=0.5]{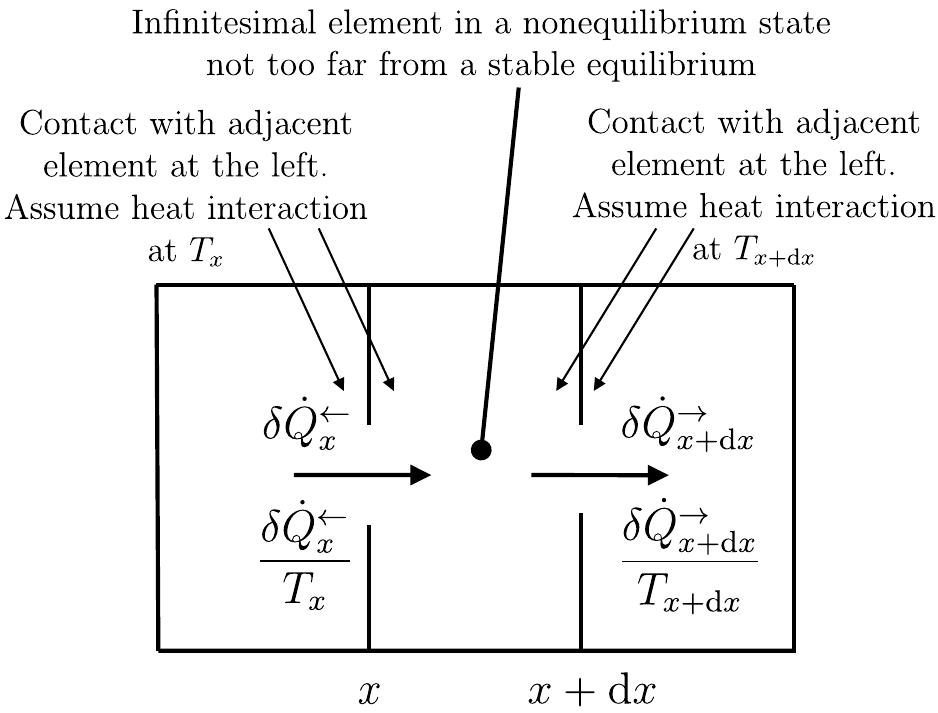}
			\caption{\label{HeatLamina}One-dimensional heat-transfer model, based on heat interactions, of an infinitesimal element of a continuum in contact with adjacent small volumes at slightly different temperatures. The infinitesimal element is in a nonequilibrium state not too far from stable equilibrium.	The attraction toward equilibrium resulting from the dissipative part of its internal  dynamics  produces spontaneous generation of entropy.}
	\end{center}
\end{figure}
	
	For simplicity, consider the one-dimensional case as schematized in Figure \ref{HeatLamina}, and the small volume between the surfaces at $x$ and $x+\diff x$, where the assigned local temperatures are $T$ and $T+\diff T$, respectively. Assuming only heat interactions, the small volume receives energy $\delta\dot Q_x^\al$ and entropy $\delta\dot Q_x^\al/T_x$ through the contact interface at temperature $T_x$ and gives away energy $\delta\dot Q^\ar_{x+\diff x}$ and entropy $\delta\dot Q^\ar_{x+\diff x}/T_{x+\diff x}$ through the one at $T_{x+\diff x}$.	The energy and entropy balances for the small volume are
		\begin{equation}
		\frac{\diff (\delta E)}{\diff t}=\delta\dot Q_x^\al-\delta\dot Q^\ar_{x+\diff x}
		\qquad
		\frac{\diff  (\delta S)}{\diff t}=\frac{\delta\dot Q_x^\al}{T_x}-\frac{\delta\dot Q^\ar_{x+\diff x}}{T_{x+\diff x}}+\delta\Sdotirr
	\end{equation}

	For example, if it is in a steady-state, i.e., its energy $\delta E$ and $\delta S$ remain constant in time,  the energy balance implies $\delta\dot Q^\ar_{x+\diff x}=\delta\dot Q_x^\al$ and the entropy balance may be rewritten as
	\begin{equation}
		\delta\Sdotirr= \delta\dot Q_x^\al \left(\frac{1}{T_{x+\diff x}}-\frac{1}{T_{x}}\right)= \delta\dot Q_x^\al \frac{\diff (1/T_x)}{\diff x}\diff x=-\delta\dot Q_x^\al \frac{1}{T_x^2}\frac{\diff T_x}{\diff x}\diff x
	\end{equation}
where we used the continuity of the temperature field  to write $T_{x+\diff x}=T_x+(\diff T_x/\diff x)\diff x$.
The condition $\delta \Sdotirr\ge 0$ must of course be always satisfied. In fact, Fourier's law of heat conduction assumes $\dot \delta\dot Q_x^\al=-kA\diff T_x /\diff x$, with $k$  positive and representing thermal conductivity and $A$ the surface area of the contact interfaces at $x$ and $x+\diff x$ where the heat interactions occur. Dividing by $A\diff x$ and denoting the entropy production per unit volume by $\sigma = \delta\Sdotirr/(A\diff x)$ and the $x$-component of the heat flux vector by $q_x''= \delta\dot Q_x^\al/A$, the entropy production density takes the well-known equivalent forms
	\begin{equation}
		\sigma=q_x''\,\frac{\diff (1/T_x) }{\diff x}=- q_x''\frac{1}{T_x^2}\frac{\diff T_x}{\diff x}= k\frac{1}{T_x^2}\Big(\frac{\diff T_x}{\diff x}\Big)^2 = \frac{(q_x'')^2}{kT_x^2}
	\end{equation}
	At steady state, the small volume maintains its entropy constant by releasing through the contact interface at $x+\diff x$ exactly the sum of the entropy it generates by irreversibility and the entropy it receives from the contact interface at $x$. The small volume generates entropy due to irreversibility because its state is steady-state but nonequilibrium, near but not coinciding with the stable equilibrium state at temperature $T_x$. Therefore, maintaining it at steady-state requires to balance  the competition between its internal dynamics, which would spontaneously push it towards equilibrium, and the heat interactions with adjacent small volumes, which keep it in disequilibrium.

	\section{Energy and entropy balances and Clausius inequalities for closed systems}\label{Clausius_inequalities}
	
	\ind{energy balance!for closed system}\ind{entropy balance!for closed system}\ind{energy!balance for closed system}\ind{entropy!balance for closed system}A system subjected only to interactions that do not transfer amounts of constituents is  called a \textit{closed system}.\ind{closed system} If the modes of interaction are only work and heat, the energy and entropy balances take the forms
	\begin{align}
		E^A_2 - E^A_1 &=  \smallsum_i W_i^{A\al}-\smallsum_j Q_j^{A\ar}&S^A_2 - S^A_1 &=-\smallsum_j Q_j^{A\ar}/T_{Q_j}+ S^A_{\rm irr}\\
		\diff E^A &=  \smallsum_i\delta W_i^{A\al}-\smallsum_j \delta  Q_j^{A\ar}&\diff S^A &= -\smallsum_j\delta Q_j^{A\ar}/T_{Q_j} + \delta S^A_{\rm irr} \\
		\diff E^A/\diff t &=  \smallsum_i\dot W_i^{A\al}-\smallsum_j \dot Q_j^{A\ar} & \diff S^A /\diff t&=-\smallsum_j\dot Q_j^{A\ar}/T_{Q_j}+ \dot S^A_{\rm irr}
	\end{align}
	For a cyclic process ($E_2-E_1=0$ and $S_2-S_1=0$) or at steady state ($\diff E^A/\diff t=0$ and $\diff S^A /\diff t=0$)  they entail  the following special forms of  Clausius inequalities (\ref{ClausiusInequalitiesNonwork})
	\begin{equation}\label{ClausiusInequalitiesHeat}
		\sum_j\left.\frac{ Q_j^{A\ar}}{T_{Q_j}}\right|_{\substack{\rm cyclic\\ \rm process}}\ge 0\qquad\qquad \sum_j\left.\frac{\dot Q_j^{A\ar}}{T_{Q_j}}\right|_{\substack{\rm steady\\ \rm state}}\ge 0	 
	\end{equation}
	
	As another special case, for historical rather than practical reasons, consider a temporal sequence of processes for system $A$ that takes it from state $A_1$ at time $t_1$ to state $A_2$ at time $t_2$. At each step of the sequence, i.e., in the interval from $t$ to $t+\diff t$, the system experiences only work interactions of magnitudes $\smallsum_i\delta W_i^{A\ar}=\smallsum_i\dot W_i^{A\ar}(t)\diff t$ and heat interactions of magnitude $\smallsum_j\delta Q_j^{A\al}=\smallsum_j\dot Q_j^{A\al}(t)\diff t$ at  temperatures $T_{Q_j}(t)$ which may all vary with time. 
	By integrating the energy and entropy balances from time $t_1$ to time $t_2$, we obtain
	\begin{align}
		E_2 - E_1 &=  \!\!\int_{t_1}^{t_2}\!\!\smallsum_i\dot W_i^{A\al}(t)\diff t- \int_{t_1}^{t_2}\!\! \smallsum_j\dot Q_j^{A\ar}(t)\diff t \\
		S_2 - S_1 &= -\int_{t_1}^{t_2}\! \smallsum_j\frac{\dot Q_j^{A\ar}(t)}{T_{Q_j}(t)}\diff t + \int_{t_1}^{t_2}\! \Sdotirr(t) \diff t
	\end{align}
	If the process is cyclic, i.e., the final state $A_2$ coincides with the initial state $A_1$, then $E_2-E_1=0$ and $S_2-S_1=0$. The energy balance yields $\oint \smallsum_j\dot Q_j^{A\ar}(t)\, \diff t = \oint \smallsum_i\dot W_i^{A\al}(t)\, \diff t$ and the entropy balance, using the condition $\Sdotirr(t)\ge 0$, yields the relation known as the \textit{Clausius inequality}\ind{Clausius!inequality}
	\begin{equation}\label{ClausiusInequality}
		\oint_{t_1}^{t_2}  \sum_j\frac{\dot Q_j^{A\ar}(t)}{T_{Q_j}(t)}\, \diff t \ge 0
	\end{equation}
	where the symbol $\oint$ serves as reminder that the relation is valid only if the process is cyclic. 
	
	\section{Non-work interactions with exchanges of volume}\label{Exchange_V}

\begin{figure}[!ht]
	\begin{center}
			\includegraphics[scale=0.42]{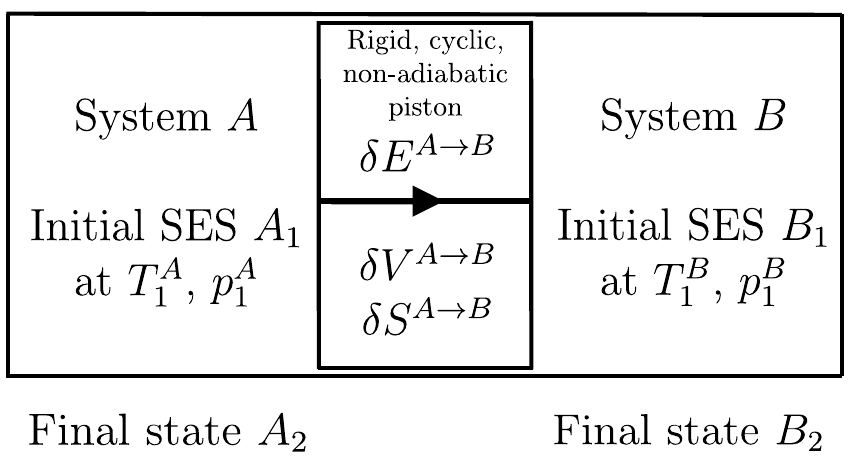}
			\caption{\label{HeatVolumeNoTest}Systems $A$ and $B$ are initially in stable equilibrium states and interact directly without other effects through a moving piston, exchanging energy, volume, and entropy. Note that $\delta V^{A\to B}>0$ when the piston moves to the left. Such an interaction can occur only if Rel.~\ref{SscamcaloreV} is satisfied. }
	\end{center}
\end{figure}

	In this section we consider  two systems, $A$ and $B$ (Figure \ref{HeatVolumeNoTest}), initially in stable equilibrium states with different temperatures $T_1^A$ and $T_1^B$, and different pressures $p_1^A$ and $p_1^B$. They interact with each other (and nothing else) in such a way as to exchange an amount of energy equal to $\delta E^{A\to B}$ but, differently from the cases considered this far, they can also exchange volume through a sliding piston that remains rigid and (except for position) returns to its initial state.

	To the energy and entropy balances we must add the volume balance
	\begin{align}
		\diff E^A&= -\delta E^{A\to B}&\quad 	\diff V^A&= -\delta V^{A\to B}&\quad
		\diff S^A&= -\delta S^{A\to B}+\delta\Sirr^A&\quad\delta\Sirr^A&\ge 0\\
		\diff E^B&= \delta E^{A\to B}&\quad 	\diff V^B&= \delta V^{A\to B}&\quad
		\diff S^B&= \delta S^{A\to B}+\delta\Sirr^B&\quad\delta\Sirr^B&\ge 0
	\end{align}
	Moreover, the maximum entropy principle together with the Taylor series expansion of the fundamental relation imply, for either system, the inequalities
	\begin{equation}\label{volumeexchangedS}
	\diff S \le (\diff E + p_1\diff V)\big/T_1 + \diff^2_{E,V} S_{\rm SES}\big/2+\cdots \le (\diff E + p_1\diff V)\big/T_1
	\end{equation}
	because the second differentials of the fundamental relations 	are non-positive by the conditions on stability,
	\begin{equation}
	\diff^2_{E,V} S_{\rm SES}= \left.\frac{\partial^2 S_{\rm SES}}{\partial E^2}\right|_{E_1,V_1}\!\!\!\!(\diff E)^2+2\left.\frac{\partial^2 S_{\rm SES}}{\partial E\partial V}\right|_{E_1,V_1}\!\!\!\!\diff E\diff V+\left.\frac{\partial^2 S_{\rm SES}}{\partial V^2}\right|_{E_1,V_1}\!\!\!\!(\diff V)^2\le 0
	\end{equation}
 The first strict equality in Rel.~\ref{volumeexchangedS} holds when the system ends in a stable equilibrium state, and the second when the second differential is zero, e.g., for thermal reservoirs with variable amounts of constituents, for which  temperature and pressure have the same values for all stable equilibrium states.
	
	Combining these relations, by eliminating $\diff E^A$, $\diff V^A$, $\diff S^A$, $\diff E^B$, $\diff V^B$, $\diff S^B$, and using $\delta\Sirr^A\ge 0$, $\delta\Sirr^B\ge 0$, yields
	\begin{equation}
	-\delta S^{A\to B}  \le -(\delta E^{A\to B}+p_1^A\,\delta V^{A\to B}) \big/T_1^A \qquad 
	\delta S^{A\to B}  \le (\delta E^{A\to B}+p_1^B\,\delta V^{A\to B}) \big/T_1^B
	\end{equation}
	which together become 
	\begin{equation}
	\frac{\delta E^{A\to B}+p_1^A\,\delta V^{A\to B}}{T_1^A} \le \delta S^{A\to B}\le \frac{\delta E^{A\to B}+p_1^B\,\delta V^{A\to B}}{T_1^B} \label{SscamcaloreV}
	\end{equation}
	Again, for the new set of conditions, these inequalities set lower and upper bounds to the range of values that the entropy transfer must and can take for given transfers of energy and volume.

	Also here, in the limiting case where $A$ and $B$ start almost in mutual equilibrium ($T_1^A\to T_1^B$ and $p_1^A\to p_1^B$) such range of values shrinks to a single value
	\begin{equation}
	\delta S^{A\to B}= \frac{\delta E^{A\to B}+p_1\,\delta V^{A\to B}}{T_1} \label{SscamcaloreVmse}
	\end{equation}
	In the particular case of an adiabatic piston, whereby $\delta S^{A\to B}=0$ so the interaction is work, denoting the energy transfer by $\delta W^{A\to B}$ instead of $\delta E^{A\to B}$ and using the identity $\delta V^{A\to B}=-\delta V^{A\gets B} $  the relation becomes
	\begin{equation}
	\delta W^{A\to B}= p_1\,\delta V^{A\gets B} \label{WorkVmse}
	\end{equation}

But, under these conditions,  a work interaction ($\delta S^{A\to B}=0$) between $A$ and $B$ is also possible if $p_1^A\ne p_1^B $ provided the energy and volume transfers obey the conditions 
	\begin{equation}
	\frac{\delta W^{A\to B}+p_1^A\,\delta V^{A\to B}}{T_1^A} \le 0\le \frac{\delta W^{A\to B}+p_1^B\,\delta V^{A\to B}}{T_1^B} \label{workVcondition}
	\end{equation}
	For positive temperatures, these imply $-p_1^B\,\delta V^{A\to B}\le \delta W^{A\to B}\le -p_1^A\,\delta V^{A\to B}$ so that  $\delta V^{A\gets B}\ge 0$ (the piston in Figure \ref{HeatVolumeNoTest} moves to the right) requires $p_1^A\ge p_1^B$,
	\begin{equation}\label{workweightexample}
	p_1^B\le \frac{\delta W^{A\to B}}{\delta V^{A\gets B}} \le p_1^A
	\end{equation}

\begin{figure}[!ht]
	\begin{center}
		\includegraphics[scale=0.42]{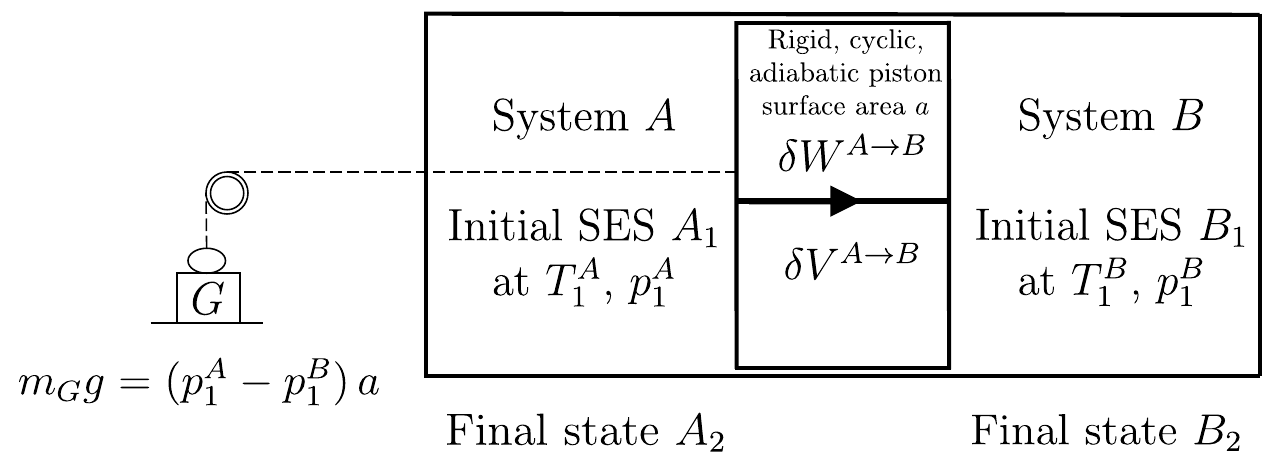}
		\caption{\label{WorkPistonWeight}Systems $A$ and $B$  are initially in stable equilibrium states and interact through an adiabatic piston of surface area $a$ attached to a weight of mass $m_G=(p_1^A-p_1^B)\,a/g$ chosen so as to balance exactly the different initial pressures applied on the two sides of the piston. When $\delta V^{A\gets B}>0$, the piston moves to the right and lifts the weight. }
	\end{center}
\end{figure}

For example, consider the setup sketched in Figure \ref{WorkPistonWeight}, where  a weight $G$ of mass $m_G=(p_1^A-p_1^B)\,a/g$ is attached to the piston (of surface area $a$) so as to balance exactly the different pressures exerted  on its two sides. When $\delta V^{A\gets B}>0$, the piston moves to the right and lifts the weight. The above relations do not hold at this stage because the assumption that systems $A$ and $B$ interact directly without other effects is not satisfied. In fact, it is a weight process for the composite system $AB$ and the work done is $\delta W^{AB\to G}=(p_1^A-p_1^B)\,\delta V^{A\gets B}$.

	But if we return the weight to its initial height by giving its energy back to $AB$ by  means of fixed-piston weight processes for $A$ and $B$, then the assumption is fulfilled and the net final effect is a work interaction between $A$ and $B$ (with no net external effects) accomplished via the piston and the weight. If we denote by $\alpha$ the fraction of $\delta W^{AB\to G}$ given  to $B$ ($0\le\alpha\le 1$) and $1-\alpha$ that given to $A$, the net works for $A$ and $B$ are $\delta W^{A\to}=p_1^A\,\delta V^{A\gets B}-(1-\alpha)\,\delta W^{AB\to G}=\delta W^{B\gets}=p_1^B\,\delta V^{A\gets B}+\alpha\,\delta W^{AB\to G}$ or, equivalently,   $\delta W^{A\to B}=\alpha\,p_1^A+(1-\alpha)\,p_1^B$, which upon varying $\alpha$ fills the entire range of values allowed by Relation \ref{workweightexample}. Of course, the work interactions with the weight leave $A$ and $B$ in nonequilibrium states and therefore the systems will spontaneously  relax toward stable equilibrium  thus generating entropy. 
	
	\section{Non-work interactions with exchanges of  constituents}\label{Exchange_n}
	
\begin{figure}[!ht]
		\begin{center}
			\includegraphics[scale=0.42]{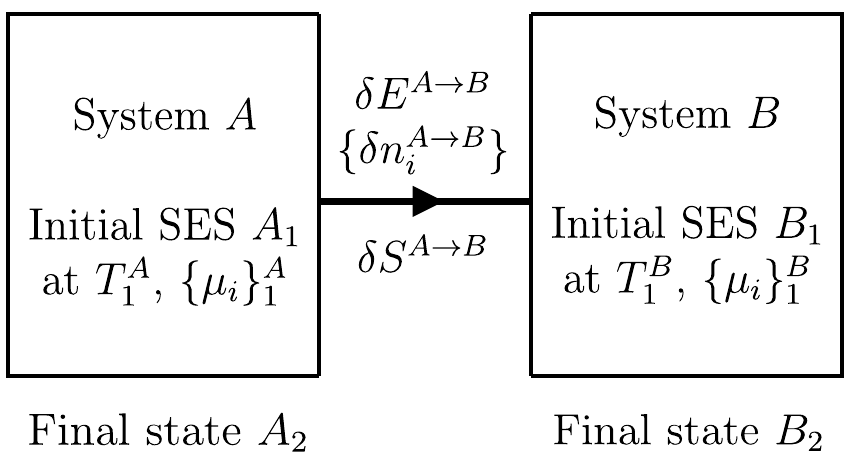}
			\caption{\label{HeatDiffusionNoTest}Systems $A$ and $B$ start in stable equilibrium states and interact directly without other effects by  exchanging energy, entropy, and amounts of constituents, without exchange of volume. Such an interaction can occur only if Rel.~\ref{SscamcaloreN} is satisfied.  }
		\end{center}
	\end{figure}

	In this section we consider  two systems, $A$ and $B$ (Figure \ref{HeatDiffusionNoTest}), initially in stable equilibrium states with different temperatures $T_1^A$ and $T_1^B$, and different chemical potentials $\{\mu_i\}_1^A$ and $\{\mu_i\}_1^B$, where $\{\mu_i\}_1^A$ is shorthand for $\mu_1|_1^A$, $\mu_2|_1^A$,\dots,$\mu_r|_1^A$ and similarly for $\{\mu_i\}_1^B$.  They interact with each other (and nothing else) in such a way that the overall exchange of energy is $\delta E^{A\to B}$ and in addition there are also exchanges of amounts of constituents, $\{\delta n_i^{A\to B}\}$, through one or more apertures that open when  the interaction begins and close immediately after, leaving the volumes unchanged.

	To the energy and entropy balances we must add a  balance for each type of constituent,
	\begin{align}
		\diff E^A&= -\delta E^{A\to B}&\quad 	\diff n_i^A&= -\delta n_i^{A\to B}&\quad
		\diff S^A&= -\delta S^{A\to B}+\delta\Sirr^A&\quad\delta\Sirr^A&\ge 0\\
		\diff E^B&= \delta E^{A\to B}&\quad 	\diff n_i^B&= \delta n_i^{A\to B}&\quad
		\diff S^B&= \delta S^{A\to B}+\delta\Sirr^B&\quad\delta\Sirr^B&\ge 0
	\end{align}
	The maximum entropy principle together with the Taylor series expansion of the fundamental relation imply, for either system, the inequalities
	\begin{equation}
	\diff S \le (\diff E - \smallsum_i \mu_i|_1\diff n_i)\big/T_1 + \diff^2_{E,\bn} S_{\rm SES}\big/ 2+\cdots \le (\diff E - \smallsum_i \mu_i|_1\diff n_i)/T_1
	\end{equation}
	because all second differentials of the fundamental relations are non-positive by the conditions on stability. The first strict equality holds when the system ends in a stable equilibrium state, and the second when the second differential is zero, e.g., for thermal reservoirs with variable amounts of constituents, for which  temperature and  chemical potentials have the same values for all stable equilibrium states.
	
	Combining these relations by eliminating $\diff E^A$, $\diff n_i^A$, $\diff S^A$, $\diff E^B$, $\diff n_i^B$, $\diff S^B$,  using $\delta\Sirr^A\ge 0$, $\delta\Sirr^B\ge 0$, and proceeding like in the previous sections, yields
	\begin{equation}
	\frac{\delta E^{A\to B}- \smallsum_i \mu_i|_1^A\,\delta n_i^{A\to B}}{T_1^A} \le \delta S^{A\to B}\le \frac{\delta E^{A\to B}- \smallsum_i \mu_i|_1^B\,\delta n_i^{A\to B}}{T_1^B} \label{SscamcaloreN}
	\end{equation}
	Again, for the new set of conditions, these inequalities set lower and upper bounds to the range of values that the entropy transfer can take for given transfers of energy and constituents. 
		Also here, in the limiting case where $A$ and $B$ start almost in mutual equilibrium ($T_1^A\to T_1^B$ and $\{\mu_i\}_1^A\to\{\mu_i\}_1^B$), such range of values shrinks to the single value
	\begin{equation}
	\delta S^{A\to B}= \frac{\delta E^{A\to B}- \smallsum_i \mu_i|_1\,\delta n_i^{A\to B}}{T_1} \label{SscamcaloreNmse}
	\end{equation}
	
		\section{Non-work interactions with exchanges of volume and constituents}\label{Exchange_V_n}

\begin{figure}[!ht]
	\begin{center}
			\includegraphics[scale=0.42]{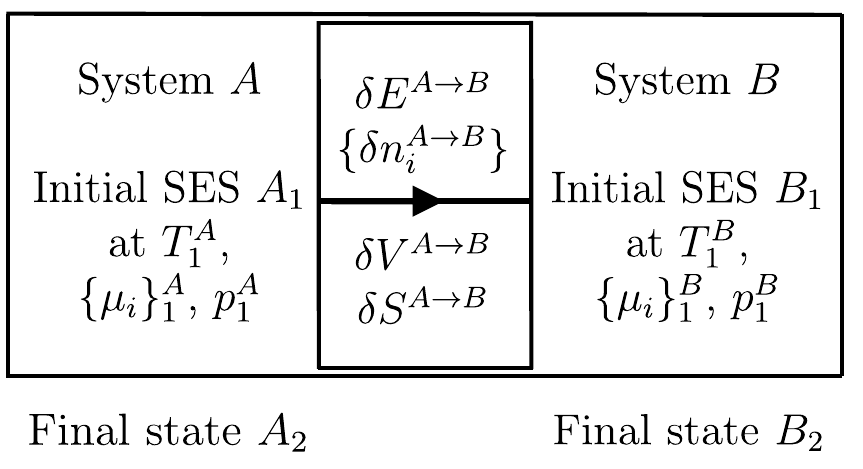}
			\caption{\label{HeatDiffusionVolumeNoTest}Systems $A$ and $B$ start in stable equilibrium states and interact directly without other effects by  exchanging energy, entropy, amounts of constituents, and volume. Such an interaction can occur only if Rel.~\ref{SscamcaloreNV} is satisfied. }
	\end{center}
\end{figure}

	If volume is also exchanged, as sketched in Figure \ref{HeatDiffusionVolumeNoTest} the interaction is possible when
	\begin{equation}
	\frac{\delta E^{A\to B}+p_1^A\,\delta V^{A\to B}- \smallsum_i \mu_i|_1^A\,\delta n_i^{A\to B}}{T_1^A} \le \delta S^{A\to B}\le \frac{\delta E^{A\to B}+p_1^B\,\delta V^{A\to B}- \smallsum_i \mu_i|_1^B\,\delta n_i^{A\to B}}{T_1^B} \label{SscamcaloreNV}
	\end{equation}
	In the limiting case where $A$ and $B$ start almost in mutual equilibrium ($T_1^A\to T_1^B$, $p_1^A\to p_1^B$,  and $\{\mu_i\}_1^A\to\{\mu_i\}_1^B$), such range of values reduces to the single value
	\begin{equation}
	\delta S^{A\to B}= \frac{\delta E^{A\to B}+p_1\,\delta V^{A\to B}- \smallsum_i \mu_i|_1\,\delta n_i^{A\to B}}{T_1} \label{SscamcaloreNmseP}
	\end{equation}

	\section{Heat-and-diffusion interactions. Definition}\label{Heat_and_diffusion}
	
	The above results lead to the following important definition of another special limiting class of non-work interactions between systems initially in stable equilibrium states. Since it can be viewed as a generalized form of heat interaction combined with the diffusion of constituents, we  call it \textit{heat-and-diffusion interaction}.\ind{interaction!heat-and-diffusion}\ind{heat-and-diffusion!definition}  As for heat, the conceptual and practical definition of heat-and-diffusion hinges on its complete distinguishability from work. The procedure is a simple extension of the discussion in Section \ref{Heat}, so we can skip most details and go directly to the results. 
	
\begin{figure}[!ht]
		\begin{center}
			\includegraphics[scale=0.42]{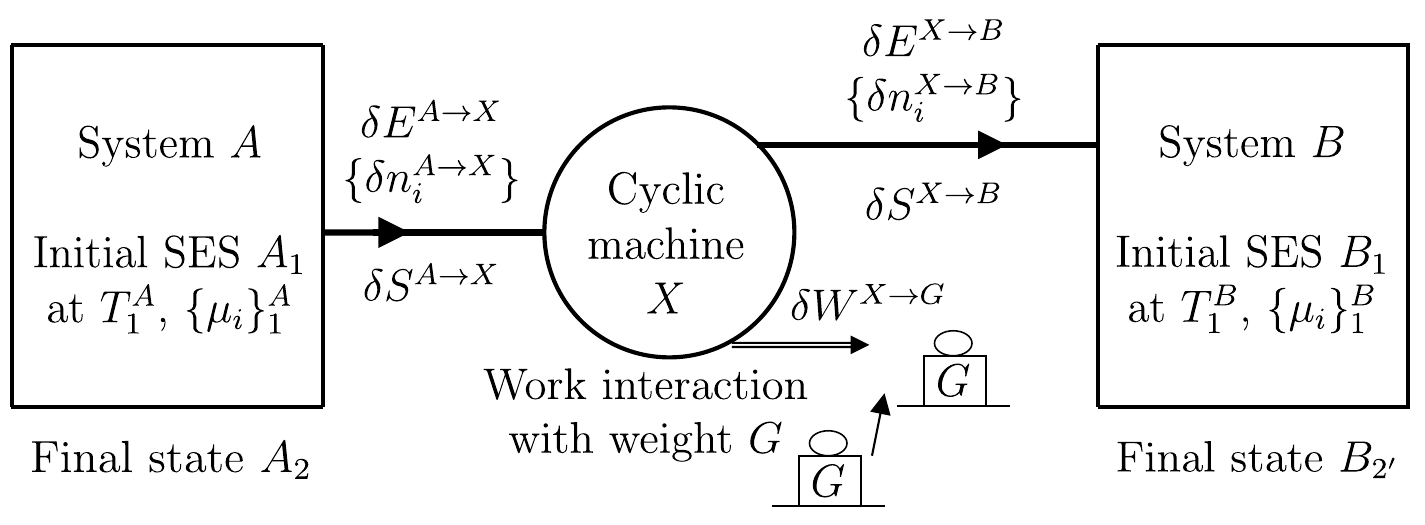}
			\caption{\label{HeatDiffusionTest1}The cyclic machine $X$ interposed between interacting systems $A$ and $B$ intercepts the energy,  constituents, and entropy they exchange and attempts to channel as much energy as possible into lifting a weight $G$. This lifting  becomes impossible in the limit as $T_1^A\to T_1^B$ and $\mu_{i1}^A\to \mu_{i1}^B$ for every $i$. These limiting conditions define the non-work interaction between $A$ and $B$ that we call a heat-and-diffusion interaction.}
		\end{center}
	\end{figure}

	Consider first the case of Figure \ref{HeatDiffusionNoTest} (no volume exchange). Let us interpose a cyclic machine $X$ between interacting systems $A$ and $B$ with the purpose to intercept the energy,  constituents, and entropy they exchange and  attempt to channel as much energy as possible into lifting a weight $G$, as sketched in  Figure \ref{HeatDiffusionTest1}. If the machine is successful, the subsequent step, like in Figure \ref{HeatTest2}, is a weight process in which system $B$ receives  the energy temporarily stored by lifting the weight, so that the end final effect is the same as in Figure \ref{HeatDiffusionNoTest}, but we can say that part of the energy transferred from $A$ to $B$ is clearly identifiable as work.

	By writing balances of energy, entropy, and constituents, and using the maximum entropy principle, like we have done above to obtain Rels.~\ref{WXtoG} and \ref{WXtoGmax}, we can easily show that the maximum work that machine $X$ can transfer to the weight is
	\begin{equation}
	\delta W_{\rm max}^{X\to G} = \bigg(1 - \frac{T_1^B}{T_1^A}\bigg)\delta E^{A\to X}  + \sum_i \bigg(\mu_i|_1^A \, \frac{T_1^B}{T_1^A}-\mu_i|_1^B \bigg) \delta n_i^{A\to B}
	\end{equation}
	This result is interesting in itself, because it generalizes the famous Carnot expression for the maximum work. But here, its importance is in showing that it is only in the limit as $T_1^A\to T_1^B$ and $\{\mu_i\}_1^A\to\{\mu_i\}_1^B$ for every $i$ that the maximum work vanishes. This limit defines the heat-and-diffusion mode of interaction. We have already proved that in this limit  the exchanges of energy, entropy, and constituent are uniquely related by Eq.~\ref{SscamcaloreNmse},
	\begin{equation}\label{SEnrelationInHeatAndDiffusion}
	\delta S^{A\to B}= \frac{\delta E^{A\to B}- \smallsum_i \mu_i|_1\,\delta n_i^{A\to B}}{T_1} 
	\end{equation}
	which of course may be viewed as a generalization of the famous relation  $\delta S^{A\to B}= \delta E^{A\to B}\big/T_1$ (Eq.~\ref{entropyHeat}) which holds for  simple heat interactions. 
	
	However, it would be misleading to think of the entire energy transfer  as ``heat.'' In fact only part of $\delta E^{A\to B}$  and part of $\delta S^{A\to B}$ can be interpreted as associated to heat. To this end, let us note that, in general, the chemical potentials may be written as  $\mu_i=h_i -Ts_i$ where $h_i=\hparder{H}{n_i}{T,p,\bn'}$ and $s_i=\hparder{S}{n_i}{T,p,\bn'}$ are the partial enthalpy\ind{partial enthalpy} and  partial entropy\ind{partial entropy} of constituent $i$.\footnote{The enthalpy $H=E+p\,V=H(S,p,\bn)$ is defined by the Legendre transform  with respect to  variable $V$ of the fundamental relation in energy form, $E=E(S,V,\bn)$. It can be shown that within the simple-system approximation \cite[Par.17.8]{GyftopoulosBeretta1991}, valid only for large amounts, $H=\sum_i n_ih_i$ and $S=\sum_i n_is_i$, where $h_i=\hparder{H}{n_i}{T,p,\bn'}=h_i(T,p,\bn)$ and $s_i=\hparder{S}{n_i}{T,p,\bn'}=s_i(T,p,\bn)$, so that the terms  $\smallsum_i h_i|_1\,\delta n_i^{A\to B}$ and $\smallsum_i s_i|_1\,\delta n_i^{A\to B}$ can be interpreted as the enthalpy and  entropy ``carried'' by the exchanged constituents as they cross from one system to the other. } Therefore, Eq.~\ref{SEnrelationInHeatAndDiffusion} may be rewritten as
	\begin{equation}\label{SEnrelationInHeatAndDiffusion1}
	\delta Q^{A\to B}\equiv \delta E^{A\to B}- \smallsum_i h_i|_1\,\delta n_i^{A\to B}=\left(\delta S^{A\to B}-\smallsum_i s_i|_1\,\delta n_i^{A\to B}\right)\,T_1 
	\end{equation}
	where the first equality defines what is  called \textit{measurable heat} \cite[Par.III.3]{DeGrootMazur}.\ind{measurable heat}\ind{heat!in heat-and-diffusion} Equivalently, we may write  this as  the generalization of Eq.~\ref{entropyHeat}, 
	\begin{equation}\label{SEnrelationInHeatAndDiffusion2}
	\delta S^{A\to B}-\smallsum_i s_i|_1\,\delta n_i^{A\to B}=\frac{\delta Q^{A\to B}}{T_1}= \frac{\delta E^{A\to B}- \smallsum_i h_i|_1\,\delta n_i^{A\to B}}{T_1}
	\end{equation}
	which shows that the temperature $T_1$ at which the heat-and-diffusion interaction occurs is \textit{not} equal to the ratio $\delta E^{A\to B}\big/\delta S^{A\to B}$ of the entire energy transfer to the entire entropy transfer, but it is  equal to the ratio $\big(\delta E^{A\to B}- \smallsum_i h_i|_1\,\delta n_i^{A\to B}\big) \big/\big(\delta S^{A\to B}-\smallsum_i s_i|_1\,\delta n_i^{A\to B}\big)$ of only the portions of   energy and entropy transfers not directly associated with the diffusion of constituents. In other words, the measurable heat $\delta Q^{A\to B}$ in a heat-and-diffusion interaction is the difference between the overall energy transfer $\delta E^{A\to B}$ and the enthalpy transfer $\smallsum_i h_i|_1\,\delta n_i^{A\to B}$, which in turn can be viewed as the energy transfer due in part to the internal energy carried by the diffused constituents and in part to the pulsion work needed to move them against their local partial pressures.\footnote{This is by analogy with the notion of bulk flow interaction \cite[Ch.22]{GyftopoulosBeretta1991}, whereby the energy transfer is due in part to the internal energy carried by the bulk displacement of  constituents and in part to the pulsion work needed to move them against the local pressure.}
	
	Similarly to Eq.~\ref{SscamcaloreNmse}, for the case of Figure \ref{HeatDiffusionVolumeNoTest} in which also volume is exchanged, we may use the identity $\mu_i=h_i -Ts_i$ in Eq.~\ref{SscamcaloreNmseP} and define the measurable heat $\delta Q^{A\to B}$ so that the energy and entropy transfers can be written as follows
		\begin{align}\label{SEnrelationInHeatAndDiffusion3}
 \delta E^{A\to B}&=	\delta Q^{A\to B}+ \smallsum_i h_i|_1\,\delta n_i^{A\to B}+p_1\,\delta V^{A\al B}\\
\delta S^{A\to B}&= \frac{\delta Q^{A\to B}}{T_1}+\smallsum_i s_i|_1\,\delta n_i^{A\to B} 
	\end{align}

	\section{Availabilities with respect to various kinds of thermal reservoirs}

\begin{figure}[!ht]
		\begin{center}
			\includegraphics[scale=0.42]{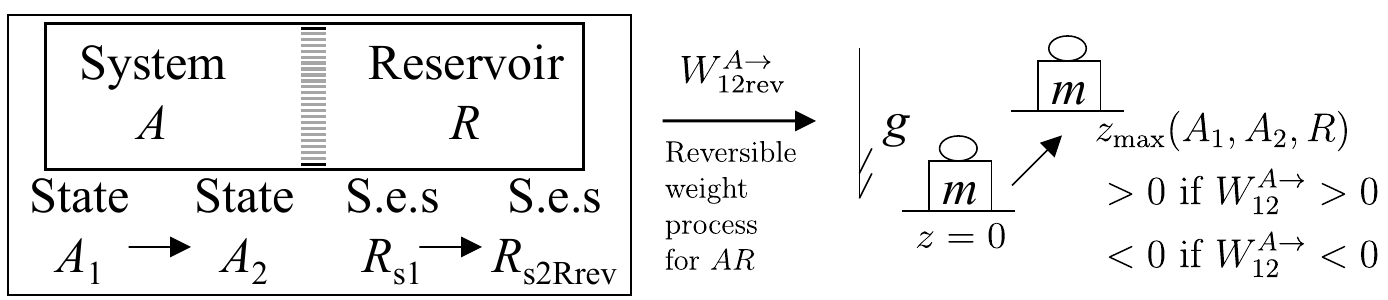}
			\caption{\label{AvailableEnergiesSetup}Schematic setup for the definition of available energy with respect to different types of thermal reservoir depending on whether volume and/or constituents can be exchanged or not between the system $A$ and the thermal reservoir $R$  (fixed $V$ and $\bn$;  variable $V$ and fixed $\bn$;  fixed $V$, variable $n_i$, fixed $\bn'$; variable $V$ and $\bn$).}
		\end{center}
	\end{figure}

	We now return to the concept of available energy, previously introduced in the context of a system $A$ interacting with a thermal reservoir $R$ through a weight process for $AR$.  In that setting, the maximum amount of energy that can be transferred to a weight is achieved when the process is reversible, and when the system ends in a state of mutual stable equilibrium with the reservoir. The reservoir's role as source or sink of entropy is indispensable to achieve reversibility of the weight process for $AR$. More generally, one may prescribe two arbitrary states of the system, say $A_1$ and $A_2$, as sketched in Fig.~\ref{AvailableEnergiesSetup} and ask a different question: given access to interaction with a specified thermal reservoir, what is the maximum mechanical work that can be extracted during the transition from $A_1$ to $A_2$, or, if work must be supplied, what is the minimum amount required? Questions of this kind are answered systematically by combining energy and entropy balances for the composite system with the fundamental relation of the reservoir.

	In the simplest case, the thermal reservoir is characterized by fixed volume and composition. Its fundamental relation (Eq.~\ref{FundRelR}) is linear, expressing proportionality between its energy and  entropy changes, and a fixed temperature $T_R$. 	
	In practical applications, however, the class of systems that can effectively play the role of thermal reservoirs is broader. Large portions of the environment --- such as the atmosphere, oceans, lakes, or rivers --- may exchange not only energy and entropy but also volume and matter with the system of interest.  
	When the exchanged amounts are small compared to the size of the reservoir, such systems may still be idealized as thermal reservoirs, albeit with different sets of constrained and unconstrained variables. Each choice leads to a distinct fundamental relation for the reservoir and, correspondingly, to a distinct definition of available energy.
	
	The notion of thermal reservoir may be conveniently generalized by relaxing the conditions of fixed volume and fixed amounts of constituents while maintaining the defining  condition that in any of its stable equilibrium states it is in mutual  equilibrium with a given system $C$ in a fixed given state $C_R$, provided that in addition to energy and entropy systems $C$ and $R$  can exchange also volume (if variable for $R$) and the  amounts of constituents that are variable for $R$, if any.  With this definition, the maximum entropy principle implies (recall Fig.~\ref{MEPconditions})  that not only the temperature $T_R$ is the same for all the stable equilibrium states of the reservoir, but also the pressure $p_R$ (if it has variable volume) and the chemical potential $\mu_{iR}$ of every constituent with variable amount.

	In this section, we first review the classical case of available energy with respect to a thermal reservoir with fixed volume and fixed amounts of constituents. We then extend the analysis to four relevant classes of reservoirs, distinguished by whether volume and/or constituents may be exchanged between the system $A$ and the reservoir $R$: (i) fixed $V$ and fixed $\bn$; (ii) variable $V$ and fixed $\bn$; (iii) fixed $V$, variable $n_i$, and fixed $\bn'$; and (iv) variable $V$ and variable $\bn$. In each case, we identify and define a corresponding ``availability function''  whose difference between states $A_1$ and $A_2$   provides a precise measure of the maximum work that can be obtained, or the minimum work that must be supplied, for the prescribed change of state, independently of whether $A_1$ and $A_2$ are equilibrium states or not. This availability function possesses an absolute minimum at the state of mutual equilibrium between the system and the reservoir, a feature that has important consequences for stability, concavity of fundamental relations, and response to perturbations.
	
	\subsection{Reservoir with fixed volume and  amounts. Helmholtz availability function vs Helmholtz free energy}
	
	With reference to Fig.~\ref{AvailableEnergiesSetup}, assume that the thermal reservoir has fixed volume and  amounts. Before imposing the reversibility of the weight process for $AR$, the energy and entropy balances are
		\begin{equation}\label{ESbalanceR}
			(E^A_2-E^A_1)+(E^R_2-E^R_1)=-W^{A\to }_{12}  \qquad
		 \quad(S^A_2-S^A_1)+(S^R_2-S^R_1)=S_{\text{gen}}
		\end{equation}

Using the fundamental relation of $R$, $E^R_2-E^R_1=T_R\,(S^R_2-S^R_1) $, to eliminate  $(E^R_2-E^R_1)$ and $(S^R_2-S^R_1)$ from Eqs.~\ref{ESbalanceR} yields
	\begin{equation}\label{WR}
W^{A\to }_{12}=E^A_1-E^A_2-T_R\,(S^A_1-S^A_2)-T_R\,S_{\text{gen}}=
W^{A\to }_{12\text{rev}}-T_R\,S_{\text{gen}}
		\end{equation}
	where we note that by defining the ``Helmholtz availability function'' $\Gamma$ and recalling the definition of the available energy $\Omega^R$
		\begin{equation}\label{Gamma}
		\Gamma^A = E^A-T_R \,S^A \qquad \qquad (\Omega^R)^A=E^A-E^A_R- T_R\,(S^A-S^A_R)=\Gamma^A-\Gamma^A_R
	\end{equation}
we can express the optimal work as
	\begin{equation}\label{WrevR}
W^{A\to }_{12\text{rev}}=(\Omega^R)^A_1-(\Omega^R)^A_2=\Gamma^A_1-\Gamma^A_2
		\end{equation}

	\begin{figure}[!ht]
		\begin{center}
				\includegraphics[scale=0.42]{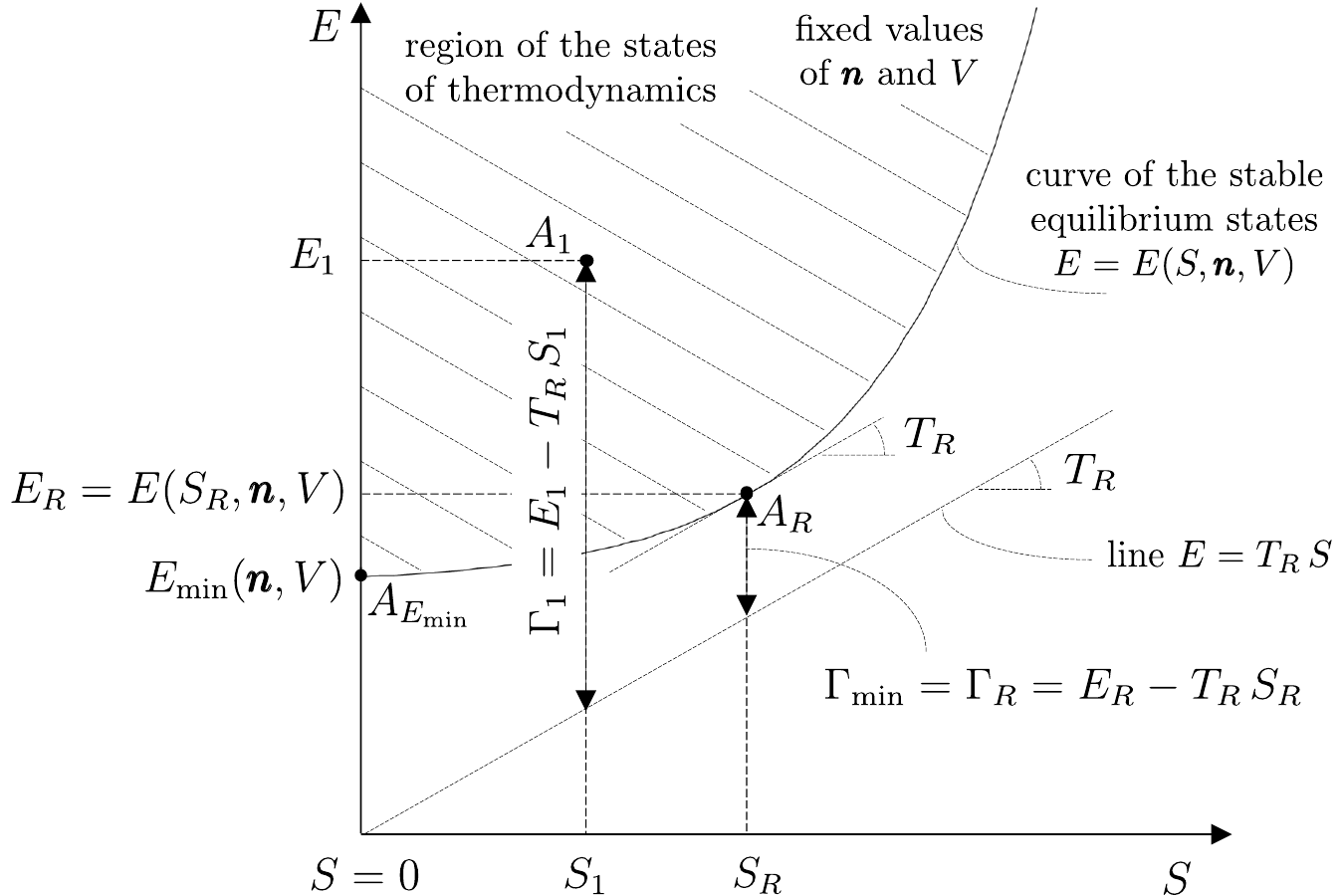}	
				\caption{\label{AvailabilityFunction}Representation on the $E$--$S$ diagram of the Helmholtz availability function $\Gamma=E-T_R\,S$ of a system $A$ with respect to a thermal reservoir $R$ with fixed volume $V$ and amounts $\bn$, showing graphically that the available energy $\left(\Omega^R\right)^A_1 =\Gamma^A_1-\Gamma^A_R$ and that $\Gamma > \Gamma_R$ for any state where $A$ is not in mutual equilibrium with $R$. Thus, the minimum value $\Gamma_R$ is achieved only at state $A_R$, where $\Gamma_R=F_R$, the Helmholtz free energy.}
		\end{center}
	\end{figure}
	
	Figure \ref{AvailabilityFunction} gives a geometrical representation of the Helmholtz availability function $\Gamma$ on the $E$--$S$ diagram of system $A$. It also shows graphically that $\Gamma$  possesses an absolute   minimum at state $A_R$, where $A$ and $R$ are in mutual equilibrium and hence $T_R^A=T_R$. 
	
	We finally note that in state $A_R$,  $\Gamma^A_R=F^A_R$ where $F=E-T\,S$ is the \textit{Helmholtz free energy}.\footnote{The Helmholtz free energy is  defined --- only for stable equilibrium states --- by the Legendre transform of the fundamental relation $E=E(S,V,\bn)$ with respect to the variable $S$. It follows that its ``natural independent variables'' are $T$, $V$, $\bn$, i.e., $F=F(T,V,\bn)$, and its differential can be written as $\diff F= -S\diff T-p\diff V+\bmu\cdot\diff\bn$.} But it is important to note that for all the other  stable equilibrium states, $\Gamma \ne F$, and while the Helmholtz availability function $\Gamma$ is defined also for nonequilibrium states, the Helmholtz free energy $F$ is not.\footnote{In the quantum thermodynamics literature, often  the  Helmholtz availability function $E-T_R\,S$ is called  ``free energy'' and denoted by the symbol $F$ (see, e.g. \cite{Crooks_2012}), potentially  generating some confusion. }
	
The observation that  $\Gamma$ has an absolute minimum at state $A_R$ can be expressed by writing that 
		\begin{equation}\label{GammaStability}\Gamma_1^A-\Gamma_R^A > 0 \qquad\text{for every state $A_1\ne A_R$	 with the same $V$ and $\bn$'s}	\end{equation} 
From this we can derive useful stability conditions, equivalent to those we already derived from the maximum entropy principle to prove  the concavity of the fundamental relation. 
 For example, choose $A_1$ to be the stable equilibrium state with the same values of $V$ and $\bn$ as state $A_R$ but with entropy $S_1^A=S^A_R+\diff S$. Then, Rel.~\ref{GammaStability} together with the Taylor expansion of fundamental relation for $A$, $E_1^A=E_A(S_1^A,V,\bn)$,  imply the general  condition	
 \begin{equation}
 	\Gamma_1^A-\Gamma_R^A= E_A(S^A_R+\diff S,V,\bn)-T_R\,(S^A_R+\diff S)-(E^A_R-T_R\,S^A_R)={\scriptstyle\frac{1}{2}}\diff^2E^A|_{V,\bn} + \cdots >0 	\end{equation} 
which in turn implies the stability condition $ \diff^2E^A|_{V,\bn} \ge 0$.

\subsection{Reservoir with variable volume and fixed amounts. Gibbs availability function vs Gibbs free energy}

	Again with reference to Fig.~\ref{AvailableEnergiesSetup}, assume now that the thermal reservoir has variable volume and fixed amounts and that $A$ and $R$ can exchange volume but no constituents. Before imposing the reversibility of the weight process for $AR$, the  energy,  entropy, and volume balance equations and the reservoir's fundamental relation  are
\begin{equation}\label{ESbalanceRV}
	(E^A_2-E^A_1)+(E^R_2-E^R_1)=-W^{A\to }_{12}  \qquad
(S^A_2-S^A_1)+(S^R_2-S^R_1)=S_{\text{gen}}
\end{equation}
\begin{equation}\label{VbalanceRV} (V^A_2-V^A_1)+(V^R_2-V^R_1)=0  \qquad E^R_2-E^R_1=T_R\,(S^R_2-S^R_1)-p_R\,(V^R_2-V^R_1)
\end{equation}
Eliminating $(V^R_2-V^R_1)$, $(E^R_2-E^R_1)$ and $(S^R_2-S^R_1)$ from these equations yields
\begin{equation}\label{WRV}
W^{A\to }_{12}=E^A_1-E^A_2-T_R(S^A_1-S^A_2)+p_R(V^A_1-V^A_2)-T_R\,S_{\text{gen}}=
W^{A\to }_{12\text{rev}}-T_RS_{\text{gen}}
\end{equation}
and by defining the ``Gibbs availability function'' $\Phi$ and the available energy $\Omega^{R_V}$
\begin{equation}\label{Phi}
	\Phi^A = E^A-T_R\, S^A +p_R\,V^A 
\end{equation}
	\begin{equation}\label{OmegaRV}
		  (\Omega^{R_V})^A = E^A-E^A_R- T_R\,(S^A-S^A_R)+ p_R\,(V^A-V^A_R)=\Phi^A-\Phi^A_R
\end{equation}
we can express the optimal work as
\begin{equation}\label{WrevRV}
W^{A\to }_{12\text{rev}}=(\Omega^{R_V})^A_1-(\Omega^{R_V})^A_2=\Phi ^A_1-\Phi ^A_2
\end{equation}
Also  $\Phi $ possesses an absolute minimum at state $A_R$, where $A$ and $R$ are in mutual equilibrium and hence $T_R^A=T_R$ and $p_R^A=p_R$. Moreover, in state $A_R$,  $\Phi^A_R=G^A_R$ where $G=E-T\,S+p\,V$ is the \textit{Gibbs free energy}.\footnote{The Gibbs free energy is  defined --- only for stable equilibrium states --- by the Legendre transform of the fundamental relation $E=E(S,V,\bn)$ with respect to both variables $S$ and $V$. Its ``natural independent variables'' are $T$, $p$, $\bn$, i.e., $G=G(T,p,\bn)$, and its differential can be written as $\diff G= -S\diff T+V\diff p+\bmu\cdot\diff\bn$.} But it is important to note that for all the other  stable equilibrium states, $\Phi\ne G$, and while the Gibbs availability function $\Phi$ is defined also for nonequilibrium states, the Gibbs free energy $G$ is not.

The stability condition that follows from the observation that  $\Gamma$ has an absolute minimum at state $A_R$ is 
\begin{equation}\label{PhiStability}\Phi_1^A-\Phi_R^A > 0 \qquad\text{for every state $A_1\ne A_R$	 with the same  $\bn$'s}		\end{equation} 

\subsection{Reservoir with fixed volume and variable amount for one constituent only.  Osmotic availability function vs osmotic free energy }

Assume now that the thermal reservoir has fixed volume and variable amount  constituent $i$ only and that $A$ and $R$ can exchange  constituents of type $i$ (through a semi-permeable rigid membrane) but no other type of constituents nor volume. Before imposing the reversibility of the weight process for $AR$, the  energy,  entropy, and constituent $i$ balance equations and the reservoir's fundamental relation  are
\begin{equation}\label{ESbalanceRni}
	(E^A_2-E^A_1)+(E^R_2-E^R_1)=-W^{A\to }_{12}  \qquad
	(S^A_2-S^A_1)+(S^R_2-S^R_1)=S_{\text{gen}}
\end{equation}
\begin{equation}\label{nibalanceRni} (n_{i2}^A-n^A_{i1})+(n^R_{i2}-n^R_{i1})=0   \qquad  E^R_2-E^R_1=T_R\,(S^R_2-S^R_1)+\mu_{iR}\,(n^R_{i2}-n^R_{i1})
\end{equation}
Eliminating $(n^R_{i2}-n^R_{i1})$, $(E^R_2-E^R_1)$ and $(S^R_2-S^R_1)$ from these equations yields
\begin{equation}\label{WRni}
W^{A\to }_{12}=E^A_1-E^A_2-T_R\,(S^A_1-S^A_2)+\mu_{iR}\,(n_{i1}^A-n^A_{i2})-T_R\,S_{\text{gen}}=
W^{A\to }_{12\text{rev}}-T_R\,S_{\text{gen}}
\end{equation}
and by defining the ``osmotic availability function'' $\Phi$ and the available energy $\Omega^{R_{n_i}}$
\begin{equation}\label{Upsilon}
	\Upsilon= E-T_R\, S -\mu_{iR}\,n_i 
\end{equation}
\begin{equation}\label{OmegaRni}
(\Omega^{R_{n_i}})^A = E^A-E^A_R- T_R\,(S^A-S^A_R)+ \mu_{iR}\,(n_i^A-n^A_{iR})=\Upsilon^A-\Upsilon^A_R
\end{equation}
we can express the optimal work as
\begin{equation}\label{WrevRni}
W^{A\to }_{12\text{rev}}=(\Omega^{R_{n_i}})^A_1-(\Omega^{R_{n_i}})^A_2=\Upsilon^A_1-\Upsilon^A_2
\end{equation}
Also  $\Upsilon $ possesses an absolute minimum at state $A_R$, where $A$ and $R$ are in mutual equilibrium and hence $T_R^A=T_R$ and $\mu_{iR}^A=\mu_{iR}$. Moreover, in state $A_R$,  $\Upsilon^A_R=\Eu^{iA}_R$ where
 \begin{equation}\label{Eui}
 	\Eu^{i}=E-T\,S-\mu_i\,n_i
 \end{equation}
 is the \textit{osmotic free energy}.\footnote{The osmotic free energy is  defined --- only for stable equilibrium states --- by the Legendre transform of the fundamental relation $E=E(S,V,\bn)$ with respect to both variables $S$ and $n_i$. Its ``natural independent variables'' are $T$, $V$, $\mu_i$, $\bn'$, i.e., $\Eu^{i}=\Eu^{i}(T,V,\mu_i,\bn')$, and its differential can be written as $\diff \Eu^{i}= -S\diff T-p\diff V-n_i\diff \mu_i+\bmu'\cdot\diff\bn'$.} But it is important to note that for all the other  stable equilibrium states, $\Upsilon\ne \Eu^{i}$, and while the osmotic availability function $\Upsilon$ is defined also for nonequilibrium states, the osmotic free energy $\Eu^{i}$ is not.

The stability condition that follows from the observation that  $\Upsilon$ has an absolute minimum at state $A_R$ is 
\begin{equation}\label{UpsilonStability}\Upsilon_1^A-\Upsilon_R^A > 0 \qquad\text{for every state $A_1\ne A_R$	 with the same $V$ and $\bn'$'s}	\end{equation}

\subsection{Reservoir with variable volume and variable amounts for all constituents.  Hill availability function vs Hill free energy }

Finally, assume  that the thermal reservoir has variable volume and variable amounts for all  constituents and that $A$ and $R$ can exchange volume as well as all types of constituents. Before imposing the reversibility of the weight process for $AR$, the  energy,  entropy, volume and constituent balance equations and the reservoir's fundamental relation  are
\begin{equation}\label{ESbalanceRnV}
	(E^A_2-E^A_1)+(E^R_2-E^R_1)=-W^{A\to }_{12}  \qquad
	(S^A_2-S^A_1)+(S^R_2-S^R_1)=S_{\text{gen}}
\end{equation}
\begin{equation}\label{nVbalanceRnv}  (V^A_2-V^A_1)+(V^R_2-V^R_1)=0 \qquad (n_{i2}^A-n^A_{i1})+(n^R_{i2}-n^R_{i1})=0\quad \forall i
\end{equation}
\begin{equation}
	 E^R_2-E^R_1=T_R\,(S^R_2-S^R_1)-p_R\,(V^R_2-V^R_1) +{\textstyle\sum_i}\mu_{iR}\,(n^R_{i2}-n^R_{i1})
\end{equation}
Eliminating $(V^R_2-V^R_1)$, $(E^R_2-E^R_1)$, $(S^R_2-S^R_1)$ and all $(n^R_{i2}-n^R_{i1})$'s yields
\begin{equation}\label{WRnV}
W^{A\to }_{12}=E^A_1-E^A_2-T_R(S^A_1-S^A_2)+p_R(V^A_1-V^A_2)+{\textstyle\sum_i}\mu_{iR}\,(n_{i1}^A-n^A_{i2})-T_R\,S_{\text{gen}}
\end{equation}
and by defining the ``Hill availability function'' $\Xi$ and the available energy $\Omega^{R_{V,n}}$
\begin{equation}\label{Xi}
	\Xi = E-T_R\, S +p_R\,V -{\textstyle\sum_i}\mu_{iR}\,n_i
\end{equation}
\begin{equation}\label{OmegaRnV}
	(\Omega^{R_{V,\bn}})^A = E^A-E^A_R- T_R\,(S^A-S^A_R)+ p_R\,(V^A-V^A_R)-{\textstyle\sum_i}\mu_{iR}\,(n_i^A-n^A_{iR})
\end{equation}
we can express the optimal work as
\begin{equation}\label{WrevRnV}
	W^{A\to }_{12\text{rev}}=(\Omega^{R_{V,\bn}})^A_1-(\Omega^{R_{V,\bn}})^A_2=\Xi ^A_1-\Xi ^A_2
\end{equation}
Also  $\Xi $ possesses an absolute minimum at state $A_R$, where $A$ and $R$ are in mutual equilibrium and hence $T_R^A=T_R$, $p_R^A=p_R$,  $\mu_{iR}^A=\mu_{iR}$ $\forall i$. Moreover, in state $A_R$,  $\Xi^A_R=\Eu^{A}_R$ where 
\begin{equation}\label{Eu}
	\Eu=E-T\,S+p\,V-\bmu\cdot\bn
\end{equation} is the \textit{Hill  free energy}.\footnote{The Hill free energy (also Euler free energy) is  defined --- only for stable equilibrium states --- by the Legendre transform of the fundamental relation $E=E(S,V,\bn)$ with respect to all its variables. Its ``natural independent variables'' are $T$, $p$, $\bmu$,  i.e., $\Eu=\Eu(T,p,\bmu)$, and its differential can be written as $\diff \Eu= -S\diff T+V\diff p-\bn\cdot\diff \bmu$. Note that, in general (in particular for small systems), $T$, $p$, and $\bmu$ are independent. They become interdependent within the simple-system model (large amounts) where the assumption of extensivity is equivalent to setting $\Eu=0$. } \footnote{We name these properties in honor of T.L.~Hill, a pioneer of the field of ``nanothermodynamics'' \cite{Hill1964,Hill2001}, because of their direct relation with Hill's ``subdivision potential'' \cite{Chamberlin2015,bedeaux2024}. However, we choose the symbol $\Eu$ for the Hill free energy  to connect with our previous naming it the Euler free energy \cite{Beretta2024MITOCW}, justified by the fact that within the simple-system model approximation (large amounts of constituents, macroscopic limit) the Hill free energy  vanishes, $\Eu=0$, leading to the well-known Euler relation, $E=TS-pV+\mu\cdot\bn$, from which follows extensivity, the fact that $T$, $p$, and $\bmu$ are not independent, and the phase rule.  }  But it is important to note that for all the other  stable equilibrium states, $\Xi\ne \Eu$, and while the Hill availability function $\Xi$ is defined also for nonequilibrium states, the Hill free energy $\Eu$ is not. It is also worth noting that setting $\Eu=0$ yields the \textit{Euler relation},   $E=TS-pV+\mu\cdot\bn$, which however does not hold in general but only within the simple-system model approximation (large amounts of constituents, macroscopic limit). In particular, for few-particle systems (nanothermodynamics) the Hill free energy $\Eu$ does not vanish. It plays an important role in determining how specific properties depend on the amounts of constituents (lack of extensivity) (see Sec.~\ref{nonextensivity}), and  it is related to the minimum work of partitioning and the maximum work that can be obtained by removing partitions (see Sec.~\ref{partitioning}). 

The stability condition that follows from the observation that  $\Xi$ has an absolute minimum at state $A_R$ is 
\begin{equation}\label{XiStability}\Xi_1^A-\Xi_R^A > 0 \qquad\text{for every state $A_1\ne A_R$}		\end{equation}

\section{Work of partitioning, Hill free energy, and subdivision potential}\label{partitioning}

The partitioning of a system into subsystems is a fundamental operation that clarifies the distinction between additivity and extensivity. Macroscopic and mesoscopic treatments often take for granted the ``simple-system model approximation''~\cite{GyftopoulosBeretta1991, BerettaGyftopoulosSS_2015}, whereby a system in a stable equilibrium state can be subdivided into a composite of smaller subsystems without energetic or entropic cost. This approximation is a cornerstone of the continuum hypothesis required to define spatial fields in many nonequilibrium frameworks; it amounts to neglecting wall rarefaction, surface effects, finite-size constraints, and changes in the spectrum of accessible states introduced by physical partitions.
	
For systems with few particles, however, this approximation fails. Such effects introduce non-negligible energetic and entropic contributions associated with the subdivision of a system into subsystems. Here, it is vital to emphasize that ``subsystems'' refer to well-defined systems (separable and uncorrelated) --- a fundamental condition for their individual energies and entropies to be formally defined.

	As part of our rigorous operational construction, which remains valid for both few- and many-particle systems, we must explicitly account for the work required to establish the boundaries that define a partition. As we shall see, the Hill (Euler) free energy quantifies precisely this contribution and is therefore referred to as the \textit{subdivision potential}. Its appearance in our derivation highlights that a thermodynamic description applicable to systems of arbitrary scale must account for deviations from extensivity rather than assuming them away from the outset --- a necessity made even more explicit in the following section.

We consider a system initially in a stable equilibrium state and examine the thermodynamic cost of partitioning it into $\lambda$ compartments. Each compartment is required to contain equal amounts of cconstituents, equal volume, and equal entropy, so that the partitioned system as a whole is again in a stable equilibrium state. Since the overall energy and entropy of the system are conserved in a reversible weight process, the transition from the unpartitioned to the partitioned configuration can, in principle, be carried out reversibly, and therefore admits a well-defined minimum work requirement.

\begin{figure}[!ht]
	\begin{center}
			\includegraphics[scale=0.52]{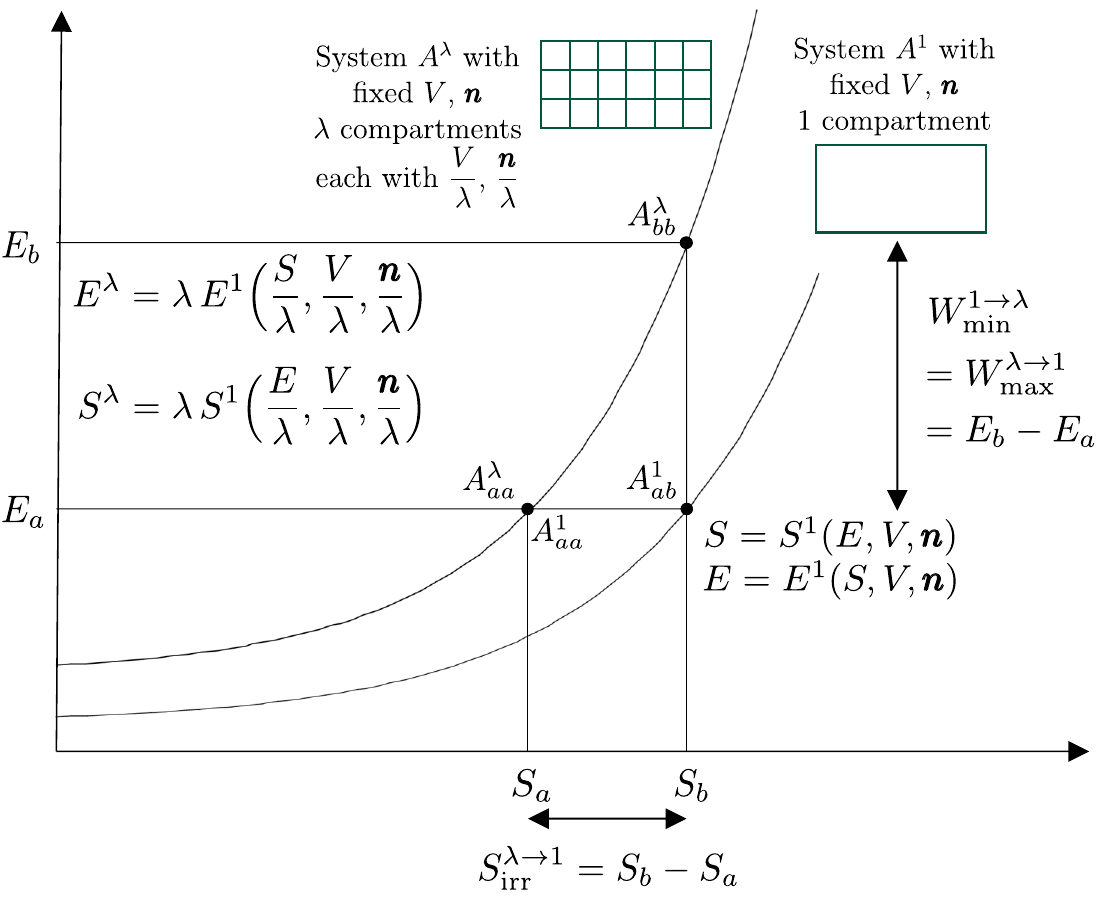}	
			\caption{\label{MinWofPartitioning}Representation on the $E$--$S$ diagram of the minimum work of partitioning, the maximum work from removing partitions, and the entropy generation by removing partitions.  In Section \ref{quantumexamples},  for the example of  distinguishable but identical point particles in the ideal-gas limit,  we derive the explicit expressions   (Eqs.~\ref{separationGas}) $S_\text{irr}^{\lambda\to 1}=n\Boltz\ln\lambda$ and $ W_\text{min}^{1\to\lambda} = W_\text{max}^{\lambda\to 1}=\frac{3}{2}(\lambda^{2/3}-1)n\Boltz T_{ab} $.}
	\end{center}
\end{figure}

The energy–entropy representation in Fig.~\ref{MinWofPartitioning} provides a clear geometric interpretation of the many ways this process can be realized. Introducing partitions changes the definition of the system and, with it, the locus of stable equilibrium states. Immediately after the partitions are removed, the $\lambda$-compartment system $A^\lambda$ initially in the stable equilibrium $A^\lambda_{aa}$  with energy $E_a$ and entropy $S_a$ finds itself transformed into a new single-compartment system $A^1$ in a nonequilibrium state, $A^1_{aa}$, even though its energy and entropy are unchanged. If left to itself, the state of $A^1$ would evolve spontaneously at constant energy $E_a$ toward the new stable equilibrium state $A^1_{ab}$ with entropy $S_b$, generating entropy by irreversibility $S^{\lambda\to 1}_\text{irr}=S_b-S_a$. Conversely, if the partitions are removed or introduced through a sufficiently rapid and controlled reversible weight process, it is possible to extract or supply the corresponding adiabatic availability, following a path of constant entropy.

From this viewpoint, the minimum work required to introduce $\lambda$ partitions is equal to the maximum work that can be extracted when the $\lambda$ compartments are reversibly merged into a single-compartment system. 
 This work is given by the difference between the energy $E_b$ of the partitioned configuration with entropy $S_b$ and the energy $E_a$ of the unpartitioned system evaluated at the same  entropy, volume, and amounts of constituents. By examining how this work changes when the number of compartments is varied by one unit, we identify a quantity naturally interpreted as the work associated with adding or removing a single partition. As shown below, this quantity coincides with the Hill (or Euler) free energy of each compartment of the partitioned system, thereby providing a direct physical interpretation of Hill’s free energy as the energetic cost of subdivision in systems with few particles.

It is noteworthy that in general the concavity of the fundamental relation $S=S^1(E,V,\bn)$ in all its variables (and the convexity of its positive-temperature energy form $E=E^1(S,V,\bn)$) implies its subadditivity, i.e., the following inequalities, which justify the relative positioning of the stable-equilibrium-state curves in  Fig.~\ref{MinWofPartitioning},
\begin{equation}\label{minWseparation}
 S^{\lambda}={\lambda}\,S^1\Big(\frac{E}{\lambda},\frac{V}{\lambda},\frac{\pmb{n}}{\lambda}\Big) < S^1(E,V,\pmb{n}) \qquad
 E^{\lambda}={\lambda}\,E^1\Big(\frac{S}{\lambda},\frac{V}{\lambda},\frac{\pmb{n}}{\lambda}\Big)>E^1(S,V,\pmb{n})
\end{equation} 
The minimum work of partitioning into $\lambda$ identical compartments in identical stable equilibrium states is equal to the maximum work that can be obtained  by removing the partitions. They obtain in  reversible weight processes that connect states $A^1_{ab}$ and $A^\lambda_{bb}$ in one or the other direction, 
\begin{equation}\label{minWseparation2}
	W_{\text{min}}^{1\to \calN} =W_{\text{max}}^{ \calN \to 1}=E^\lambda_{bb}-E^1_{ab}
	={ \calN}\,E^1\Big(\frac{S_b}{ \calN},\frac{V}{ \calN},\frac{\bn}{ \calN}\Big)-E^1(S_b,V,\bn) > 0
\end{equation} 

It is interesting to compute the minimum work to increment  $\lambda$ by one and the maximum work to decrement $\lambda$ by one
\begin{equation}
	W_{\text{min}}^{\calN \to\calN+1}=\frac{W_{\text{min}}^{1\to \calN+1}-W_{\text{min}}^{1\to \calN}}{(\calN+1)-\calN}\qquad\text{and}\qquad
	W_{\text{max}}^{\calN \to\calN-1}=\frac{W_{\text{max}}^{1\to \calN}-W_{\text{max}}^{1\to \calN-1}}{\calN-(\calN-1)}
\end{equation} 
For $\lambda$ sufficiently large, we can approximate these incremental ratios by 
\begin{equation}
	W_{\text{min}}^{\calN \to\calN+1}\approx  \frac{\partial W_{\text{min}}^{1\to \calN} }{\partial\calN} \qquad\text{and}\qquad
	W_{\text{max}}^{\calN \to\calN-1}\approx \frac{\partial W_{\text{max}}^{\calN\to 1} }{\partial\calN}
\end{equation} 
where the partial derivatives of Eq.~\ref{minWseparation2} (for $S_b=S$)  are given by
\begin{align}
	\frac{\partial W_{\text{min}}^{1\to \calN} }{\partial\calN}&= \frac{\partial W_{\text{max}}^{\calN\to 1} }{\partial\calN}=E^1\Big(\frac{S}{ \calN},\frac{V}{ \calN},\frac{\bn}{ \calN}\Big)+\calN\,T^1\Big(\frac{S}{ \calN},\frac{V}{ \calN},\frac{\bn}{ \calN}\Big)\Big(-\frac{S}{\calN^2}\Big)\nonumber \\ &\ \ \  -\calN\,p^1\Big(\frac{S}{ \calN},\frac{V}{ \calN},\frac{\bn}{ \calN}\Big)\Big(-\frac{V}{\calN^2}\Big)+\calN\,\bmu^1\Big(\frac{S}{ \calN},\frac{V}{ \calN},\frac{\bn}{ \calN}\Big)\cdot\Big(-\frac{\bn}{\calN^2}\Big)\nonumber\\
	&= E^1\Big(\frac{S}{ \calN},\frac{V}{ \calN},\frac{\bn}{ \calN}\Big)-\frac{S}{\calN}\,T^1\Big(\frac{S}{ \calN},\frac{V}{ \calN},\frac{\bn}{ \calN}\Big) +\frac{V}{ \calN}\,p^1\Big(\frac{S}{ \calN},\frac{V}{ \calN},\frac{\bn}{ \calN}\Big)-\bmu^1\Big(\frac{S}{ \calN},\frac{V}{ \calN},\frac{\bn}{ \calN}\Big)\cdot\frac{\bn}{\calN}\nonumber\\ 
	&=\Eu^1\Big(\frac{S}{ \calN},\frac{V}{ \calN},\frac{\bn}{ \calN}\Big)
\end{align}
Thus we see that the Hill (Euler) free energy $\Eu = E-TS+pV-\bmu\cdot\bn$ for one of the $\calN$ compartments equals (for large $\lambda$) the optimal work to increase or decrease $\lambda$ by one.

\section{Stable equilibrium  properties in the absence of extensivity}\label{nonextensivity}

For positive-temperature stable equilibrium states the following relations hold in general regardless of the system's size
\begin{align}
	&s=\frac{S}{n}=\frac{1}{n}S(nu,nv,n\by)=s(u,v,\by,n) & \Big(\frac{\partial s}{\partial n}\Big)_{u,v,\by}&=\frac{1}{n^2}\frac{\Eu}{T}\\
	&e=\frac{E}{n}=\frac{1}{n}E(ns,nv,n\by) =e(s,v,\by,n)& \Big(\frac{\partial e}{\partial n}\Big)_{s,v,\by}&=-\frac{1}{n^2}\Eu\\
	&f=\frac{F}{n}=\frac{1}{n}F(T,nv,n\by)=f(T,v,\by,n) & \Big(\frac{\partial f}{\partial n}\Big)_{T,v,\by}&=-\frac{1}{n^2}\Eu\\
	&g=\frac{G}{n}=\frac{1}{n}G(T,p,n\by)=g(T,p,\by,n) & \Big(\frac{\partial g}{\partial n}\Big)_{T,p,\by}&=-\frac{1}{n^2}\Eu\\
		&h=\frac{H}{n}=\frac{1}{n}H(ns,p,n\by)=h(s,p,\by,n) & \Big(\frac{\partial h}{\partial n}\Big)_{s,p,\by}&=-\frac{1}{n^2}\Eu
\end{align}
where $n=\sum_i  n_i$ is the overall amount of constituents, $\by$ denotes the mole fractions ($y_i=n_i/n$), and  $s$, $e$,  $f$, $g$, $h$  respectively represent \textit{specific molar} entropy, energy, Helmholtz free energy, Gibbs free energy, enthalpy. We see that in general specific molar properties are not independent of $n$. By contrast, a defining requirement for extensivity is that specific properties be independent of $n$, which these relations  clearly show can happen only approximately for large $n$, when the specific molar Hill free energy $\Eu/n$ is finite, or exactly when $\Eu=0$. 

The following relations also hold in general, i.e., in the absence of modeling assumptions that imply extensivity. When $\Eu=0$ they all reduce to well-known relations of macroscopic equilibrium thermodynamics.
\begin{align}
		 S&=\sum_{i=1}^rn_i\,s_i - \Big(\frac{\partial \Eu}{\partial T}\Big)_{p,\pmb{n}} & \overset{ n  \text{ large}}{\longrightarrow
	} & & S&=\sum_{i=1}^rn_i\,s_i\\
	 E&=\sum_{i=1}^rn_i\,e_i +\Eu - T\Big(\frac{\partial \Eu}{\partial T}\Big)_{p,\pmb{n}}-p\Big(\frac{\partial \Eu}{\partial p}\Big)_{T,\pmb{n}} & \overset{ n  \text{ large}}{\longrightarrow
	} & & E&=\sum_{i=1}^rn_i\,e_i\\
	 F&=\sum_{i=1}^rn_i\,f_i +\Eu -p\Big(\frac{\partial \Eu}{\partial p}\Big)_{T,\pmb{n}} \hphantom{-p\Big(\frac{\partial \Eu}{\partial p}\Big)_{T,\pmb{n}}} & \overset{ n  \text{ large}}{\longrightarrow
	} & & F&=\sum_{i=1}^rn_i\,f_i\\
	 G&=\sum_{i=1}^rn_i\,g_i +\Eu=\sum_{i=1}^rn_i\,\mu_i +\Eu \hphantom{\Big(\frac{\partial \Eu}{\partial p}\Big)_{T,\pmb{n}}} & \overset{ n  \text{ large}}{\longrightarrow
	} & & G&=\sum_{i=1}^rn_i\,\mu_i\\
	 H&=\sum_{i=1}^rn_i\,h_i +\Eu - T\Big(\frac{\partial \Eu}{\partial T}\Big)_{p,\pmb{n}}\hphantom{-p\Big(\frac{\partial \Eu}{\partial p}\Big)_{T,\pmb{n}}} & \overset{ n  \text{ large}}{\longrightarrow
	} &  &H&=\sum_{i=1}^rn_i\,h_i\\
V&=\sum_{i=1}^rn_i\,v_i+\Big(\frac{\partial \Eu}{\partial p}\Big)_{T,\pmb{n}} & \overset{ n  \text{ large}}{\longrightarrow
} & & V&=\sum_{i=1}^rn_i\,v_i\\
0&=	\sum_{i=1}^rn_i\,\mu_{i,j}+\Big(\frac{\partial \Eu}{\partial n_j}\Big)_{T,p,\pmb{n}'_j} & \overset{ n  \text{ large}}{\longrightarrow
} &  & 0&=\sum_{i=1}^rn_i\,\mu_{i,j}\label{DuhemMargules}
\end{align}
where $s_i$, $e_i$,  $f_i$, $g_i$, $h_i$, $v_i$, $\mu_{i,j}$ are respectively the \textit{$i$-th constituent partial molar} entropy, energy, Helmholtz free energy, Gibbs free energy, enthalpy, volume, and chemical potentials  defined in terms of the Gibbs free energy $G=E-TS+pV=G(T,p,\bn)$ (Legendre transform of $E=E(S,V,\bn)$ with respect to $S$ and $V$) and the chemical potentials as follows
\begin{align}
s_i&= -\Big(\frac{\partial \mu_i}{\partial T}\Big)_{p,\pmb{n}}=-\Big(\frac{\partial^2 G}{\partial T\partial n_i}\Big)_{p,\pmb{n}'_i}=\Big(\frac{\partial S}{\partial n_i}\Big)_{T,p,\pmb{n}'_i}= s_i(T,p,n\pmb{y}) \\
e_i&=\Big(\frac{\partial E}{\partial n_i}\Big)_{T,p,\pmb{n}'_i}=\mu_i+T\,s_i-p\,v_i= e_i(T,p,n\pmb{y}) \\
f_i&=\Big(\frac{\partial F}{\partial n_i}\Big)_{T,p,\pmb{n}'_i}= f_i(T,p,n\pmb{y}) =\mu_i -p\,v_i= f_i(T,p,n\pmb{y})\\
g_i& =\mu_i=\Big(\frac{\partial G}{\partial n_i}\Big)_{T,p,\pmb{n}'_i}=\mu_i(T,p,n\pmb{y})\\
h_i&=\Big(\frac{\partial H}{\partial n_i}\Big)_{T,p,\pmb{n}'_i}= h_i(T,p,n\pmb{y}) =T\,s_i+\mu_i=\Big(\frac{\partial (\mu_i/T)}{\partial (1/T)}\Big)_{p,\pmb{n}}\\
v_i&=\Big(\frac{\partial \mu_i}{\partial p}\Big)_{T,\pmb{n}}=\Big(\frac{\partial^2 G}{\partial p\partial n_i}\Big)_{T,\pmb{n}'_i}=\Big(\frac{\partial V}{\partial n_i}\Big)_{T,p,\pmb{n}'_i}= v_i(T,p,n\pmb{y}) \\
\mu_{i,j}	&= \Big(\frac{\partial \mu_i}{\partial n_j}\Big)_{T,p,\pmb{n}'_j}=\Big(\frac{\partial^2 G}{\partial n_j\partial n_i}\Big)_{T,p,\pmb{n}'_{ij}}=\Big(\frac{\partial \mu_j}{\partial n_i}\Big)_{T,p,\pmb{n}'_{ij}}= \mu_{i,j}(T,p,n\pmb{y})= \mu_{j,i}(T,p,n\pmb{y})
\end{align}
Eq.~\ref{DuhemMargules} for large $n$ is the \textit{ Duhem-Margules relation}. For large $n$ the dependences on $(T,p,n\by)$ reduce to $(T,p,\by)$, and the partial properties become independent of $n$.

\section{Stability conditions and 	LeChatelier-Braun principle}

	The inequalities  that obtain from the stability conditions \ref{GammaStability}, \ref{PhiStability}, \ref{UpsilonStability}, and \ref{XiStability}, give body to the general \textit{LeChatelier-Braun theorem} (or \textit{principle}). 	
For example, from the general condition $ \diff^2E^A|_{V,\bn} \ge 0$ we have seen that for normal systems at stable equilibrium
\begin{equation} \Big(\frac{{\partial }^2S}{\partial E^2}\Big)_{V,\bn} \le 0\ \Rightarrow\  \Big(\frac{{\partial }T}{\partial E}\Big)_{V,\bn}\ge 0\qquad \Big(\frac{{\partial }^2E}{\partial S^2}\Big)_{V,\bn} \ge 0\ \Rightarrow\  \Big(\frac{{\partial }T}{\partial S}\Big)_{V,\bn}\ge 0
		\end{equation} 
Combined with the idea that $T$ is an escaping tendency for energy, we may interpret this as follows. 
	
 If we change a stable equilibrium state to another  with higher energy (or entropy), the temperature increases, hence enhancing the systems' tendency to give energy (or entropy) away. The increase of temperature can be interpreted as an attempt of the system to counteract the externally imposed increase of energy (or entropy) by enhancing its own tendency to give energy (and entropy) away. 
	
If the system is initially in mutual equilibrium with a reservoir $R$, an injection (subtraction) of energy pushes its  state away from mutual equilibrium, but the consequent increase (decrease) of its temperature, away from the initial $T_R$, favors a spontaneous process whereby the system exchanges energy (and entropy) with $R$ so as to return toward mutual equilibrium.
	
	\subsection{Mathematical basis of the LeChatelier-Braun principle}

Assume we have a function $P=P(x,y,\bz  )$ where $P$, $x$, $y$ are additive exchangeable properties (such as $S$, $E$, $V$, the $n_i$'s or linear combinations of them, such as $\Gamma$, $\Phi$, $\Upsilon$, and $\Xi$) and $P$ is subject to a stability condition $\diff^2 P|_{\bz }\ge 0$ or $\diff^2 P|_{\bz }\le 0$. We write its first and second partial differentials  (constant $\bz$) with the following notation
	\[ \diff P|_{\bz  }=P_{,x}\, \diff x+ P_{,y}\diff y \qquad P_{,x}=\Big(\frac{\partial P}{\partial x}\Big)_{y,\bz  }=P_{,x}(x,y,\bz  ) \qquad  P_{,y}=\Big(\frac{\partial P}{\partial y}\Big)_{x,\bz  }=P_{,y}(x,y,\bz  ) \]
	\[ \diff^2 P|_{\bz  }= [\diff x\ \diff y]\left[\begin{matrix} \displaystyle P_{,xx} &\!\!\displaystyle\!\! P_{,xy}\\  \displaystyle P_{,xy} & \displaystyle\!\!\! P_{,yy}\\ \end{matrix}\right] \left[\begin{matrix} \displaystyle \diff x\\   \displaystyle \diff y\\ \end{matrix}\right]= P_{,xx}\,(\diff x)^2+2P_{,xy}\diff x\diff y+P_{,yy}\,(\diff y)^2\]
	\[  P_{,xx}=\Big(\frac{\partial^2 P}{\partial x^2}\Big)_{y,\bz  }=\Big(\frac{\partial P_{,x}}{\partial x}\Big)_{y,\bz  } \qquad P_{,yy}=\Big(\frac{\partial^2 P}{\partial y^2}\Big)_{x,\bz  }=\Big(\frac{\partial P_{,y}}{\partial y}\Big)_{x,\bz  } \]
	\[  P_{,xy}=\Big(\frac{\partial^2 P}{\partial x\partial y}\Big)_{\bz  }=\Big(\frac{\partial P_{,y}}{\partial x}\Big)_{y,\bz  }=\Big(\frac{\partial P_{,x}}{\partial y}\Big)_{x,\bz  }=\Big(\frac{\partial^2 P}{\partial y\partial x}\Big)_{\bz }=P_{,yx}\]
	The quadratic form can be rewritten (check by substitution) in the two canonical forms 
	\begin{align*} \diff^2 P|_{\bz }&= P_{,xx}\,\Big(\diff x + \frac{P_{,xy}}{P_{,xx}}\diff y\Big)^2+\lambda_{y}\,(\diff y)^2 &\text{where}\qquad  & \lambda_{y}=P_{,yy}-\Big( \frac{P_{,xy}}{P_{,xx}}\Big)^2 P_{,xx} \\ \ &=  \lambda_{x}\,(\diff x)^2+P_{,yy}\,\Big(\diff y + \frac{P_{,xy}}{P_{,yy}}\diff x\Big)^2 &\text{where}\qquad  & \lambda_{x}=P_{,xx}-\Big( \frac{P_{,xy}}{P_{,yy}}\Big)^2 P_{,yy}
	\end{align*}
	Therefore, since the stability conditions must hold for arbitrary $\diff x$ and $\diff y$,
	\[\diff^2 P|_{\bz }\ge 0 \ \Rightarrow\  \left\{\begin{matrix}P_{,xx}\ge \lambda_x\ge 0 \\ P_{,yy}\ge \lambda_y\ge 0\end{matrix}\right.\quad\text{whereas}\quad \diff^2 P|_{\bz }\le 0 \ \Rightarrow\  \left\{\begin{matrix}P_{,xx}\le \lambda_x\le 0 \\ P_{,yy}\le \lambda_y\le 0\end{matrix}\right.\]
	Whether $\diff^2 P|_{\bz }$ is positive semidefinite or negative semidefinite, we can use the properties of Jacobians to write
	\[ 0\le \text{det(Hess}(P))
	= \left|\begin{matrix} \displaystyle P_{,xx} &\!\!\displaystyle\!\! P_{,xy}\\  \displaystyle P_{,xy} & \displaystyle\!\!\! P_{,yy}\\ \end{matrix}\right|  = 
	\frac{\partial (P_{,x},P_{,y})}{\partial (x,y)}=\frac{\partial (P_{,x},P_{,y})}{ }\,\frac{ }{\partial (x,y)}\]
	\[ =\left\{ \begin{array}{l}\displaystyle\frac{\partial (P_{,x},P_{,y})}{\partial (P_{,x},y)}\,\frac{\partial (P_{,x},y)}{\partial (x,y)}=\Big(\frac{\partial P_{,y}}{\partial y}\Big)_{P_{,x}} \Big(\frac{\partial P_{,x}}{\partial x}\Big)_y= \Big(\frac{\partial P_{,y}}{\partial y}\Big)_{P_{,x}}\, P_{,xx}\ge 0\\ \ \\ \displaystyle   \frac{\partial (P_{,x},P_{,y})}{\partial (x,P_{,y})}\,\frac{\partial (x,P_{,y})}{\partial (x,y)}=\Big(\frac{\partial P_{,x}}{\partial x}\Big)_{P_{,y}} \Big(\frac{\partial P_{,y}}{\partial y}\Big)_x= \Big(\frac{\partial P_{,x}}{\partial x}\Big)_{P_{,y}}\,P_{,yy}\ge 0 \end{array}\right. \]
	\noindent Therefore, we can rewrite  $\lambda_{x}$ and $\lambda_{y}$ as 
	\[\lambda_{x}=P_{,xx}-\Big( \frac{P_{,xy}}{P_{,yy}}\Big)^2 P_{,yy}=\frac{\text{det(Hess}(P))}{P_{,yy} } = \Big(\frac{\partial P_{,x}}{\partial x}\Big)_{P_{,y}} \] \[ \lambda_{y}=P_{,yy}-\Big( \frac{P_{,xy}}{P_{,xx}}\Big)^2 P_{,xx}=\frac{\text{det(Hess}(P))}{P_{,xx}} = \Big(\frac{\partial P_{,y}}{\partial y}\Big)_{P_{,x}} \]
	\noindent so that, finally, the stability conditions become
	\[\diff^2 P|_{\bz }\ge 0  \Rightarrow \left\{\begin{matrix}\displaystyle\Big(\frac{\partial P_{,x}}{\partial x}\Big)_{y}\ge \Big(\frac{\partial P_{,x}}{\partial x}\Big)_{P_{,y}}\ge 0 \\ \displaystyle\Big(\frac{\partial P_{,y}}{\partial y}\Big)_{x}\ge \Big(\frac{\partial P_{,y}}{\partial y}\Big)_{P_{,x}}\ge 0\end{matrix}\right.\qquad \diff^2 P|_{\bz }\le 0  \Rightarrow  \left\{\begin{matrix}\displaystyle\Big(\frac{\partial P_{,x}}{\partial x}\Big)_{y}\le \Big(\frac{\partial P_{,x}}{\partial x}\Big)_{P_{,y}}\le 0 \\ \displaystyle\Big(\frac{\partial P_{,y}}{\partial y}\Big)_{x}\le \Big(\frac{\partial P_{,y}}{\partial y}\Big)_{P_{,x}}\le 0\end{matrix}\right.\]
	
	\subsection{LeChatelier-Braun principle}
	
	We may interpret these inequalities as follows. To fix ideas, assume $P=E$ and $x=S$, $y=V$, $\bz =\bn $, so that $P_{,x}=T$ and $P_{,y}=-p$. The corresponding stability condition is  $ \diff^2E|_{\bn} \ge 0$ which implies the conditions\footnote{Condition \ref{cpcv} entails the general inequality $C_p\ge C_V\ge 0$ where $C_V$ and $C_p$ are the heat capacities at constant volume and pressure. Condition \ref{kskT} entails the general inequality $\kappa_T\ge \kappa_S\ge 0$ where $\kappa_T$ and $\kappa_S$ are the isothermal and isoentropic compressibilities.}
	\begin{align}
		\Big(\frac{\partial P_{,x}}{\partial x}\Big)_{y}&\ge \Big(\frac{\partial P_{,x}}{\partial x}\Big)_{P_{,y}}\ge 0&\text{that is}& & \Big(\frac{\partial T}{\partial S}\Big)_{V}&\ge \Big(\frac{\partial T}{\partial S}\Big)_{p}\ge 0\label{cpcv}\\
		\Big(\frac{\partial P_{,y}}{\partial y}\Big)_{x}&\ge \Big(\frac{\partial P_{,y}}{\partial y}\Big)_{P_{,x}}\ge 0&\text{that is}& & -\Big(\frac{\partial p}{\partial V}\Big)_{S}&\ge -\Big(\frac{\partial p}{\partial V}\Big)_{T}\ge 0\label{kskT}
	\end{align}
	
	\noindent\textbf{Assertion 1}. 
	If a system initially in mutual equilibrium with a thermal reservoir $R$ is perturbed to a neighboring stable equilibrium state in which the value of an additive property $x$ is changed to $x+\diff x$, the system responds by changing the conjugate potential $P_{,x}$ in the direction that increases (decreases) the escaping tendency of $x$ when $x$ is increased (decreased). As a consequence, the system tends to oppose the imposed exchange of $x$ by favoring a spontaneous exchange with $R$ that acts in the opposite direction, thereby tending to restore mutual equilibrium.
	
	\noindent\textbf{Assertion 2}. 
	The magnitude of this response depends on how many mutual equilibrium conditions are disrupted by the perturbation. A perturbation that constrains the system to maintain a fixed value of another additive property $y$ produces a stronger response,
	$\big(\partial P_{,x}/\partial x\big)_{y}\diff x$,
	than a perturbation that constrains the system to maintain a fixed value of the conjugate potential $P_{,y}$,
	$\big(\partial P_{,x}/\partial x\big)_{P_{,y}}\diff x$.
	In general, the system’s counterreaction is stronger when the perturbation breaks a larger number of mutual equilibrium conditions.

\section{\label{quantumexamples}Entropy and uncertainty in quantum models}

The following discussion of Quantum Thermodynamics serves as a concrete demonstration of the operational construction  developed in the first part of this work for one of the nonequilibrium frameworks listed in Section~\ref{state}. In the quantum regime, the identification of the thermodynamic state and its properties is particularly challenging, as many classical heuristic partitions of energy transfer into work and heat are currently the subject of active investigation~\cite{Binder2018,Davoudi2024}. There are numerous interesting results that stem from the application of the concepts discussed so far to the study of additional properties defined within the framework of quantum-theoretical models. Our framework provides a self-consistent logical scaffolding that remains independent of specific modeling choices for dissipative processes and may help to clarify currently controversial definitions. This discussion also allows us to explore fundamental aspects such as the interpretation of entropy as a measure of uncertainty and the origin of the ideal-gas model.

The description of equilibrium and nonequilibrium states within the scope of \emph{atomic and quantum theory} is based on two fundamental observations of quantum theory: the quantization of energy levels and the irreducible need for probabilities in the description of the states of a system.

\subsection{Energy levels. Quantization}

The value of a property of a system in a given state (any state) generally does not coincide with the result of a single act of measurement of that property. In fact, contrary to what was suggested for simplicity in Sec.~\ref{properties}, a single act of measurement is not sufficient to determine the value of the measured property. The measurement procedure that defines the property must be repeated on identical replicas of the system, all prepared in the same way, and the full statistics of the collected measurement outcomes must be analyzed.

To fix ideas, let us consider the measurement procedure that defines the property energy. When applied to a system described within the context of classical modeling, repeated applications of the procedure to identical and identically prepared replicas always yield the same numerical value, namely the value of the energy. By contrast, an essential feature of quantum modeling is that, when the procedure is applied repeatedly to identical and identically prepared systems, the outcomes of individual measurement acts are, in general, \emph{unpredictable}\footnote{In principle, in addition to the conditions of separability and statistical independence implicit in our use of the word ``system,'' one should also ensure a clear distinction between intrinsic quantum uncertainties and classical statistical mixing, according to the notion of ``homogeneity'' of the preparation scheme. This refinement goes beyond our scope here, but is explained, for example, in \cite{park_1968_nature}. For a technical discussion see also \cite[Secs.~4--5]{RayBeretta2025}.} yet statistically structured. The outcomes occur with consistently repeatable frequencies among different numerical values, the \emph{energy levels} of the system, which we denote by \(\epsilon_j\).

The set \(\{\epsilon_j\}\) of possible energy levels constitutes a characteristic of the system, called the \emph{energy spectrum}, which depends only on the amounts of constituents \(\bn\) and the parameters \(\bbeta\). The spectrum may be continuous, discrete, or partly discrete and partly continuous.\footnote{For simplicity, but without significant loss of generality, we restrict attention to systems with a \emph{discrete} energy spectrum.}

A similar situation holds for any other property whose measurement procedure is defined by mechanics (particle position, momentum, angular momentum, magnetic dipole moment, etc.). To determine the (quantum) state of a system, it is therefore necessary to collect the full measurement statistics for a ``quorum'' (a complete and independent set) of such mechanical properties. This procedure is known as \emph{quantum tomography} \cite{park_1979,park_1971_general,Dariano_2003}.

There are other types of properties, such as adiabatic availability, available energy with respect to a reservoir, and entropy, which --- being defined through measurement procedures that involve the determination of a maximum obtainable quantity (e.g., the maximum energy transferable to an external weight) or conditions on the final state of a weight process --- do not admit direct measurement procedures with a meaningful interpretation of individual measurement acts. The determination of the values of properties of this kind therefore requires a full tomography.

The existence of properties, such as energy, with a \emph{discrete} spectrum of values accessible through individual measurement acts is a distinctive feature of quantum theory, referred to as \emph{quantization}. Discrete spectra are generally associated with modes of the electromagnetic field and with the internal degrees of freedom (rotational, vibrational, electronic, and magnetic) of atoms and molecules, aggregates of interacting atoms and molecules, and crystalline lattices. Continuous spectra are instead typically associated with the translational degrees of freedom of free particles, atoms, and molecules not confined to finite regions of space.

Molecular theory provides methods for deriving expressions for the set \(\{\epsilon_j\}\) of energy levels as a function of the system structure and of the atoms and molecules that compose it, or equivalently as a function of the amounts of constituents and the parameters,
\begin{equation}
	\{\epsilon_j\}=\{\epsilon_j(\bn,\bbeta) \}
	\label{livellichiuso}
\end{equation}
We will present two examples of such expressions below; however, the methods used to calculate them lie beyond the scope of the present overview.

\subsection{Quantum probabilities. Uncertainty}

In addition to quantization, the aspect that most sharply contrasts with the determinism on which classical physics was based is the unpredictability of the outcome of a single measurement act. This unpredictability reflects an intrinsic indeterminacy in the state of systems and is a defining discovery of modern physics, its technological applications, and the various philosophical interpretations that have emerged over the last century and remain under discussion.\footnote{``God does not play dice!'' declared Albert Einstein --- one of the founders of quantum theory --- expressing his profound dissatisfaction with the impossibility of predicting which energy level a system will yield in a given measurement act. Today, this feature is widely accepted as a fact of physical reality, although the dissatisfaction expressed by Einstein is still shared by some.}

The indeterminacy inherent in the state of every system does not prevent the rigorous definition of the state itself, nor does it require abandoning deterministic equations of motion that describe temporal evolution in accordance with the principle of causality, another cornerstone of classical physics.

Indeed, while the outcome of a single measurement act is not predictable, the limiting frequency \(f(\epsilon_j,N)\) with which each energy level \(\epsilon_j\) occurs in a sequence of \(N\) measurement acts on identical replicas of the system, all prepared in the same state, is perfectly predictable when \(N\) is sufficiently large.\footnote{For systems with a continuous spectrum, the predictable quantity is the frequency \(f(\epsilon,N)\,d\epsilon\) with which individual measurement acts yield values between \(\epsilon\) and \(\epsilon+d\epsilon\).} It is therefore possible to define, for each energy level \(\epsilon_j\), a property denoted by \(p_{\epsilon_j}\), called the \emph{probability that a single measurement act yields the energy level} \(\epsilon_j\), defined by the limit
\begin{equation}
	p_{\epsilon_j}=\lim_{N\to\infty} f(\epsilon_j,N) 
\end{equation}
The value of this property becomes part of the set of quantities that define the state of the system. Evidently,
\begin{equation}
	\sum_{\{\epsilon_j\}} p_{\epsilon_j}=1 
\end{equation}
where the sum extends over all values in the energy spectrum.\footnote{Similarly, for a continuous spectrum one defines a probability density \(p_\epsilon\) such that
	\begin{equation}
		p_\epsilon\,d\epsilon=\lim_{N\to\infty} f(\epsilon,N)\,d\epsilon 
	\end{equation}
	With both discrete and continuous spectra, the normalization condition becomes
$\sum_{\{\epsilon_j\}} p_{\epsilon_j} + \int p_\epsilon\,d\epsilon = 1 $.}

\subsection{Energy}

The energy measurement procedure must be repeated $N$ times on identical replicas of the system, all prepared identically, with $N$ sufficiently large (in the infinite limit), until the mean value of the energy levels provided by individual measurement acts weighted by their respective frequencies,
\begin{equation} E(N)= \sum_{\{\epsilon_j\}} f(\epsilon_j,N)\, \epsilon_j \end{equation} 
stabilizes and becomes insensitive to further repetitions of the measurement, i.e., becomes independent of $N$.
The value $E$ obtained in this way is
the result of the energy measurement procedure and, formally, is given by the relation
\begin{equation} E= \sum_{\{\epsilon_j\}}\lim_{N\to\infty} f(\epsilon_j,N)\, \epsilon_j =\sum_{\{\epsilon_j\}}p_{\epsilon_j}\epsilon_j \end{equation} 
equal to the average of the energy levels weighted by their respective probabilities (expectation value).\footnote{With both discrete and  continuous spectra, the value is $ E= \lim_{N\to\infty} f(\epsilon_j,N) \,\epsilon_j =\sum_{\{\epsilon_j\}}p_{\epsilon_j}\epsilon_j +\int p_{\epsilon}\epsilon \,d\epsilon  $. } In terms of the probabilities $p_{\epsilon_j}$ we may also compute the dispersion $\sigma_E$ of the energy measurement results around the mean value, defined by
\begin{equation} \sigma^2_E= \sum_{\{\epsilon_j\}}p_{\epsilon_j}(\epsilon_j -E)^2\end{equation} 

Now that we have introduced the probabilities $p_{\epsilon_j}$, it might seem legitimate to say that ``if as a result of a single measurement act, the system provides the energy value $\epsilon_j$, then it means that `that' was the value of the energy before the measurement act, i.e., the system was in a state with energy $\epsilon_j$.'' However, if this were true, we could also say that ``since, in general, other measurement acts performed on identical replicas of the system prepared identically provide different values (in the set $\{\epsilon_j\}$), it means that these identical replicas of the system were prepared in different states,'' and consequently, the probabilities $p_{\epsilon_j}$ would not satisfy the definition of properties since they would characterize not the states of the system but uncertainties introduced by the (inhomogeneity of the) preparation method, which with some hidden stochastic rule ``chooses'' to prepare the system in this or that state.

What has been said for energy can be repeated for other properties. For example, for a particle with translational degrees of freedom confined in a container, the components $v_x$, $v_y$, and $v_z$ of the velocity vector. Although not always, it is generally possible to measure two (or more) properties simultaneously in a single measurement act, obtaining two (or more) `responses' from the system for each act, for example, an energy level $\epsilon_j$ and three velocity levels $(v_x)_{n_x}$, $(v_y)_{n_y}$, $(v_z)_{n_z}$ for the three velocity components. It is therefore possible for the same energy level $\epsilon_j$ to emerge in different measurement acts together with different combinations of possible levels of other properties measurable simultaneously. For each $\epsilon_j$, the number of different combinations of this type that can occur is a characteristic of the system called {\it degeneracy} or {\it multiplicity of the energy level $\epsilon_j$},  denoted by $g_{\epsilon_j}$ and which is a function, in addition to the level itself, of the amounts of constituents and parameters,
\begin{equation} g_{\epsilon_j}= g_{\epsilon_j} (\bn,\bbeta) \end{equation}

\subsection{Entropy}

In general, quantum theory provides a well-defined mathematical representation of the system's state and, with it, an explicit expression of entropy valid for all states (equilibrium and nonequilibrium). This representation requires the introduction of mathematical concepts beyond the scope of these brief notes. However, for systems with a discrete energy spectrum a quite broad subclass of states\footnote{In terms of technical quantum-thermodynamics jargon, it is the subclass of states for which the density operator $\rho$, which represents the state of the system,  commutes with the Hamiltonian operator $H$, whose eigenvalues are the energy levels $\epsilon_j$, i.e., the states for which $[\rho,H]=0$. This condition implies that operators $H$ and $\rho$ share a common set of eigen-projectors $P_{\epsilon_j}$ such that $ g_{\epsilon_j} =\Tr(P_{\epsilon_j})$ and the spectral expansions of $\rho$ and $H$ may be written as $H=\sum_j \epsilon_j P_{\epsilon_j}$ and $\rho=\sum_j p_{\epsilon_j} P_{\epsilon_j} $. Outside this subclass of states, the more general explicit expression for the entropy is $S=-\Boltz\Tr(\rho\ln\rho)$, which clearly  reduces to Eq.~\ref{Squant} when  $[\rho,H]=0$.} which includes the stable equilibrium states,  the explicit expression of entropy reduces to the following
\begin{equation} S= -\Boltz  \sum_{\{\epsilon_j\} } p_{\epsilon_j} \ln (p_{\epsilon_j}/ g_{\epsilon_j})\label{Squant}\end{equation} 
where $k$ is the Boltzmann constant,\ind{Boltzmann, constant}\ind{Boltzmann constant} $\Boltz =1.380649\times 10^{-23}$ J/K, and \{$p_{\epsilon_j}$\} and \{$g_{\epsilon_j}$\} are the probabilities and multiplicities of all energy levels.

In fact, Equation \ref{Squant} represents a measure of the {\it breadth of the probability distribution} \{$p_{\epsilon_j}$\}. For example, in the particular case of a state with probability $p_{\epsilon_i}=1$ for the energy level $\epsilon_i$ (with multiplicity $g_{\epsilon_i}$) and zero for all other levels ($p_{\epsilon_j}=0$ for all $j\ne i$), the entropy is given by \begin{equation} S=\Boltz \,\ln g_{\epsilon_i}\label{Sgi}\end{equation}  and is therefore higher the greater the multiplicity of the only level. Another notable particular case is the state with $M$ equiprobable energy levels ($p_{\epsilon_j}=1/M$ for $j=$1, 2, ..., $M$) for a system with nondegenerate energy levels ($g_{\epsilon_j}=1$ for every $j$); the entropy is given by
\begin{equation} S=\Boltz \,\ln M\end{equation} 
and is therefore higher the greater the number $M$ of equiprobable (nondegenerate) levels.

The broader the probability distribution, the higher the entropy, and the greater the uncertainty about the outcome of the next measurement act. It is in this sense (and limited to the subclass of states for which the expression \ref{Squant} holds) that entropy can be interpreted as an indicator of the uncertainty of the outcomes of individual measurement acts; uncertainty, be aware, that for a given state of the system cannot be eliminated in any way, is irreducible, being intrinsic in the nature of the state itself. It can also be called {\it disorder}:\ind{disorder and entropy} the disorder with which measurement results emerge. The most ordered situation is the one that always provides the same value; it corresponds to the narrowest possible probability distribution, and if the energy level is nondegenerate, entropy (Equation \ref{Sgi}) is zero. However, the opposite is not true since not all states of mechanics belong to the subclass for which the entropy expression is given by Equation \ref{Squant}. Therefore, it is not correct to conclude (and indeed it is not true) that states of mechanics, having zero entropy, are all ordered, in the sense of being free of indeterminacy. Most of them still exhibit uncertainty in the outcomes of individual measurement acts. To emphasize the fact that there are irreducible uncertainties \cite{park_1968_nature} even in states with zero entropy, the term {\it indeterminacy}\ind{indeterminacy and entropy} has been introduced for the states of mechanics.

The most disorderly conceivable situation for a given system (with $M$ nondegenerate levels\footnote{In the case of degenerate levels, the most disorderly situation corresponds to the following probability distribution and, hence, the following values of entropy and energy,
	\begin{equation} p_{\epsilon_j}=\frac{g_{\epsilon_j}}{\sum_i g_{\epsilon_i}}\qquad S=\Boltz \,\ln\sum_i g_{\epsilon_i}\qquad E= \frac{\sum_j g_{\epsilon_j}\epsilon_j }{\sum_i g_{\epsilon_i}}\end{equation}  }) is the state that provides all possible levels with equal probability,
\begin{equation} p_{\epsilon_j}=\frac{1}{M}\qquad S=\Boltz \,\ln M \qquad E=\frac{1}{M}\sum_j\epsilon_j \end{equation}  
where we have also indicated the corresponding values $S$ and $E$ of entropy and energy.

\subsection{Stable Equilibrium States}
All stable equilibrium states belong to the subclass for which the expression \ref{Squant} for entropy holds. From the principle of maximum entropy (Section \ref{maxEntropyPrinciple}), we know that among all states with energy $E$, amounts $\bn$, and parameters $\bbeta$, the stable equilibrium state is the one with the maximum entropy. The corresponding probability distribution \{$p_{\epsilon_j}$\} is therefore the solution to the following constrained maximization problem
\begin{equation} \max_{\{p_{\epsilon_j}\}}\ S= -\Boltz  \sum_{\{\epsilon_j\} } p_{\epsilon_j} \ln \frac{p_{\epsilon_j}}{ g_{\epsilon_j}}\quad {\rm subject\ to\ the\ constraints\ }\quad \sum_{\{\epsilon_j\} } p_{\epsilon_j}=1 \quad {\rm and}\quad \sum_{\{\epsilon_j\} } p_{\epsilon_j}\epsilon_j =E \label{maxSq}\end{equation} 
namely,\footnote{Applying the {\it method of Lagrange multipliers}
	\begin{equation} \max_{\{p_{\epsilon_j}\}}\ L (p_{\epsilon_j},\lambda_1, \lambda_E)\big|_{\epsilon_j, g_{\epsilon_j},E} = -\Boltz  \sum_j p_{\epsilon_j} \ln (p_{\epsilon_j}/ g_{\epsilon_j})+\lambda_1\Big( \sum_i p_{\epsilon_i}-1\Big) +\lambda_E\Big( \sum_i p_{\epsilon_i}\epsilon_i -E\Big)\end{equation} 
	Setting $\big(\partial L/\partial  p_{\epsilon_j}\big)_{ g_{\epsilon_j},\lambda_1, \lambda_E }=0$, we obtain $p_{\epsilon_j}= g_{\epsilon_j}\exp\left(\lambda_1/\Boltz -1+\lambda_E \epsilon_j/\Boltz  \right)$ and, using the constraint $\sum_j p_{\epsilon_j}=1$ to obtain the value of $\lambda_1$, we find \begin{equation} p_{\epsilon_j}= \frac{g_{\epsilon_j}\exp (\lambda_E \epsilon_j/\Boltz )}{\sum_i g_{\epsilon_i}\exp (\lambda_E \epsilon_i/\Boltz )}\end{equation}  
	Substituting into the expressions for energy and entropy, we find
	\begin{equation} E = \frac{\sum_j g_{\epsilon_j}\epsilon_j \exp (\lambda_E \epsilon_j/\Boltz )}{\sum_i g_{\epsilon_i}\exp (\lambda_E \epsilon_i/\Boltz )}\qquad S=-\lambda_E\,E+\Boltz \ln\sum_i g_{\epsilon_i}\exp (\lambda_E \epsilon_i/\Boltz )\end{equation}  and, applying the definition of temperature and remembering that $\epsilon_j=\epsilon_j(\bn,\bbeta)$ and $ g_{\epsilon_j}= g_{\epsilon_j} (\bn,\bbeta)$, \begin{equation} \frac{1}{T} = \vparder{S}{E}{\bn,\bbeta}=\vparder{S}{E}{\{\epsilon_j\}, \{g_{\epsilon_j}\}}= -\lambda_E \end{equation} }
\begin{equation} p_{\epsilon_j}= \frac{g_{\epsilon_j}\exp (- \epsilon_j/\Boltz T)}{\sum_i g_{\epsilon_i}\exp (- \epsilon_i/\Boltz T)}\label{pepsilon}\end{equation} 
where $T$ is the temperature. For a lower and upper bounded energy spectrum, $-1/T$ can range between $-\infty$ and $+\infty$ and this distribution yields an $E$--$S$ diagram as shown in Fig.~\ref{ESdiagram_NegativeT}.

\subsection{Third law}

The ground-energy stable equilibrium states has energy  $E=E_{min}= \epsilon_{j_{\rm min}}$, inverse temperature $-1/T= -\infty$, and entropy  $S=\Boltz \ln g_{j_{\rm min}}$, thus its temperature is zero (third law) but its entropy is zero only if the ground-state energy level of the system is nondegenerate.

\subsection{Ergotropy versus adiabatic availability}

We have seen that adiabatic availability is the maximum amount of energy that a system can transfer to a weight in a weight process, and that its realization requires a reversible weight process in which the system ends in a stable equilibrium state. Reversibility of the weight process requires that the initial and final entropies be equal. From Eq.~\ref{Squant} it is clear that, in principle, there are many ways in which the entropy can be kept constant while changing the probabilities. Adiabatic availability therefore requires that the initial probabilities $p_{\epsilon_j}(t_1)$ change in time, while keeping the value of the entropy constant, so as to reach the unique distribution $p_{\epsilon_j}(t_2)$ that satisfies Eq.~\ref{pepsilon} and such that $S(t_2)=S(t_1)$.

However, as first noted in \cite[Secs.~3.2--3.4]{hatsopoulos_1976_unifieda}, if the state of the system is assumed to obey a strictly unitary equation of motion, it can be shown that the values of the probabilities $p_{\epsilon_j}$ cannot be changed, but can only be rearranged so as to decrease the energy while keeping the entropy unchanged, by manipulating their order with respect to that of the set $\{\epsilon_j\}$ of energy levels. This observation effectively defines a special class of reversible weight processes, called CCP (cyclic change in parameters) unitary processes. Starting from any initial state with probabilities $p_{\epsilon_j}(t_1)$, the largest amount of energy that can be extracted through a reversible weight process in this class is obtained by a unitary process that rearranges the probabilities so that the final set of values $p_{\epsilon_j}(t_2)$ consists of exactly the same values as initially, but ordered oppositely to the increasing order of the energy levels, i.e.,
$p_{\epsilon_1}(t_2)>p_{\epsilon_2}(t_2)>\cdots>p_{\epsilon_k}(t_2)>\cdots$
for $\epsilon_1<\epsilon_2<\cdots<\epsilon_k<\cdots$.

The resulting final energy $E(t_2)$ is the minimum among all states compatible with the initial probabilities, but it is generally higher than the energy of the stable equilibrium state with entropy equal to $S(t_1)$. Consequently, the extractable energy $E(t_1)-E(t_2)$, which in recent years has been called \emph{ergotropy} \cite{Allahverdyan_2004}, is in general smaller than the adiabatic availability of the initial state.

This difference has a clear operational meaning: it quantifies the portion of the adiabatic availability that is inaccessible when the dynamics is restricted to unitary, probability-distribution-preserving transformations, and therefore cannot drive the system to the stable equilibrium state of equal entropy.

From the thermodynamic viewpoint adopted here, this limitation reflects the absence of internal reversible mechanisms capable of reshaping the probability distribution beyond unitary rearrangements. The gap between ergotropy and adiabatic availability thus provides a direct measure of the role of internal dynamics in enabling the full conversion of available energy, and highlights the distinction between idealized, dynamically constrained processes and fully reversible thermodynamic weight processes.

\subsection{Stable-equilibrium partition function}

 Defining the so-called {\it partition function},\ind{partition function}
\begin{equation} {\cal Q}(T, \{\epsilon_j\}, \{g_{\epsilon_j}\})=\sum_i g_{\epsilon_i}\exp (- \epsilon_i/\Boltz T)\label{partition}\end{equation} 
 the following expressions for probabilities, energy, and entropy, are easily verified
\begin{align} p_{\epsilon_j}&= -\Boltz T\,\frac{\partial \ln{\cal Q}}{\partial \epsilon_j}=\frac{g_{\epsilon_j}\exp (- \epsilon_j/\Boltz T)}{ {\cal Q}} \\
E&=  \Boltz T^2\,\frac{\partial \ln{\cal Q}}{\partial T}=\frac{\sum_i g_{\epsilon_i}\epsilon_i \exp (- \epsilon_i/\Boltz T)}{ {\cal Q}} \label{EQ}\\
S&=  \Boltz  \,\frac{\partial T\ln{\cal Q}}{\partial T}=\frac{E}{T}+\Boltz \ln{\cal Q} \label{SQ}\\
\sigma_E^2&= k^2_\text{B} T^3\,\frac{\partial^2 T\ln{\cal Q}}{\partial T^2}=-\Boltz \vparder{{}^2S}{E^2}{\{\epsilon_j\}, \{g_{\epsilon_j}\}}^{-1}\ge 0 \label{sigmaEQ}
\end{align}
Note that Equations \ref{EQ} and \ref{SQ} have the form $E=E(T, \{\epsilon_j\}, \{g_{\epsilon_j}\})$ and $S=S(T, \{\epsilon_j\}, \{g_{\epsilon_j}\})$, defining implicitly, through the parameter $T$, the fundamental relation of the system $S=S(E,\bn,\bbeta)=S(E, \{\epsilon_j(\bn,\bbeta)\}, \{g_{\epsilon_j} (\bn,\bbeta)\})$
from which, as we have seen, all the other properties defined for stable equilibrium states can be derived.  Rel.~\ref{sigmaEQ} confirms the concavity of the entropy versus energy relation.

\subsection{Harmonic Oscillator}
As a first example, consider a system consisting of a single harmonic oscillator with frequency $\nu$. The energy levels are quantized. In addition to the minimum energy level of $h\nu/2$, the others are separated by intervals all equal to $h\nu$, where $h= 6.6260\times 10^{-34}$ J\,s is the  Planck constant,
\begin{equation} \epsilon_j=\Big(j+\frac{1}{2}\Big)h\nu\quad {\rm with}\ j\ {\rm an\ integer}\ge 0 \end{equation} 
so that the partition function (Eq.~\ref{partition}), is
\begin{align} {\cal Q}&= \sum_{j=0}^\infty \exp\bigg(\!-\Big(j+\frac{1}{2}\Big)\frac{h\nu}{\Boltz T}\bigg)=  \frac{\exp (-h\nu/2\Boltz T)}{1-\exp(-h\nu/2\Boltz T)} \end{align}
from which it follows, for example,
\begin{align} p_{\epsilon_j}&=  \left[1-\exp(-h\nu/2\Boltz T)\right] \exp (-jh\nu/\Boltz T)\\
E&= h\nu\Big(\frac{1}{2}+\frac{1}{\exp(h\nu/\Boltz T)-1}\Big)\\
S&= \Boltz \bigg(\frac{h\nu/\Boltz T}{ \exp(h\nu/\Boltz T)-1}-\ln\big[1-\exp(-h\nu/\Boltz T)\big]\bigg)\\
\sigma^2_E&= k^2_\text{B}T^2\frac{(h\nu/2\Boltz T)^2}{\sinh^2(h\nu/2\Boltz T)} \label{CVvibraz}
\end{align}

\subsection{Single Structureless Particle Confined in a Box}

As a second example, consider a system consisting of a single particle of mass $m$ without internal structure and therefore endowed only with translational degrees of freedom, confined in a parallelepiped-shaped container with sides $\ell_1$, $\ell_2$, $\ell_3$ (and volume $V=\ell_1\ell_2\ell_3$). Also in this case, the energy levels are quantized. They are given by the relation
\begin{equation}  \epsilon_{j_1,j_2,j_3}=\epsilon_{j_1}+\epsilon_{j_2}+\epsilon_{j_3}=\frac{h^2}{8m}\Big(\frac{j_1^2}{\ell_1^2}+\frac{j_2^2}{\ell_2^2}+\frac{j_3^2}{\ell_3^2}\Big) \quad {\rm where}\  j_1,j_2,j_3 = 1,2,3\dots,\infty  >0\end{equation} 
The maximum entropy principle implies that the stable equilibrium state probability distribution \{$p_{j_1,j_2,j_3}$\} is given by the solution of the  constrained maximization problem
$ \max_{\{p_{j_1,j_2,j_3}\}}\ S= -nR \sum_{\pmb{j}} p_{j_1,j_2,j_3} \ln p_{j_1,j_2,j_3}$ subject to $\sum_{{\pmb{j}} } p_{j_1,j_2,j_3}=1$ and $ \sum_{{\pmb{j}} } p_{j_1,j_2,j_3}\epsilon_{j_1,j_2,j_3} =E $
where $\sum_{{\pmb{j}} }=\sum_{j_1=1 }^\infty\sum_{j_2=1 }^\infty\sum_{j_3=1 }^\infty $.
Assigning the Lagrange multiplier $1/\Boltz T$ to the energy constraint, and recalling that for one particle $n=1/N_\text{Av}$ and  $nR=R/N_\text{Av}=\Boltz $, we find the stable equilibrium state distribution
\begin{equation}p_{j_1,j_2,j_3}= \frac{\exp (-\epsilon_{j_1,j_2,j_3}/\Boltz T)}{{\cal Q}}=p_{j_1}p_{j_2}p_{j_2}=\frac{\exp (-\epsilon_{j_1}/\Boltz T)}{{\cal Q}_1}\frac{\exp (-\epsilon_{j_2}/\Boltz T)}{{\cal Q}_2}\frac{\exp (-\epsilon_{j_3}/\Boltz T)}{{\cal Q}_3}\end{equation} 
where we define the ``directional partition functions'' 
${\cal Q}_i=\sum_{j_i=1}^\infty \exp (- \epsilon_{j_i}/\Boltz T)$
so that the partition function, given by Relation \ref{partition}, becomes
\begin{equation} {\cal Q}={\cal Q}_1{\cal Q}_2{\cal Q}_3 \qquad\text{with}\qquad {\cal Q}_i=\sum_{j_i=1}^\infty \exp\Big(-\frac{h^2}{8m\Boltz T}\frac{j_i^2}{\ell_i^2}\Big)\qquad\text{for }i=1,2,3
\end{equation}
It easy to verify that the probabilities and all the properties can be obtained from derivatives of the ${\cal Q}_i$'s and that $T=\big(\partial E/\partial S\big)_{\ell_1, \ell_2, \ell_3}$, i.e., the Lagrange multiplier indeed represents  the temperature. For $i=1,2,3$, we have the relations
\begin{equation} p_{j_i}=-\Boltz T\,\Big(\frac{\partial \ln{\cal Q}_i}{\partial \epsilon_{j_i}}\Big)_{T}\qquad  E=E_1+E_2+E_3\qquad E_i=\sum_{j_i } p_{j_i}\epsilon_{j_i}= \Boltz T^2\,\Big(\frac{\partial \ln{\cal Q}_i}{\partial T}\Big)_{\ell_i}\end{equation} \begin{equation} S=S_1+S_2+S_3\qquad S_i= -\Boltz  \sum_{j_i} p_{j_i} \ln p_{j_i}= \Boltz  \,\Big(\frac{\partial T\ln{\cal Q}_i}{\partial T}\Big)_{\ell_i}=\frac{E_i}{T}+\Boltz \ln{\cal Q}_i \end{equation}
\begin{equation} \diff \ln{\cal Q}_i=\frac{E_i}{\Boltz T^2}\diff T+\frac{2E_i}{\Boltz T}\frac{\diff \ell_i}{\ell_i} \qquad \Big(\frac{\partial \ln{\cal Q}_i}{\partial T}\Big)_{\ell_i}=\frac{ E_i}{\Boltz  T^2}\qquad \Big(\frac{\partial  \ln{\cal Q}_i}{\partial \ln\ell_i}\Big)_{T}=\frac{ 2E_i}{\Boltz  T}\end{equation}
\begin{equation}S_i= \frac{E_i}{T}+\Boltz \ln{\cal Q}_i\qquad \diff S_i =\frac{1}{T}\diff E_i+ \frac{2E_i}{T}\frac{\diff \ell_i}{\ell_i} \end{equation}
We may  define the ``directional pressure'' $\pi_i$ representing the change in the directional energy $ E_i$ at constant $S_i$ due to a partial change in the volume $V=\ell_1\ell_2\ell_3$ obtained by changing only the side length $\ell_i$, while keeping the other two side lengths fixed, so that $\partial V/V=\partial \ell_i/\ell_i$,
\begin{equation} \pi_i=-\Big(\frac{\partial E_i}{\partial V} \Big)_{S,\pmb{\ell}_i'}=-\frac{1}{V}\Big(\frac{\partial E_i}{\partial \ln\ell_i} \Big)_{S,\pmb{\ell}_i'} =\frac{2E_i}{V} \qquad E= E_1+E_2+E_3=(\pi_1+\pi_2+\pi_3)\frac{V}{2}\end{equation}
The directional and overall stable-equilibrium-state Gibbs relations rewrite, in general, as
\begin{equation}  \diff E_i=T\diff S_i -2E_i \frac{\diff \ell_i}{\ell_i} =T\diff S_i -\pi_i V \frac{\diff \ell_i}{\ell_i} \end{equation}
\begin{equation} \diff E=T\diff S-\pi_1V\frac{\diff \ell_1}{\ell_1}-\pi_2V\frac{\diff \ell_2}{\ell_2}-\pi_3V\frac{\diff \ell_3}{\ell_3}
\end{equation}

\subsection{Ideal gas equation of state for the single particle in a box}

For `practical' values of $m$, $\ell_1$, $\ell_2$, $\ell_3$, and  $T$, the values of $ h^2/8m\Boltz T\ell_i^2$ are typically much smaller than one,\footnote{In terms of the  de Broglie wavelength $\Lambda_\text{dB}=h/\sqrt{2\pi m\Boltz T}$ this condition may be expressed as $\Lambda_\text{dB}\ll \ell_i $. Eq.~\ref{QiApprox}  then rewrites as $ {\cal Q}_i\approx \ell_i/\Lambda_\text{dB}$.}  and therefore the sum in ${\cal Q}_i$  can be approximated by an integral,
\begin{equation} {\cal Q}_i= \sum_{j_i=1}^\infty {\rm e}^{-\frac{h^2}{8m\Boltz T}\frac{j_i^2}{\ell_i^2}}\approx \int_0^\infty {\rm e}^{-\frac{h^2}{8m\Boltz T}\frac{x^2}{\ell_i^2}}\diff x = \left(\frac{2\pi m \Boltz T\ell_i^2}{h^2}\right)^{1/2}\label{QiApprox}\qquad \Big(\frac{\partial \ln{\cal Q}_i}{\partial T}\Big)_{\ell_i}\approx\frac{1}{2T}\end{equation}
\begin{equation}  {\cal Q}= {\cal Q}_1 {\cal Q}_2 {\cal Q}_3\approx \left(\frac{2\pi m \Boltz TV^{2/3}}{h^2}\right)^{3/2} \end{equation}
and, therefore, the overall partition function becomes independent of the details $\ell_1$, $\ell_2$, $\ell_3$ of the shape of the container, given the same  volume $V=\ell_1\ell_2\ell_3$.
As a result,  in this practical limit of  large $T$ for a given $V$, so that the approximation $  TV^{2/3}\gg \frac{ h^2}{8m\Boltz }$ holds, the directional energies and the directional pressures become independent of direction, and  we have
\begin{equation} E_i\approx \frac{1}{2}\Boltz T  \qquad S_i\approx\frac{1}{2}\Boltz \Big(1+ \ln\frac{2\pi m \Boltz T\ell_i^2}{h^2}\Big)\qquad \pi_i\approx\frac{\Boltz T}{V}\label{pressureidealgasi}\end{equation}
	\begin{equation}E\approx \frac{3}{2}\Boltz T \qquad  S\approx\frac{3}{2}\Boltz \Big(1+ \ln\frac{2\pi m \Boltz TV^{2/3}}{h^2}\Big)\qquad
		p=-\vparder{E}{V}{S}=T\vparder{S}{V}{E}\approx\frac{\Boltz T}{V} \label{pressureidealgas}\end{equation}
	\begin{equation} \diff E\approx T\diff S-\Boltz T\Big(\frac{\diff \ell_1}{\ell_1}+\frac{\diff \ell_2}{\ell_2}+\frac{\diff \ell_3}{\ell_3}\Big)=T\diff S-\Boltz T\frac{\diff V}{V}=T\diff S-p\diff V\end{equation}
where, in the last of Eqs.~\ref{pressureidealgas}, we have evaluated the pressure using its definition, Eq.~\ref{defp}. 
The result obtained is valid, as seen, if $ h^2/8mkTV^{2/3}\ll 1$ and shows that the \textit{equation of state} (relation between $T$, $p$, and $V$ for the stable equilibrium states) of a single particle ($n=1$ molecule) confined in the container of volume $V$  is
\begin{equation} pV=\Boltz T \qquad\text{or, equivalently for }n=1\text{ molecule,}\qquad pV=nRT \end{equation} 
which is the well-known \textit{ideal-gas equation of state}. Eqs.~\ref{pressureidealgasi} and \ref{pressureidealgas} also show that the overall energy $E$ is equally partitioned into the three directional contributions $E_i$ (equipartition theorem).

\subsection{Ideal gas equation of state for $n$ distinguishable but identical point particles in a box}

The ideal gas equation of state is a good approximation also for the stable equilibrium states of a system consisting of many point (structureless) particles of mass $m$ confined in a box at relatively high temperatures and low pressures, in which at any given instant of time only a negligibly small fraction of the particles are  close to one another in the collision range where intermolecular forces are strong, whereas every other particle (the vast majority) feels negligible intermolecular forces and, hence, behaves like a single particle in a box, essentially not feeling the presence  of  other particles except during the negligible time it spends colliding with them. In these limiting conditions (ideal-gas limit), the properties of  $n$ distinguishable but identical particles in the box of volume $V$ can be approximated by those of a composite of $n$ separable and independent identical systems each consisting of a single-particle in a box of volume $V$. From the additivity of energy and entropy for composites of separable and independent systems, it follows that
\begin{equation}E\approx \frac{3}{2}n\Boltz T \qquad  S\approx\frac{3}{2}n\Boltz [1+ \ln (c TV^{2/3})]\qquad
	p\approx\frac{n\Boltz T}{V} \qquad c=\frac{2\pi m \Boltz}{h^2} \label{pressureidealgasN}\end{equation}

We may use these relations to compute the minimum work of partitioning and the entropy of partition removal with reference to Figure \ref{MinWofPartitioning}. Let $T_{ab}$ be the temperature of  stable equilibrium state $A^1_{ab}$ of the single-compartment system $A^1$ with $n$ particles in volume $V$. By the first of Eqs.~\ref{pressureidealgasN},  the energy is $E^1_{ab}= \frac{3}{2}n\Boltz T_{ab}$ and the entropy $S^1_{ab}=  \frac{3}{2}n\Boltz [1+ \ln(c T_{ab}V^{2/3})] $. For the  $\lambda$-compartments system $A^\lambda$, each compartment has $n/\lambda$ particles in volume $V/\lambda$. If they are all in mutual equilibrium at temperature $T_{ab}$, Eqs.~\ref{pressureidealgasN} imply that each of them has energy  $\frac{3}{2}\frac{n}{\lambda}\Boltz T_{ab}$ and entropy $\frac{3}{2}\frac{n}{\lambda}\Boltz [1+ \ln(c T_{ab}(V/\lambda)^{2/3}]$, so that $E^\lambda_{aa} = \frac{3}{2}n\Boltz T_{ab}=E^\lambda_{ab}$, $T_{aa}=T_{ab}$ and $S^\lambda_{aa} =\frac{3}{2}n\Boltz [1+ \ln(c T_{ab}(V/\lambda)^{2/3})]  $. Also, they imply that the state $A^\lambda_{bb}$, identified by the condition that  $S^\lambda_{bb}=\frac{3}{2}n\Boltz [1+ \ln(c T_{bb}(V/\lambda)^{2/3})]=S^\lambda_{ab}$, has temperature such that $T_{bb}(V/\lambda)^{2/3}= T_{ab}V^{2/3}$, i.e., $T_{bb}= T_{ab}\lambda^{2/3}$. Therefore, with reference to Figure \ref{MinWofPartitioning}, for distinguishable but identical point particles in the ideal-gas limit,  the entropy of partition removal and the minimum work of partitioning are given by
\begin{equation} S_\text{irr}^{\lambda\to 1}=S^1_{ab}-S^\lambda_{aa} =n\Boltz\ln\lambda \qquad   W_\text{min}^{1\to\lambda} = W_\text{max}^{\lambda\to 1}= E^\lambda_{bb}-E^1_{ab}=\frac{3}{2}(\lambda^{2/3}-1)n\Boltz T_{ab}   \label{separationGas}
\end{equation}

\subsection{Variable amounts of constituents. Open system model}

So far in these brief quantum theory notes, the amounts of constituents $\bn$ have been considered fixed. To model the stable equilibrium states of systems with many particles in the simple-system approximation (i.e., when the effects of adding or removing partitions are negligible, and correlations are rapidly erased by dissipation) it is possible to consider  the amounts $\bn$ as variable, behaving like  normal properties, such as energy, with their respective quantum uncertainties. This allows to model a system open to exchanges of constituents.

For a system with a single type of constituents, the result of measuring the number of particles is an integer, denoted by $z$, with $0\le z <\infty$, generally unpredictable. But repeating the measurement on a large number of identical replicas of the system, all identically prepared, we can compute the average value of the number of particles as well as the dispersion of measurement results around the mean value  \begin{equation} n = \sum_{\{z\}}\lim_{N\rightarrow\infty} f(z,N)\, z =\sum_{z=0}^\infty p_z \, z \qquad\text{and}\qquad  \sigma^2_n= \sum_{z}p_z\,(z -n)^2 \end{equation} 

In general, if we simultaneously measure the energy and amounts of all $r$ constituents of a multi-constituent system, the result of a single measurement will be the set of $r+1$ values $z_1$, $z_2$,\dots, $z_r$, $\epsilon_j$, or, more succinctly, $\bz $,  $\epsilon_j$, where $\epsilon_j$ belongs to the set of possible  values $\{ \epsilon_j(\bz ,\bbeta)\}$  compatible with the values $\bbeta$ of the parameters and the measured numbers of particles $\bz$. The functions $\epsilon_j(\bz ,\bbeta) $ are the ones already defined by Rel.~\ref{livellichiuso} for the closed system, with the fixed $\bn$ replaved by the variable $\bz $. 
The joint probability of obtaining from a single measurement both amounts and energy values $\bz $ and  $\epsilon_j$ is then defined. It is a property that we can indicate with the symbol $p_{\bz  ,\epsilon_j}$ or $p_{z_1,z_2,\dots,z_r ,\epsilon_j}$. The values of energy and amounts of constituents are then given by the relations
\begin{equation}  E= \sum_{\{\epsilon_j\}}p_{\epsilon_j}\epsilon_j = \sum_{\{\bz  ,\epsilon_j\}}p_{\bz ,\epsilon_j}\epsilon_j \qquad\text{and}\qquad
n_i= \sum_{\{z_i\} }p_{z_i}z_i = \sum_{\{\bz ,\epsilon_j\} }p_{\bz  ,\epsilon_j}z_i \end{equation}
where we have also defined the marginal probabilities $ p_{\epsilon_j}$ and $p_{z_i}$ connected with measurements of only the energy or only the amount of constituent $i$,
\begin{equation} p_{\epsilon_j} =  \sum_{\{\bz  \}}p_{\bz ,\epsilon_j} \qquad\qquad
p_{z_i} =  \sum_{\{\bz ',\epsilon_j\}}p_{\bz ,\epsilon_j} \end{equation} 
where $\bz '$ denotes the set $\{z_1$,\dots, $z_{(i-1) }$, $z_{(i+1) }$,\dots, $z_r$\}.

Also in this case, for a quite broad subclass of states that includes the stable equilibrium states of systems with discrete energy spectra, the explicit expression for the entropy is reduced to the following
\begin{equation} S= -\Boltz \sum_{\{\bz  ,\epsilon_j\} } p_{\bz  ,\epsilon_j} \ln (p_{\bz  ,\epsilon_j}/ g_{\epsilon_j})\label{Sgrand}\end{equation} 
With a procedure similar to what seen above, we obtain the stable-equilibrium  probability distribution
\begin{equation} p_{\bz  ,\epsilon_j}= \frac{g_{\epsilon_j}\exp (\bz \cdot\bmu/\Boltz T - \epsilon_j/\Boltz T)}{\sum_{\{\bz  ,\epsilon_j\}} g_{\epsilon_i}\exp (\bz \cdot\bmu/\Boltz T - \epsilon_i/\Boltz T)}\end{equation} 
where $T$ is the temperature, $\bz \cdot\bmu=z_1\mu_1+\dots+ z_r\mu_r$, and $\mu_i$ is the total potential of constituent $i$. 
The {\it partition function} 
\begin{equation} {\cal Q}(T,\bmu, \{\bz \},\{\epsilon_j\}, \{g_{\epsilon_j}\})=\sum_{\{\bz  ,\epsilon_j\}} g_{\epsilon_i}\exp (\bz \cdot\bmu/\Boltz T - \epsilon_i/\Boltz T)\label{grandpartition}\end{equation} 
is all that is needed to compute marginal probabilities, energy, entropy, and amounts using the relations
\begin{equation} p_{\epsilon_j}=-\Boltz T\,\frac{\partial \ln{\cal Q}}{\partial \epsilon_j} \qquad p_{z_i}=\frac{\Boltz T}{\mu_i}\,\frac{\partial \ln{\cal Q}}{\partial z_i} \qquad
n_i = kT\,\frac{\partial \ln{\cal Q}}{\partial \mu_i}\end{equation}
\begin{equation}
E = \Boltz T^2\,\frac{\partial \ln{\cal Q}}{\partial T}+\sum_i\mu_i n_i\qquad
S= \Boltz \,\frac{\partial T\ln{\cal Q}}{\partial T}=\frac{E}{T}-\frac{\sum_i\mu_i n_i }{T}+\Boltz\ln{\cal Q}\label{SgrandQ}\end{equation}
Assuming  the compatible energy values depend only on volume $V$, $\{ \epsilon_j(\bz ,V)\}$, the pressure is given by the mean value of their negative variation with volume, 
\begin{equation}
	p = -\vparder{E}{V}{S,\bn}= \Boltz T\,\vparder{ \ln{\cal Q}}{ V}{T,\bmu}=\Boltz T\,\sum_{\{\epsilon_j\}}\frac{\partial \ln{\cal Q}}{\partial \epsilon_j} \frac{\partial \epsilon_j}{\partial V}=-\sum_{\{\epsilon_j\}}p_{\epsilon_j}\frac{\partial \epsilon_j}{\partial V}\label{pgrandQ}\end{equation}
and the Hill (Euler) free energy is 

\begin{equation}
	\Eu = E-TS+pV-\bmu\cdot\bn =  pV- \Boltz T \ln{\cal Q} =-\Boltz T\,\vparder{ V\ln{\cal Q}}{ V}{T,\bmu}  \label{EuQ}\end{equation}

\section{Conclusions}

This paper has presented a unified and operationally grounded exposition of the elementary foundations of thermodynamics in which all concepts are defined independently of system size, extensivity, and equilibrium assumptions. By introducing entropy as a property of all states and by distinguishing clearly between stable equilibrium and nonequilibrium states, the formulation provides a logically consistent basis for thermodynamic reasoning across scales.

The energy–entropy diagram has been shown to offer a powerful geometric framework for understanding availability, irreversibility, and energy conversion limits, while the analysis of entropy transfer in non-work interactions has led to precise definitions of heat and heat-and-diffusion interactions relevant to mesoscopic and continuum nonequilibrium theories. From this analysis, Clausius inequalities and the Clausius statement of the second law emerge naturally in forms valid beyond equilibrium.

The perspective advanced in this paper reinforces the view that thermodynamics is not a theory limited to macroscopic, extensive systems in equilibrium, but a universal physical framework applicable to all systems and all states. By avoiding assumptions of extensivity at the foundational level, defining entropy and energy operationally beyond equilibrium, and analyzing entropy transfer in non-work interactions, the theory retains both logical coherence and broad applicability. Extensivity, equilibrium, and macroscopic behavior emerge as special cases rather than prerequisites. In this sense, thermodynamics appears not as a phenomenology tied to scale, but as a general structure governing the evolution and interaction of physical systems, from macroscopic energy technologies to few-particle and mesoscopic regimes.


\begin{thebibliography}{99}
	
	\bibitem[Fermi(1937)]{Fermi1956}
	Fermi, E.
	\newblock {\em Thermodynamics}; Prentice-Hall: New York, NY, USA,  1937.
	\newblock Reprinted by Dover, New York, NY, USA, 1956.
	
	\bibitem[Callen(1985)]{Callen1985}
	Callen, H.B.
	\newblock {\em Thermodynamics and an Introduction to Thermostatistics}, 2 ed.;
	Wiley: New York, NY, USA,  1985.
	
	\bibitem[Van~Wylen and Sonntag(1978)]{VanWylenSonntag1978}
	Van~Wylen, G.J.; Sonntag, R.E.
	\newblock {\em Fundamentals of Classical Thermodynamics}, 2 ed.; Wiley: New
	York, NY, USA,  1978.
	
	\bibitem[Moran and Shapiro(1988)]{MoranShapiro1988}
	Moran, M.J.; Shapiro, H.N.
	\newblock {\em Fundamentals of Engineering Thermodynamics}; Wiley: New York,
	NY, USA,  1988.
	
	\bibitem[Feynman et~al.(1963)Feynman, Leighton, and Sands]{Feynman1963}
	Feynman, R.P.; Leighton, R.B.; Sands, M.
	\newblock {\em The Feynman Lectures on Physics, Vol. I: Mainly Mechanics,
		Radiation, and Heat}; Addison--Wesley: Reading, MA, USA,  1963.
	
	\bibitem[Tisza(1966)]{Tisza1966}
	Tisza, L.
	\newblock {\em Generalized Thermodynamics}; MIT Press: Cambridge, MA, USA,
	1966.
	
	\bibitem[Landau and Lifshitz(1980)]{LandauLifshitz1980}
	Landau, L.D.; Lifshitz, E.M.
	\newblock {\em Statistical Physics, Part I}, 3rd ed.; Pergamon Press: Oxford, UK,
	1980.
	\newblock Revised by E. M. Lifshitz and L. P. Pitaevskii; translated by J. B.
	Sykes and M. J. Kearsley.
	
	\bibitem[Hatsopoulos and Keenan(1965)]{HatsopoulosKeenan1965}
	Hatsopoulos, G.N.; Keenan, J.H.
	\newblock {\em Principles of General Thermodynamics}; Wiley: New York, NY, USA,
	1965.
	
	\bibitem[Keenan(1941)]{Keenan1941}
	Keenan, J.H.
	\newblock {\em Thermodynamics}; Wiley: New York, NY, USA,  1941.
	\newblock Reprinted by MIT Press, Cambridge, MA, USA, 1970.
	
	\bibitem[Guggenheim(1967)]{Guggenheim1967}
	Guggenheim, E.A.
	\newblock {\em Thermodynamics}, 7 ed.; North-Holland: Amsterdam, The
	Netherlands,  1967.
	
	\bibitem[Gyftopoulos and Beretta(1991)]{GyftopoulosBeretta1991}
	Gyftopoulos, E.P.; Beretta, G.P.
	\newblock {\em Thermodynamics: Foundations and Applications}; Macmillan: New
	York, NY, USA,  1991.
	\newblock Reprinted by Dover, Mineola, NY, USA, 2005.
	
	\bibitem[Zanchini and Beretta(2010)]{ZanchiniBeretta2010}
	Zanchini, E.; Beretta, G.P.
	\newblock Removing Heat and Conceptual Loops from the Definition of Entropy.
	\newblock {\em International Journal of Thermodynamics} {\bf 2010}, {\em
		13},~67--76.
	\newblock {\url{https://doi.org/10.1063/1.2979048}}.
	
	\bibitem[Beretta and Zanchini(2011)]{BerettaZanchini2011}
	Beretta, G.P.; Zanchini, E.
	\newblock Rigorous and General Definition of Thermodynamic Entropy. In {\em
		Thermodynamics}; Letcher, T.M., Ed.; InTech: Rijeka, Croatia,  2011.
	\newblock {\url{https://doi.org/10.5772/13371}}.
	
	\bibitem[Carath\'eodory(1909)]{Caratheodory1909}
	Carath\'eodory, C.
	\newblock Untersuchungen \"uber die Grundlagen der Thermodynamik.
	\newblock {\em Mathematische Annalen} {\bf 1909}, {\em 67},~355--386.
	\newblock {\url{https://doi.org/10.1007/BF01450409}}.
	
	\bibitem[Lieb and Yngvason(1999)]{LiebYngvason1999}
	Lieb, E.H.; Yngvason, J.
	\newblock The Physics and Mathematics of the Second Law of Thermodynamics.
	\newblock {\em Physics Reports} {\bf 1999}, {\em 310},~1--96.
	\newblock {\url{https://doi.org/10.1016/S0370-1573(98)00082-9}}.
	
	\bibitem[Lieb and Yngvason(2013)]{LiebYngvason2013}
	Lieb, E.H.; Yngvason, J.
	\newblock The Entropy Concept for Non-Equilibrium States.
	\newblock {\em Proceedings of the Royal Society A} {\bf 2013}, {\em
		469},~20130408.
	\newblock {\url{https://doi.org/10.1098/rspa.2013.0408}}.
	
	\bibitem[Lieb and Yngvason(2014)]{LiebYngvason2014}
	Lieb, E.H.; Yngvason, J.
	\newblock Entropy Meters and the Entropy of Non-Extensive Systems.
	\newblock {\em Proceedings of the Royal Society A} {\bf 2014}, {\em
		470},~20140192.
	\newblock {\url{https://doi.org/10.1098/rspa.2014.0192}}.
	
	\bibitem[Zanchini and Beretta(2014)]{ZanchiniBeretta2014}
	Zanchini, E.; Beretta, G.P.
	\newblock Recent Progress in the Definition of Thermodynamic Entropy.
	\newblock {\em Entropy} {\bf 2014}, {\em 16},~1547--1570.
	\newblock {\url{https://doi.org/10.3390/e16031547}}.
	
	\bibitem[Hill(1964)]{Hill1964}
	Hill, T.L.
	\newblock {\em Thermodynamics of Small Systems}; W. A. Benjamin: New York, NY,
	USA,  1964.
	\newblock Reprinted by Dover, Mineola, NY, USA, 1994.
	
	\bibitem[Hill(2001)]{Hill2001}
	Hill, T.L.
	\newblock A Different Approach to Nanothermodynamics.
	\newblock {\em Nano Letters} {\bf 2001}, {\em 1},~273--275.
	\newblock {\url{https://doi.org/10.1021/nl010027w}}.
	
	\bibitem[Chamberlin(2015)]{Chamberlin2015}
	Chamberlin, R.V.
	\newblock The Big World of Nanothermodynamics.
	\newblock {\em Entropy} {\bf 2015}, {\em 17},~52--73.
	\newblock {\url{https://doi.org/10.3390/e17010052}}.
	
	\bibitem[Chamberlin and Lindsay(2024)]{Chamberlin2024}
	Chamberlin, R.V.; Lindsay, S.M.
	\newblock Nanothermodynamics: There’s Plenty of Room on the Inside.
	\newblock {\em Nanomaterials} {\bf 2024}, {\em 14},~1828.
	\newblock {\url{https://doi.org/10.3390/e17010052}}
	
		\bibitem[Bedeaux et~al.(2024)Bedeaux, Kjelstrup, and Schnell]{bedeaux2024}
	Bedeaux, D.; Kjelstrup, S.; Schnell, S.K.
	\newblock {\em Nanothermodynamics: Theory and applications}; World Scientific,
	2024.
	
	\bibitem[Beretta and Zanchini(2019)]{BerettaZanchini2019}
	Beretta, G.P.; Zanchini, E.
	\newblock New Definitions of Thermodynamic Temperature and Entropy Not Based on
	the Concepts of Heat and Thermal Reservoir.
	\newblock {\em Atti della Accademia Peloritana dei Pericolanti} {\bf 2019},
	{\em 97},~A1, 1--28.
	\newblock {\url{https://doi.org/10.1478/AAPP.97S1A1}}.
	
	\bibitem[Beretta(2024)]{Beretta2024MITOCW}
	Beretta, G.P.
	\newblock 2.43 Advanced Thermodynamics.
	\newblock MIT OpenCourseWare,  2024.
	\newblock Spring 2024,
	\newblock {\url{https://ocw.mit.edu/courses/2-43-advanced-thermodynamics-spring-2024/}}.
	
	\bibitem[Margenau(1950)]{Margenau1950}
	Margenau, H.
	\newblock {\em The Nature of Physical Reality: A Philosophy of Modern Physics};
	McGraw-Hill: New York, USA,  1950.
	\newblock Reprinted by Ox Bow Press, 1970.
	
	\bibitem[Feynman(1964)]{Feynman1964great}
	Feynman, R.P.
	\newblock The Great Conservation Principles.
	\newblock Messenger Lectures series, 'The Character of Physical Law', Cornell
	University,  1964.
	\newblock Recorded by BBC.
	
	
	
	\bibitem[Pavelka et al.(2018)]{Pavelka2018}
	Pavelka, M.; Klika, V.; Grmela, M.
	\newblock {\em Multiscale Thermo-Dynamics: Introduction to GENERIC}.
	\newblock De Gruyter: Berlin {\bf 2018}.
	
	\bibitem[Ottinger(2005)]{Ottinger2005}
	Öttinger, H.~C.
	\newblock {\em Beyond Equilibrium Thermodynamics}.
	\newblock Wiley: Hoboken, NJ, USA {\bf 2005}.
	
	\bibitem[Grmela and Pavelka(2026)]{Grmela2026}
	Grmela, M.; Pavelka, M.
	\newblock Comparison of some geometric frameworks for dissipative evolution in multiscale non-equilibrium thermodynamics.
	\newblock {\em Journal of Non-Equilibrium Thermodynamics} {\bf 2026}, {\em in press}.
	\newblock {\url{https://doi.org/10.48550/arXiv.2512.05168}}.
	
	\bibitem[Gyarmati(1970)]{Gyarmati1970}
	Gyarmati, I.
	\newblock {\em Non-equilibrium thermodynamics} (Vol. 184).
	\newblock Berlin: Springer {\bf 1970}.
	
	\bibitem[Ván and Kovács(2020)]{Van2020}
	Ván, P.; Kovács, R.
	\newblock Variational principles and nonequilibrium thermodynamics.
	\newblock {\em Phil. Trans. R. Soc. A} {\bf 2020}, {\em 378},~20190178.
	\newblock {\url{https://doi.org/10.1098/rsta.2019.0178}}.
	
	\bibitem[Gibbs(1902)]{Gibbs1902}
	Gibbs, J.~W.
	\newblock {\em Elementary Principles in Statistical Mechanics}.
	\newblock New York: Charles Scribner's Sons {\bf 1902}.
	
	\bibitem[Tolman(1938)]{Tolman1938}
	Tolman, R.~C.
	\newblock {\em The Principles of Statistical Mechanics}.
	\newblock Oxford: Oxford University Press {\bf 1938}.
	
	\bibitem[Pathria and Beale(2021)]{Pathria2021}
	Pathria, R.~K.; Beale, P.~D.
	\newblock {\em Statistical Mechanics}.
	\newblock 4th ed.; Elsevier {\bf 2021}.
	
	\bibitem[Jaynes(1957)]{Jaynes1957}
	Jaynes, E.~T.
	\newblock Information Theory and Statistical Mechanics.
	\newblock {\em Physical Review} {\bf 1957}, {\em 106},~620--630.
	\newblock {\url{https://doi.org/10.1103/PhysRev.106.620}}.
	
	\bibitem[Parrondo et al.(2015)]{Parrondo2015}
	Parrondo, J.~M.~R.; Horowitz, J.~M.; Sagawa, T.
	\newblock Thermodynamics of Information.
	\newblock {\em Nature Physics} {\bf 2015}, {\em 11},~131--139.
	\newblock {\url{https://doi.org/10.1038/nphys3230}}.
	
	\bibitem[Seifert(2012)]{Seifert2012}
	Seifert, U.
	\newblock Stochastic thermodynamics, fluctuation theorems and molecular machines.
	\newblock {\em Rep. Prog. Phys.} {\bf 2012}, {\em 75},~126001.
	\newblock {\url{https://doi.org/10.1088/0034-4885/75/12/126001}}.
	
	\bibitem[Sekimoto(2010)]{Sekimoto2010}
	Sekimoto, K.
	\newblock {\em Stochastic Energetics}.
	\newblock Berlin: Springer-Verlag {\bf 2010}.
	
	\bibitem[Risken(1996)]{Risken1996}
	Risken, H.
	\newblock {\em The Fokker-Planck Equation: Methods of Solution and Applications}.
	\newblock 2nd ed.; Springer: Berlin/Heidelberg, Germany {\bf 1996}.
	
	\bibitem[Onsager(1931)]{Onsager1931}
	Onsager, L.
	\newblock Reciprocal Relations in Irreversible Processes. I.
	\newblock {\em Physical Review} {\bf 1931}, {\em 37},~405--426.
	\newblock {\url{https://doi.org/10.1103/PhysRev.37.405}}.
	
	\bibitem[De Groot and Mazur(1962)]{DeGroot1962}
	De Groot, S.~R.; Mazur, P.
	\newblock {\em Non-Equilibrium Thermodynamics}.
	\newblock Amsterdam: North-Holland Publishing Company {\bf 1962}.
	
	\bibitem[Ziegler(1983)]{Ziegler1983}
	Ziegler, H.
	\newblock {\em An Introduction to Thermomechanics: Field Theory and Variational Principles}.
	\newblock 2nd ed.; North-Holland: Amsterdam, The Netherlands {\bf 1983}.
	
	
	\bibitem[Cahn and Hilliard(1958)]{Cahn1958}
	Cahn, J.~W.; Hilliard, J.~E.
	\newblock Free energy of a nonuniform system. I. Interfacial free energy.
	\newblock {\em J. Chem. Phys.} {\bf 1958}, {\em 28},~258--267.
	\newblock {\url{https://doi.org/10.1063/1.1744102}}.
	
	\bibitem[Antanovskii(1996)]{Antanovskii1996}
	Antanovskii, L.~K.
	\newblock Microscale theory of surface tension.
	\newblock {\em Physical Review E} {\bf 1996}, {\em 54},~6285--6290.
	\newblock {\url{https://doi.org/10.1103/PhysRevE.54.6285}}.
	
	\bibitem[Ván(2003)]{Van2003}
	Ván, P.
	\newblock Weakly nonlocal irreversible thermodynamics.
	\newblock {\em Annalen der Physik} {\bf 2003}, {\em 12},~146--173.
	\newblock {\url{https://doi.org/10.1002/andp.200310002}}.
	
	\bibitem[Mauri(2013)]{Mauri2013}
	Mauri, R.
	\newblock {\em Non-Equilibrium Thermodynamics in Multiphase Flows}.
	\newblock Springer: Dordrecht, The Netherlands {\bf 2013}.
	
	\bibitem[Malvern(1969)]{Malvern1969}
	Malvern, L.~E.
	\newblock {\em Introduction to the Mechanics of a Continuous Medium}.
	\newblock Upper Saddle River, NJ: Prentice-Hall {\bf 1969}.
	
	\bibitem[Truesdell and Noll(2004)]{Truesdell2004}
	Truesdell, C.; Noll, W.
	\newblock {\em The Non-Linear Field Theories of Mechanics}.
	\newblock 3rd ed.; Berlin: Springer-Verlag {\bf 2004}.
	
	\bibitem[Gurtin et al.(2010)]{Gurtin2010}
	Gurtin, M.~E.; Fried, E.; Anand, L.
	\newblock {\em The Mechanics and Thermodynamics of Continua}.
	\newblock Cambridge University Press: Cambridge, UK {\bf 2010}.
	
	\bibitem[Berezovski and Ván(2017)]{Berezovski2017}
	Berezovski, A.; Ván, P.
	\newblock {\em Internal Variables in Thermoelasticity}.
	\newblock Springer: Cham, Switzerland {\bf 2017}.
	
	\bibitem[Jou et al.(2010)]{Jou2010}
	Jou, D.; Casas-Vázquez, J.; Lebon, G.
	\newblock {\em Extended Irreversible Thermodynamics}.
	\newblock 4th ed.; Dordrecht: Springer {\bf 2010}.
	
	\bibitem[Mazur(1998)]{Mazur1998}
	Mazur, P.
	\newblock Mesoscopic nonequilibrium thermodynamics; irreversible processes and fluctuations.
	\newblock {\em Physica A} {\bf 1998}, {\em 261},~59--73.
	\newblock {\url{https://doi.org/10.1016/S0378-4371(98)00366-9}}.
	
	\bibitem[Rubi(2012)]{Rubi2012}
	Rubi, J.~M.
	\newblock Mesoscopic thermodynamics.
	\newblock {\em Physica Scripta} {\bf 2012}, {\em T151},~014027.
	\newblock {\url{https://doi.org/10.1088/0031-8949/2012/T151/014027}}.
	
	\bibitem[Anisimov and Longo(2024)]{Anisimov2024}
	Anisimov, M.~A.; Longo, G.
	\newblock {\em Mesoscopic Thermodynamics for Scientists and Engineers}.
	\newblock London: Academic Press {\bf 2024}.
	
	\bibitem[Muller and Ruggeri(1998)]{Muller1998}
	Müller, I.; Ruggeri, T.
	\newblock {\em Rational Extended Thermodynamics}.
	\newblock New York: Springer {\bf 1998}.
	
	\bibitem[Ruggeri and Sugiyama(2015)]{Ruggeri2015}
	Ruggeri, T.; Sugiyama, M.
	\newblock {\em Rational Extended Thermodynamics beyond the Monatomic Gas}.
	\newblock Cham: Springer {\bf 2015}.
	
	\bibitem[Cercignani(1988)]{Cercignani1988}
	Cercignani, C.
	\newblock {\em The Boltzmann Equation and Its Applications}.
	\newblock New York: Springer-Verlag {\bf 1988}.
	
	\bibitem[Struchtrup(2005)]{Struchtrup2005}
	Struchtrup, H.
	\newblock {\em Macroscopic Transport Equations for Rarefied Gases}.
	\newblock Springer: Heidelberg, Germany {\bf 2005}.
	
	\bibitem[Prigogine(1967)]{Prigogine1967}
	Prigogine, I.
	\newblock {\em Introduction to Thermodynamics of Irreversible Processes}.
	\newblock 3rd ed.; New York: Interscience Publishers {\bf 1967}.
	
	\bibitem[Beretta et al.(2012)]{Beretta2012}
	Beretta, G.~P.; Keck, J.~C.; Janbozorgi, M.; Metgalchi, H.
	\newblock The rate-controlled constrained-equilibrium approach to far-from-local-equilibrium thermodynamics.
	\newblock {\em Entropy} {\bf 2012}, {\em 14},~92--130.
	\newblock {\url{https://doi.org/10.3390/e14010092}}.
	
	\bibitem[Kondepudi and Prigogine(2014)]{Kondepudi2014}
	Kondepudi, D.; Prigogine, I.
	\newblock {\em Modern Thermodynamics: From Heat Engines to Dissipative Structures}.
	\newblock 2nd ed.; Chichester: John Wiley \& Sons {\bf 2014}.
	
	\bibitem[Alicki and Lendi(2007)]{Alicki2007}
	Alicki, R.; Lendi, K.
	\newblock {\em Quantum Dynamical Semigroups and Applications}.
	\newblock Berlin: Springer-Verlag {\bf 2007}.
	
	\bibitem[Deffner and Jarzynski(2013)]{Deffner2013}
	Deffner, S.; Jarzynski, C.
	\newblock Information processing and the second law of thermodynamics: An inclusive, Hamiltonian approach.
	\newblock {\em Physical Review X} {\bf 2013}, {\em 3},~041003.
	\newblock {\url{https://doi.org/10.1103/PhysRevX.3.041003}}.
	
	\bibitem[Binder et al.(2018)]{Binder2018}
	Binder, F.; Correa, L.~A.; Gogolin, C.; Anders, J.; Adesso, G., Eds.
	\newblock {\em Thermodynamics in the Quantum Regime}.
	\newblock Cham: Springer {\bf 2018}.
	
	
	
	
	
	
	\bibitem[Ray and Beretta(2025)]{RayBeretta2025}
	Ray, R.K.; Beretta, G.P.
	\newblock No-Signaling in Steepest Entropy Ascent: A Nonlinear, Non-Local,
	Non-Equilibrium Quantum Dynamics of Composite Systems Strongly Compatible
	with the Second Law.
	\newblock {\em Entropy} {\bf 2025}, {\em 27}.
	\newblock {\url{https://doi.org/10.3390/e27101018}}.
	
	\bibitem[Zanchini(1986)]{Zanchini1986}
	Zanchini, E.
	\newblock On the Definition of Extensive Property Energy by the First Postulate
	of Thermodynamics.
	\newblock {\em Foundations of Physics} {\bf 1986}, {\em 16},~923--935.
	\newblock {\url{https://doi.org/10.1007/BF00765339}}.
	
	\bibitem[Zanchini(1988)]{Zanchini1988}
	Zanchini, E.
	\newblock Thermodynamics: Energy of closed and open systems.
	\newblock {\em Il Nuovo Cimento B} {\bf 1988}, {\em 101},~453--465.
	\newblock {\url{https://doi.org/10.1007/BF02828923}}.
	
	\bibitem[Pekola and Karimi(2024)]{PekolaKarimi2024}
	Pekola, J.P.; Karimi, B.
	\newblock Heat Bath in a Quantum Circuit.
	\newblock {\em Entropy} {\bf 2024}, {\em 26}.
	\newblock {\url{https://doi.org/10.3390/e26050429}}.
	
	\bibitem[Beretta and Gyftopoulos(2015)]{BerettaGyftopoulos2015}
	Beretta, G.P.; Gyftopoulos, E.P.
	\newblock What is the Third Law?
	\newblock {\em Journal of Energy Resources Technology} {\bf 2015}, {\em
		137},~021004.
	\newblock {\url{https://doi.org/10.1115/1.4026380}}.
	
	\bibitem[De~Groot and Mazur(1962)]{DeGrootMazur}
	De~Groot, S.R.; Mazur, P.
	\newblock {\em Non-Equilibrium Thermodynamics}; North-Holland: Amsterdam, NL,
	1962.
	\newblock Reprinted by Dover, 1984.
	
	\bibitem[Sivak and Crooks(2012)]{Crooks_2012}
	Sivak, D.A.; Crooks, G.E.
	\newblock Near-Equilibrium Measurements of Nonequilibrium Free Energy.
	\newblock {\em Phys. Rev. Lett.} {\bf 2012}, {\em 108},~150601.
	\newblock {\url{https://doi.org/10.1103/PhysRevLett.108.150601}}.
	

	
	\bibitem[Beretta and Gyftopoulos(2015)]{BerettaGyftopoulosSS_2015}
	Beretta, G.P.; Gyftopoulos, E.P.
	\newblock What is a Simple System?
	\newblock {\em Journal of Energy Resources Technology} {\bf 2015}, {\em
		137},~021007.
	\newblock {\url{https://doi.org/10.1115/1.4026383}}.
	
	\bibitem[Davoudi et al.(2024)]{Davoudi2024}
	Davoudi, Z.; Jarzynski, C.; Mueller, N.; Oruganti, G.; Powers, C.; Yunger Halpern, N.
	\newblock Quantum thermodynamics of nonequilibrium processes in lattice gauge theories.
	\newblock {\em Physical Review Letters} {\bf 2024}, {\em 133},~250402.
	\newblock {\url{https://doi.org/10.1103/PhysRevLett.133.250402}}.
	
	
	\bibitem[Park(1968)]{park_1968_nature}
	Park, J.L.
	\newblock Nature of {{Quantum States}}.
	\newblock {\em American Journal of Physics} {\bf 1968}, {\em 36},~211--226.
	\newblock {\url{https://doi.org/10.1119/1.1974484}}.
	
	\bibitem[Band and Park(1979)]{park_1979}
	Band, W.; Park, J.L.
	\newblock Quantum State Determination: Quorum for a Particle in One Dimension.
	\newblock {\em American Journal of Physics} {\bf 1979}, {\em 47},~188--191.
	\newblock {\url{https://doi.org/10.1119/1.11870}}.
	
	\bibitem[Park and Band(1971)]{park_1971_general}
	Park, J.L.; Band, W.
	\newblock A General Theory of Empirical State Determination in Quantum Physics:
	{{Part I}}.
	\newblock {\em Foundations of Physics} {\bf 1971}, {\em 1},~211--226.
	\newblock {\url{https://doi.org/10.1007/BF00708608}}.
	
	\bibitem[D'Ariano et~al.(2003)D'Ariano, Paris, and Sacchi]{Dariano_2003}
	D'Ariano, G.M.; Paris, M.G.; Sacchi, M.F.
	\newblock Quantum Tomography; Elsevier,  2003; Vol. 128, {\em Advances in
		Imaging and Electron Physics}, pp. 205--308.
	\newblock {\url{https://doi.org/10.1016/S1076-5670(03)80065-4}}.
	
	\bibitem[Hatsopoulos and Gyftopoulos(1976)]{hatsopoulos_1976_unifieda}
	Hatsopoulos, G.N.; Gyftopoulos, E.P.
	\newblock A Unified Quantum Theory of Mechanics and Thermodynamics. {{Part
			IIa}}. {{Available}} Energy.
	\newblock {\em Foundations of Physics} {\bf 1976}, {\em 6},~127--141.
	\newblock {\url{https://doi.org/10.1007/BF00708955}}.
	
	\bibitem[Allahverdyan et~al.(2004)Allahverdyan, Balian, and
	Nieuwenhuizen]{Allahverdyan_2004}
	Allahverdyan, A.E.; Balian, R.; Nieuwenhuizen, T.M.
	\newblock Maximal Work Extraction from Finite Quantum Systems.
	\newblock {\em Europhysics Letters} {\bf 2004}, {\em 67},~565.
	\newblock {\url{https://doi.org/10.1209/epl/i2004-10101-2}}.
	
\end{thebibliography}

\begingroup
\setcounter{secnumdepth}{-1} 
\reftitle{References}
\endgroup

\end{document}